%% file: main.tex
\documentclass[a4paper,fleqn]{cas-sc}

\usepackage[numbers, sort&compress]{natbib}

\def\tsc#1{\csdef{#1}{\textsc{\lowercase{#1}}\xspace}}
\tsc{WGM}
\tsc{QE}
\tsc{EP}
\tsc{PMS}
\tsc{BEC}
\tsc{DE}

\input{preamble}

\begin{document}
\let\WriteBookmarks\relax
\def\floatpagepagefraction{1}
\def\textpagefraction{.001}
\shorttitle{Reference-free Human-Object Interaction Editing}
\shortauthors{J.T. Hoe et~al.}

\title [mode = title]{Reference-free Human-Object Interaction Editing}                      


\cortext[1]{Corresponding authors}
\affiliation[1]{organization={School of Electrical and Electronic Engineering, Nanyang Technological University},
                addressline={50 Nanyang Ave}, 
                postcode={639798}, 
                country={Singapore}}
                
\affiliation[2]{organization={School of Intelligent Systems Engineering, Sun Yat-sen University},
                postcode={518107}, 
                postcodesep={}, 
                city={Shenzhen Campus},
                state={Guangdong Province},
                country={China}}
                
\affiliation[3]{organization={College of Mechanical and Electronic Engineering, Nanjing Forestry University},
                city={Nanjing},
                postcode={210037}, 
                state={Jiangsu Province}, 
                country={China}}
                
\affiliation[4]{organization={Faculty of Computer Science and Information Technology, Universiti Malaya},
                city={Kuala Lumpur},
                postcode={50603}, 
                country={Malaysia}}

\affiliation[5]{organization={College of Engineering and Computer Science, VinUniversity},
                city={Hanoi},
                postcode={12426}, 
                country={Vietnam}}
\author[1]{Jiun Tian Hoe}[type=author,
                        auid=000,
                        bioid=1,
                        ]
\ead{jiuntian001@e.ntu.edu.sg}

\credit{Conceptualization, Methodology, Software, Validation, Formal analysis, Investigation, Data Curation, Writing - Original Draft, Visualization} 

\author[1,2]{Weipeng Hu}[]
\cormark[1]
\ead{huwp7@mail.sysu.edu.cn}
\credit{Methodology, Validation, Data Curation, Writing - Review \& Editing}

\author[2]{Wei Zhou}[]
\ead{zhouw75@mail3.sysu.edu.cn}
\credit{Software, Data Curation, Visualization} 

\author[3]{Chao Xie}[]
\ead{chaoxie@njfu.edu.cn}
\credit{Software, Data Curation, Visualization}

\author[1]{Ziwei Wang}[]
\ead{ziwei.wang@ntu.edu.sg}
\credit{Methodology, Validation, Writing - Review \& Editing, Supervision}

\author[1]{Xudong Jiang}[]
\ead{exdjiang@ntu.edu.sg}
\credit{Methodology, Writing - Review \& Editing, Supervision}

\author[1,5]{Yap-Peng Tan}[]
\ead{eyptan@ntu.edu.sg}
\ead{yp.t@vinuni.edu.vn}
\credit{Conceptualization, Writing - Review \& Editing, Supervision, Funding acquisition}

\author[4,5]{Chee Seng Chan}[]
\cormark[1]
\ead{cs.chan@um.edu.my}
\credit{Conceptualization, Methodology, Validation, Formal analysis, Data Curation, Writing - Review \& Editing, Supervision}



\input{sec/0_abstract}




\maketitle

\input{sec/1_intro}
\input{sec/2_relatedwork}
\input{sec/3_method}
\input{sec/4_experiment}

\input{sec/5_conclusion}

\FloatBarrier
\appendix
\input{sec/X_suppl}

\printcredits

\bibliographystyle{elsarticle-num}

\bibliography{main.bib}


\Needspace*{12\baselineskip}
\bio{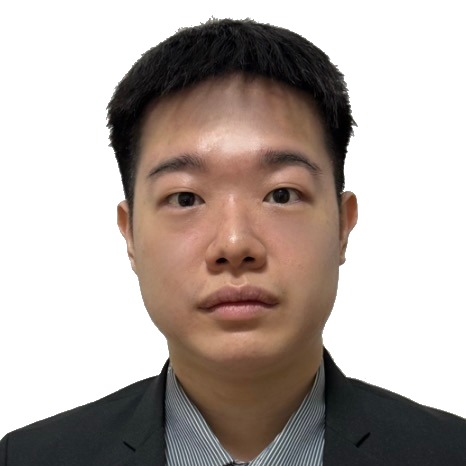}
\textbf{Jiun Tian Hoe} is a Ph.D. candidate with the Centre for Advanced Robotics Technology Innovation (CARTIN) in the School of Electrical and Electronic Engineering, Nanyang Technological University, Singapore, where he is advised by Prof. Yap-Peng Tan and Prof. Xudong Jiang. He received the Bachelor of Computer Science degree from the Faculty of Computer Science and Information Technology, Universiti Malaya, Malaysia, in 2022. His research interests include computer vision, image generation, image editing, world models and generative models. He regularly serves as a reviewer for top-tier computer vision and artificial intelligence venues, including CVPR, ICCV, ECCV, NeurIPS, BMVC, and AAAI.
\endbio

\Needspace*{12\baselineskip}
\bio{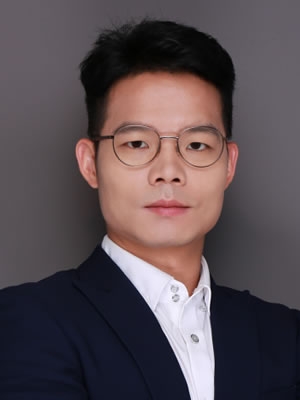}
\textbf{Weipeng Hu} received the Ph.D. degree in Electronics and Information Technology from Sun Yat-sen University, Guangzhou, China, in 2022. From 2022 to 2026, he was a Research Fellow with the Centre for Advanced Robotics Technology Innovation (CARTIN) in the School of Electrical and Electronic Engineering, Nanyang Technological University, Singapore. He is currently an Associate Professor with the School of Intelligent Systems Engineering, Sun Yat-sen University, Shenzhen, China. His research interests include computer vision, intelligent perception and understanding, and multimodal information processing. He has published over 30 papers in top-tier journals and conferences. Dr. Hu serves as an Area Chair for IEEE ICME 2026 and BMVC 2026, and regularly reviews for prestigious venues including IEEE T-PAMI, T-IP, T-IFS, T-MM, T-CSVT, and CVPR.
\endbio

\Needspace*{12\baselineskip}
\bio{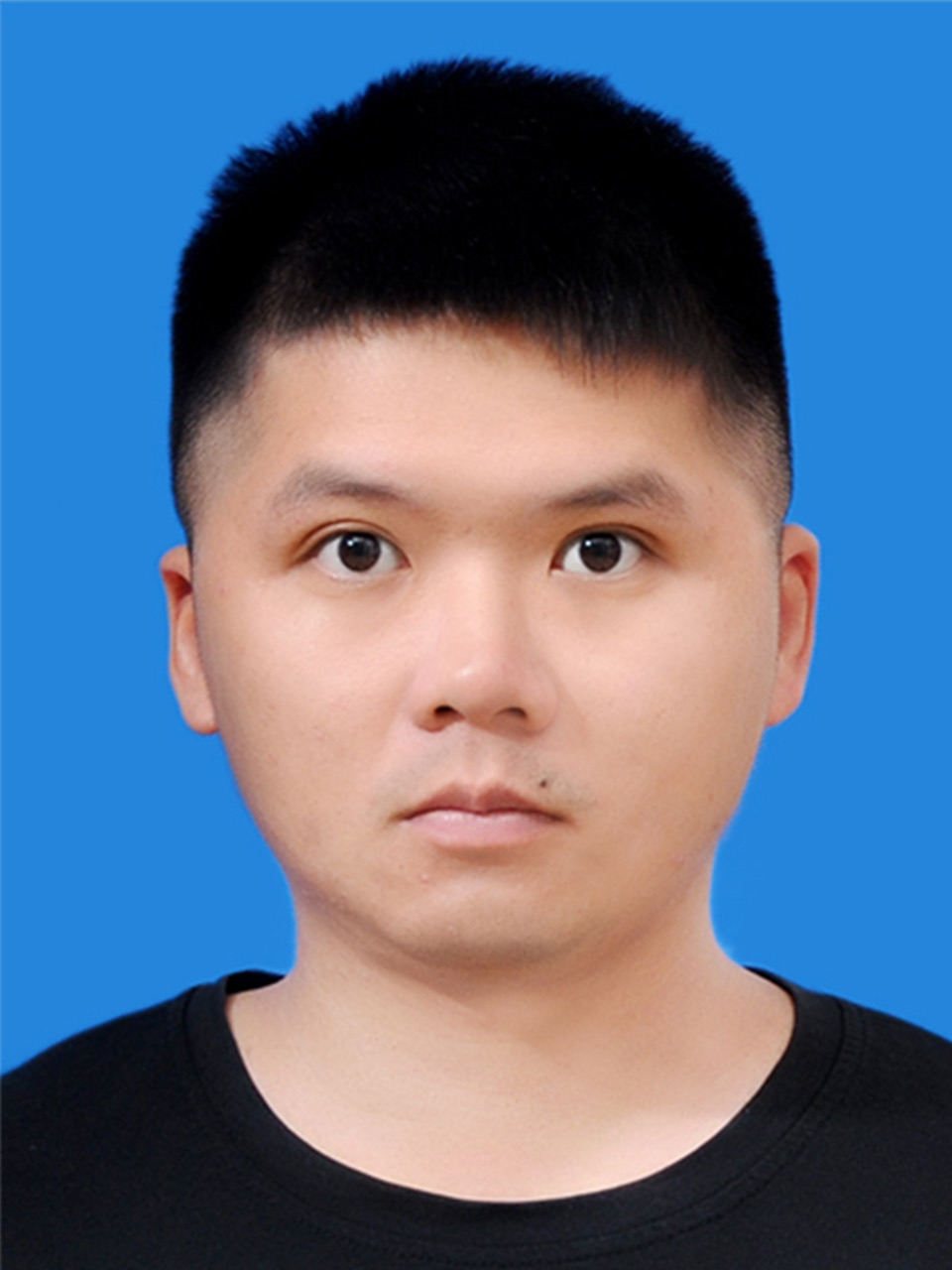}
\textbf{Wei Zhou} received the M.E. degree from the Beijing University of Posts and Telecommunications, Beijing, China, in 2020, and the Ph.D. degree from the School of Electronics and Information Technology, Sun Yat-sen University, Guangzhou, China, in 2024. His research interests include computer vision, pattern recognition, and deep learning, with a specific focus on multi-label image classification, image captioning, weakly-supervised temporal action localization, and image generation and editing.
\vspace{5em}
\endbio

\Needspace*{12\baselineskip}
\bio{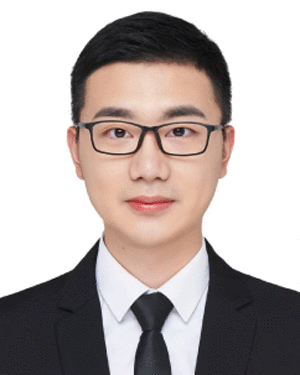}
\textbf{Chao Xie} received the Ph.D. degree from Southeast University, Nanjing, China, in 2018. From July 2023 to July 2024, he was a Visiting Scholar with the School of Electrical and Electronic Engineering, Nanyang Technological University, Singapore. He is currently an Associate Professor with the College of Mechanical and Electronic Engineering, Nanjing Forestry University, Nanjing. His current research interests include image processing and deep learning.
\vspace{5em}
\endbio

\Needspace*{12\baselineskip}
\bio{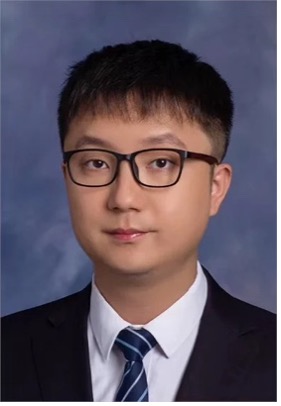}
\textbf{Ziwei Wang} is currently an assistant professor in School of Electrical and Electronic Engineering, Nanyang Technological University and the director of Perception and embodied INtElligence (PINE) Lab. Before joining NTU, he was a postdoc fellow in Robotics Institute, Carnegie Mellon University, with Prof. Changliu Liu. He received the Ph.D and the B.S degrees from the Department of Automation, Tsinghua University in 2023 and the Department of Physics, Tsinghua University in 2018 respectively. His research goal is to design foundation models (FMs) for robotic manipulation. He has published over 50 scientific papers in TPAMI, IJCV, RAL, CVPR, ICCV, ECCV, NeurIPS, CoRL, IROS and ICRA. He serves as a regular reviewer member for a variety of conferences and journals.
\vspace{3em}
\endbio

\Needspace*{12\baselineskip}
\bio{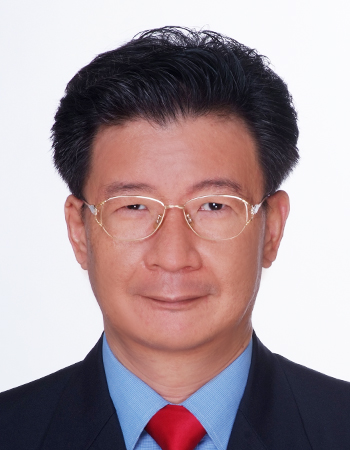}
\textbf{Xudong Jiang} (Fellow, IEEE) received the BE and PhD degrees from the University of Electronic Science and Technology of China (UESTC), Chengdu, China, and the PhD degree from Helmut Schmidt University, Hamburg, Germany. From 1998 to 2004, he was with the Institute for Infocomm Research, A*STAR, Singapore, as a lead scientist, and the head of the Biometrics Laboratory. He joined Nanyang Technological University (NTU), Singapore, as a faculty member, in 2004, where he served as the director of the Centre for Information Security from 2005 to 2011. He is currently a professor with the School of EEE, NTU and serves as the director of the Centre for Information Sciences and Systems of School of EEE, NTU. He holds 7 patents and has authored about 300 papers with more than 60 papers in the IEEE journals and more than 30 papers in top conferences CVPR/NeurIPS/ICCV/ECCV/ICLR/AAAI. He served as IFS TC member of the IEEE Signal Processing Society from 2015 to 2017, associate editor for IEEE Signal Processing Letter from 2014 to 2018 and associate editor for IEEE Transactions on Image Processing from 2016 to 2020. Currently, he serves as senior area editor for IEEE Transactions on Image Processing and editor-in-chief for IET Biometrics. His research interests include image processing, pattern recognition, computer vision, machine learning, and biometrics.
\endbio

\Needspace*{12\baselineskip}
\bio{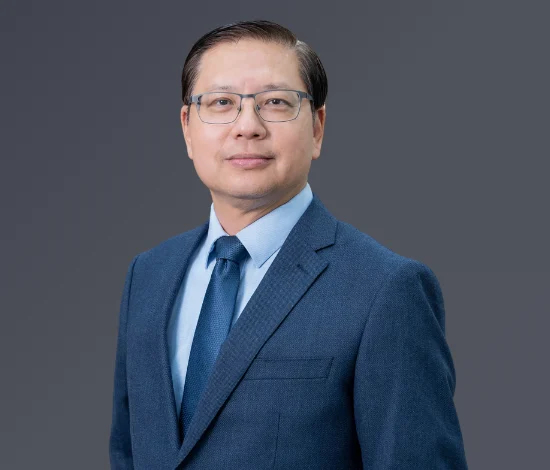}
\textbf{Yap-Peng Tan} (Fellow, IEEE) received the B.S. degree in electrical engineering from National Taiwan University, Taipei, Taiwan, in 1993, and the M.A. and Ph.D. degrees in electrical engineering from Princeton University, Princeton, New Jersey, in 1995 and 1997, respectively. He is currently the Provost at VinUniversity, Hanoi, Vietnam, and a Full Professor with the School of Electrical and Electronic Engineering, Nanyang Technological University, Singapore. His research interests include image and video processing, computer vision, pattern recognition, and data analytics. 
Prof. Tan has served as an Associate Editor for IEEE Transactions on Circuits and Systems for Video Technology, IEEE Signal Processing Letters, IEEE Transactions on Multimedia, and IEEE Access. He was also an editorial board member for the EURASIP Journal on Advances in Signal Processing and the EURASIP Journal on Image and Video Processing. His conference leadership includes serving as General Co-Chair for ICME 2010 and VCIP 2015, and Technical Program Co-Chair for ICME 2015 and ICIP 2019.
\endbio

\Needspace*{12\baselineskip}
\bio{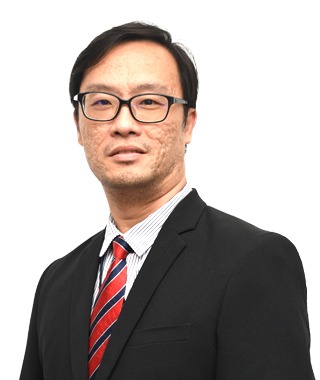}
\textbf{Chee Seng Chan} (Senior Member, IEEE) is a Full Professor at the Faculty of Computer Science and Information Technology, Universiti Malaya, and Affiliate Faculty at VinUniversity. His work spans computer vision, machine learning, and trustworthy artificial intelligence. He leads an active research team whose work has been published in leading international venues such as CVPR, NeurIPS etc. He is among the key architects of ILMU, Malaysia’s first sovereign multilingual large language model, and has played an important role in advancing locally grounded, strategically important AI systems ready for real-world deployment. His contributions include Ryt AI, one of the earliest regulator-approved, language-centric AI systems for banking, and MalayMMLU, the first benchmark for evaluating AI proficiency in Bahasa Melayu. He has received several recognitions, including the Top Research Scientists Malaysia (TRSM) Award, the Young Scientists Network of the Academy of Sciences Malaysia (YSN-ASM), and the Hitachi Research Fellowship. He has held leadership roles in major international conferences, including IEEE VCIP 2013, ACPR 2015, IEEE MMSP 2019, and ACM MMAsia 2025 and 2026. From 2020 to 2022, he served as Undersecretary for the Division of Data Strategic and Foresight at Ministry of Science, Technology and Innovation (MOSTI).
\endbio

\end{document}

%% file: preamble.tex
\usepackage{multirow}
\usepackage{needspace}
\usepackage{amsthm}
\usepackage{amsmath,amsfonts}
\usepackage{duckuments}
\usepackage{xr}
\usepackage{xr-hyper}
\usepackage[capitalize]{cleveref}
\usepackage{listings}
\usepackage{tcolorbox}
\usepackage{url}

\definecolor{codegreen}{rgb}{0,0.6,0}
\definecolor{codegray}{rgb}{0.5,0.5,0.5}
\definecolor{codepurple}{rgb}{0.58,0,0.82}
\definecolor{backcolour}{rgb}{0.95,0.95,0.92}
\lstdefinestyle{mystyle}{
  backgroundcolor=\color{backcolour},   commentstyle=\color{codegreen},
  keywordstyle=\color{magenta},
  numberstyle=\tiny\color{codegray},
  stringstyle=\color{codepurple},
  basicstyle=\ttfamily\footnotesize,
  breakatwhitespace=false,         
  breaklines=true,                 
  captionpos=b,                    
  keepspaces=true,                 
  numbers=left,                    
  numbersep=5pt,                  
  showspaces=false,                
  showstringspaces=false,
  showtabs=false,                  
  tabsize=2
}
\newcommand{\code}[1]{\mbox{
    \ttfamily
    \tcbox[
        on line,
        boxsep=0pt, left=4pt, right=4pt, top=2pt, bottom=1.5pt,
        toprule=0pt, rightrule=0pt, bottomrule=0pt, leftrule=0pt,
        oversize=0pt, enlarge left by=0pt, enlarge right by=0pt,
        colframe=white, colback=black!12,
        height=1.0\baselineskip 
    ]{#1}
}}
\crefname{lstlisting}{Listing}{Listings} 
\Crefname{lstlisting}{Listing}{Listings}

\crefname{appsec}{Appendix Section}{Appendix Sections}
\Crefname{appsec}{Appendix Section}{Appendix Sections}

\crefname{appfig}{Appendix Figure}{Appendix Figures}
\Crefname{appfig}{Appendix Figure}{Appendix Figures}
\crefname{apptab}{Appendix Table}{Appendix Tables}
\Crefname{apptab}{Appendix Table}{Appendix Tables}

\tcbset{
  colback=blue!5,
  colframe=blue!50!black,
  boxsep=3pt,
  left=6pt,right=6pt,top=4pt,bottom=4pt,
  arc=2pt,
}

\makeatletter
\newcommand{\specialurl}[1]{%
  \begingroup%
  \g@addto@macro{\UrlBreaks}{\UrlOrds}%
  \url{#1}%
  \endgroup%
}
\makeatother

\newtheorem{theorem}{Theorem}[section]
\newtheorem{lemma}[theorem]{Lemma}
\newcommand{\iebench}{\textup{\rmfamily\scshape IEBench}}
\newcommand{\iebenchext}{\textup{\rmfamily\scshape IEBench-L}}

\usepackage{subcaption}
\usepackage{placeins}
\usepackage{lipsum}

%% file: sec/0_abstract.tex
\begin{abstract}
This paper presents \textsc{InteractEdit}, a novel framework for reference-free Human-Object Interaction (HOI) editing that tackles the challenging task of transforming an existing interaction in an image into a new, desired interaction while preserving the identities of the subject and object. Unlike prior image editing tasks such as attribute manipulation, object replacement or style transfer, HOI editing involves complex spatial, contextual, and relational dependencies inherent in HOI. Existing methods often overfit to the source image structure, limiting adaptability to the substantial structural modifications demanded by the new interactions. To address this, \textsc{InteractEdit} disassembles each scene into subject, object, and background components to disentangle intricate HOI relationship, and introduces a selective inversion strategy combined with Selective-Rank Adaptation (SeRA) to leverage pretrained interaction priors while learning visual identity from the source image. This enables a balanced trade-off between interaction editing and identity preservation. We also introduce \iebench, a new benchmark for HOI editing, and a new metric that jointly evaluates the trade-off between successful interaction editing and identity preservation. Extensive experiments show that \textsc{InteractEdit} outperforms 23 existing methods, providing a strong baseline for future HOI editing research. Code and the dataset: \url{https://jiuntian.github.io/InteractEdit/}.
\end{abstract}

%% file: sec/1_intro.tex
\section{Introduction}\label{sec:intro}
Human–Object Interaction (HOI) understanding plays a key role in computer vision, supporting a wide range of applications from activity recognition~\cite{morais2021learning} and augmented-reality overlays~\cite{yang2024lemon} to robotics~\cite{gao2024coohoi}. While HOI \emph{recognition} and \emph{generation} have been well studied~\cite{li2024diffusionhoi,luo2024sichoi,hoe2024interactdiffusion,hu2025personahoi}, the task of \emph{editing an existing interaction within a single image} remains largely unexplored despite its clear commercial potential. 
In industries such as film and media, modifying a character’s action, for example from \textit{riding a horse} to \textit{kissing a horse} can save an entire re-shoot. Similarly, in e-commerce and advertising, updating a product visual, {\it e.g.}\ changing an athlete from \textit{holding} to \textit{drinking} a sports drink, can significantly reduce production time and expenses. These scenarios demand a careful trade-off between modifying the HOI and maintaining subject and object identity, all within a coherently adapted visual context. 

\input{assets/figure/teaser}

Achieving this balance is non-trivial, as people, objects, and scene context are tightly interdependent and must remain visually coherent after editing. Yet, most modern image editors~\cite{shuai2024diffusion-edit-survey} focus on manipulating attributes, objects, or styles under a fixed layout assumption. This causes them to be ill-suited for interaction-level editing. \textbf{Interaction editing, by contrast, requires substantial structural changes}, such as reposing bodies, relocating objects and sometimes modifying the background, while preserving subject and object identity. For example, transforming \textit{a person blowing a cake} into \textit{eating the cake} (see \cref{fig:teaser}) involves altering the action, adjusting body posture, and repositioning the object, all without disrupting scene coherence or visual likeness. Moreover, existing methods are not designed to operate without reference images or paired supervision \cite{xu2025hoiedit}. This gap highlights the need for a new task formulation that supports flexible, identity-preserving edits directly from a single image and a target prompt. 

To this end, we define the \emph{reference-free HOI editing} task, which addresses four key challenges:
\begin{enumerate}[a)]
\item \textbf{Intricate interaction relationships}: subjects, objects, and actions are tightly coupled spatially \& semantically;
\item \textbf{Rigid image structure}: existing editors rigidly preserve the original structure, limiting their ability to deform pose and layout changes required for new interactions;
\item \textbf{Pretrained-prior corruption}: unconstrained inversion risks overwriting pretrained interaction priors, thereby reducing generalizability to new target interactions;
\item \textbf{Missing standard dataset and metrics}: lack of a standard benchmark for a fair and reproducible comparison.
\end{enumerate}

To tackle these challenges, we propose \textsc{InteractEdit} (\cref{fig:arch}), a novel framework that requires only a single source image and a target prompt. First, the image is decomposed into subject, object, and background tokens, facilitating the disentanglement of intricate HOI relationship. To support reference-free editing, it is essential to retain the model’s pretrained understanding of interactions. We introduce \emph{Selective Inversion}, which preserves the pretrained Query weights~\(W^{Q}\) and adapts only the Key and Value weights~\(W^{K},W^{V}\), thereby maintaining the original attention “question space” while allowing controllable interaction modifications. Building on this, we propose \emph{Selective-Rank Adaptation} (SeRA) to guide each attention weight update within a structured, low-rank subspace, with each direction modulated by a learnable gating scalar. The goal of this design is to retain the pretrained action vocabulary while enabling identity-aware adaptations, striking a balance between interaction transformation and subject and object identity consistency.

To support a fair and reproducible evaluation, we also introduce \iebench, a dedicated benchmark for HOI editing. It systematically assesses HOI editability across a diverse set of source–target interaction pairs and reports performance using the Editability–Identity score (\text{\textsc{ei}}), a new metric that jointly quantifies editing success and identity preservation.

\input{assets/figure/arch}

As a summary, our contributions are:
\begin{enumerate}[(i)]
\item We formally define the reference-free HOI editing task and introduce \iebench, a comprehensive HOI benchmark, featuring diverse evaluation scenarios, as well as a new metric (\textit{i.e.}, Editability–Identity score (\text{\textsc{ei}})) that jointly assesses HOI editability and identity preservation.
\item We present \textsc{InteractEdit}, a novel framework for editing HOI from a single image. Specifically, it disassembles HOI components and combines \emph{Selective Inversion} with SeRA to preserve pretrained action priors while enabling identity-consistent interaction edits.
\item Experiments show state-of-the-art performance where \textsc{InteractEdit} achieves the best balance between HOI editability and identity consistency, establishing a strong foundation for future research in HOI editing.
\end{enumerate}

\noindent A preliminary version of this work was previously disseminated as an arXiv preprint as ``InteractEdit: Zero-Shot Editing of Human-Object Interactions in Images'' \cite{hoe2025interactedit} prior to this submission.

%% file: assets/figure/teaser.tex


\begin{figure}
    \centering
    \includegraphics[width=1\linewidth]{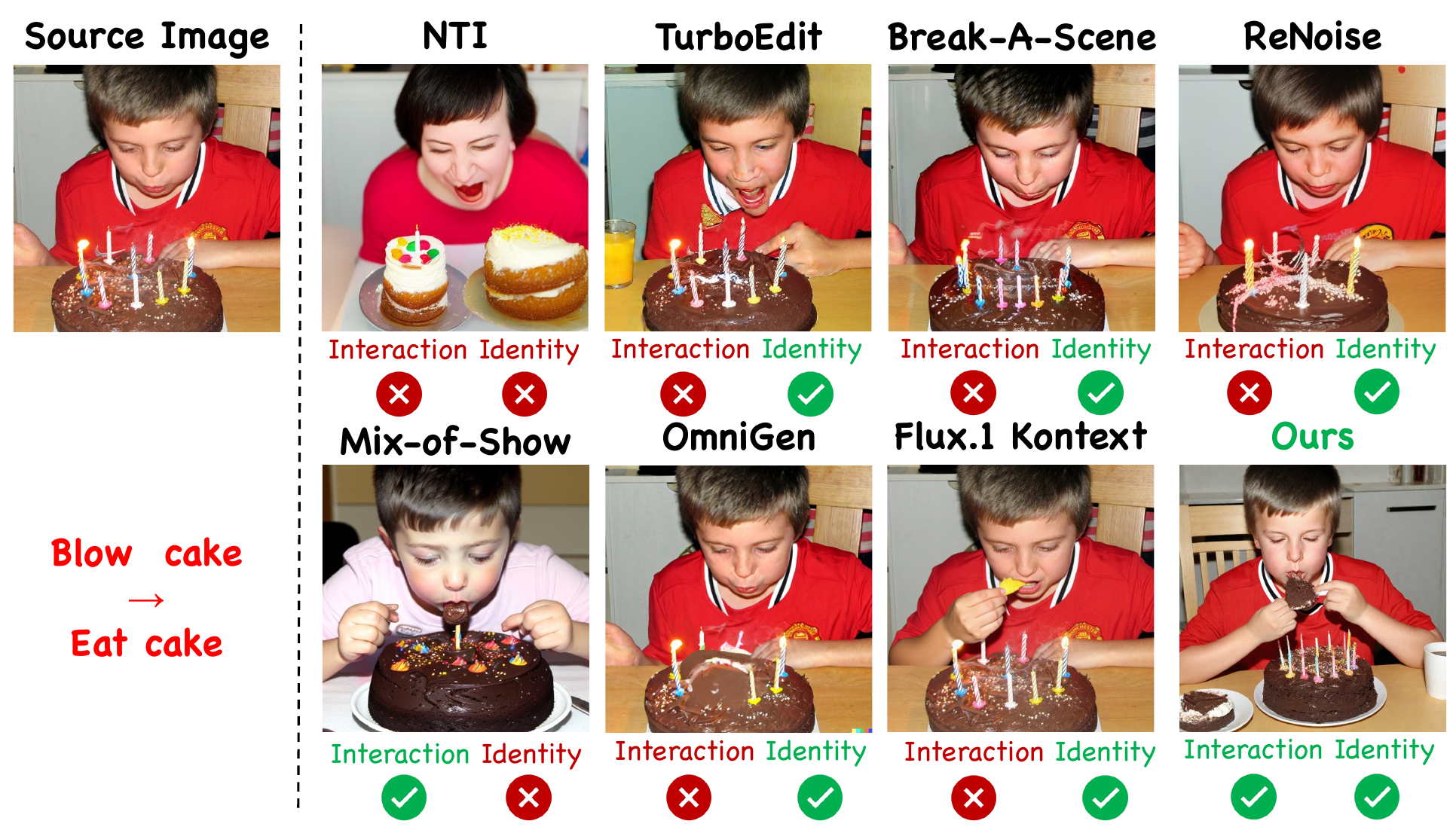}
    \caption{Sample results of editing Human-Object Interaction in the source image (left). Existing methods overly preserve the structure, making interaction edits ineffective.}
    \label{fig:teaser}
\end{figure}


%% file: assets/figure/arch.tex
\begin{figure}
\centering
\includegraphics[width=0.95\linewidth]{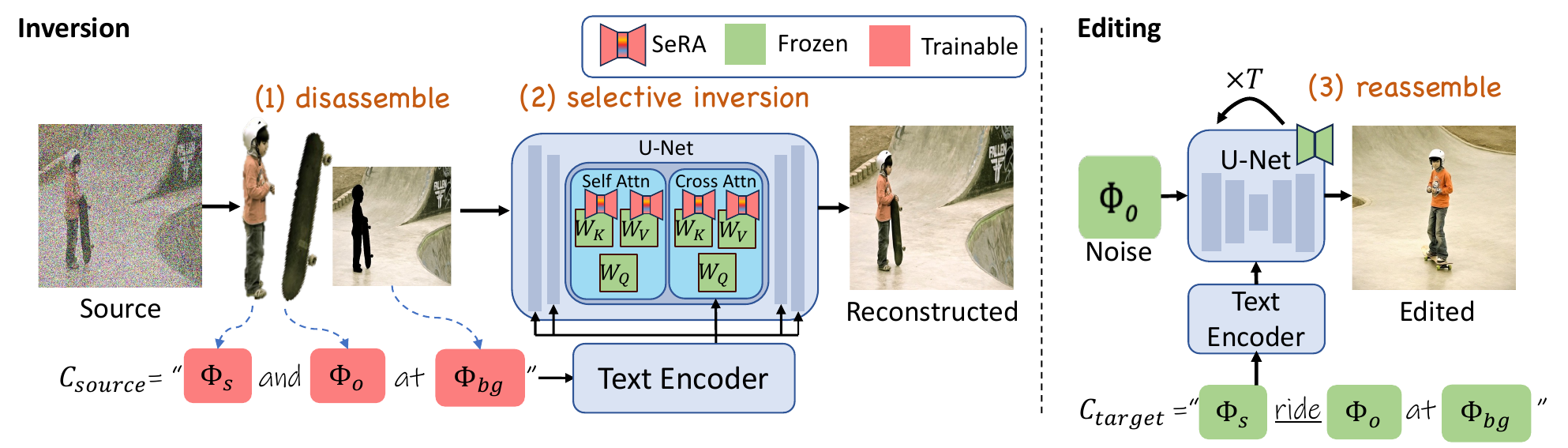}
\caption{Overview of the {\rmfamily\scshape InteractEdit} framework. HOI components are disassembled into subject, object, and background descriptors during inversion (\cref{subsec:disassemble-hoi}). SeRA regularization enables non-rigid edits by capturing essential identity features while suppressing overfitting to structural details (\cref{subsec:lora}). Selective inversion preserves pretrained interaction priors while adapting to the source image’s identity (\cref{subsec:selective-training}). Editing reassembles these components with the target interaction, using SeRA weights delta to guide the diffusion model (\cref{subsec:hoi-editing}).}  
\label{fig:arch}
\end{figure}

%% file: sec/2_relatedwork.tex
\section{Related Work}\label{sec:related_work}

\subsection{Human-Object Interactions}
Human-Object Interactions (HOIs) are typically represented as triplets $\left\langle \text{human, action, object} \right\rangle$ with corresponding bounding boxes for the human and the object. HOI detection aims to localize these entities and classify their interactions. Recent works \cite{zhang2023pvic,fang2023hodn,luo2024sichoi, ehotr, ehotr+, pmeqc2026, HAN2026114765, ascformer2026} have improved generalization, particularly for rare and unseen interactions.
HOI generation \cite{hoe2024interactdiffusion,jianglin2024record,ge2024invertedinteractionhoi,hu2025personahoi,cha2025verbdiff,taghipour2025b2b,arpe} is a recent field focused on generating images from given HOI triplets. InteractDiffusion \cite{hoe2024interactdiffusion} employs diffusion models to generate images matching the specified interactions and bounding boxes. While valuable for data augmentation and creative applications, HOI generation does not address edit interactions in existing images, where the goal is to modify the interaction while preserving original identity.
Our proposed framework fills this gap by enabling interaction editing in images. It modifies existing HOI while preserving the visual integrity of the original subject and object.

\subsection{Diffusion Models} 
Diffusion Models \cite{ddpm2020,song2021scoresde} generate high-quality images by iteratively denoising random noise and have proven more stable than GANs \cite{diffusionbeatsgan2021}. Conditional diffusion models further guide generation using text prompts \cite{glide2021,dalle2-2022,podell2023sdxl}, layout \cite{gligen2023,layoutdiffusion2023,xie2023boxdiff,phung2024attentionrefocusing,zheng2025semantic}, personalized embeddings \cite{ye2023ipadapter} and visual conditions \cite{controlnet2023,mo2023freecontrol}. Text-conditioned diffusion is itself a form of multimodal learning, where heterogeneous signals must be aligned and fused rather than merely concatenated \cite{mnerevidential2025,unimodallabels2026}. While recent advances enable HOI-aware generation \cite{hoe2024interactdiffusion}, editing existing human-object interactions while maintaining identity coherence remains largely unexplored. To bridge this gap, we leverage text-conditioned diffusion models, which are uniquely suited for flexible, controllable image manipulation.

\subsection{Image Editing Diffusion Models}\label{subsec:image-editing-diffusion}
Image Editing Diffusion Models \cite{revise,tftdiffusion} span single- and multi-subject customization, local editing, text-guided editing, and image translation~\cite{shuai2024diffusion-edit-survey}. Subject customization~\cite{shi2024instantbooth,li2024blipdiffusion,liu2023cones2,han2023svdiff,wang2024msdiffusion,ma2024subjectdiffusion,ng2024partcraft,wang2024instantid,shi2024foodfusion,jiang2024animediff} generates or modifies images with specific subjects using a small set of reference images. Local editing~\cite{epstein2023selfguidance,mou2024diffeditor,shi2024dragdiffusion,mou2023dragondiffusion} includes object change \cite{yang2024personinplace} (adding, removing, replacing objects), attribute modification (color or texture), spatial transformation \cite{shi2024dragdiffusion,liu2024dragnoise} (translation, scaling, warping), and inpainting \cite{wang2023imageneditor}. Text-guided image editing \cite{brooks2022instructpix2pix,tumanyan2023pnp,kawar2023imagic,geng2024instructdiffusion,hu2024instructimagen,tsaban2023ledits,brack2024ledits++,song2024dac} leverages natural language prompts to direct image modification, enabling a range of edits. Image translation \cite{meng2022sdedit,jiang2024scedit} involves transforming images across domains (e.g., from sketches to photorealistic images) or applying style transfers. Mainstream customization methods typically require multiple images per subject, limiting their practical use. Break-A-Scene \cite{avrahami2023breakascene} showed that multiple visual concepts can be learned from a single image, which is crucial for real-world image editing where multiple reference images per concept are rarely available.

Although some of the above methods share building blocks with InteractEdit (e.g., concept decomposition, textual inversion, LoRA adaptation), they exhibit failure modes specific to HOI editing and serve different purposes from ours. DreamBooth-style personalization \cite{ruiz2023dreambooth} adapts model weights but tends to overfit to the source image's structural information (original pose and spatial arrangement), making it unable to adapt to a new target interaction that requires large non-rigid change. Textual Inversion \cite{gal2022textualinversion} is too capacity-limited, as a few learned tokens cannot faithfully capture the appearance identity of both subject and object simultaneously. Break-A-Scene \cite{avrahami2023breakascene} decomposes concepts for multi-concept generation from scratch, rather than for preserving identities across a non-rigid interaction change. In contrast, InteractEdit explicitly separates these concerns: a subject-object-background decomposition disentangles the deeply coupled identities in an HOI image, while adapting model weights provides the representational capacity of DreamBooth without its structural overfitting. Furthermore, unlike prior parameter-freezing strategies (e.g., ControlNet \cite{controlnet2023}, IP-Adapter \cite{ye2023ipadapter}) that freeze parameters as a lightweight conditioning shortcut, our Selective Inversion is an interaction-aware design that follows directly from the algebraic structure of cross-attention (see Sec.~\ref{subsec:selective-training}): freezing $W_Q$ preserves the pretrained interaction-aware spatial routing, while adapting $W_K, W_V$ binds the new identity tokens to the prompt.

\subsection{Interaction Editing}
Interaction editing is fundamentally distinct. While conventional editing modifies object attributes or positions within a largely static structure, interaction editing aims to change only the interaction while preserving subject and object identity. Person in Place \cite{yang2024personinplace} places human onto a given background image rather than editing an existing interaction, making it unsuitable for direct comparison to HOI editing tasks. Recent advanced instructional image editing methods \cite{wei2024omniedit,labs2025flux1kontext} excel at attributes or style manipulation but struggle with the flexible, non-rigid changes required by HOI edits. The concurrent HOIEdit \cite{xu2025hoiedit} tackles HOI editing using multiple reference images of target interactions, making it less practical where such references are unavailable.

In contrast, our method enables \textit{reference-free} HOI editing by leveraging the model's pretrained interaction knowledge, removing the need for reference images of target interactions. This makes our approach both practical and efficient. To the best of our knowledge, this is the \textit{\textbf{first work}} to systematically address reference-free interaction editing.

\subsection{Low-Rank Adaptation} LoRA \cite{hu2022lora} has been widely used in image editing for efficient fine-tuning \cite{song2024dac}, concept customization \cite{gu2024mixofshow}, and style transfer \cite{frenkel2024blora}. We extend LoRA with Selective-Rank Adaptation (SeRA), introducing per-rank gating to steer the inversion process more effectively. Unlike LoRA, our SeRA modulates each rank direction to capture essential identity attributes for reconstruction while suppressing overfitting to structural details. This adaptive mechanism is essential for flexible non-rigid interaction edits, a core requirement for effective reference-free HOI editing as pursued in this work.

%% file: sec/3_method.tex
\section{Method}\label{sec:method}
Given an input image $I_{\text{source}}$ depicting an initial HOI as a triplet $\left \langle s, i_{\text{original}}, o \right\rangle$, where $s$ is subject, $o$ is object and $i_{\text{original}}$ is their original interaction, our goal is to semantically modify this interaction to a new, specified interaction $i_{\text{target}}$. This requires balancing the preservation of subject and object identity with sufficient structural adaptability in editing the interaction and adapting the background as necessary.
The resulting image $I_{\text{target}}$ should depict the new HOI $\left \langle s, i_{\text{target}}, o \right\rangle$ while maintaining realism and coherence.

\subsection{Preliminary}

Image editing generally involves two stages: \textit{inversion} and \textit{editing}. The inversion algorithm $F_{\text{inv}}$ encodes an input image into a set of \textbf{descriptors} $\mathbf{\Phi}$ that capture the key identity and structural information needed for reconstruction. 
These descriptors are then used in the editing stage to generate the desired content. Formally, the inversion is defined as:
\begin{equation}
    \mathbf{\Phi} = F_{\text{inv}}(I_{\text{source}}, C_{\text{source}}),
\end{equation}
where $C_{\text{source}}$ represents the original text prompt.
During the editing stage, the edit function $F_{\text{edit}}$ integrates $\mathbf{\Phi}$ into the base model to produce an image that retains visual identity while adhering to the new instruction:
\begin{equation}\label{eq:f_edit}
I_\text{target} = F_{\text{edit}}(\mathbf{\Phi}, C_{\text{target}}),
\end{equation}
where $C_{\text{target}}$ is the target prompt.


Break-A-Scene \cite{avrahami2023breakascene} is an inversion technique that bridges Textual Inversion \cite{gal2022textualinversion} and DreamBooth \cite{ruiz2023dreambooth} for single-image multi-concept extraction. Technically, given an input image $I_{\text{source}}$ and a set of $N$ masks $\{M_n\}^N_{n=1}$ representing distinct concepts, it extracts the token descriptor $\{\Phi_n\}^N_{n=1}$, which can be incorporated into the target text prompts to generate new instances of these concepts. Additionally, it learns weight deltas $\Delta W$ to retain the source image information to improve reconstruction. 
While effective at disentangling static visual concepts, Break-A-Scene \cite{avrahami2023breakascene} struggles with non-rigid edits like HOI editing, which require modifying the spatial relationships between subjects and objects, as well as altering the subject’s pose and the object’s placement. Our approach, described next, enhances the Break-A-Scene with a regularized inversion process designed to better support non-rigid HOI edits.

\subsection{InteractEdit: Overall framework}\label{subsec:framework}

We innovatively design \textsc{InteractEdit} (see \cref{fig:arch}) as a \textit{reference-free} HOI-editing framework that leverage on pretrained interaction priors to guide target edits without relying on external references. The core idea is to retain the model’s existing HOI understanding while enabling flexible transformation. In other words, \textsc{InteractEdit} performs an inversion that learns weight deltas $\Delta W$ to encode the source image’s identity while preserving the model’s pretrained interaction priors. Specifically, we introduce \emph{Selective Inversion} and Selective-Rank Adaptation (SeRA), which are designed to constrain updates to identity subspaces. This helps to retain the subject and the object identity while still allowing the targeted HOI transformation. The next section presents the decomposition of HOI to disentangle intricate HOI relationship, followed by a description of how \emph{Selective Inversion} and SeRA are designed to leverage interaction priors and retain subject and object identity. Finally, we outline the overall HOI editing process. 



\subsection{Disassemble Human-Object Interaction}\label{subsec:disassemble-hoi}
Editing HOI in images is challenging due to the tightly coupled relationships between the human, object, and background. Directly editing these interdependent components often leads to one of two critical issues: either the intended interaction fails to transform, or the visual identity is lost.
To address this issue, we \textbf{decompose} each scene into three elements: human (subject), object, and background, treating them as three distinct concepts, and represent them as a set of learnable token descriptors $\Phi_s$, $\Phi_o$, and $\Phi_\text{bg}$, following Break-A-Scene \cite{avrahami2023breakascene}. These embeddings are incorporated into text prompts and optimized during inversion. Each concept is associated with a mask $M_n$, where $n \in {s, o, {bg}}$, which defines its spatial region within the image.
At each optimization step, a subset of $k \leq N$ concepts is randomly selected, and a corresponding text prompt $C_{\text{source}}$ is constructed. 
For example, if the subject and background are chosen, the text prompt takes the form: ``a photo of [$\Phi_s$] and [$\Phi_\text{bg}$].''

A masked version of standard diffusion loss \cite{avrahami2023breakascene} is then applied, ensuring that only the pixels covered by the selected concept masks contribute to optimization:
\begin{equation}\label{eq:l_rec}
    \mathcal{L}_\text{rec} = \mathbb{E}_{\mathbf{z},\epsilon\sim\mathcal{N}(0,1),t} \Big[
    \lVert \epsilon \odot M_U - \epsilon_\theta(\mathbf{z}_t,t,C_\text{source}) \odot M_U \rVert^2_2 
    \Big]
\end{equation}
where $\mathbf{z}_t$ is the noisy latent at timestep $t$, $C_\text{source}$ is the text prompt, $M_U$ is the union of the corresponding masks, $\epsilon$ is the added noise, and, $\epsilon_\theta(\cdot)$ is the denoising network.

To further disentangle different concepts, an additional loss is imposed on the cross-attention maps to explicitly align each concept token $\Phi_n$ with its corresponding spatial region defined by the mask $M_n$, where $n \in {s, o, {bg}}$ (subject, object, and background). The cross-attention loss is defined as:
\begin{equation}\label{eq:l_attn}
    \mathcal{L}_\text{attn}=\mathbb{E}_{\mathbf{z},t,n}\left[ \lVert \text{CA}_\theta(\Phi_n,\mathbf{z}_t) - M_{n} \rVert ^2_2 \right],
\end{equation}
where $\text{CA}_\theta(\Phi_n,\mathbf{z}_t)$ is the cross-attention map between the token $\Phi_n$ and the noisy latent $\mathbf{z}_t$. Cross-attention maps are extracted at a 16×16 resolution, averaged across all corresponding U-Net layers, and normalized to the range (0,1). This regularization ensures that each learned concept token effectively corresponds to its spatial region, preventing entanglement between different concepts.

However, unconstrained updates as in \cite{avrahami2023breakascene}, these descriptors tend to overfit to the structural details of the source image, such as spatial relationships, arrangement, and pose. In other words, this causes memorization of specific configurations rather than learning generalizable representations. More critically, it corrupts the model’s interaction priors, undermining its adaptability to new interactions. As a result, the generated outputs often fail to reflect the intended modifications, posing a significant barrier to effective HOI editing (see \cref{tab:ablation} and \cref{fig:qualitative_abl} for results).


\subsection{Regularizing Inversion via SeRA}\label{subsec:lora}
\begin{figure}
    \centering
    \includegraphics[width=1\linewidth]{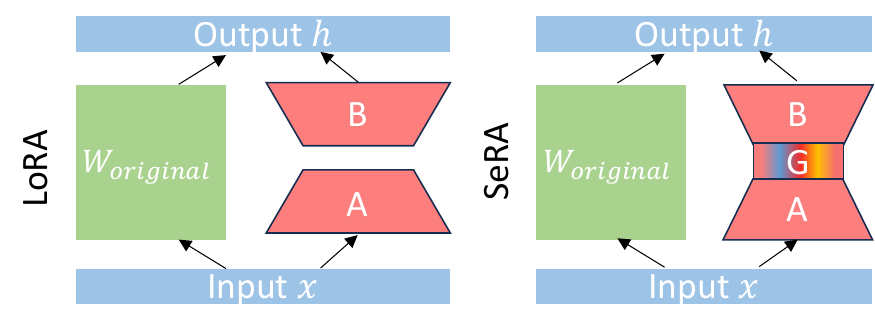}
    \caption{Selective-Rank Adaptation (SeRA): It extends LoRA by integrating a learnable gating vector $G$ between low-rank matrices $A$ and $B$, enabling selective activation of rank directions during weight updates. This constrains adaptation capacity to reduce overfitting to source-image structure while supporting adaptive interaction edits.}
    \label{fig:sera}
\end{figure}
To mitigate this, we introduce SeRA, a specialized LoRA \cite{hu2022lora} variant that integrates a learnable gating vector $G$ between the low-rank matrices $A$ and $B$ (see \cref{fig:sera}). 
Technically, the gate $G \in \mathbb{R}^{r \times r}$ modulates each update direction, allowing the model to learn how many and which rank components to activate during inversion. Combined with the disentangled representation (from \cref{subsec:disassemble-hoi}), this prevents the LoRA from overusing its capacity to memorize source-image structure, while retaining enough adaptation for HOI transformations such as pose and subject–object spatial relations.

Formally, a SeRA-adapted weight matrix $W$ is defined as:
\begin{equation}\label{eq:lora}
W = W_{\text{original}} + \Delta W,\quad \Delta W = AGB^\top,
\end{equation}
where $W \in \mathbb{R}^{d_1 \times d_2}$, $A \in \mathbb{R}^{d_1 \times r}$, $B \in \mathbb{R}^{d_2 \times r}$, $r$ is the rank hyperparameter, and $G \in \mathbb{R}^{r \times r}$ is a diagonal, learnable matrix that gates the contribution of each rank direction. In practice, we apply a sigmoid to each diagonal entry of $G$, i.e., $G \leftarrow \sigma(G)$, so that every gate lives in $(0,1)$ and softly controls each rank direction. We apply SeRA in the U-Net backbone, as detailed in the following section.

To ensure each gate in G commits to either opening or closing its corresponding subspace, rather than hovering at an indecisive midpoint, we introduce an uncertainty loss that actively repels gates from 0.5. Concretely, for a diagonal gating matrix $G\in\mathbb{R}^{r\times r}$ with entries $G_{jj}$, we define:
\begin{equation}\label{eq:l_sera}
    \mathcal{L}_{\mathrm{SeRA}} \;=\;\frac{1}{r}\sum_{j=1}^{r} G_{jj}\bigl(1 - G_{jj}\bigr)\,.
\end{equation}
Since the term $G_{jj}(1 - G_{jj})$ attains its maximum at $G_{jj}=0.5$ and falls to zero at the endpoints 0 and 1, minimizing $\mathcal{L}_{\mathrm{SeRA}}$ drives each gate away from uncertainty and toward a clear “on” or “off” state.



\subsection{Selective Inversion for Prior Retention}\label{subsec:selective-training}
The U-Net used in diffusion models comprises multiple blocks arranged in an encoder-decoder structure. Each block contains self-attention and cross-attention mechanisms to capture relationships within the image features and between the image and conditioning inputs, respectively.
In each attention layer, input features are transformed into Query ($Q$), Key ($K$) and Value ($V$) representations. The attention mechanism is described as 
   $$
   \text{Attention}(Q, K, V) = \text{Softmax}\left(\frac{QK^\top}{\sqrt{d_k}}\right)V,
   $$
where $Q = XW^Q$, $K = XW^K$, and $V = XW^V$ are linear transformations of the input $X$ using weight matrices $W^Q$, $W^K$, and $W^V$, respectively, and $d_k$ is the dimensionality of the Key vectors used for scaling. $Q$ encodes what features a token is ``looking for,'' $K$ represents what features a token offers to be attended to, and $V$ carries the actual information used to generate the output. 

In HOI editing, our goal is to leverage the pretrained model's interaction knowledge while learning to encode and maintain the visual traits of the source image. Rather than assigning trainable and frozen projections heuristically, we motivate the choice from the algebraic structure of attention. The attention operation can be factorised into a \emph{structural} factor and a \emph{content} factor:
\begin{equation}
    \underbrace{A = \mathrm{softmax}\!\left(\tfrac{QK^{\top}}{\sqrt{d_k}}\right)}_{\text{structure: \emph{where} attention is routed}},
    \qquad
    \underbrace{O = AV}_{\text{content: \emph{what} is placed there}}.
    \label{eq:attn-factorisation}
\end{equation}
Two observations follow immediately from \cref{eq:attn-factorisation}. First, $V$ does \emph{not} enter the routing matrix $A$ at all; it only modulates the \emph{content} that is read out at each spatial location. Second, $A$ is determined jointly by $Q$ and $K$: $Q$ comes from the image-side latent $X_\text{img}$ and acts as the spatial query, while $K$ comes from the text embedding $X_\text{txt}$ and provides the per-token binding key for routing. In cross-attention, the asymmetry is sharper still: $Q$ is image-side and $K, V$ are text-side, so the newly introduced concept tokens for the disassembled subject, object, and background can only enter the attention through $W^K$ and $W^V$.

This factorisation motivates a design for which projections to adapt and which to freeze during inversion:
\begin{itemize}
    \item \textbf{Train $W^V$ (content lever).} Since $V$ is the only factor in $AV$ that is not shared with $A$, adapting $W^V$ provides a direct content pathway for writing the inverted subject/object/background appearance into the output without perturbing the routing.
    \item \textbf{Train $W^K$ (binding key).} Adapting $W^K$ helps make the newly inverted concept tokens addressable by the image queries; otherwise their attention column $A[:,\text{new}] \propto \exp(Q k^\top)$ with $k = X_\text{txt} W^K$ remains generic, and the new content in $V$ is not routed to the intended spatial region.
    \item \textbf{Freeze $W^Q$ (pretrained spatial routing).} $W^Q$ produces the image-side queries that select \emph{where} text tokens attend, and the pretrained $W^Q$ already encodes an interaction-aware spatial routing prior. Freezing $W^Q$ preserves this pathway rather than fitting it to the pose and spatial layout of the single source image, which we expect to mitigate the ``structural locking'' that otherwise hinders re-routing toward the target interaction (\cref{fig:attnmap}).
\end{itemize}

Our hypothesis is therefore that freezing $W^Q$ while adapting $W^K, W^V$ should preserve the general spatial routing while still allowing the new identities to bind to the prompt. Leaving both $W^Q$ and $W^K$ trainable instead removes the constraint on $A$, so the adaptation is free to fit, and potentially \textit{memorise the source attention map}; we confirm empirically in \cref{sec:qkv-ablation} that this fully-trainable configuration performs worst. For the frozen-$W^Q$ case, \cref{lem:logit_invariance} further shows that the row-space of the attention-score matrix $S(X)$ remains contained in the row-space induced by the pretrained query map, for any update of $W^K$.

We apply the same Q-frozen/$K,V$-trainable schedule consistently to both self- and cross-attention: in self-attention the same decomposition holds with $Q, K, V$ all derived from the image latent, and freezing $W^Q$ analogously preserves the pretrained intra-image spatial relations.
These updates are parameterized with low-rank SeRA modules. Specifically, the pretrained weights ($W_\text{original}$) remain frozen, while only the SeRA parameters ($A$, $B$ and $G$) are optimized. Collectively, the SeRA-adapted weights across all U-Net blocks constitute $\Phi_{\text{SeRA}}$, capturing image-specific adjustments that faithfully reconstruct the source while retaining pretrained interaction priors.

The overall loss combines the above masked reconstruction loss (\cref{eq:l_rec}), the cross-attention alignment loss (\cref{eq:l_attn}) and the SeRA loss (\cref{eq:l_sera}):
\begin{equation}
\mathcal{L}_\text{total}=\mathcal{L}_\text{rec}+\lambda_\text{attn}\mathcal{L}_\text{attn}+\lambda_\mathrm{SeRA}\mathcal{L}_{\mathrm{SeRA}},
\end{equation}
where $\lambda_\text{attn}=0.01$ following Break-A-Scene \cite{avrahami2023breakascene} and $\lambda_\mathrm{SeRA}=0.01$.

\subsection{HOI Editing}\label{subsec:hoi-editing}
To implement selective inversion, we extract descriptors as:
\begin{equation}
    \mathbf{\Phi} = [\Phi_s,\Phi_o,\Phi_\text{bg},\Phi_\text{SeRA}] = F_{\text{inv}}(I_{\text{source}}, C_{\text{source}}),
\end{equation}
where $\Phi_s$,$\Phi_o$,$\Phi_\text{bg}$ are the descriptors representing the subject, object, and background, respectively, and $\Phi_\text{SeRA}$ denotes the SeRA weight deltas learned during inversion.

During the editing phase, we generate an image that depicts the target interaction while preserving the original identity. To do this, we reassemble the descriptor tokens ($\Phi_s$, $\Phi_o$, and $\Phi_\text{bg}$) into a new target prompt, replacing the original action $i_\text{original}$ with the target interaction $i_{\text{target}}$:
$$C_{\text{target}} = \text{``a photo of } \left[\Phi_s\right] \text{ } i_{\text{target}} \text{ } \left[\Phi_o\right] \text{ at } \left[\Phi_\text{bg}\right]\text{."}$$
Using this prompt, the diffusion model $\epsilon_\theta$, along with selectively inverted weights $\Phi_\text{SeRA}$, generates the edited image $I_{\text{target}}$ via an iterative denoising process starting from random noise $\mathbf{z}_T$. At each timestep $t=T,T-1,\cdots,1$, the model predicts the denoised image:
\begin{align}
    \mathbf{z}_{t-1} = \epsilon_{\theta}(\mathbf{z}_t, t, C_{\text{target}};~\Phi_\text{SeRA}).
\end{align}
The final image $I_\text{target}$ is obtained as $\mathbf{z}_0$. This process realizes \cref{eq:f_edit}, where $\mathbf{\Phi}$ including the descriptor tokens ($\Phi_s$, $\Phi_o$, $\Phi_\text{bg}$) and SeRA weights $\Phi_\text{SeRA}$.

In this process, the selectively inverted SeRA weights $\Phi_\text{SeRA}$ and the descriptor tokens ($\Phi_s$, $\Phi_o$, $\Phi_\text{bg}$) work together to balance preserving the visual identity from the source image and enabling sufficient flexibility for interaction edits, while the pretrained model parameters provide interaction priors. This synergy ensures the generated image reflects the new interaction $i_{\text{target}}$ as specified by $C_{\text{target}}$, without significantly compromising the original identity.

%% file: sec/4_experiment.tex
\section{Interaction Editing Benchmark (IEBench)}\label{sec:hoieditbench}
\subsection{Benchmark}
A rigorous benchmark is essential for properly evaluating HOI editing task. While HOIEdit \cite{xu2025hoiedit} made initial progress, but it has two key limitations. First, its evaluation is limited to a small-scale benchmark covering only 3 actions. Second, it relies on text-image and image-image similarity metrics, which are unreliable for assessing HOI edit correctness, as CLIP is known to underperform in the action recognition task \cite{lei2024lex}.

To address these shortcomings, we introduce \iebench\footnote{We will release this benchmark publicly.}, a new benchmark specifically designed to evaluate both intended interaction transformation and identity preservation in the HOI editing task. \iebench~is constructed from the test subset of HICO-DET dataset \cite{hicodet2018} by selecting source images and target interactions that: (i) the source image must depict a single HOI instance, (ii) the target interaction must be plausible for the object, and (iii) the target interaction must produce significant visual changes. We define criterion (iii) operationally: a $\langle$source image, target interaction$\rangle$ pair is retained only if realising the target requires the person to physically reconfigure their limbs or body relative to the object. For example, we retain \textit{hold} $\rightarrow$ \textit{eat} broccoli (the arm raises the object to the mouth) and \textit{hold} $\rightarrow$ \textit{ride} a skateboard (the board moves from the side to underfoot), while excluding near-synonymous verb swaps that leave the same limbs in the same contact with the object, as well as attribute-only edits such as object colour or texture, since a method could satisfy these without changing the interaction at all.

The resulting benchmark comprises of 28 source images covering 25 actions and 13 objects, forming a total of 100 unique $\langle$ source image, target interaction $\rangle$ pairs. The full set of actions includes: \textit{feed, make, pick up, sit on, hit, ride, walk, cut, eat at, jump, throw, dribble, smell, kick, hug, eat, hold, catch, sit at, wash, stand on, kiss, groom, carry, and lie on}. 
\Cref{tab:object_interaction_pairs} lists the complete set of objects and their paired target interactions. Each object is matched with multiple interactions to ensure diverse and meaningful interaction editing scenarios. 

\input{assets/figure/obj-act-pairs}

\subsubsection{Extended benchmark ({\let\textsc\textnormal\iebenchext})} \label{subsubsec:iebench-l}
To evaluate generalization at a larger scale and broader interaction coverage, we additionally introduce \iebenchext, an extended and \emph{more challenging} version of \iebench~constructed under the same selection criteria. \iebenchext~comprises \textbf{100 source images} (\textbf{3.6$\times$} larger), spanning \textbf{30 unique actions} (\textbf{1.2$\times$}) and \textbf{21 unique objects} (\textbf{1.6$\times$}), and forming \textbf{326} unique $\langle$source image, target interaction$\rangle$ pairs (\textbf{3.3$\times$}). Beyond scale, \iebenchext~is intentionally harder: it covers a much wider distribution of subjects, objects, and viewpoints, and incorporates rarer verb--object compositions and more demanding non-rigid pose changes that stress-test both the editability and identity-preservation aspects of HOI editing. The newly added objects and their candidate target interactions are listed in the bottom block of \cref{tab:object_interaction_pairs}. \iebench~remains a strict subset of \iebenchext, ensuring backward compatibility with all results reported on \iebench. We re-evaluate \textsc{InteractEdit} together with the top-performing baselines on \iebenchext~to verify that our conclusions hold under this larger and more demanding setting; the full results are reported in \cref{sec:iebenchext-results}.

\subsection{New Metric}
\label{subsec:new_metric}
\input{assets/figure/clip_score}
Traditional metrics like CLIP-T, which measure text–image alignment between a prompt and the generated image, are inadequate for HOI editing. CLIP \cite{clip2021} is known to struggle with verb semantics \cite{momeni2023verbsinaction,thrush2022winoground}. As shown in \cref{fig:clip_score}, very different human–horse interactions (e.g., stand, wash, ride) yield nearly identical similarity scores with the prompt “person ride horse,” demonstrating CLIP-T’s invariance to verbs. To analyze this, we further test CLIP-T’s sensitivity to action (verb) versus object mismatches. Verb mismatches (blue) yield extremely narrow margins clustered near zero, while noun mismatches (orange) produce broader margins. This confirms CLIP is primarily noun-sensitive but weak at distinguishing verbs, highlighting why CLIP-T cannot serve as a reliable HOI metric and motivating the need for our HOI-specific alternatives. Further details are given in \cref{sec:clip-score}.

To complement this benchmark, we introduce new evaluation metrics specifically designed for HOI editing task. These metrics quantify the trade-off between intended interaction transformation correctness and identity preservation.

\textbf{(i) HOI Editability, \text{\textsc{he}}} quantifies editing success by determining whether the target interaction is present in the edited image. Leveraging PViC \cite{zhang2023pvic}, a state-of-the-art HOI detector, each generated image is assigned a score of one if the target interaction is detected, and zero otherwise. The final \text{\textsc{he}} score is computed as the mean detection rate over all edited samples.

\textbf{(ii) Identity Consistency} assesses how well the subject and object identities are preserved after editing. It is deliberately region-restricted, comparing only the segmented subject and object regions, so that the large pose and layout changes required by an interaction edit are not themselves penalized.

Let $I$ be a source image and $\hat{I}$ its edited counterpart. For a concept $c \in \{s, o\}$ (subject, object), we define the region-extraction operator
\begin{equation}\label{eq:region-extraction}
    \mathcal{R}_c(I) = I \odot \text{SAM}\big(I, \text{GDINO}(I, \tau_c)\big),
\end{equation}
which grounds the concept using GroundingDINO \cite{liu2023groundingdino} with the text query $\tau_c$ (the subject or object noun of the HOI triplet), segments it with SAM \cite{kirillov2023segmentanything}, and returns the masked region. Let $f(\cdot) \in \mathbb{R}^{d}$ denote the DINOv2 \cite{oquab2023dinov2} embedding of a masked region, chosen for its strong object-centric and fine-grained representation. The per-concept identity similarity is then the cosine similarity between the source and the edited region:
\begin{equation}\label{eq:simc}
    \text{sim}_c\big(I, \hat{I}\big) = \frac{\big\langle f(\mathcal{R}_c(I)),\, f(\mathcal{R}_c(\hat{I})) \big\rangle}{\big\lVert f(\mathcal{R}_c(I)) \big\rVert_2 \cdot \big\lVert f(\mathcal{R}_c(\hat{I})) \big\rVert_2}.
\end{equation}
Each edited image is scored by the average of $\text{sim}_s$ and $\text{sim}_o$, and the final Identity Consistency score is computed as the mean over all edited samples. If GroundingDINO fails to ground concept $c$ in the edited image, we set $\text{sim}_c = 0$ rather than excluding it from the average, so that a method cannot obtain a high identity score by removing the subject or the object altogether.

\textbf{(iii) Editability-Identity Score, \text{\textsc{ei}}} quantifies the trade-off between HE score and Identity Consistency via the harmonic mean, analogous to the $F_1$ score \cite{rijsbergen1979fscore}. This formulation ensures a balanced evaluation by penalizing low performance in either dimension:
\begin{align}
    \text{\textsc{ei}} &= \frac{2 \times \text{HOI Editability} \times \text{Identity Consistency}} {\text{HOI Editability} + \text{Identity Consistency}}.
\end{align}

\textbf{(iv) Human Preference Score v2 (HPSv2)} \cite{wu2023hpsv2} is a perceptual evaluation model \cite{blindgraphattention2024,perceptionoriented2025,deepbiattention2024} trained on extensive human rankings to quantify the alignment of generated images with human aesthetic and quality preferences. We leverage this metric to assess the fine-grained visual realism of the edited HOI scenes, which automated structural metrics may overlook. While \textbf{\textsc{he}} and \textbf{\textsc{ei}} scores ensure semantic interaction correctness and identity consistency, \textbf{HPSv2} provides a necessary complementary perspective by evaluating the overall photorealism of the generated result.


\begin{figure}
    \centering
    \includegraphics[width=.95\linewidth]{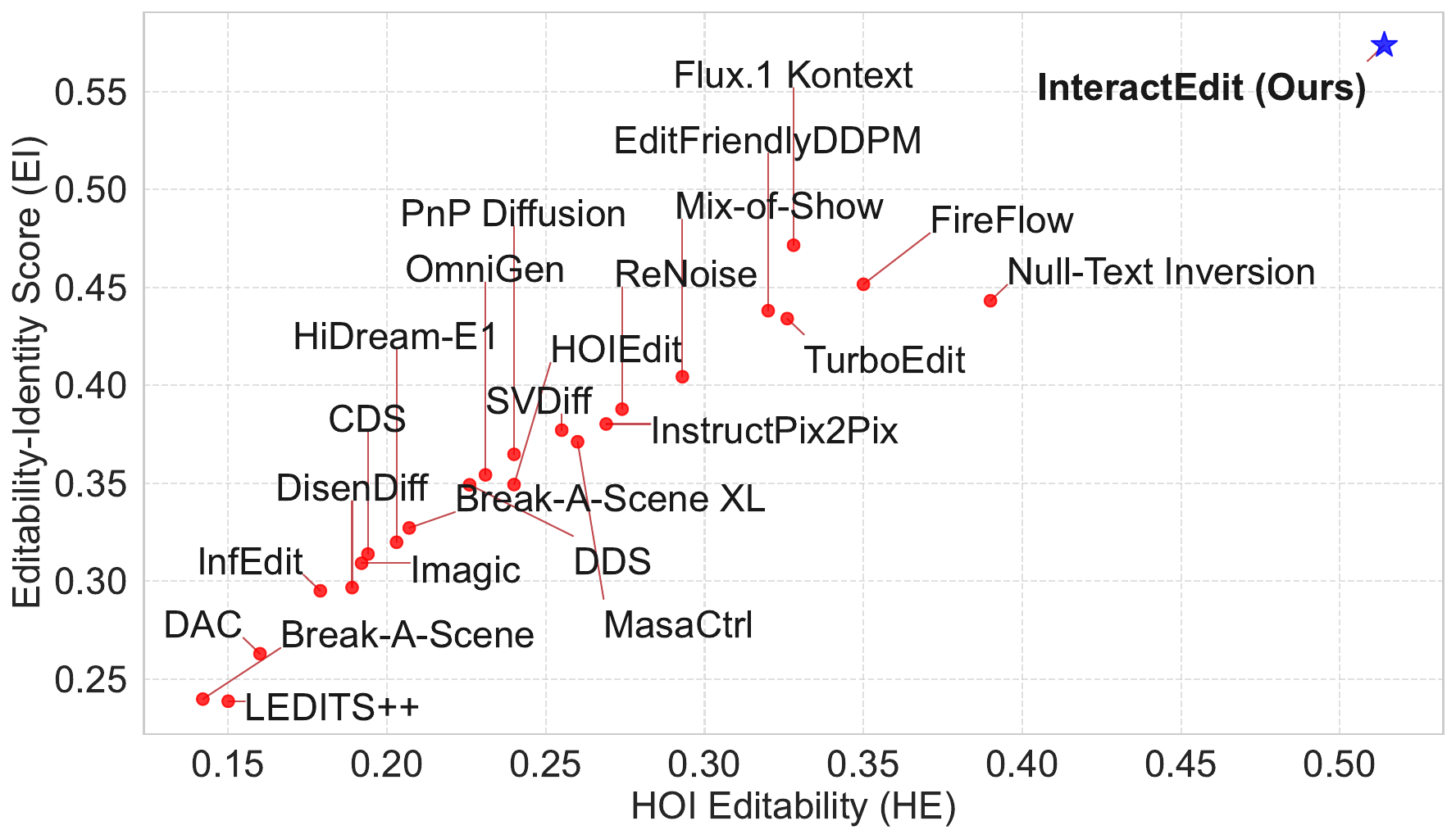}
    \caption{Our method achieves the best for both HOI editability and Editability-Identity score.}
    \label{fig:overall}
\end{figure}
\input{assets/figure/edit_perf}

\input{assets/figure/qualitative}

\section{Experiments}\label{sec:experiments}
We conducted experiments at 512x512 resolution, basing our method on Stable Diffusion XL \cite{podell2023sdxl}. Following Break-A-Scene \cite{avrahami2023breakascene}, we use a two-stage optimization approach. In the first stage, we optimize only the descriptor tokens $\Phi_s, \Phi_o, \Phi_\text{bg}$ for 1000 steps with a learning rate of 5e-4. In the second stage, we jointly optimized the SeRA weights $\Phi_\text{SeRA}$ and the descriptor token for 200 additional steps with a learning rate of 1e-4. Additional details are in \cref{sec:implementation-details}.

\subsection{Qualitative results}
\Cref{fig:qualitative} compares our edits with those of leading baselines, illustrating how each method trades off interaction change versus identity preservation. NTI and InstructPix2Pix manage to alter actions but severely distort the person or object. InfEdit, Flux.1 Kontext, and Break-A-Scene instead lock in identity perfectly yet leave poses and layouts essentially unchanged. For example, when asked to \textit{``wash the horse,''} Flux.1 Kontext still depicts the rider in the original ``ride horse'' pose, resulting in almost no interaction change.

Our approach strikes the balance: SeRA's gated low-rank updates constrain inversion to identity-relevant subspaces, avoiding overfit to the source structure. Selective inversion freezes the model's query layers so that pretrained interaction priors remain intact. This synergy unlocks substantial non-rigid edits like re-posing arms and repositioning objects, without sacrificing who or what is in the scene. The crisp ``hold cake'' and dynamic ``kiss horse'' examples best illustrate this effective trade-off (see \cref{sec:more-qualitative} for more).


\subsection{Quantitative results}
For each pair of source image and target interaction in IEBench, we generate 10 outputs using different random seeds, totalling 1000 edited images. \cref{tab:edit-perf} and \cref{fig:overall} compare our method against 23 existing baselines in terms of HOI editability and Editability-Identity scores. \textsc{InteractEdit} outperforms all others, advancing the state of HOI editing.

Our \textsc{InteractEdit} achieves an HOI Editability of \textbf{0.514}, a \textbf{31.8\%} relative gain over the previous best (NTI). On the composite \text{\textsc{ei}} score, which harmonically combines editability and identity consistency, we reach \textbf{0.573}, outperforming the next-best method (Flux.1 Kontext) by \textbf{21.7\%}. Furthermore, our framework secures the highest HPSv2 score of \textbf{0.2640}, surpassing FireFlow (0.2530).

These results demonstrate that our approach not only improves interaction accuracy but also maintains a strong trade-off between posing new actions and preserving visual identity, while also advances in perceptual quality. We attribute these gains to our SeRA-based inversion regularization, which prevents overfitting to the source image, and our \emph{Selective Inversion} strategy, which preserves target interaction priors from the pretrained model.

\input{assets/figure/edit_perf_l}

\subsection{Quantitative results on \iebenchext}\label{sec:iebenchext-results}
To verify the generalizability and the scalability beyond \iebench, we re-evaluate \textsc{InteractEdit} together with the top-performing baselines on the extended \iebenchext~benchmark (\cref{subsubsec:iebench-l}). For each $\langle$source image, target interaction$\rangle$ pair, we generate \textbf{5} samples with different random seeds, totalling $\textbf{1{,}630}$ edited images per method. \cref{tab:edit-perf-l} reports HOI Editability (\textsc{he}), Editability--Identity (\textsc{ei}), Identity Consistency, and HPSv2 on \iebenchext.

\iebenchext~is intentionally harder than \iebench~(broader object distribution, rarer HOIs, and more demanding non-rigid pose changes), providing a stricter test of HOI editing capability. Despite this, \textsc{InteractEdit} remains the top performer on the \textsc{ei} with \textbf{0.332}, \textbf{6.4\%} relative gain over the next-best InfEdit (0.312); the \textsc{he} with \textbf{0.498}, \textbf{38.7\%} relative gain over the runner-up Null-Text Inversion (0.359); and the highest HPSv2 of \textbf{0.2625} (vs.\ FireFlow's 0.2549). On Identity Consistency alone, InfEdit attains the highest score (0.831); however, this comes at the cost of editability: InfEdit's \textsc{he} of only 0.192, shows that it preserve identity largely by leaving the source pose unchanged, failing to actually realize the target interaction. 
This is precisely the trade-off \textsc{ei} is designed to capture, where the \textsc{InteractEdit} strikes the most effective balance between editability and identity preservation. 
This empirical result shows that \textsc{InteractEdit} transfers cleanly from smaller-scale \iebench~(\cref{tab:edit-perf}) to the broader, more demanding 100-image, 30-action, 21-object setting of \iebenchext.

\subsection{Ablation studies}\label{subsec:ablation}
\input{assets/figure/edit_abl}
\input{assets/figure/ab_qual_ab}

HOI editing poses significant challenges, such as overfitting to the source image and ineffective use of pretrained interaction knowledge, limiting the adaptability to target interactions. To address these, we conduct an ablation study to assess the contributions of our \emph{Selective Inversion} strategy and SeRA. The results are summarized in \cref{tab:ablation} and visualized in \cref{fig:qualitative_abl}. The baseline refers to SDXL with DreamBooth \cite{ruiz2023dreambooth} and Textual Inversion \cite{gal2022textualinversion}.

The baseline configuration greatly overfits the source image, yielding a very low \text{\textsc{ei}} (0.252) and  \text{\textsc{he}} (0.150) scores.  Simply disassembling HOI components raises \text{\textsc{ei}} to 0.327 and \text{\textsc{he}} to 0.207, showing that concept split alone is insufficient without suitable regularization. Introducing standard LoRA across all parameters boosts \text{\textsc{ei}} to 0.454 and \text{\textsc{he}} to 0.331, demonstrating that capacity control helps unlock interaction changes but still risks overwriting pretrained interaction priors. Applying SeRA, which selectively gates parameter adaptation, raises \text{\textsc{ei}} to 0.498 and \text{\textsc{he}} to 0.388, underscoring the value of constrained, direction‐wise updates that preserve pretrained priors. Adding \emph{Selective Inversion} to SeRA yields a slight uptick (\text{\textsc{ei}} = 0.508 and \text{\textsc{he}} = 0.410), confirming that freezing interaction queries enhances editability by preserving pretrained interaction knowledge. Finally, integrating disassembly with \emph{Selective Inversion} and SeRA achieves the best performance (\text{\textsc{ei}} = 0.573 and \text{\textsc{he}} =  0.514). For completeness, we also compare against AdaLoRA~\cite{zhang2023adalora}, which adaptively prunes ranks based on importance scores. As shown in \cref{tab:ablation}, AdaLoRA has good editability (\text{\textsc{he}} = 0.502) but achieves a lower \text{\textsc{ei}} (0.525) than SeRA. This is because rank pruning discards entire update directions, which can erase useful identity information, leading to less stable identity preservation. In contrast, SeRA softly regulates each rank through continuous gating, striking a more effective balance between editability and identity consistency. 

\begin{figure}
    \centering
    \includegraphics[width=0.8\linewidth]{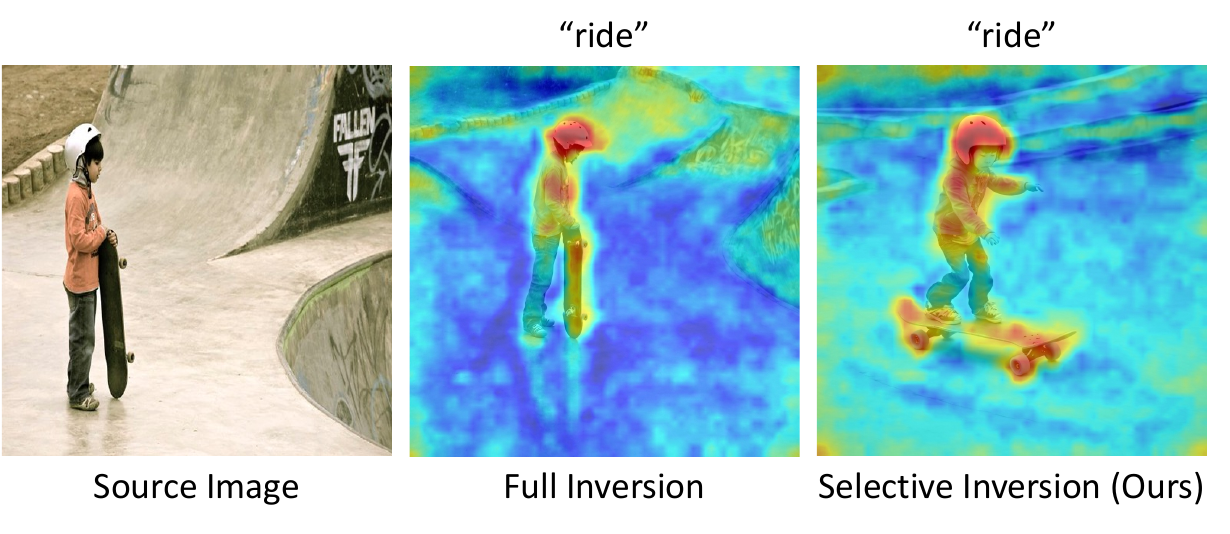}
    \caption{Cross-attention maps on the action token ``ride''. In \textbf{Full Inversion}, the model overfits to the source structure, leaving the attention map stuck on the original interaction region (holding skateboard). In our \textbf{Selective Inversion}, freezing $W_Q$ retains the pretrained ``question space,'' enabling the attention to successfully shift to the new ``riding'' semantic layout. This validates the role of $W_Q$ in preserving interaction reasoning capabilities during inversion.}
    \label{fig:attnmap}
\end{figure}

To empirically validate \textbf{Selective Inversion}, we visualize the cross-attention maps in \cref{fig:attnmap} for the target interaction token \textit{``ride''} across different inversion strategies. In the \textbf{Full Inversion} baseline, the attention map remains spatially anchored to the original \textit{``hold''} interaction region. This ``structural locking'' occurs because updating all weights causes the model to overfit to the source image's layout, thereby corrupting the pretrained interaction priors and preventing the intended transformation. In contrast, our \textbf{Selective Inversion} freezes the Query weights ($W_Q$), which preserves the model's pretrained ``question space'' and its fundamental interaction reasoning capabilities. As shown in the visualization, this constraint enables the attention map to successfully shift its focus to the new semantic structure required for the \textit{``ride''} interaction, facilitating the successful rendering of the intended edit. This empirical evidence confirms that $W_Q$ is the primary component responsible for structural retrieval and interaction logic, while updating only the Key ($W_K$) and Value ($W_V$) weights is sufficient for capturing the visual identities of the subject and object from the source image.

\begin{figure}
    \centering
    \includegraphics[width=0.8\linewidth]{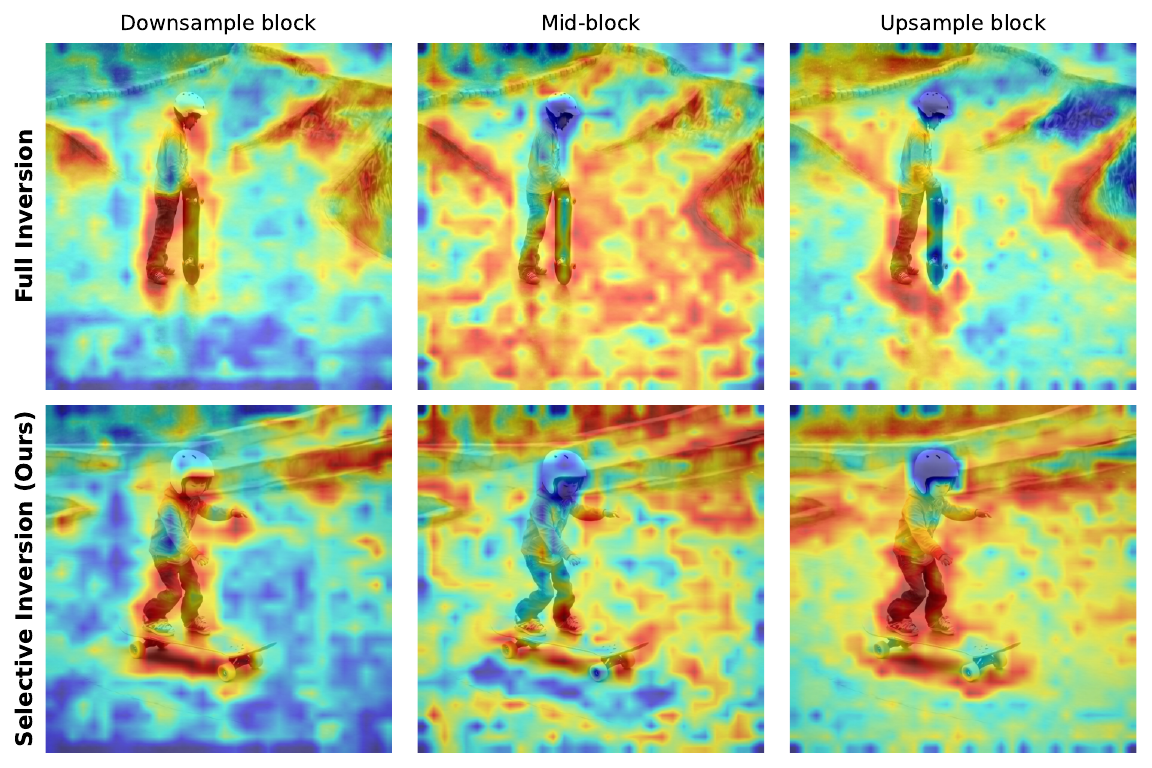}
    \caption{Cross-attention on the action token \textit{``ride''} at three U-Net depths. Under \textbf{Full Inversion} (top) the attention remains organised around the source \textit{hold} configuration, at every depth, whereas under \textbf{Selective Inversion} (bottom) it follows the target \textit{ride} configuration, covering the rider's body and the board underfoot. Warm colours denote higher attention; maps are averaged over heads and denoising timesteps.}
    \label{fig:attnmap_layers}
\end{figure}

\Cref{fig:attnmap} shows one map per method, pooled over the cross-attention layers. To confirm that the effect is not an artefact of this pooling, \cref{fig:attnmap_layers} resolves the same comparison by depth, at the downsample, mid, and upsample blocks. The pattern is consistent throughout: under Full Inversion the attention for the target action token stays organised around the source interaction layout at the downsample, mid, and upsample blocks alike, whereas under Selective Inversion it is organised around the target layout.

\input{assets/figure/qkv_ablation}

\subsubsection{Which cross-attention projection should be frozen?} \label{sec:qkv-ablation}
To verify the selective inversion derived from \cref{eq:attn-factorisation} and \cref{lem:logit_invariance} in \cref{subsec:selective-training}, we re-train \textsc{InteractEdit} under every partition of $\{W_Q, W_K, W_V\}$ into trainable and frozen projections, keeping concept disassembly and SeRA fixed. \cref{tab:qkv-ablation} reports \text{\textsc{ei}}, HOI Editability, and Identity and Background consistency on \iebench.

\textbf{(i) Training $W_Q$ and $W_K$ jointly is strictly worst.} The ``None frozen'' row drops to \text{\textsc{ei}} = 0.496 (7.7 points below ours): with both factors of $A$ trainable, the optimiser memorises the source-image attention map, exactly as predicted, and exhibits the structural locking visible in \cref{fig:attnmap}. The high Background score (0.634) reflects pixel-level source reconstruction, not interaction editing.
\textbf{(ii) Freezing $W_V$ or $W_K$ starves the inverted identity.} Freezing $W_V$ removes the only content lever in $O = AV$ and collapses Background to 0.509 --- the worst of any configuration. Freezing $W_K$ instead forces the new concept tokens to use the pretrained key projection, which has never seen these tokens, so image queries cannot selectively attend to them; \text{\textsc{ei}} settles at an intermediate 0.558. Both confirm that $W_K$ and $W_V$ are the necessary entry points for the inverted identity.
\textbf{(iii) Freezing both $W_K$ and $W_V$ over-edits.} Training only $W_Q$ attains the highest HOI Editability (0.540) by bending the routing toward generic pretrained interactions, but Identity (0.566) and Background (0.515) drop sharply: the inverted concept tokens have no path into the prompt.
\textbf{(iv) Freezing only $W_Q$ (ours) is best trade-off.} The configuration certified by \cref{lem:logit_invariance} simultaneously attains the best \text{\textsc{ei}} (0.573), Identity (0.649) and Background (0.641), while remains competitive on HOI Editability (0.514). No alternative partition dominates ours on more than a single metric.

\textbf{The main effect of $W_Q$.} Reading \cref{tab:qkv-ablation} by whether $W_Q$ is trainable isolates its contribution independently of $W_K$ and $W_V$. The four configurations that leave $W_Q$ trainable reach \text{\textsc{ei}} of 0.496, 0.528, 0.553, and 0.558 (mean 0.534), all below the 0.573 obtained by freezing it. Marginalising over what happens to $W_K$ and $W_V$, freezing $W_Q$ is therefore worth $+0.039$ \text{\textsc{ei}} on average, and no $W_Q$-trainable partition matches the frozen-$W_Q$ result even at its best.

Together with \cref{lem:logit_invariance} and the attention visualisation in \cref{fig:attnmap} and \cref{fig:attnmap_layers}, this ablation supplies the missing empirical link: the subspace preserved by freezing $W_Q$ is interaction-relevant, not merely a generic visual subspace.

\subsubsection{Sensitivity to mask quality} \label{sec:mask-sensitivity}
InteractEdit obtains its subject and object masks from an automatic SAM + GroundingDINO pipeline, raising the question of whether the method depends on the precision of this off-the-shelf segmenter. To investigate, we evaluate InteractEdit from scratch on 6 random IEBench scenes under 10 perturbed mask conditions covering two axes: (i) mask source — replacing the original mask with its tight bounding box, simulating the weakest practical prior a user could supply; and (ii) controlled geometric degradation of the SAM mask via dilation, erosion, or translation at ±5/15/30 px.
Only the subject and object masks are perturbed; the background mask is recomputed as the complement to remain consistent. Across all 10 perturbed conditions, the perturbed mean stays within approximately one standard deviation of the original baseline on every metric: HOI Editability (0.579 → 0.595 ± 0.070), Identity Consistency (0.542 → 0.520 ± 0.021), and EI (0.560 → 0.553 ± 0.036) (\cref{tab:mask-sensitivity}). We conclude that InteractEdit is \textbf{insensitive to mask quality} and that the masks act as attention localizers, not pixel-level constraints, so a bounding box from an off-the-shelf detector is a sufficient prior.

\begin{table}[ht]
    \caption{\textbf{Sensitivity of InteractEdit to mask quality.} InteractEdit retrained under 10 perturbed mask conditions (bbox-replacement and dilation/erosion/translation at ±5/15/30 px). The last row shows the grouped means. Performance stays within $\approx1\sigma$ of the original baseline across all three metrics.}
    \label{tab:mask-sensitivity}
    \begin{tabular}{l|c|rrr}
    \hline
    \textbf{Mask condition}    & \textbf{mask IoU} & \textbf{Editability-Identity (EI)} & \textbf{HOI Editability} & \textbf{Identity Consistency} \\ \hline
    Original (baseline)        & 1.00              & 0.560                                & 0.579                      & 0.542                           \\
    Tight bounding box         & 0.50              & 0.573                                & 0.671                      & 0.500                           \\
    Dilation +5 px             & 0.83              & 0.570                                & 0.618                      & 0.528                           \\
    Dilation +15 px            & 0.64              & 0.511                                & 0.513                      & 0.509                           \\
    Dilation +30 px            & 0.49              & 0.538                                & 0.605                      & 0.484                           \\
    Erosion -5 px              & 0.77              & 0.588                                & 0.645                      & 0.540                           \\
    Erosion -15 px             & 0.56              & 0.517                                & 0.526                      & 0.509                           \\
    Erosion -30 px             & 0.38              & 0.627                                & 0.724                      & 0.553                           \\
    Translation 5 px           & 0.65              & 0.540                                & 0.579                      & 0.507                           \\
    Translation 15 px          & 0.57              & 0.577                                & 0.632                      & 0.534                           \\
    Translation 30 px          & 0.68              & 0.536                                & 0.548                      & 0.524                           \\ \hline
    All perturbed (mean ± std) & 0.62              & 0.553 ± 0.036                        & 0.595 ± 0.070              & 0.520 ± 0.021                   \\ \hline
    \end{tabular}
\end{table}

\input{assets/figure/rank-reg-sweep}
\subsubsection{Sensitivity to rank and SeRA regularization strength} \label{sec:rank-reg-sweep}
We analyse the sensitivity of \textsc{InteractEdit} to the two hyperparameters introduced in \cref{subsec:lora}: the rank $r$ of the adapter (\cref{eq:lora}) and the SeRA uncertainty-loss weight $\lambda_{\textrm{SeRA}}$ (\cref{eq:l_sera}). For the rank, we sweep $r \in \{4, 8, 16, 32, 64\}$ for \emph{both} vanilla LoRA and SeRA to compare how the ungated and gated adapters scale with capacity; for the regularization, we sweep $\lambda_{\textrm{SeRA}} \in \{0.001, 0.01, 0.1, 1.0\}$ on SeRA. All variants are evaluated on \iebench~setting.

\textbf{Rank.} The SeRA-vs-LoRA comparison is governed by a single mechanism: as $r$ in \cref{eq:lora} grows, vanilla LoRA spends surplus capacity on memorising source-image structure, while SeRA's gating $G$ closes the surplus directions. At $r=4$, four rank directions are too few for either adapter to fit the source: LoRA shows slightly higher HE (0.537 vs.\ SeRA 0.508, \cref{fig:rank_reg_sweep_hoi}), but Identity is low for both (SeRA 0.585, LoRA 0.599, \cref{fig:rank_reg_sweep_id}), so neither holds an advantage. From $r=8$ onward the two methods diverge as predicted by \cref{subsec:lora}: vanilla LoRA's HE drops steadily (0.498 $\rightarrow$ 0.461, \cref{fig:rank_reg_sweep_hoi}) and its \text{\textsc{ei}} peaks at $r=32$ before degrading at $r=64$ (\textbf{0.555 $\rightarrow$ 0.541}, \cref{fig:rank_reg_sweep_f1}), while SeRA's gating absorbs the surplus, keeping HE flat ($\approx 0.51$, \cref{fig:rank_reg_sweep_hoi}) and growing \text{\textsc{ei}} monotonically (\textbf{0.551 $\rightarrow$ 0.571}, \cref{fig:rank_reg_sweep_f1}). Identity Consistency tracks closely across all ranks (\cref{fig:rank_reg_sweep_id}), so the SeRA-vs-LoRA gap is purely an editability gap.

\textbf{Regularization.} All three metrics peak at the default $\lambda_{\textrm{SeRA}} = 0.01$ (\text{\textsc{ei}} 0.571, HE 0.514, Identity 0.641) in \cref{fig:rank_reg_sweep_reg}, and both weakening ($\lambda_{\textrm{SeRA}} = 0.001$) and strengthening ($\lambda_{\textrm{SeRA}} = 0.1$) the uncertainty loss degrade \text{\textsc{ei}} and HE. The default $\lambda_{\textrm{SeRA}} = 0.01$ is therefore the optimal for all three metrics.

\subsection{Per-category failure analysis}\label{sec:failure-analysis}
\input{assets/figure/per-action}
\input{assets/figure/per-object}
To identify \emph{where} \textsc{InteractEdit} fails rather than only how often, we disaggregate the \iebenchext~results by target verb (\cref{tab:he-per-action}) and by object (\cref{tab:he-per-object}).

\textbf{(i) Verbs.} \text{\textsc{he}} varies widely across target verbs. The weakest are \textit{groom} (0.000), \textit{open} (0.000), \textit{fill} (0.100), \textit{make} (0.114), and \textit{stand on} (0.133), while the strongest are \textit{throw} (0.980), \textit{ride} (0.975), \textit{hug} (0.950), and \textit{dribble} (0.900).

\textbf{(ii) Objects.} The weakest objects are small, hand-held, or deformable: \textit{bottle} (0.167), \textit{cup} (0.200), \textit{banana} (0.240), and \textit{cake} (0.296). The strongest are larger and rigid: \textit{bench} (0.783), \textit{snowboard} (0.750), \textit{sports ball} (0.743), and \textit{skateboard} (0.733). This is consistent with the small-object failure mode described in \cref{subsec:failure-limitations}.

\subsection{User study}\label{subsec:user-study}
\begin{figure} 
    \centering
    \includegraphics[width=0.8\linewidth]{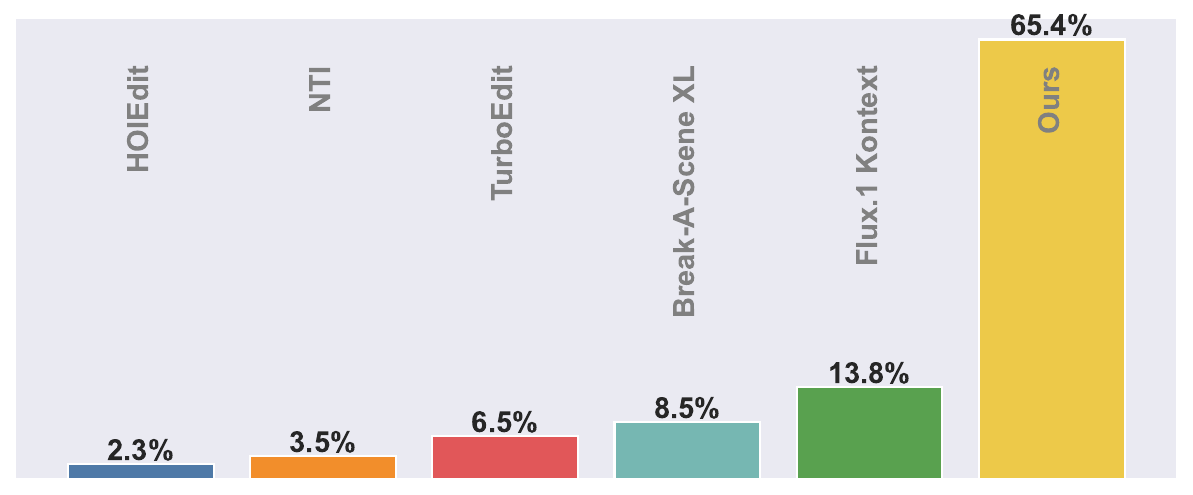}
    \caption{User preference study results. A total of 480 trials were conducted with 32 participants to assess HOI editing quality. Our method, InteractEdit, was preferred in 65.4\% of cases, significantly outperforming the next-best method, Flux.1 Kontext (13.8\%).}
    \label{fig:userstudy-overall}
\end{figure}

To assess human preference, we conducted a user study comprising a total of 480 individual trials, in which participants were asked to select the best result from six competing methods across 15 unique interaction edits, involving 32 participants in total. To ensure fairness, outputs were randomly shuffled before presentation. 


The study included NTI, TurboEdit, and Flux.1 Kontext as the strongest pre-existing interaction editing methods, HOIEdit as an HOI-specific baseline, and Break-A-Scene XL as a disentanglement-based editing baseline. Our method (\textsc{InteractEdit}) was evaluated alongside them.

As shown in \cref{fig:userstudy-overall}, 65.4\% of participants preferred \textsc{InteractEdit}. Flux.1 Kontext received 13.8\%, Break-A-Scene XL 8.5\%, TurboEdit 6.5\%, NTI 3.5\%, and HOIEdit only 2.3\%. The strong preference for our method highlights its ability to balance realism with effective interaction editing, preserving identity while enabling meaningful transformations. More information is detailed in \cref{sec:userstudy-details}.

\begin{figure}
    \centering
    \includegraphics[width=0.8\linewidth]{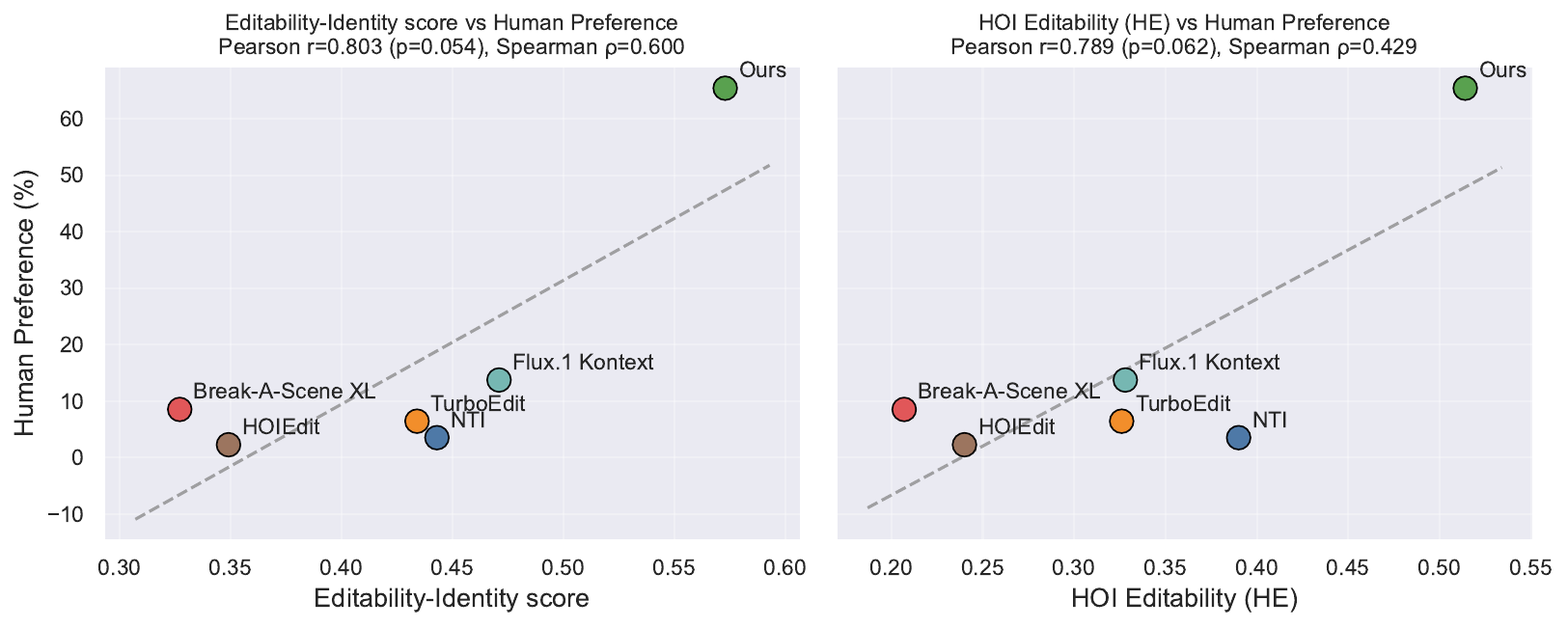}
    \caption{Correlation between automated metrics and human preference. Both HOI Editability (HE) and Editability-Identity (EI) scores show strong positive correlation with human preference rates, with EI exhibiting a slightly stronger relationship (Pearson r = 0.803, p = 0.054) than HE (r = 0.789, p = 0.062). This validates the reliability of our proposed metrics as proxies for human judgment in evaluating HOI editing quality.}
    \label{fig:metric_correlation}
\end{figure}

To verify that our automated metrics: HOI Editability (HE) and the joint Editability–Identity score (EI), are reliable proxies for human perception, we measure their agreement with the user study presented in \cref{sec:userstudy-details}. For each of the six methods, we take the metric score reported in \cref{tab:edit-perf} and the overall human preference rate (\%) in the user study. \cref{fig:metric_correlation} plots HE and EI against the human preference rate, together with a linear regression fit. Both metrics show a strong positive correlation with human judgment, with Pearson r = 0.803 (p = 0.054) for EI and r = 0.789 (p = 0.062) for HE; rank correlation follows the same trend (Spearman $\rho$ = 0.600 for EI and 0.429 for HE). EI consistently aligns more closely with human preference than HE, which is expected by design: HE evaluates only whether the target interaction is realized, while EI jointly captures editing success and identity preservation — the two qualities users implicitly trade off when comparing edited images. The strong, monotonic relationship between EI and human preference confirms that improvements in EI translate into perceivable improvements for human observers, and supports our use of EI as the primary quantitative metric throughout the paper.

%% file: assets/figure/obj-act-pairs.tex
\begin{table}[ht]
    \centering
    \caption{Objects and their candidate target interactions in \iebench~and \iebenchext. The original \iebench~objects are listed in the top block; \iebenchext~extends this set with the additional objects and actions.}
    \label{tab:object_interaction_pairs}
    \begin{tabular}{|l|l|l|}
        \hline
        \textbf{Subset} & \textbf{Object} & \textbf{Target Interactions} \\ \hline
        \multirow{13}{*}{\iebench} &Broccoli & cut, eat, hold, smell, wash \\
        & Dining table & clean, eat at, sit at \\
        & Skateboard & hold, jump, ride, sit on \\
        & Chair & hold, lie on, sit on, stand on \\
        & Book & carry, hold, read \\
        & Snowboard & hold, jump, ride \\
        & Surfboard & hold, jump, ride, sit on \\
        & Sports ball & catch, dribble, hit, hold, kick, throw \\
        & Cake & blow, cut, eat, hold, make \\
        & Horse & feed, kiss, ride, walk, wash \\
        & Dog & feed, groom, hug, walk, wash \\
        & Pizza & cut, eat, hold, make, pick up \\
        & Cat & feed, hold, hug, kiss, wash \\ \hline        
        \multirow{7}{*}{\iebenchext} & Bench & sit on, lie on \\
        & Cup & hold, drink with, sip, fill, smell \\
        & Banana & hold, eat, cut \\
        & Bottle & drink with, open, hold \\
        & Couch & sit on, lie on \\
        & Teddy bear & hold, hug, carry \\
        & Donut & hold, eat, pick up \\
        & Sandwich & hold, eat \\ \hline
    \end{tabular}
\end{table}

%% file: assets/figure/clip_score.tex
\begin{figure}
\centering
\begin{minipage}[t]{0.62\linewidth}
\vspace{0pt}
\centering
\setlength{\tabcolsep}{1pt}
\renewcommand{\arraystretch}{1}
\resizebox{\linewidth}{!}{
  \begin{tabular}{cccc}
    \multicolumn{4}{c}{\footnotesize CLIP similarity with ``person ride horse''}\\
    \footnotesize 0.2020 & \footnotesize 0.2419 & 
    \footnotesize 0.2414 & \footnotesize 0.2415 \\
    \includegraphics[width=.21\linewidth]{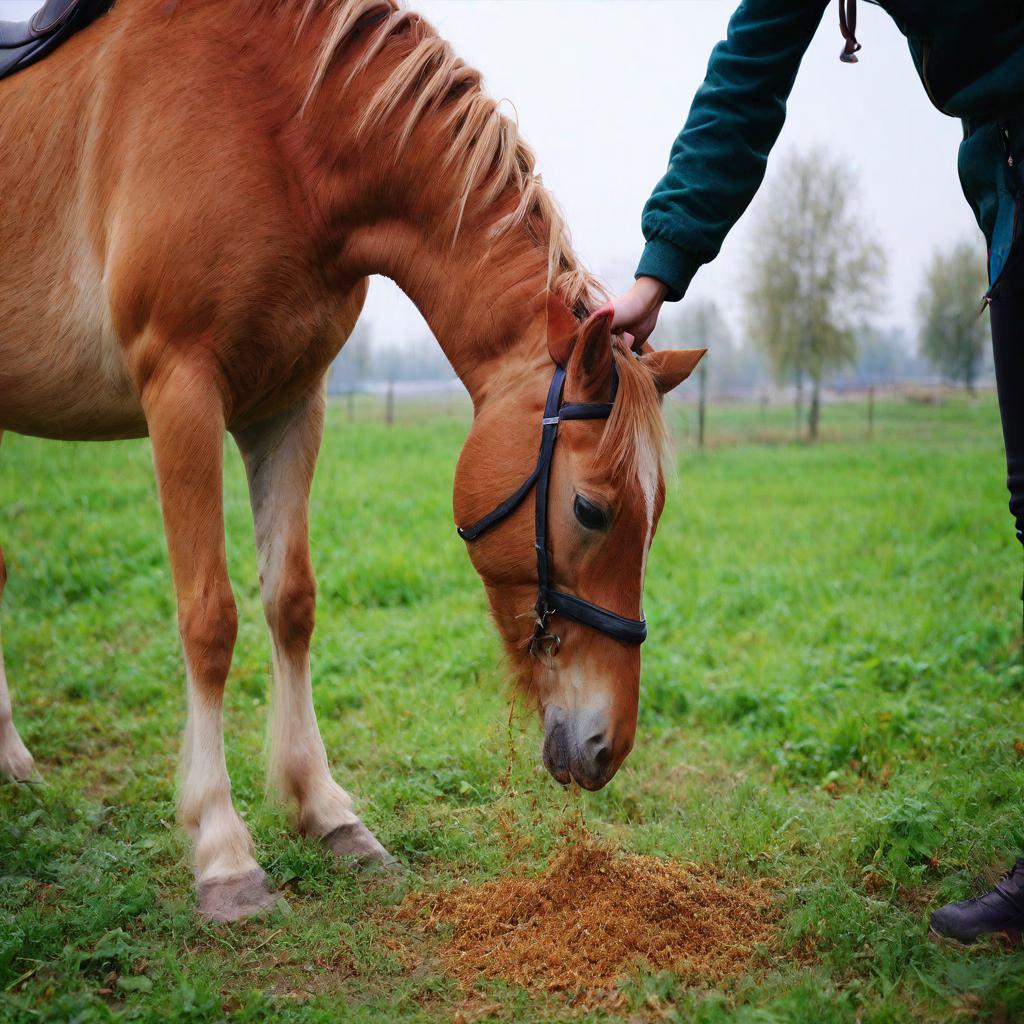} &
    \includegraphics[width=.21\linewidth]{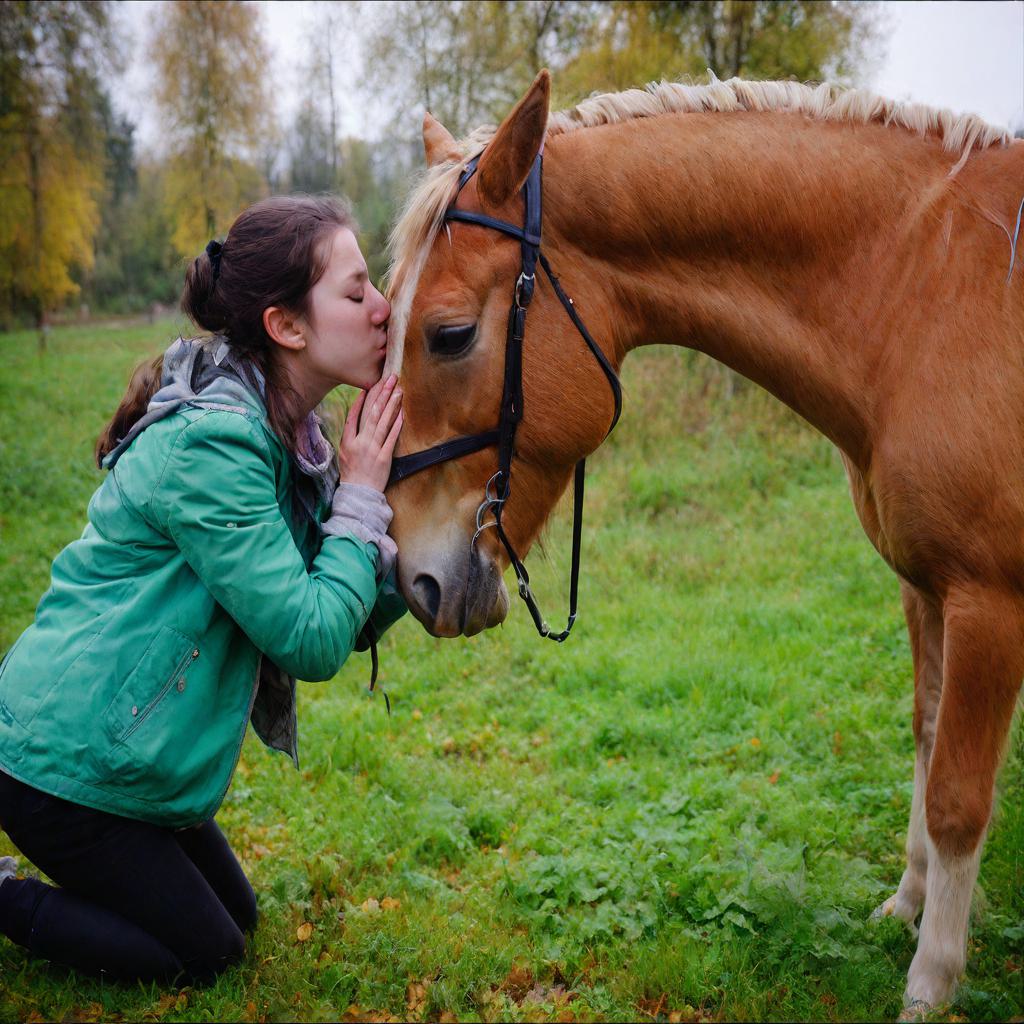} &
    \includegraphics[width=.21\linewidth]{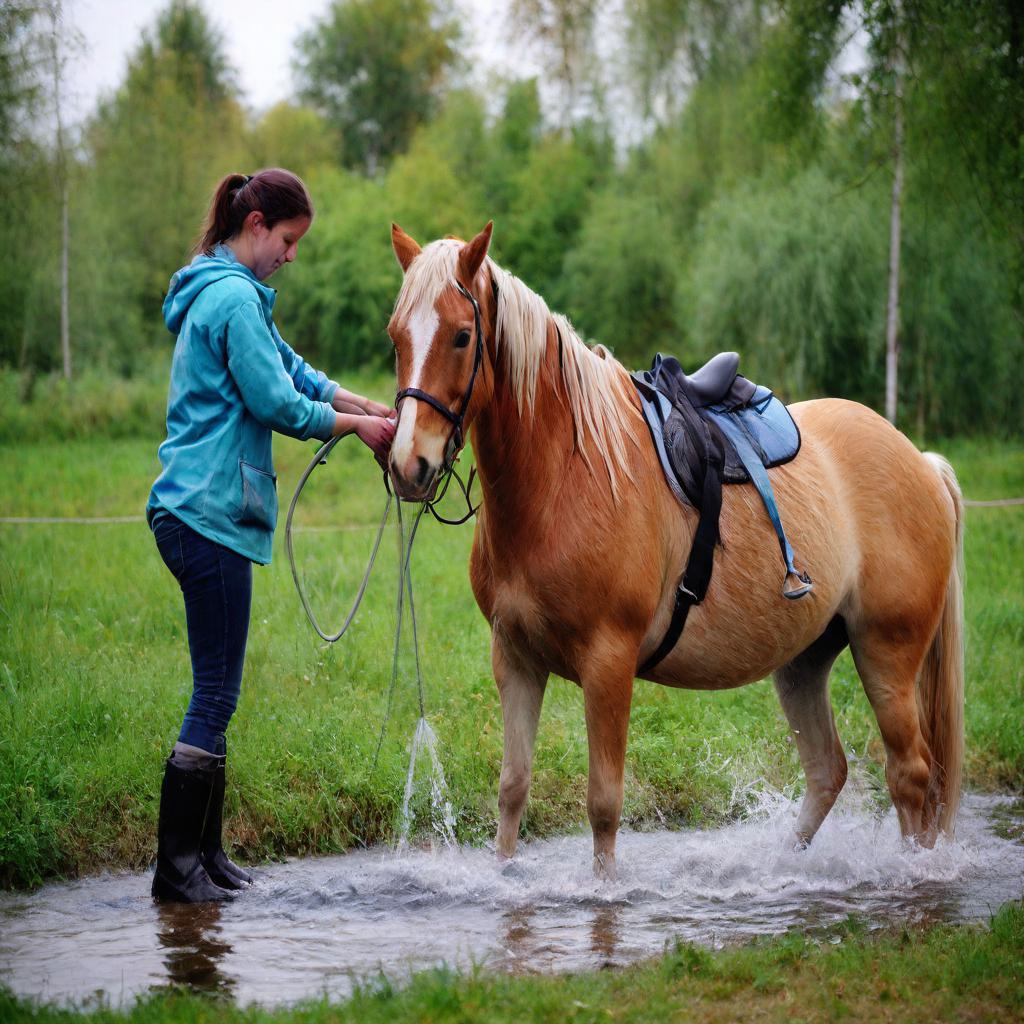} &
    \includegraphics[width=.21\linewidth]{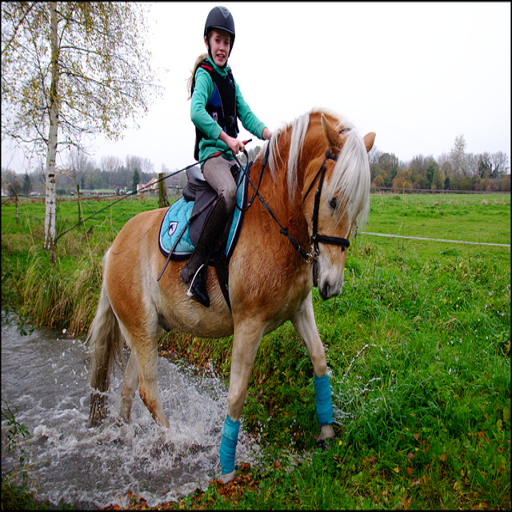}
  \end{tabular}
}
\end{minipage}%
\hfill
\begin{minipage}[t]{0.37\linewidth}
\vspace{0pt}
\centering
\includegraphics[width=\linewidth]{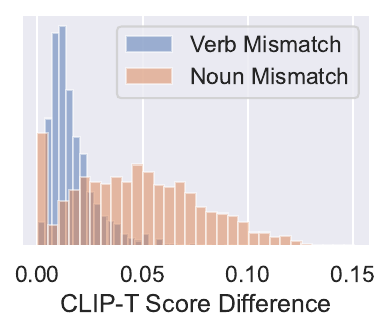}
\end{minipage}
\caption{CLIP-T score fails to distinguish verb-specific interaction differences.}
\label{fig:clip_score}
\end{figure}

%% file: assets/figure/edit_perf.tex

\begin{table}[t!]
  \caption{Comparison of {\rmfamily\scshape InteractEdit} with existing baselines on the Editability-Identity Score \text{{\rmfamily\scshape ei}}, HOI Editability and Identity Consistency using \iebench. We use HPSv2 \cite{wu2023hpsv2} as a human-preference aligned perceptual metric to evaluate fine-grained realism. The \textbf{best} and \underline{second best} are indicated.}
  \label{tab:edit-perf}
  \centering
  \begin{tabular}{l||c|c|c|c}
    \hline \hline
    Method              & \text{{\rmfamily\scshape ei}}              & HOI Editability      & Identity Consistency  & HPSv2\\ \hline \hline
    Null-Text Inversion \cite{mokady2023nti}        & 0.443          & \underline{0.390}    & 0.513                      & 0.2483           \\
    PnP Diffusion \cite{ju2024pnpinversion}         & 0.365          & 0.240                & 0.759                      & 0.2520           \\
    MasaCtrl \cite{cao2023masactrl}                 & 0.371          & 0.260                & 0.648                      & 0.2212          \\
    HOIEdit \cite{xu2025hoiedit}                    & 0.349          & 0.240                & 0.641                      & 0.2129          \\
    Imagic \cite{kawar2023imagic}                   & 0.309          & 0.192                & 0.792                      & 0.2417         \\
    CDS \cite{Nam2024cds}                           & 0.314          & 0.194                & \underline{0.820}                      & 0.2303          \\
    DAC \cite{song2024dac}                          & 0.263          & 0.160                & 0.736                      & 0.2135         \\
    LEDITS++ \cite{brack2024ledits++}               & 0.239          & 0.150                & 0.583                      & 0.2043         \\
    InstructPix2Pix \cite{brooks2022instructpix2pix}& 0.380          & 0.269                & 0.648                      & 0.2178        \\
    ReNoise \cite{garibi2024renoise}                & 0.388          & 0.274                & 0.663                      & 0.2312         \\
    TurboEdit \cite{deutch2024turboedit}            & 0.434          & 0.326                & 0.649                      & 0.2437         \\
    SVDiff \cite{han2023svdiff}                     & 0.377          & 0.255                & 0.723                      & 0.2351         \\
    EditFriendlyDDPM \cite{huberman2024editfriendly}& 0.438          & 0.320                & 0.694                      & 0.2470         \\
    DDS \cite{hertz2023dds}                         & 0.349          & 0.226                & 0.767                      & 0.2270        \\
    InfEdit \cite{xu2023infedit}                    & 0.295          & 0.179                & \textbf{0.838}                      & 0.2296       \\
    Break-A-Scene  \cite{avrahami2023breakascene}   & 0.240          & 0.142                & 0.768                      & 0.2296         \\
    Break-A-Scene XL                                & 0.327          & 0.207                & 0.779                      & 0.2367     \\
    Mix-of-Show \cite{gu2024mixofshow}              & 0.404          & 0.293                & 0.652                      & 0.2505      \\
    DisenDiff \cite{zhang2024disendiff}             & 0.297          & 0.189                & 0.689                      & 0.2234       \\
    OmniGen \cite{xiao2025omnigen}                  & 0.354          & 0.231                & 0.759                      & 0.2120         \\
    HiDream-E1 \cite{cai2025hidream}                & 0.320          & 0.203                & 0.753                      & 0.2297       \\
    Flux.1 Kontext \cite{labs2025flux1kontext}      & \underline{0.471} & 0.328             & \textbf{0.838}                      & 0.2427      \\
    FireFlow \cite{deng2024fireflow}                & 0.451          & 0.350                & 0.636                      & \underline{0.2530}         \\
    \hline
    \hline
    \textbf{InteractEdit (Ours)}                    & \textbf{0.573} & \textbf{0.514}       & 0.649                      & \textbf{0.2640}\\ \hline \hline
  \end{tabular}%
\end{table}


%% file: assets/figure/qualitative.tex
\begin{figure*}
    \centering
    \setlength{\tabcolsep}{1pt} 
    \renewcommand{\arraystretch}{1} 
    \begin{tabular}{ccccccccc}
        \footnotesize Source & \footnotesize NTI & \scriptsize InstructPix2Pix & \footnotesize InfEdit & \footnotesize OmniGen & \footnotesize Flux.1 Kontext & \footnotesize TurboEdit & \footnotesize Break-A-Scene & \footnotesize Ours\\
          \includegraphics[width=.104\linewidth]{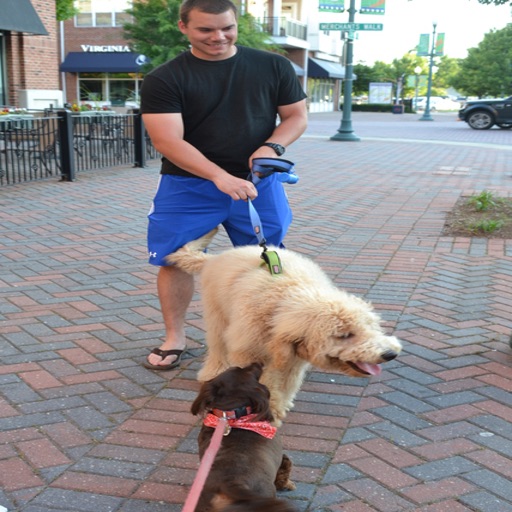}
        & \includegraphics[width=.104\linewidth]{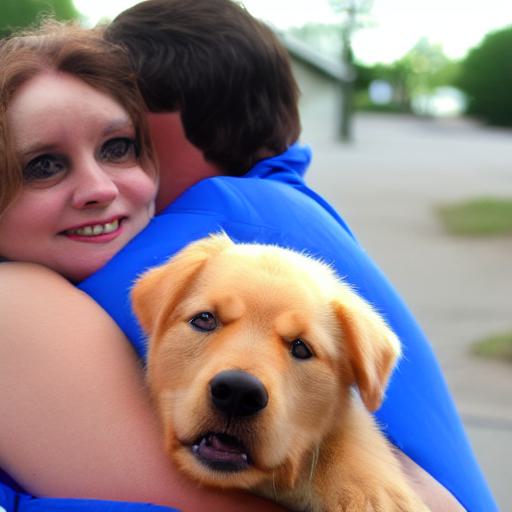}
        & \includegraphics[width=.104\linewidth]{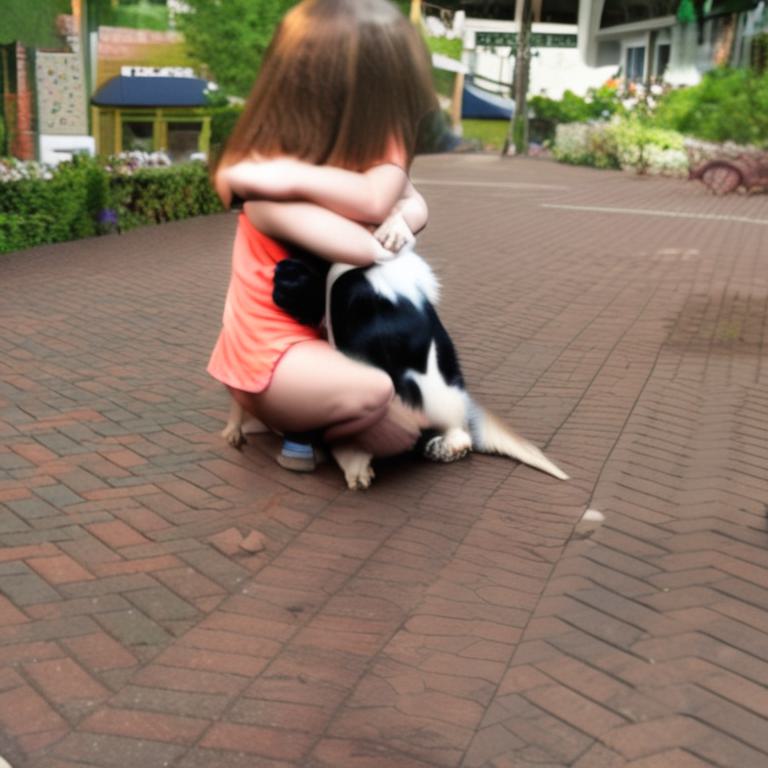}
        & \includegraphics[width=.104\linewidth]{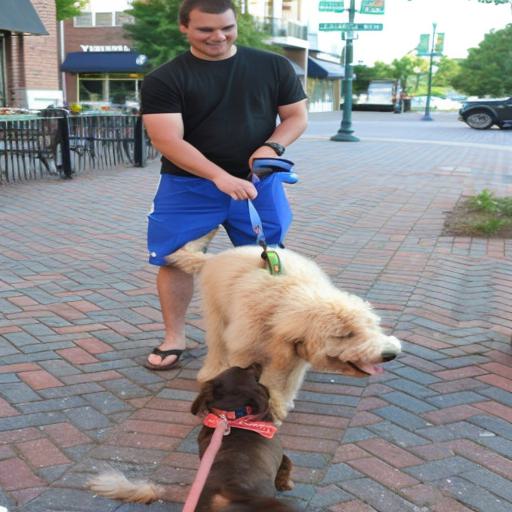}
        & \includegraphics[width=.104\linewidth]{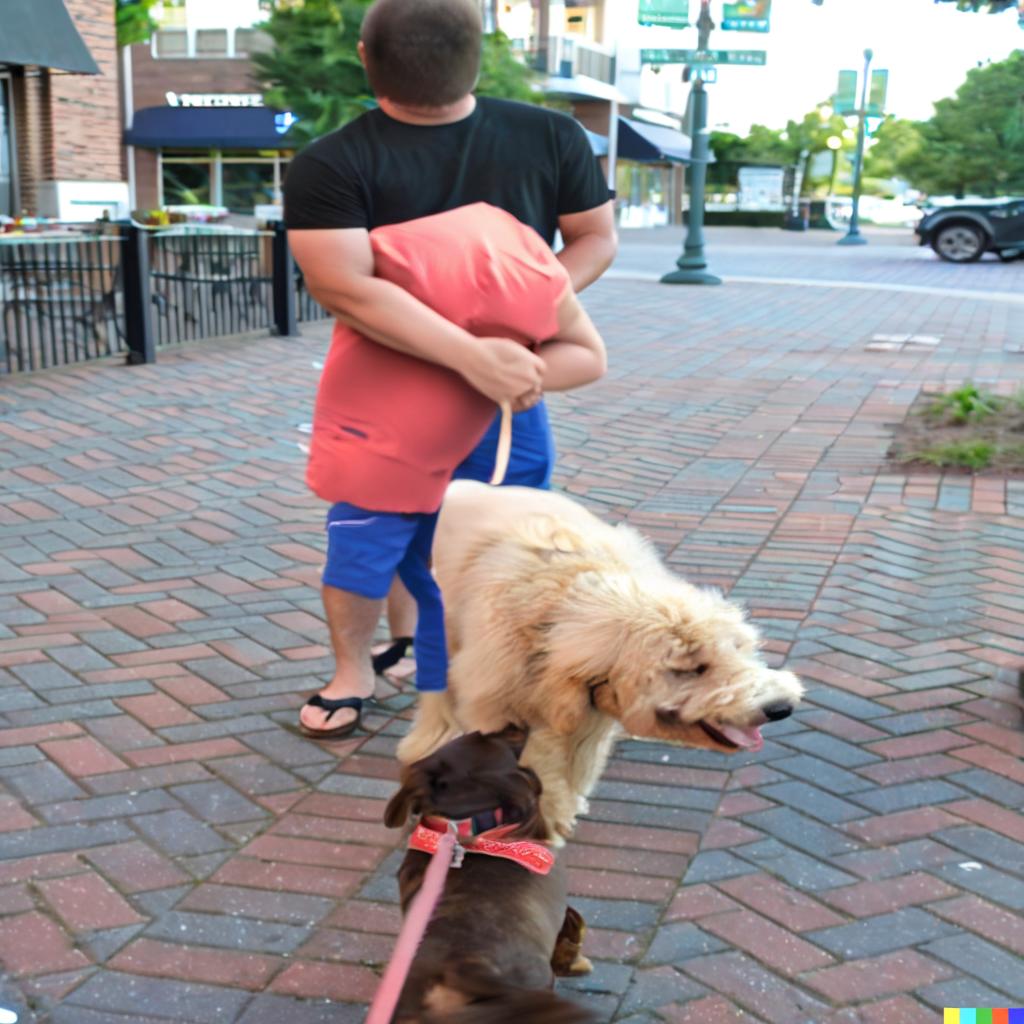}
        & \includegraphics[width=.104\linewidth]{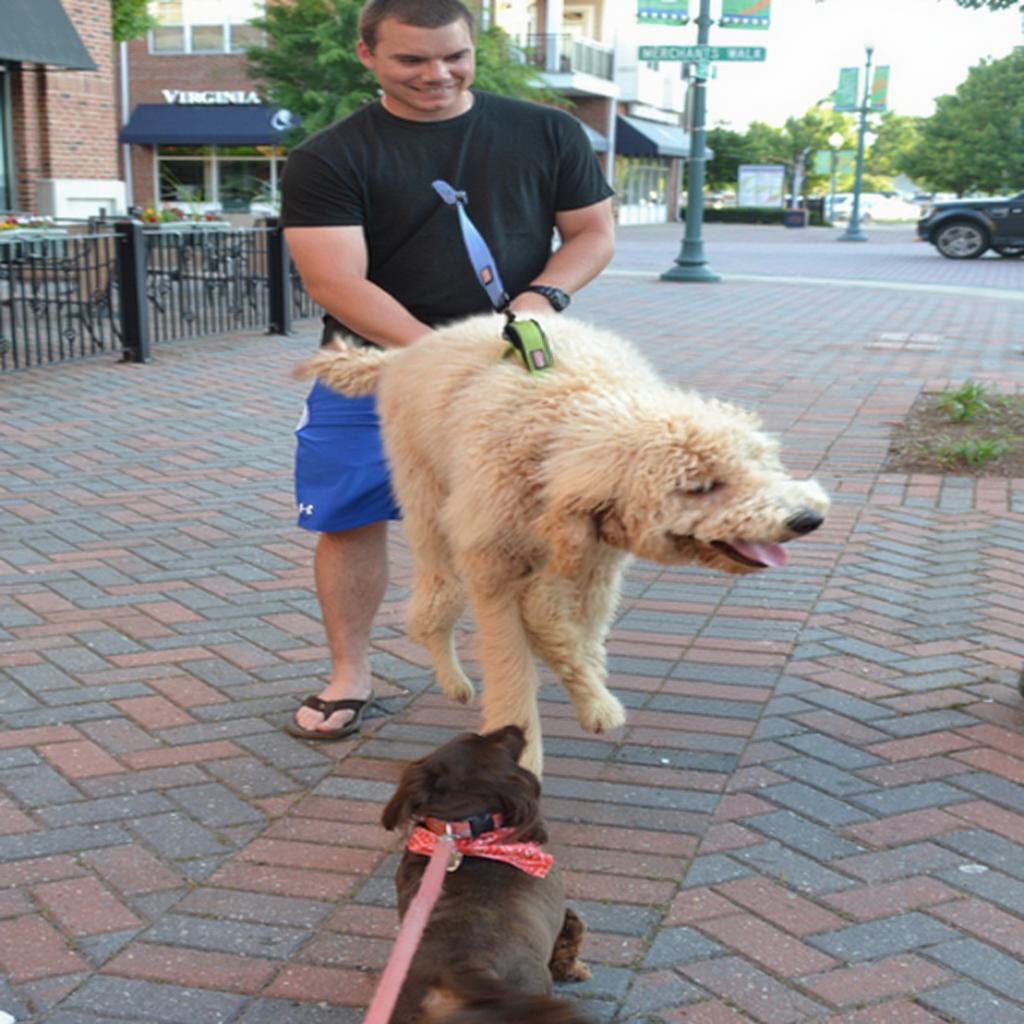}
        & \includegraphics[width=.104\linewidth]{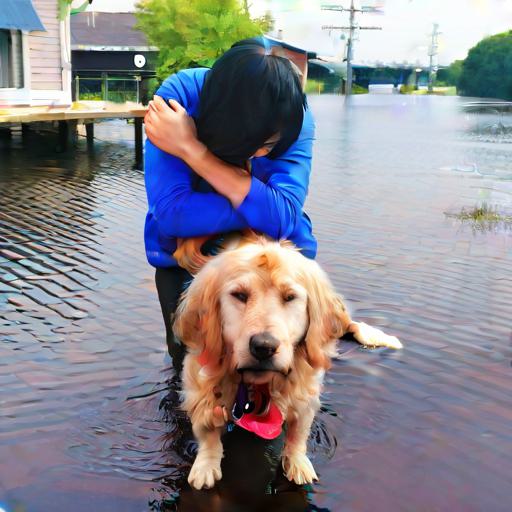}
        & \includegraphics[width=.104\linewidth]{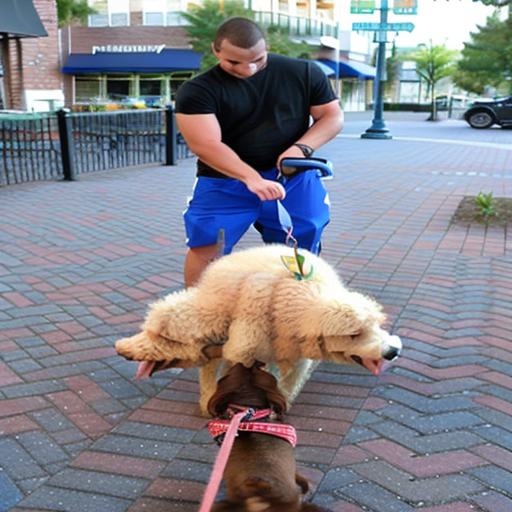}
        & \includegraphics[width=.104\linewidth]{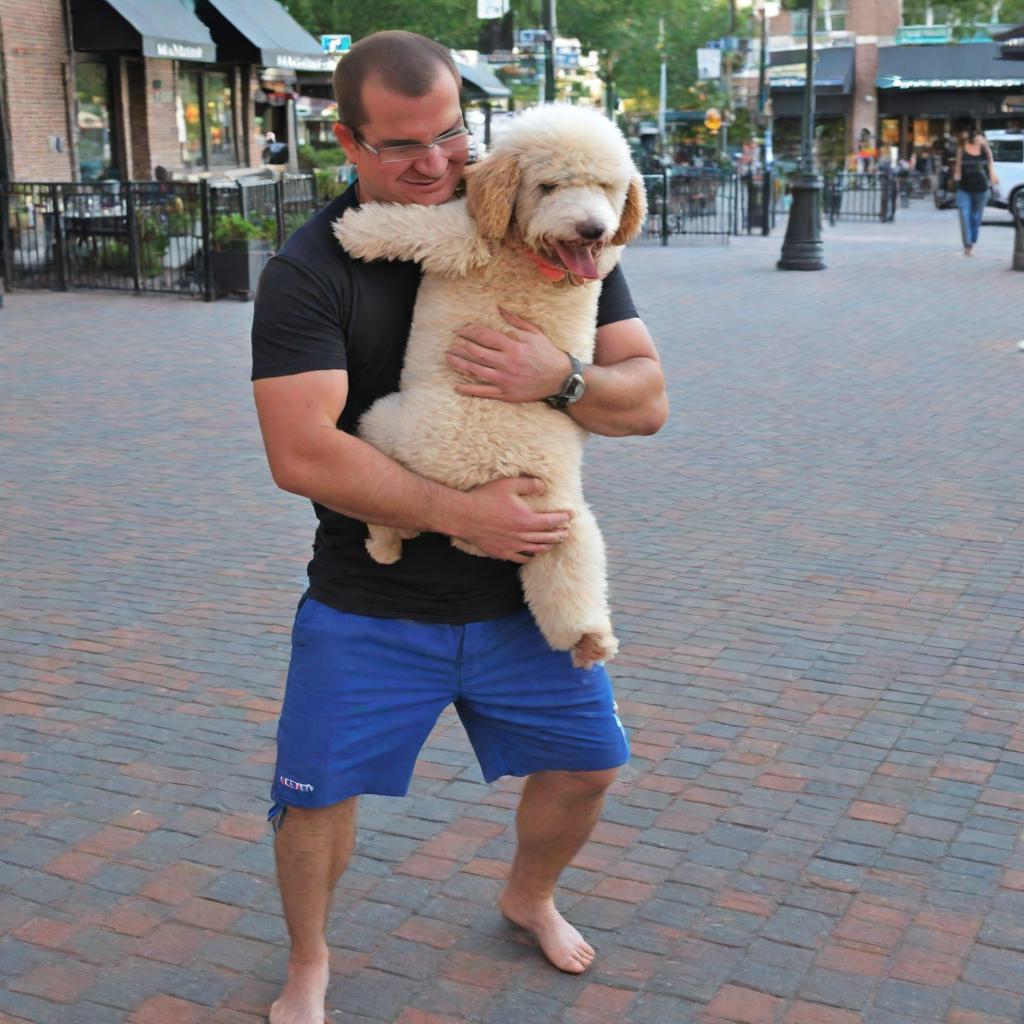}\\
        \\[-1.2em]
        walk dog & \multicolumn{8}{c}{hug dog}\\
          \includegraphics[width=.104\linewidth]{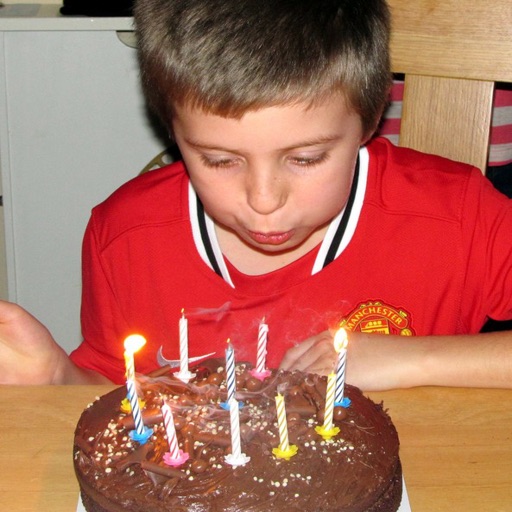}
        & \includegraphics[width=.104\linewidth]{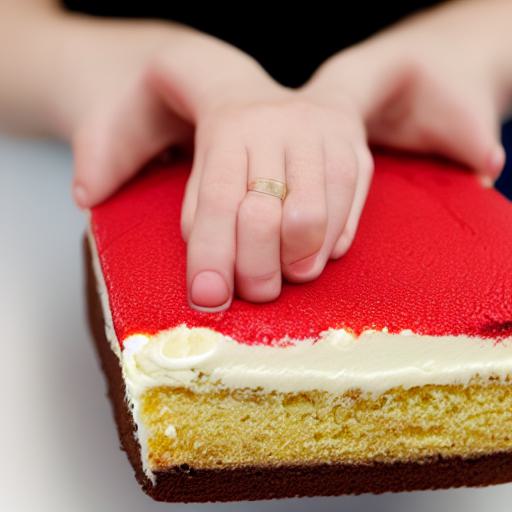}
        & \includegraphics[width=.104\linewidth]{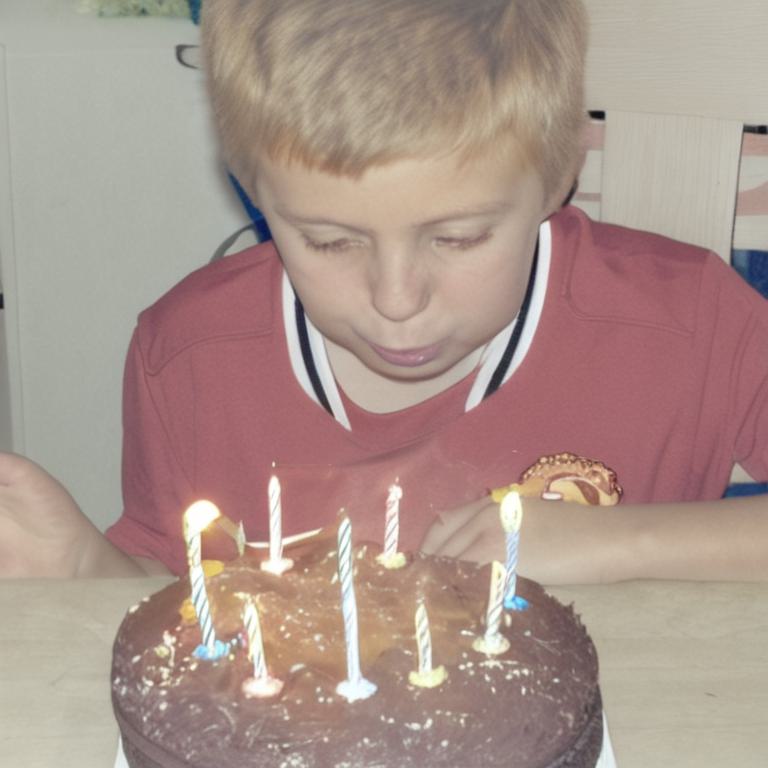}
        & \includegraphics[width=.104\linewidth]{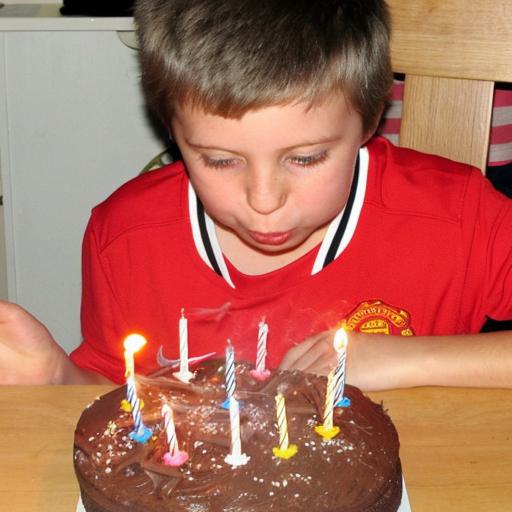}
        & \includegraphics[width=.104\linewidth]{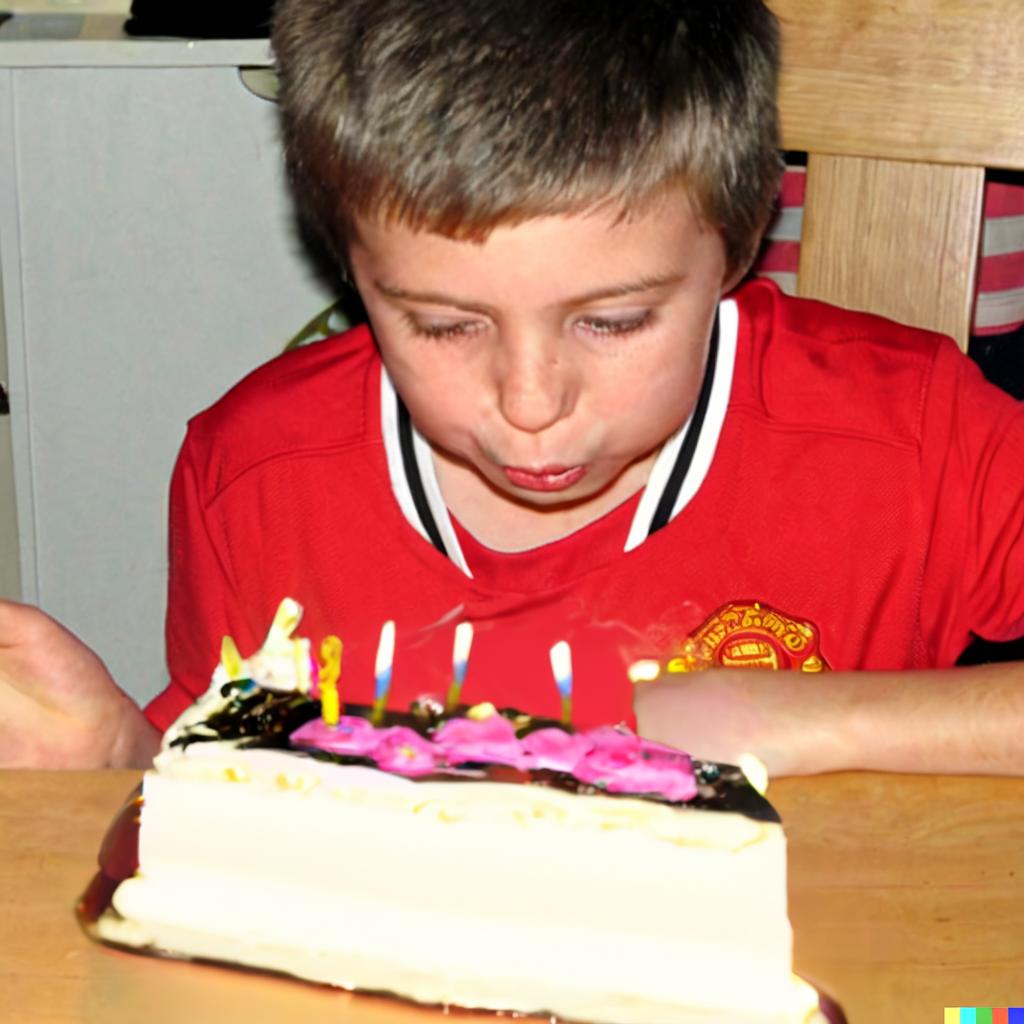}
        & \includegraphics[width=.104\linewidth]{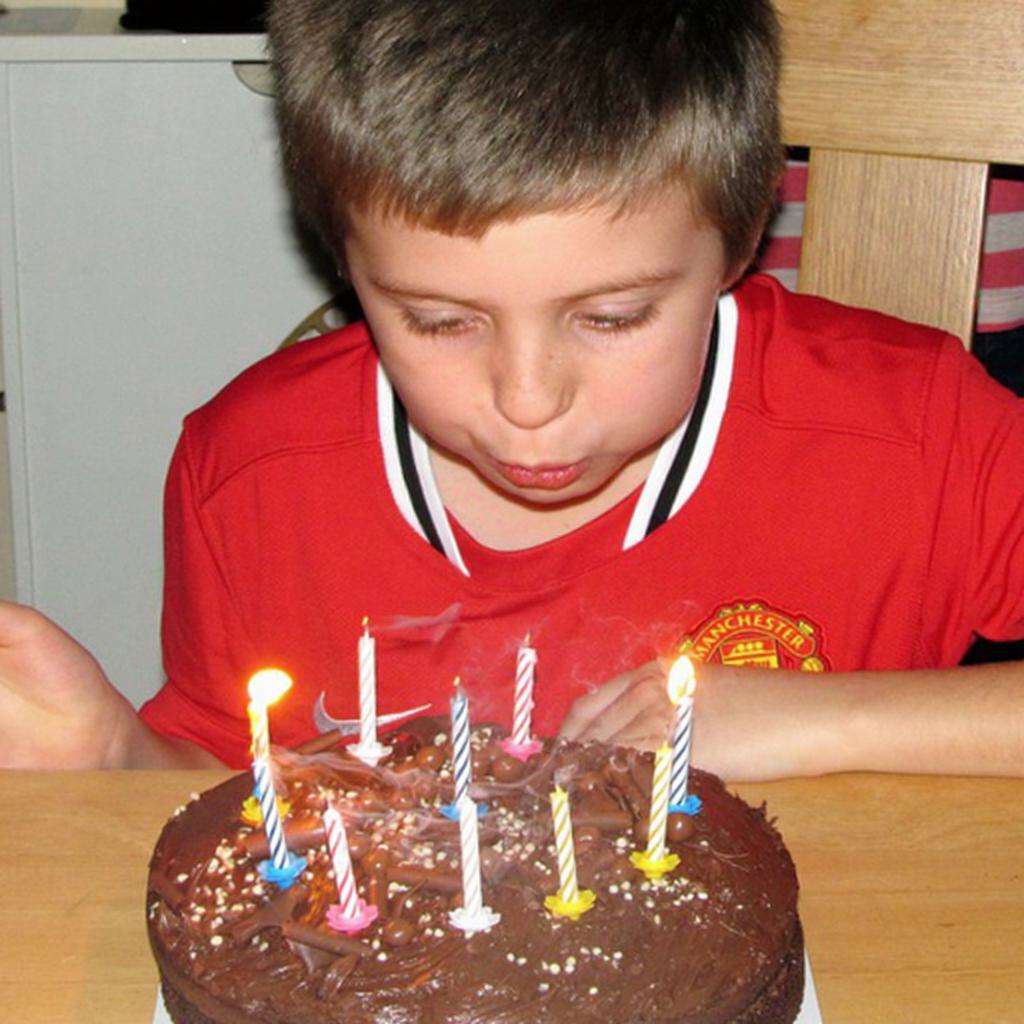}
        & \includegraphics[width=.104\linewidth]{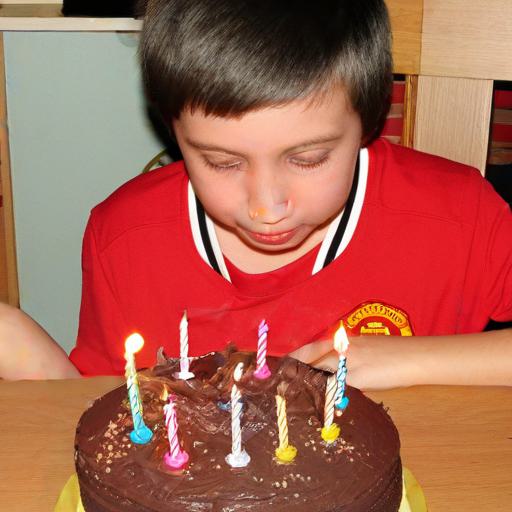}
        & \includegraphics[width=.104\linewidth]{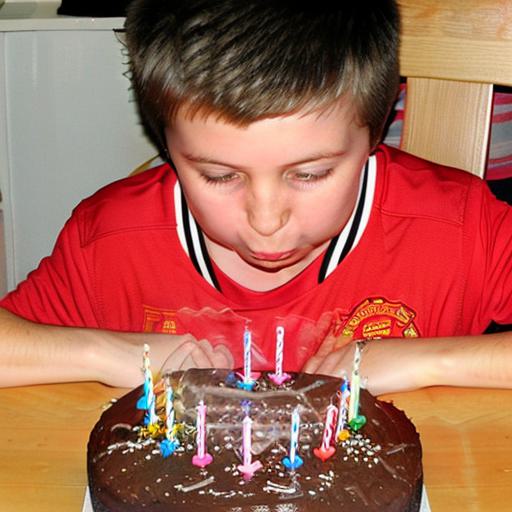}
        & \includegraphics[width=.104\linewidth]{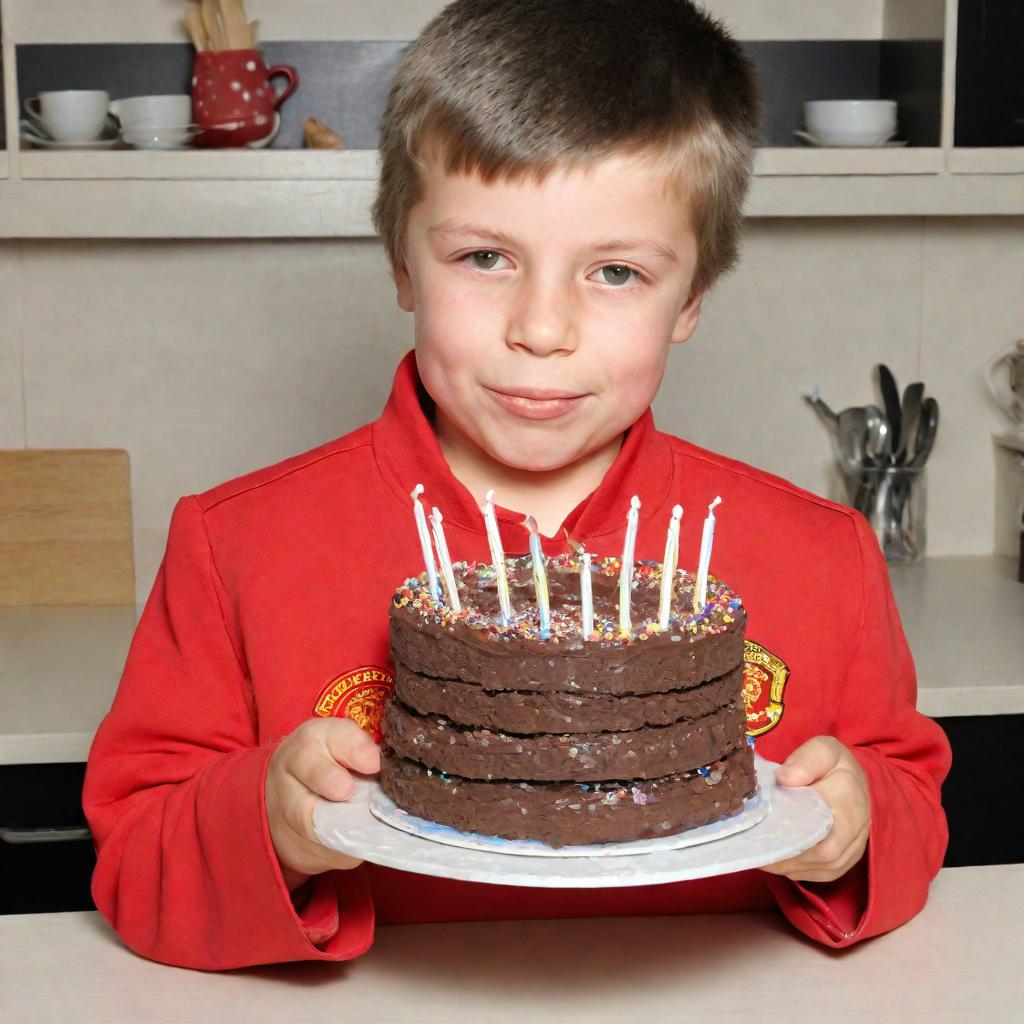}\\
        \\[-1.2em]
        blow cake & \multicolumn{8}{c}{hold cake}\\
          \includegraphics[width=.104\linewidth]{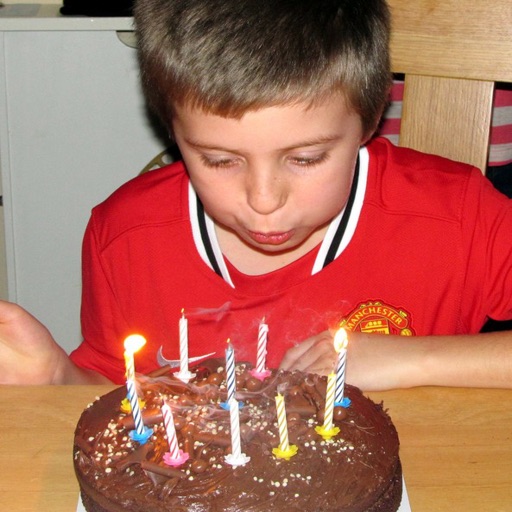}
        & \includegraphics[width=.104\linewidth]{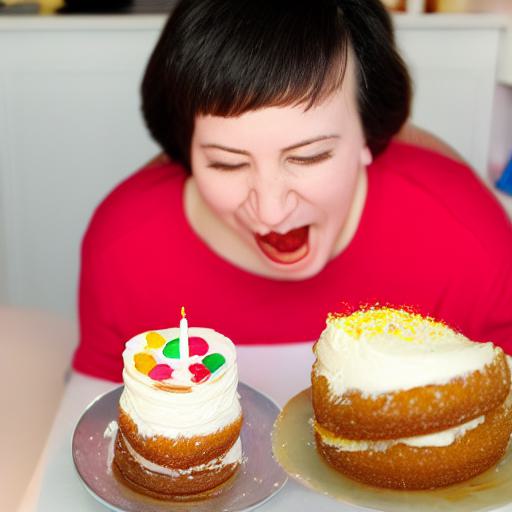}
        & \includegraphics[width=.104\linewidth]{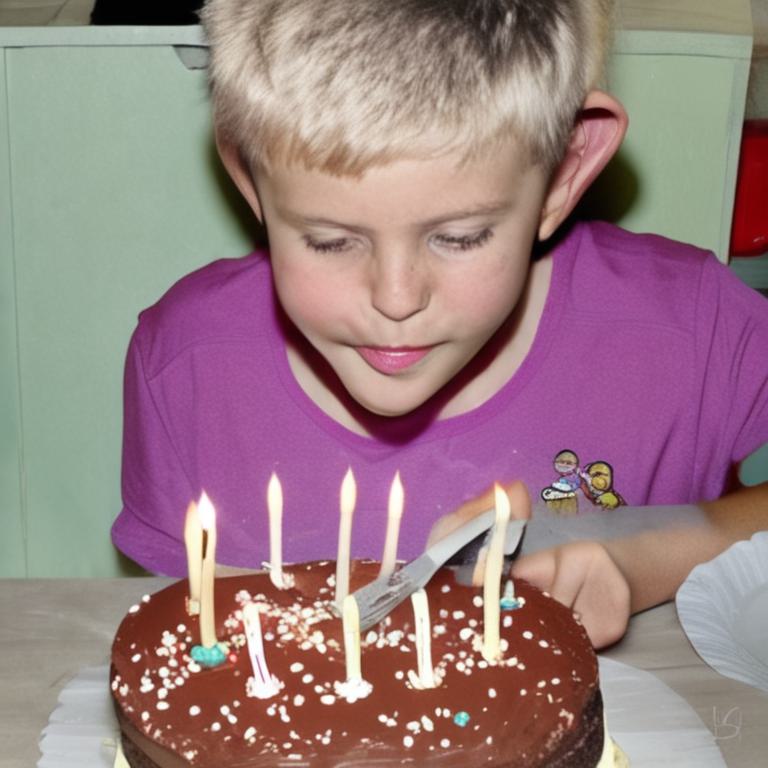}
        & \includegraphics[width=.104\linewidth]{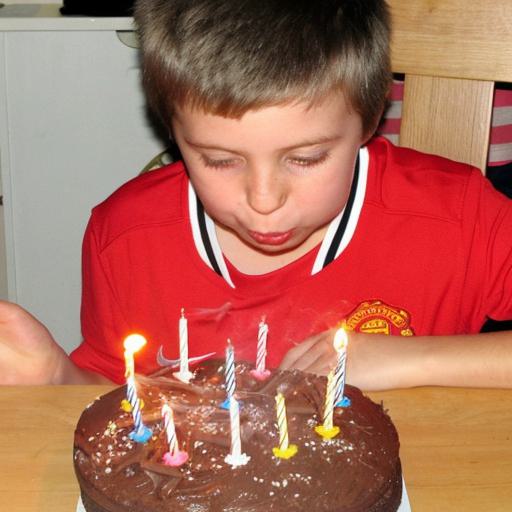}
        & \includegraphics[width=.104\linewidth]{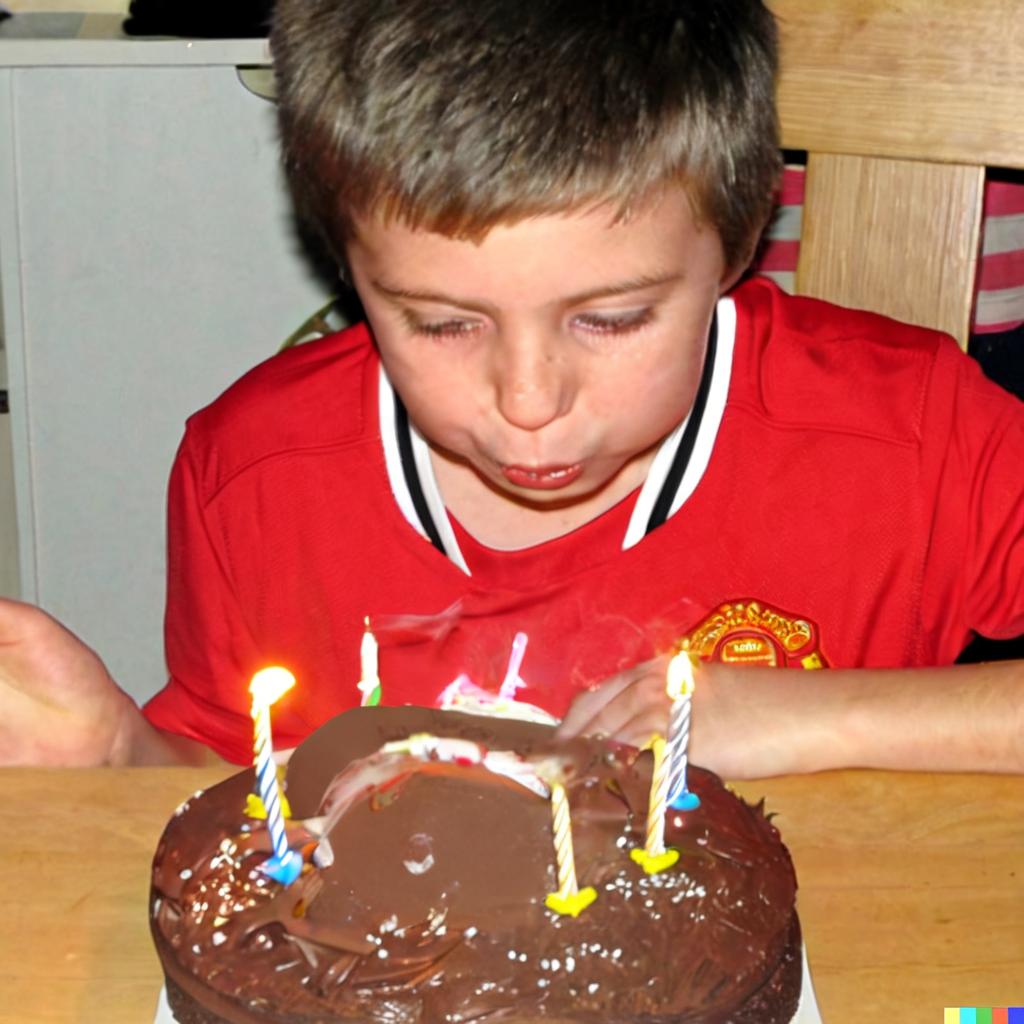}
        & \includegraphics[width=.104\linewidth]{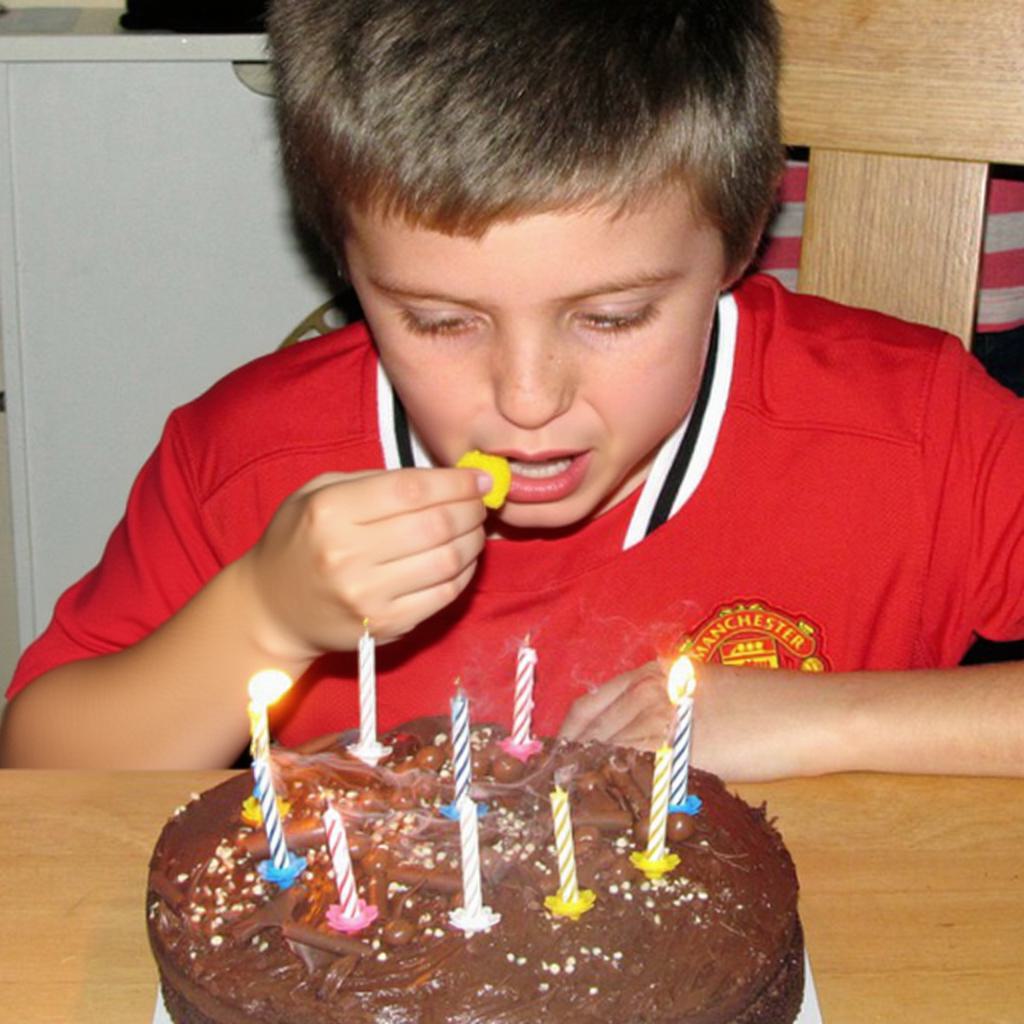}
        & \includegraphics[width=.104\linewidth]{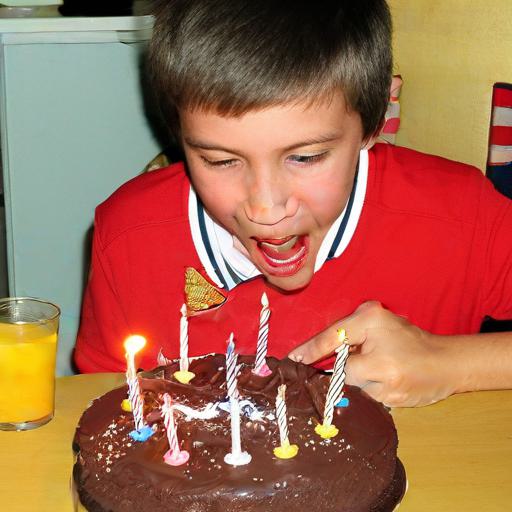}
        & \includegraphics[width=.104\linewidth]{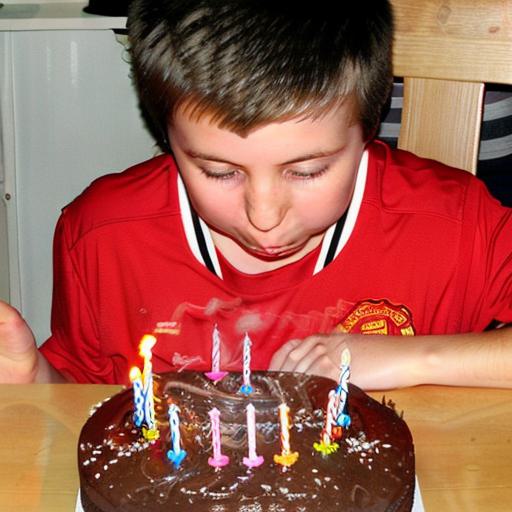}
        & \includegraphics[width=.104\linewidth]{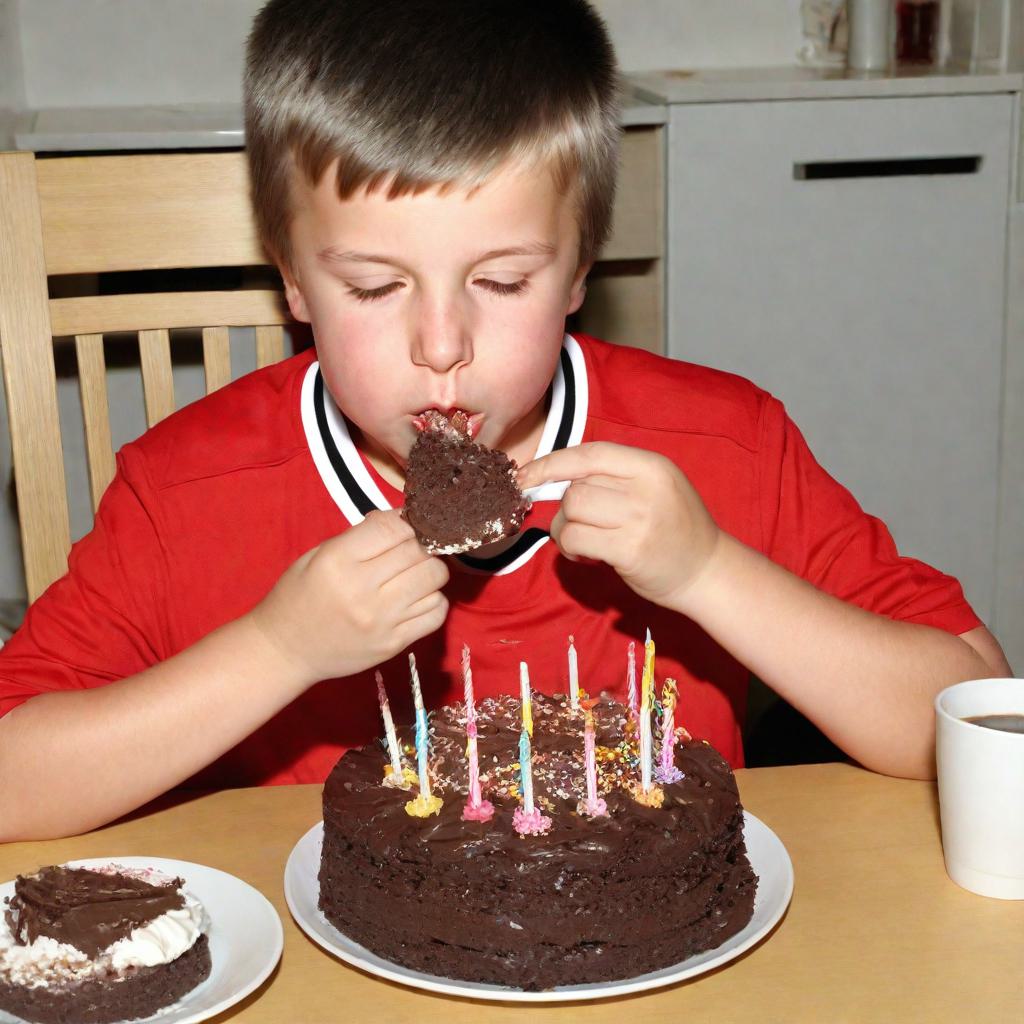}\\
        \\[-1.2em]
        blow cake & \multicolumn{8}{c}{eat cake}\\
          \includegraphics[width=.104\linewidth]{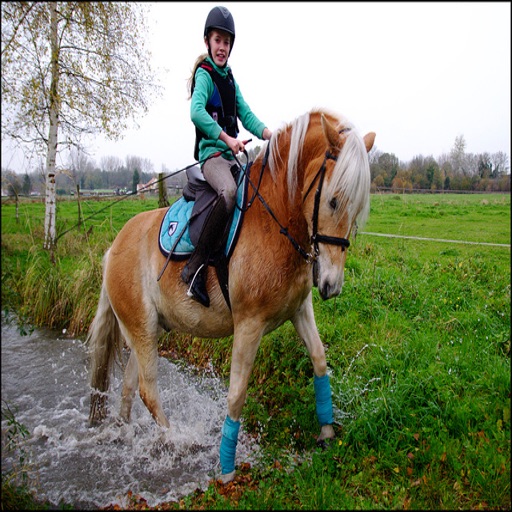}
        & \includegraphics[width=.104\linewidth]{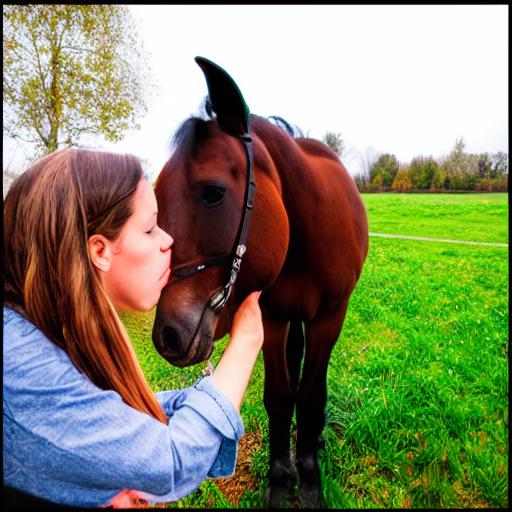}
        & \includegraphics[width=.104\linewidth]{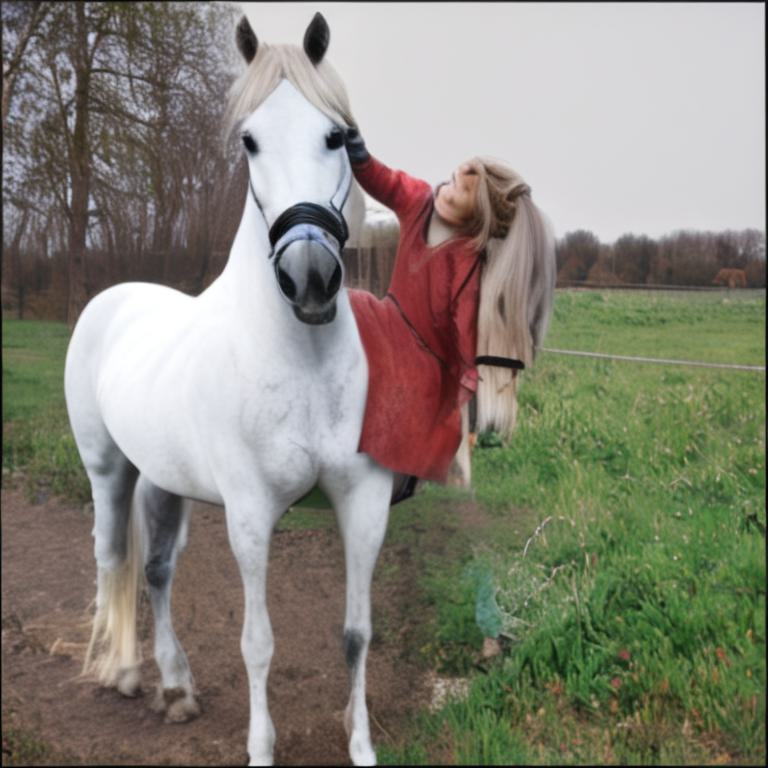}
        & \includegraphics[width=.104\linewidth]{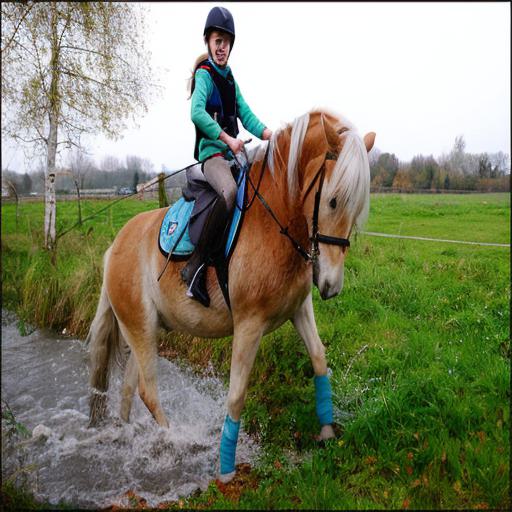}
        & \includegraphics[width=.104\linewidth]{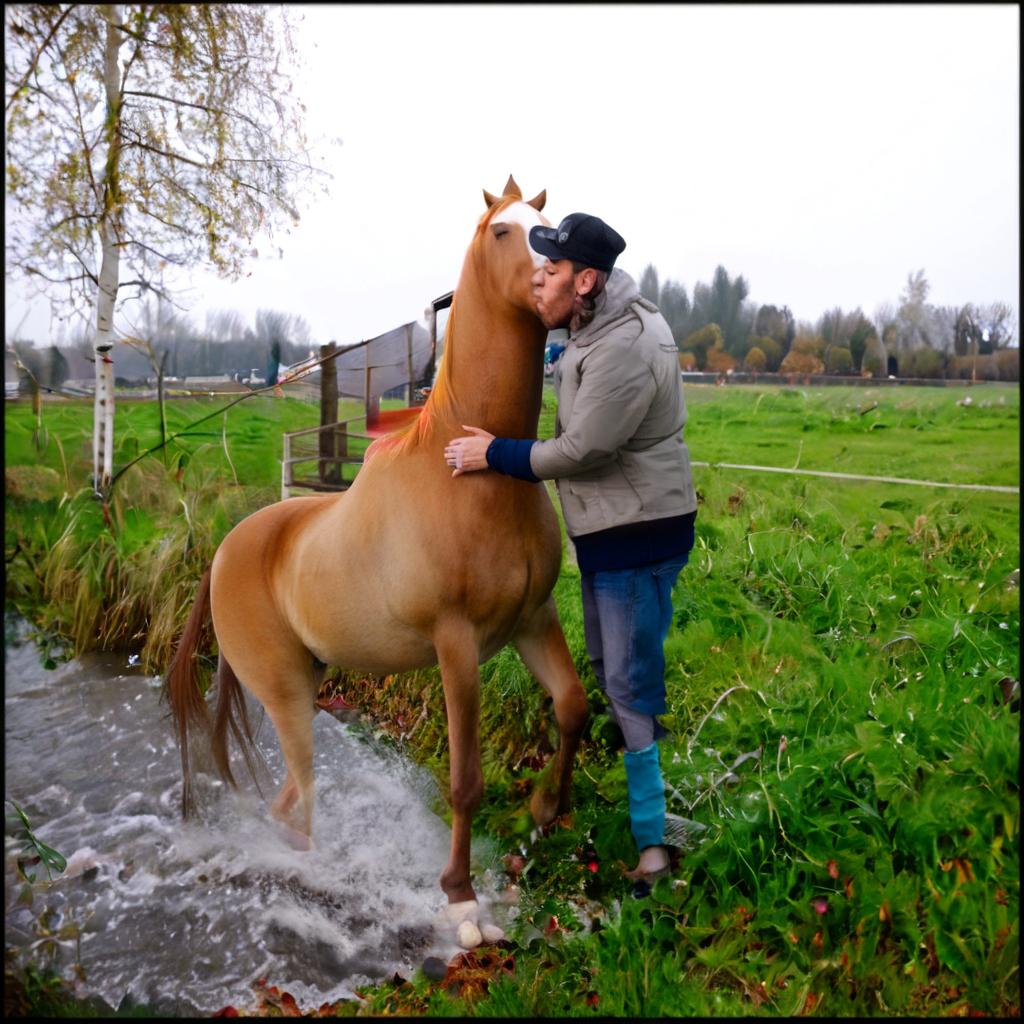}
        & \includegraphics[width=.104\linewidth]{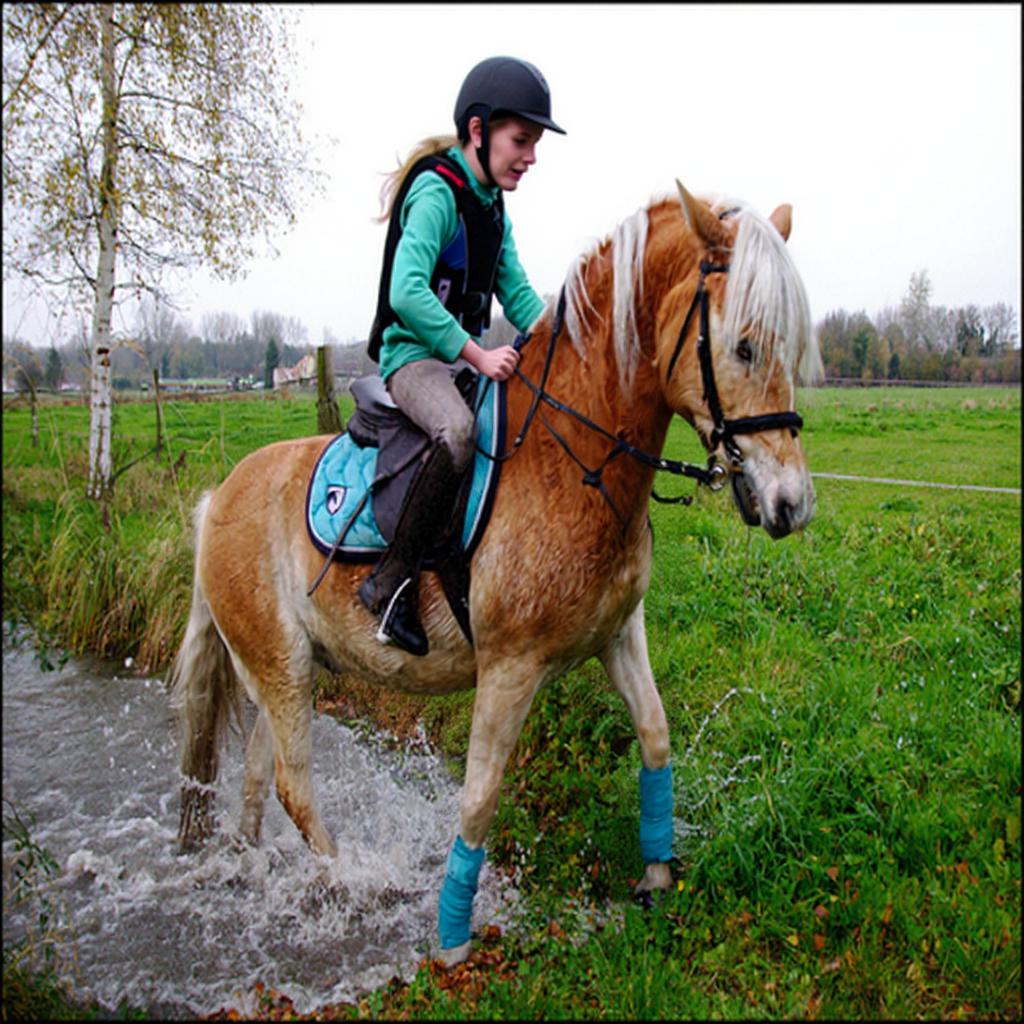}
        & \includegraphics[width=.104\linewidth]{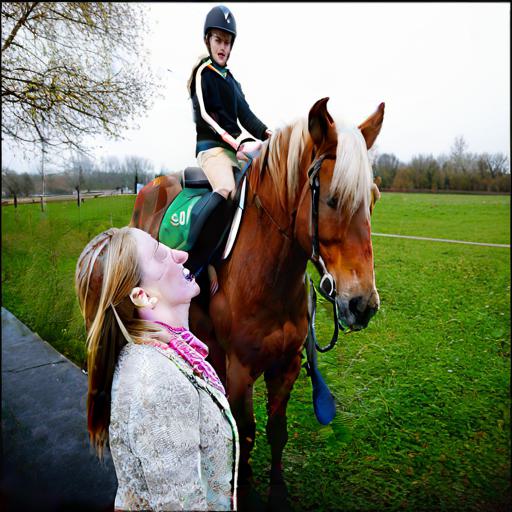}
        & \includegraphics[width=.104\linewidth]{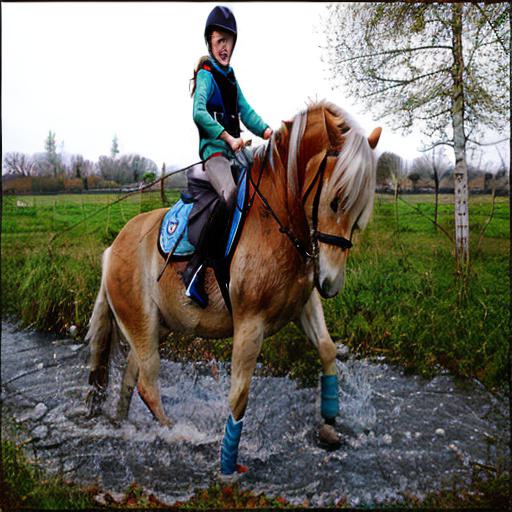}
        & \includegraphics[width=.104\linewidth]{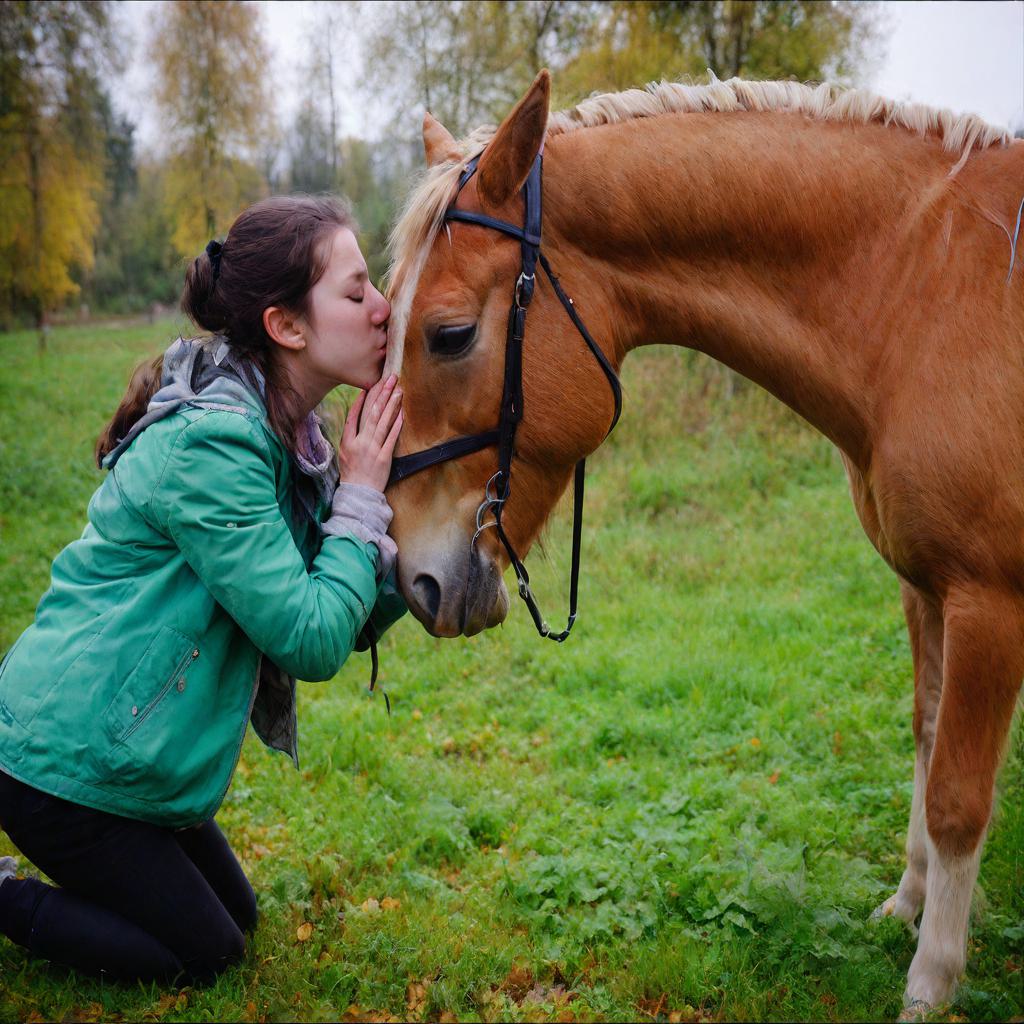}\\
        \\[-1.2em]
        {ride horse} & \multicolumn{8}{c}{kiss horse}\\
          \includegraphics[width=.104\linewidth]{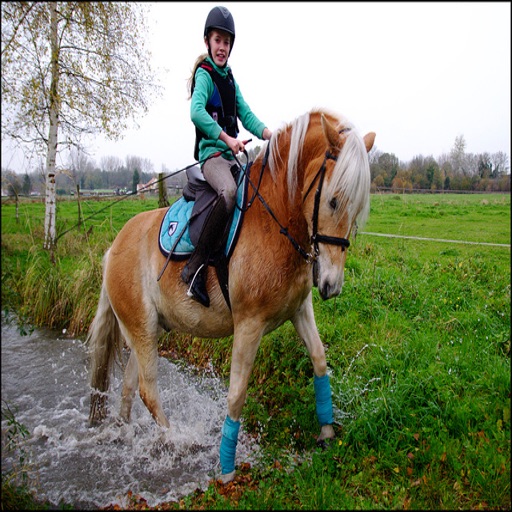}
        & \includegraphics[width=.104\linewidth]{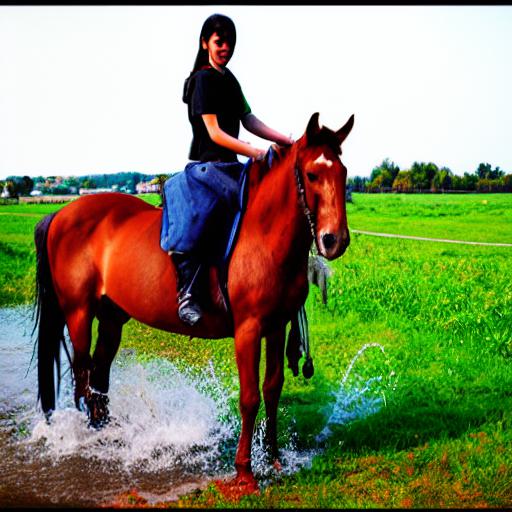}
        & \includegraphics[width=.104\linewidth]{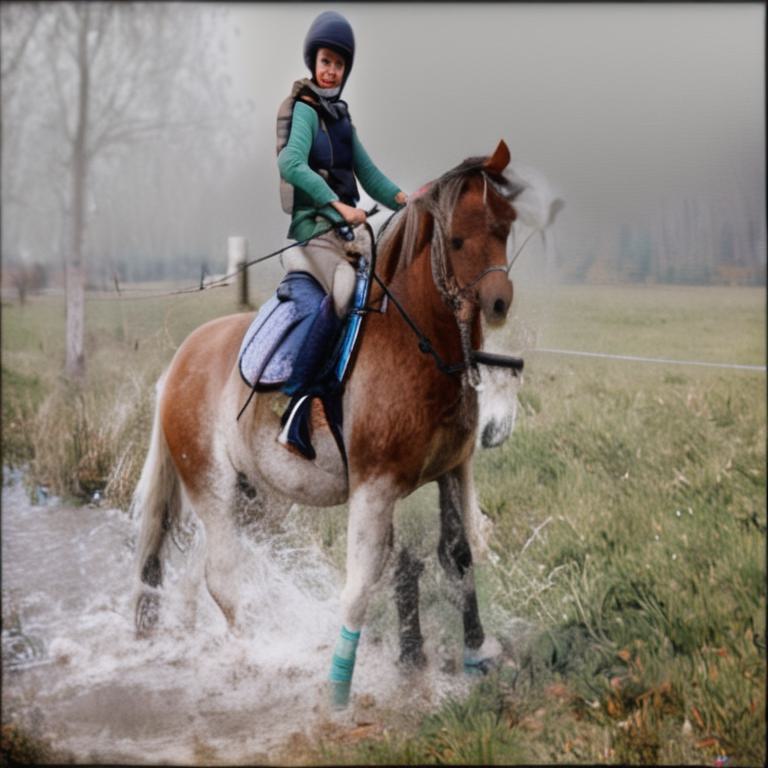}
        & \includegraphics[width=.104\linewidth]{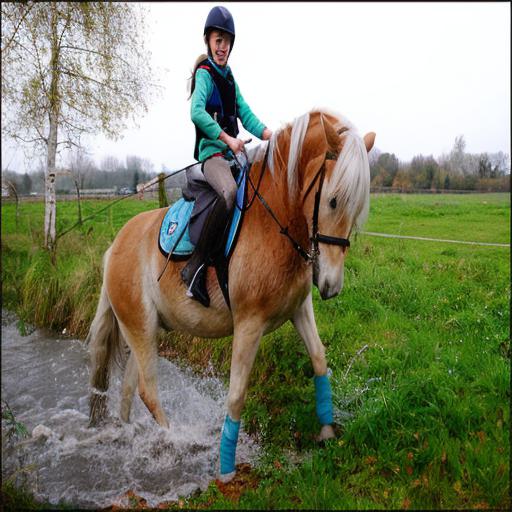}
        & \includegraphics[width=.104\linewidth]{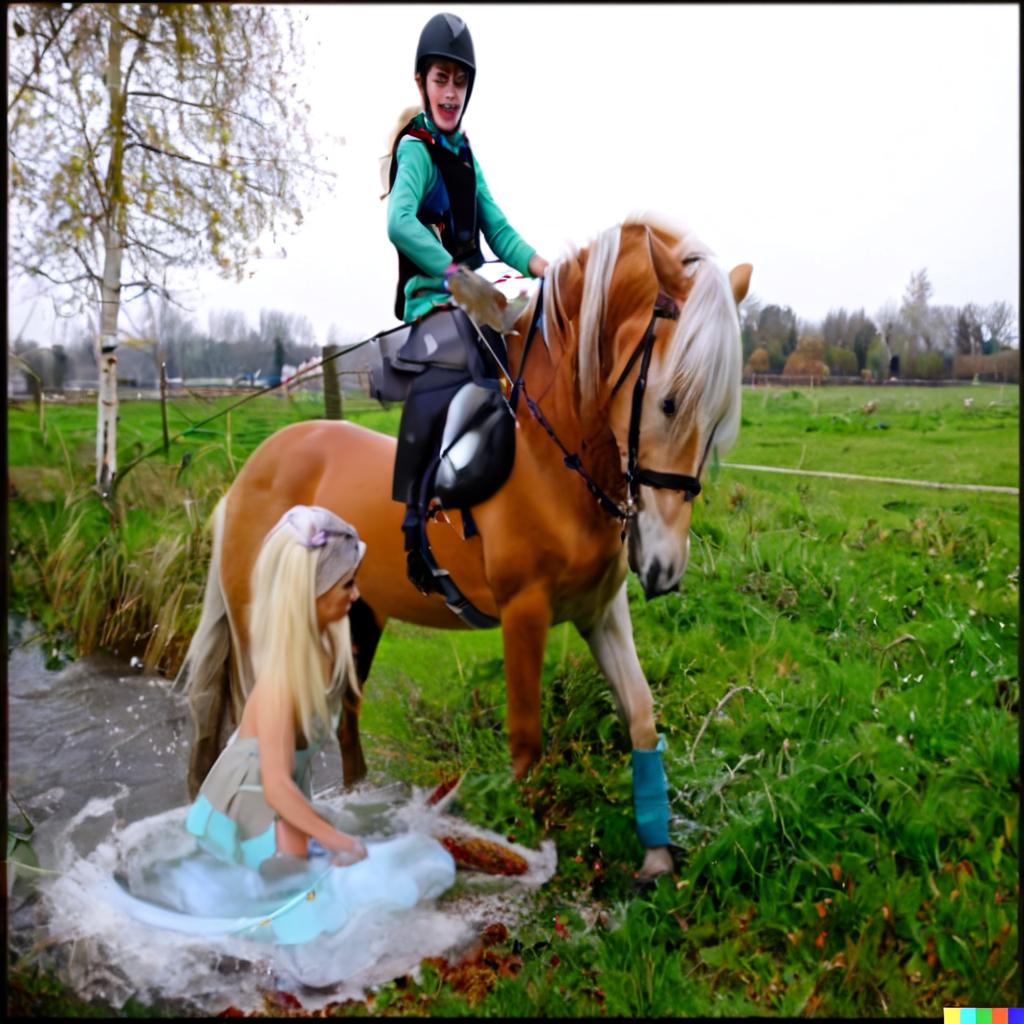}
        & \includegraphics[width=.104\linewidth]{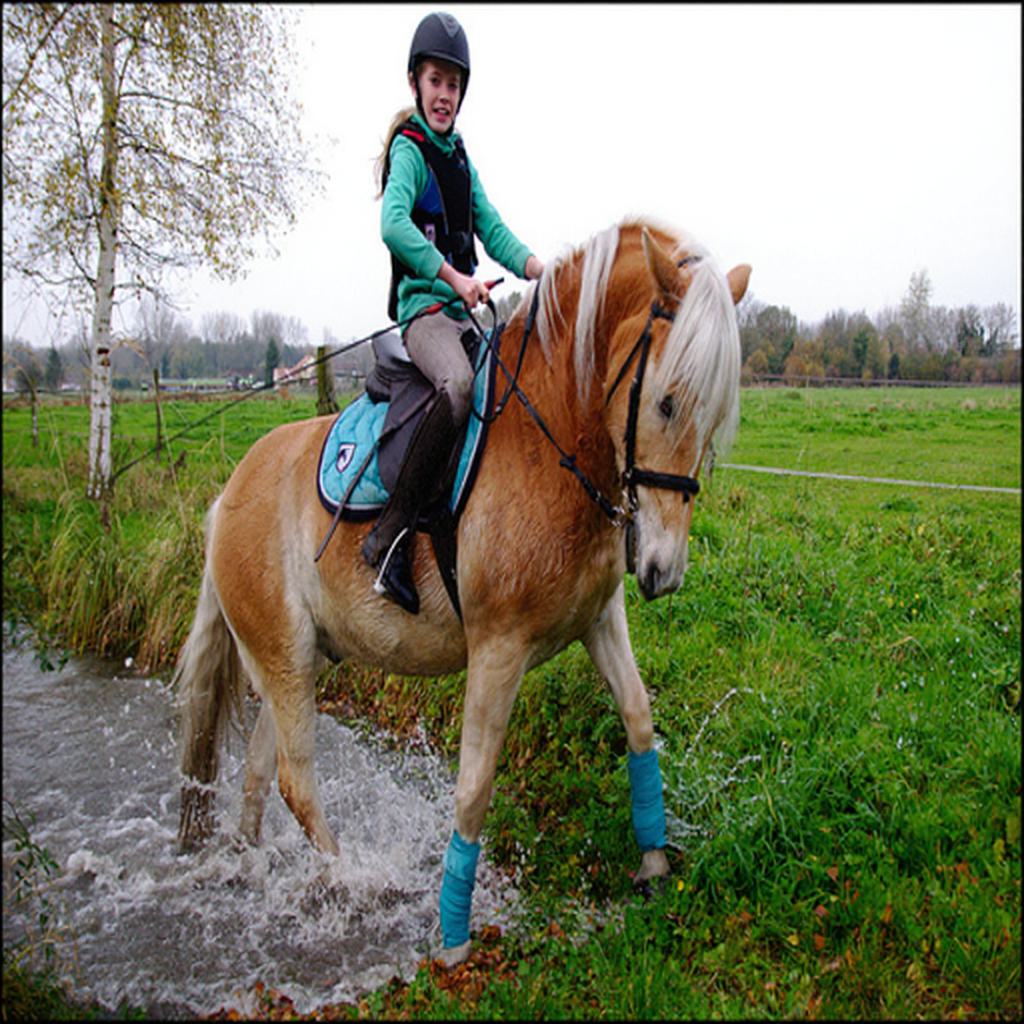}
        & \includegraphics[width=.104\linewidth]{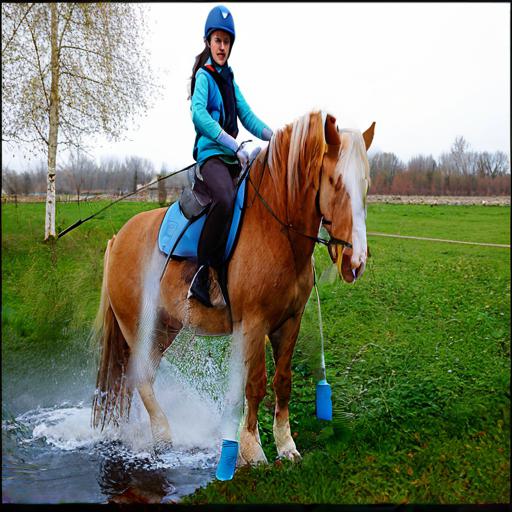}
        & \includegraphics[width=.104\linewidth]{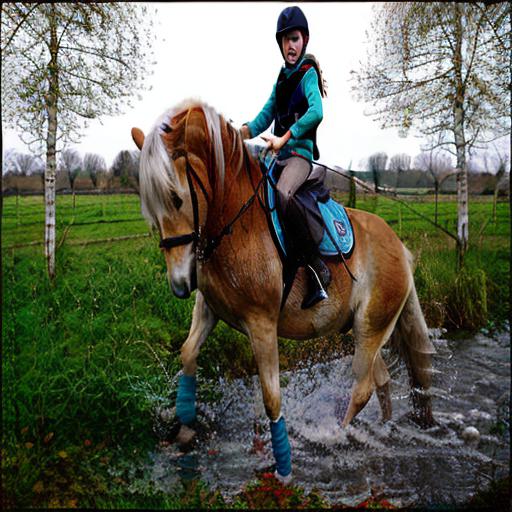}
        & \includegraphics[width=.104\linewidth]{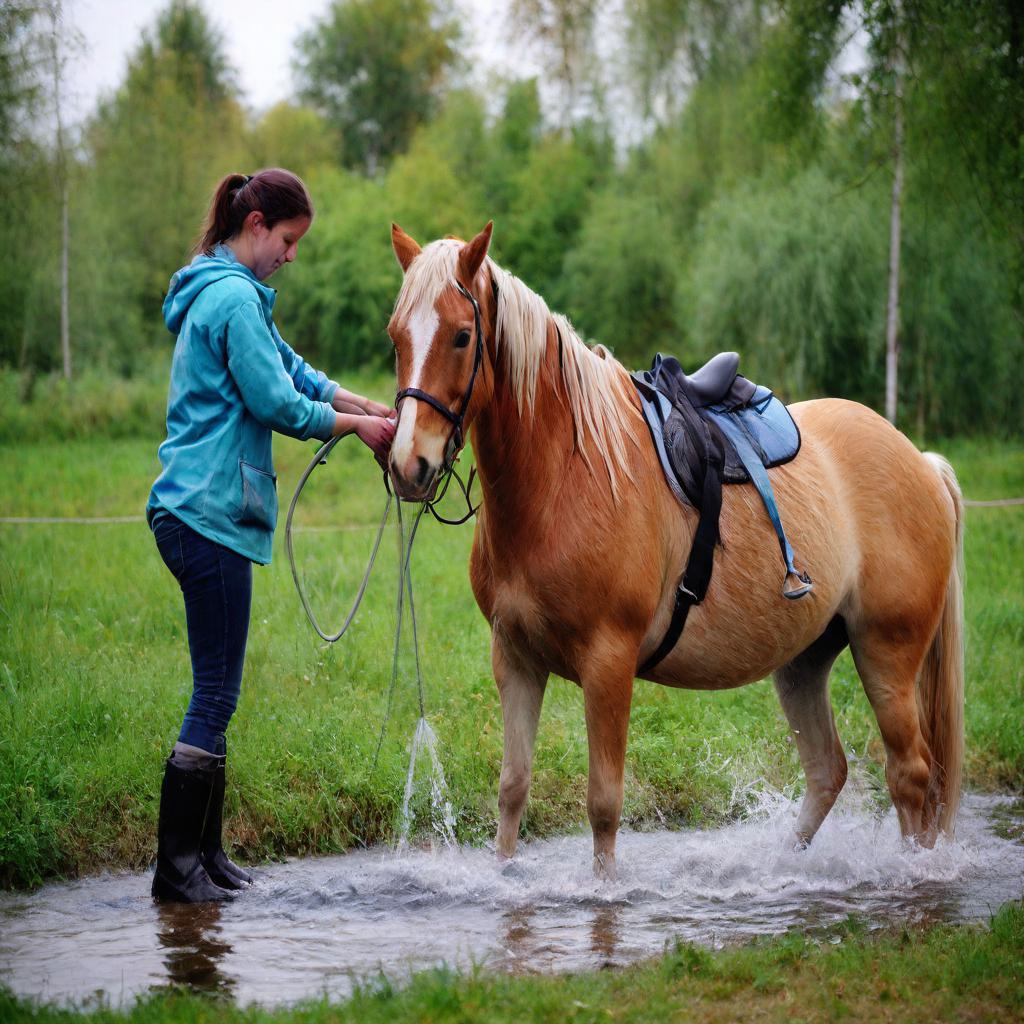}\\
        \\[-1.2em]
        {ride horse} & \multicolumn{8}{c}{wash horse}\\
          \includegraphics[width=.104\linewidth]{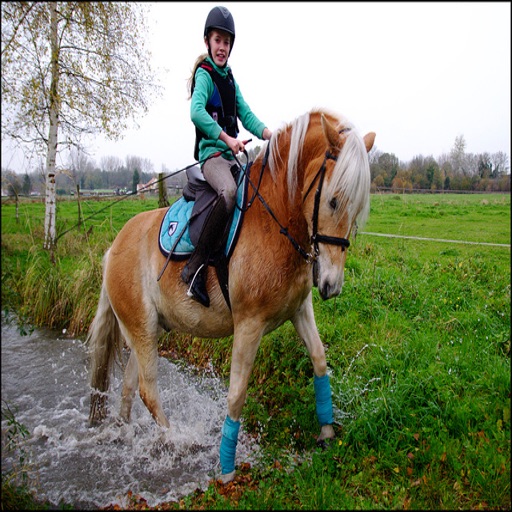}
        & \includegraphics[width=.104\linewidth]{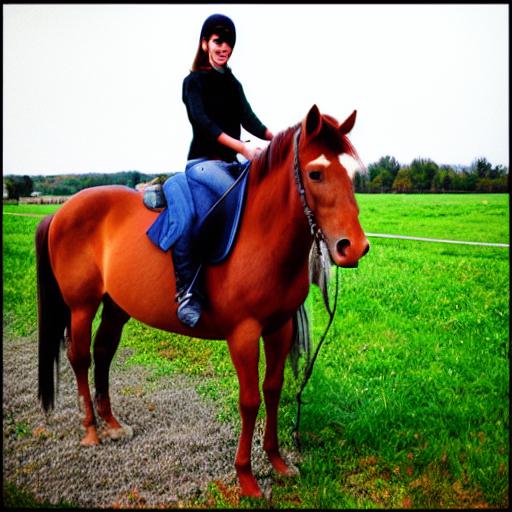}
        & \includegraphics[width=.104\linewidth]{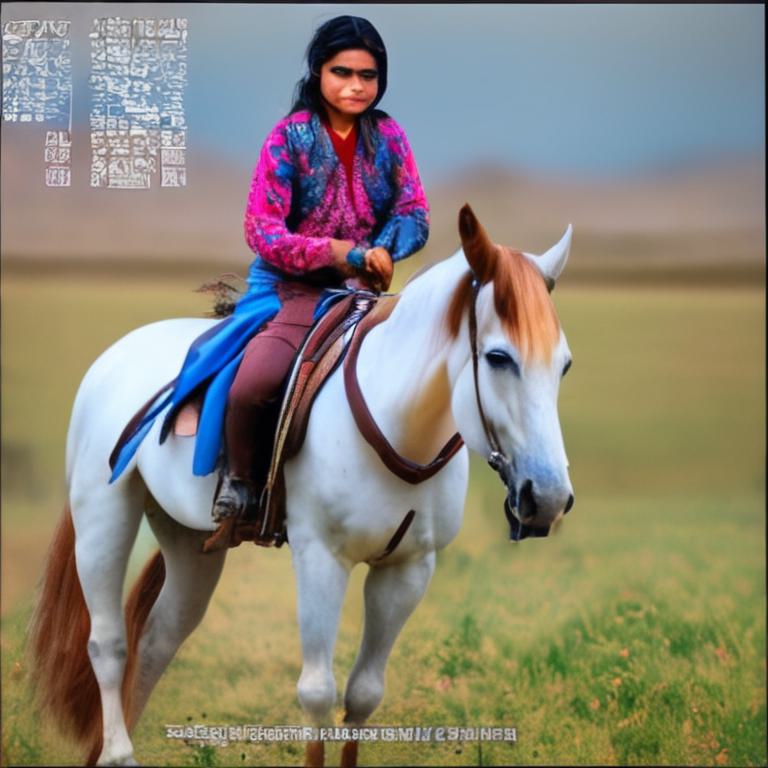}
        & \includegraphics[width=.104\linewidth]{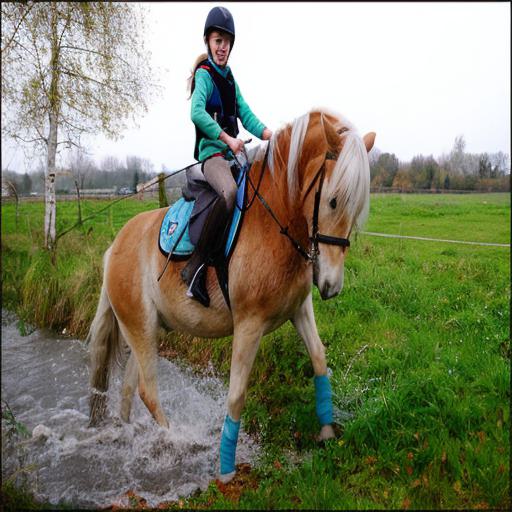}
        & \includegraphics[width=.104\linewidth]{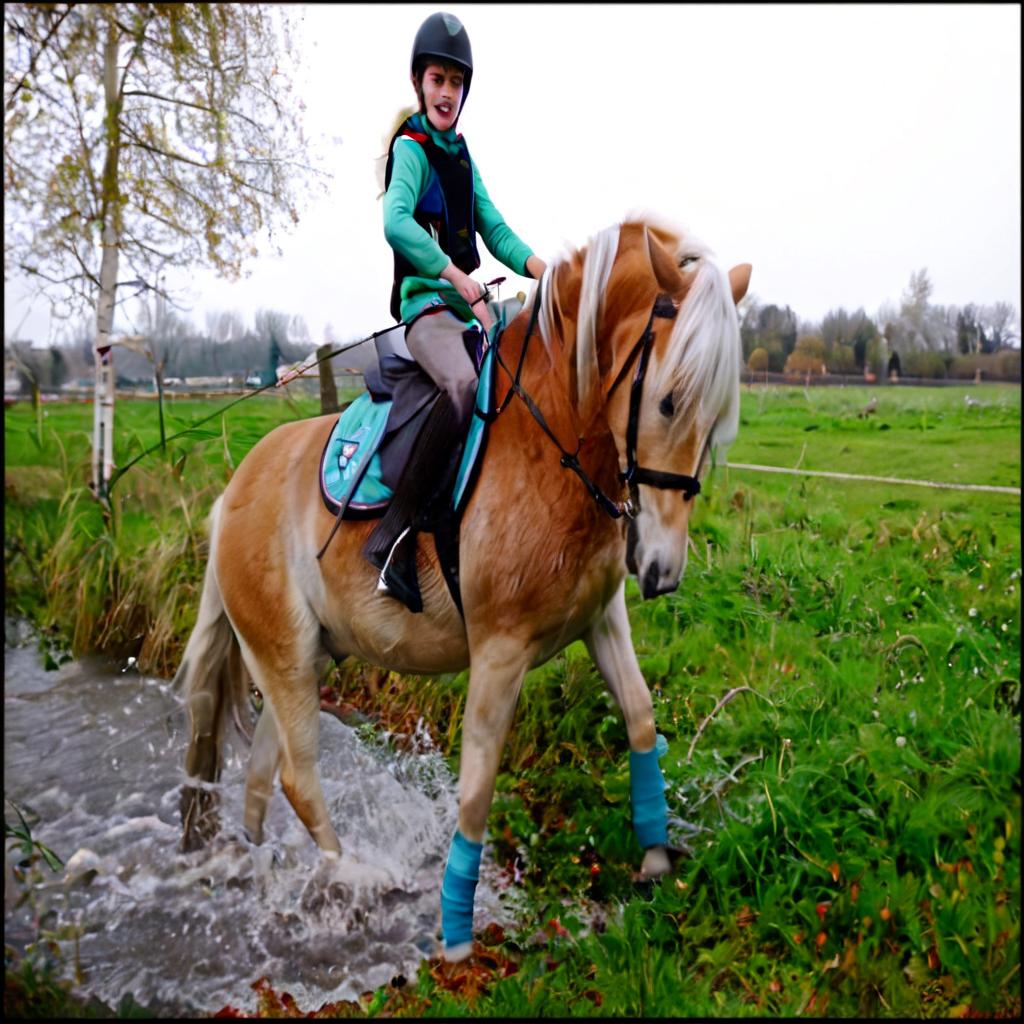}
        & \includegraphics[width=.104\linewidth]{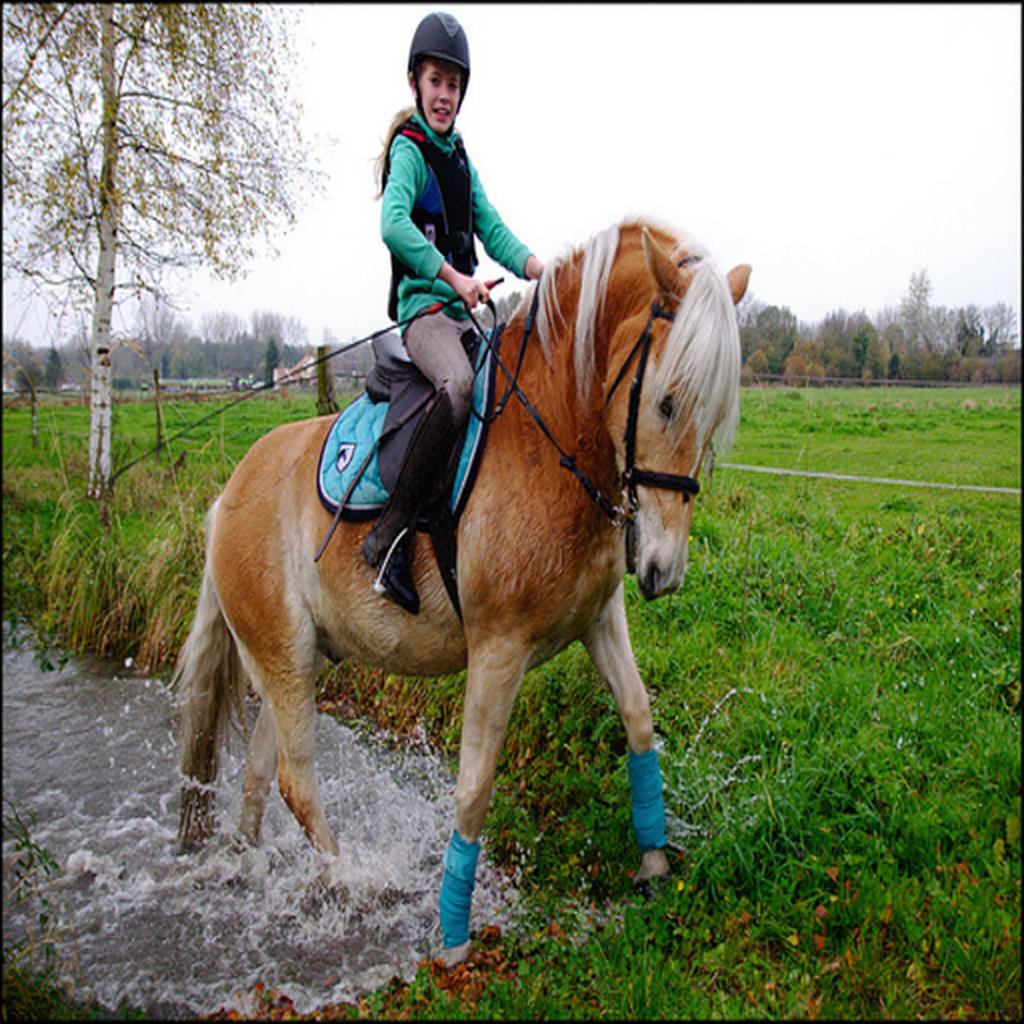}
        & \includegraphics[width=.104\linewidth]{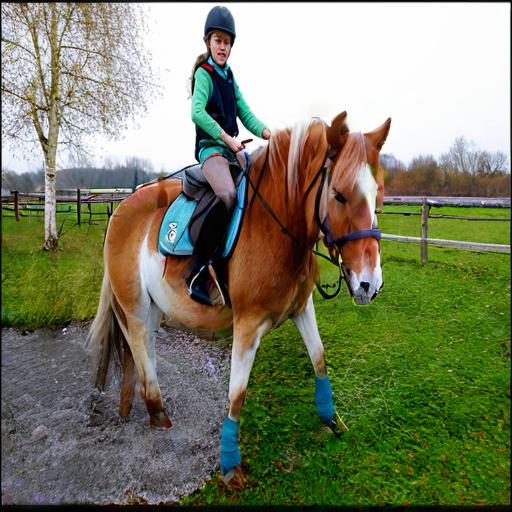}
        & \includegraphics[width=.104\linewidth]{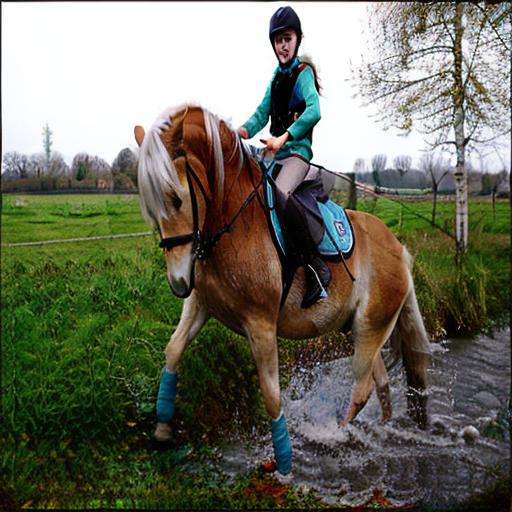}
        & \includegraphics[width=.104\linewidth]{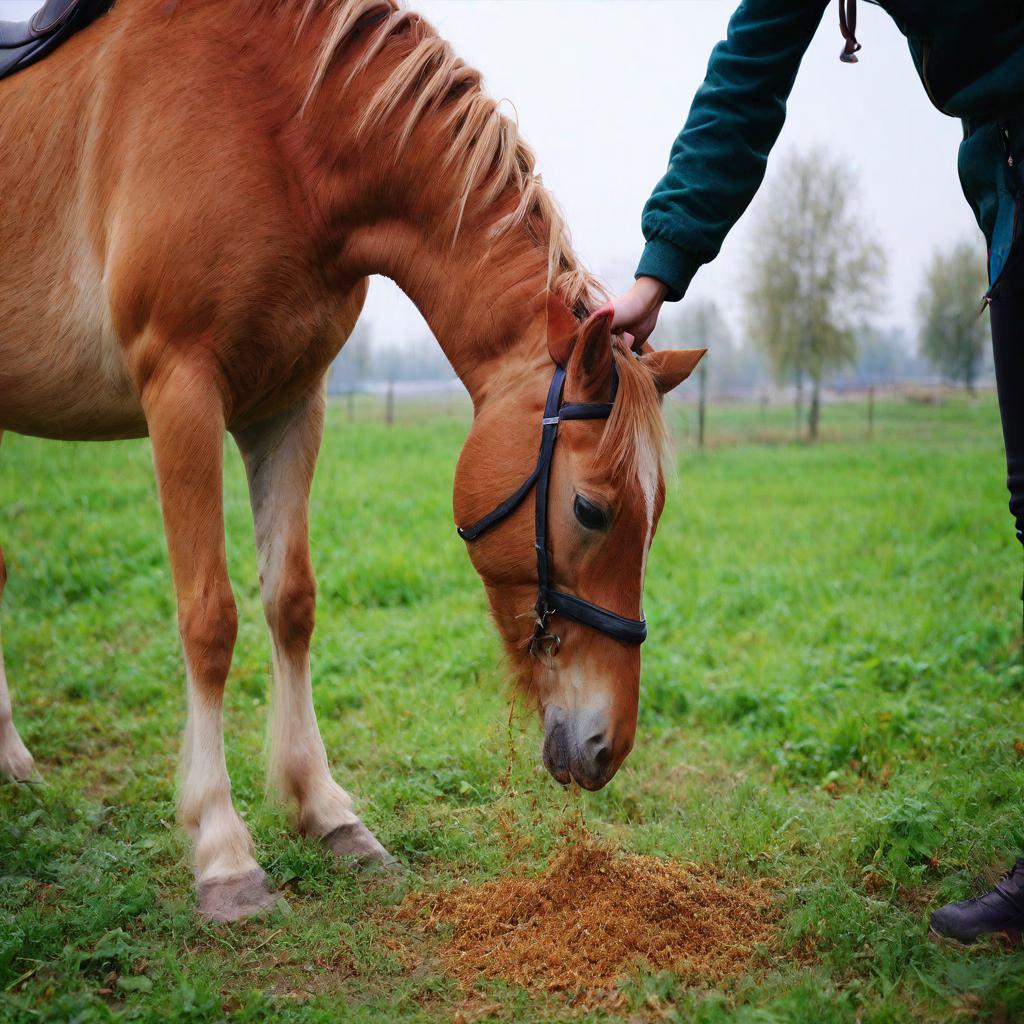}\\
        {ride horse} & \multicolumn{8}{c}{feed horse}\\
    \end{tabular}
    \caption{Qualitative comparison with SOTAs. The Source column shows the original image and its associated interaction. Our method demonstrates the most balanced results among all methods. Additional qualitative results are provided in \cref{sec:more-qualitative}.
    }
    \label{fig:qualitative}
\end{figure*}

%% file: assets/figure/edit_perf_l.tex
\begin{table}[t!]
  \caption{Comparison of {\rmfamily\scshape InteractEdit} with existing baselines on the more challenging \iebenchext~benchmark, reporting Editability--Identity Score (\text{{\rmfamily\scshape ei}}), HOI Editability, Identity Consistency, and HPSv2 \cite{wu2023hpsv2} (a human-preference aligned perceptual metric for fine-grained realism). \iebenchext~scales \iebench~to 100 source images, 33 actions, and 21 objects with rarer verb--object compositions and harder non-rigid pose changes, causing absolute scores to drop across all methods. The \textbf{best} and \underline{second best} are indicated.}
  \label{tab:edit-perf-l}
  \centering
  \begin{tabular}{l||c|c|c|c}
    \hline \hline
    Method              & \text{{\rmfamily\scshape ei}}              & HOI Editability      & Identity Consistency  & HPSv2\\ \hline \hline
    Null-Text Inversion \cite{mokady2023nti}        & 0.296          & \underline{0.359}    & 0.252                       & 0.2437           \\
    MasaCtrl \cite{cao2023masactrl}                 & 0.243          & 0.245                & 0.241                      & 0.2205          \\
    TurboEdit \cite{deutch2024turboedit}            & 0.280          & 0.299                & 0.264                      & 0.2426         \\
    EditFriendlyDDPM \cite{huberman2024editfriendly}& 0.255          & 0.258                & 0.252                      & 0.2438         \\
    InfEdit \cite{xu2023infedit}                    & \underline{0.312}          & 0.192                & \textbf{0.831}                      & 0.2332       \\
    Mix-of-Show \cite{gu2024mixofshow}              & 0.260          & 0.282                & 0.241                      & 0.2524      \\
    Flux.1 Kontext \cite{labs2025flux1kontext}      & 0.272           & 0.276             & \underline{0.269}                      & 0.2471      \\
    FireFlow \cite{deng2024fireflow}                & 0.290          & 0.337                & 0.255                      & \underline{0.2549}         \\
    \hline
    \hline
    \textbf{InteractEdit (Ours)}                    & \textbf{0.332} & \textbf{0.498}       & 0.250                      & \textbf{0.2625}\\ \hline \hline
  \end{tabular}%
\end{table}

%% file: assets/figure/edit_abl.tex
\begin{table}[t]

\caption{Ablation study of {\rmfamily\scshape InteractEdit}. Dis., SI, and SeRA represents Disassemble HOI, Selective Inversion, and Regularization with SeRA.}
\label{tab:ablation}
\centering
\begin{tabular}{l||ccc||c||c}
\hline \hline
\multirow{2}{*}{Method} & \multirow{2}{*}{Dis.} & \multirow{2}{*}{SI} & \multirow{2}{*}{SeRA} & \multirow{2}{*}{\text{{\rmfamily\scshape ei}}} & HOI            \\
                        &                       &                     &                       &                          & Editability    \\ \hline \hline
B0: Baseline            &                       &                     &                       & 0.252                    & 0.150          \\
B1: +Dis                & \checkmark            &                     &                       & 0.327                    & 0.207          \\
L1: LoRA                &                       &                     & LoRA                  & 0.454                    & 0.331          \\ \hline
S1: SeRA                &                       &                     & \checkmark            & 0.498                    & 0.388          \\
S2: SeRA +SI            &                       & \checkmark          & \checkmark            & 0.508                    & 0.410          \\ \hline 
A: AdaLoRA +SI +Dis.    & \checkmark            & \checkmark          & AdaLoRA               & 0.525                    & 0.502          \\ \hline
S3: Full (Ours)         & \checkmark            & \checkmark          & \checkmark            & \textbf{0.573}           & \textbf{0.514} \\ \hline \hline
\end{tabular}%
\end{table}

%% file: assets/figure/ab_qual_ab.tex
\begin{figure}
    \centering
    \setlength{\tabcolsep}{1pt} 
    \renewcommand{\arraystretch}{1} 
    \begin{tabular}{cccccccc}
        \footnotesize Source & \footnotesize B0 & \footnotesize B1 & \footnotesize L1 & \footnotesize S1 & \footnotesize S2 & \footnotesize A & \footnotesize Ours \\
        
        \raisebox{-0.5\height}{\includegraphics[width=.118\linewidth]{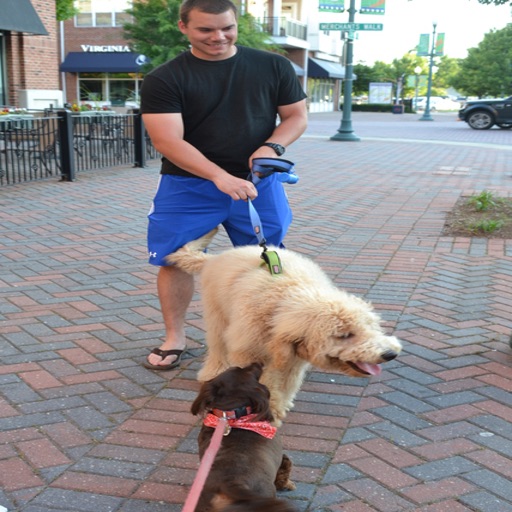}}
        & \raisebox{-0.5\height}{\includegraphics[width=.118\linewidth]{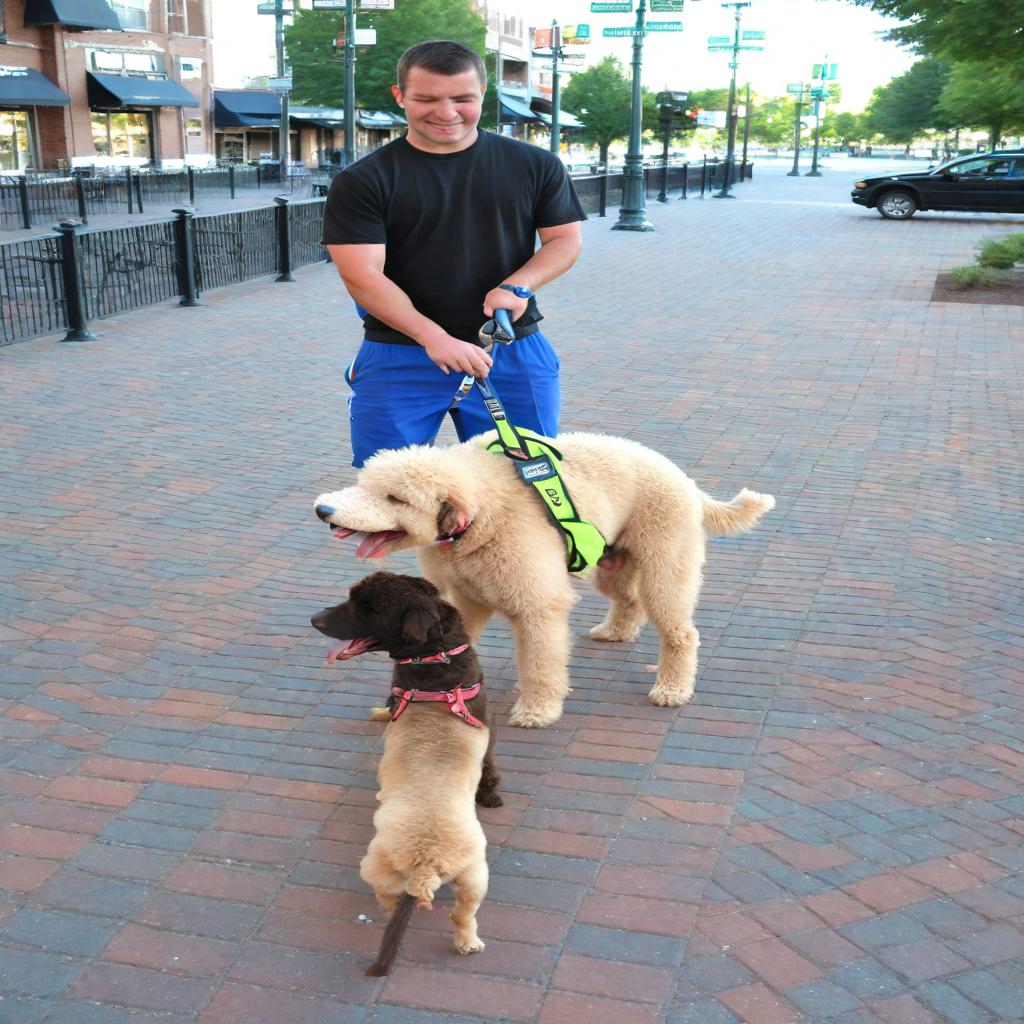}}
        & \raisebox{-0.5\height}{\includegraphics[width=.118\linewidth]{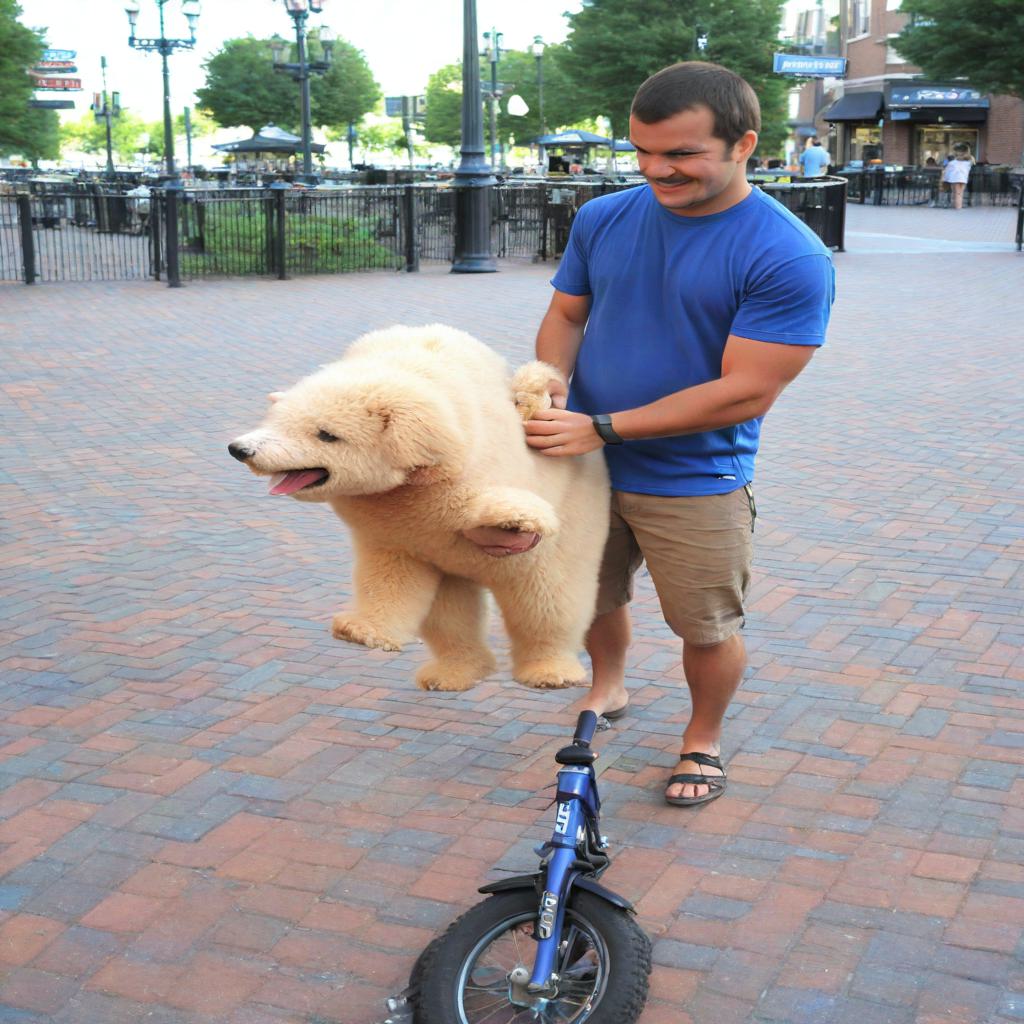}}
        & \raisebox{-0.5\height}{\includegraphics[width=.118\linewidth]{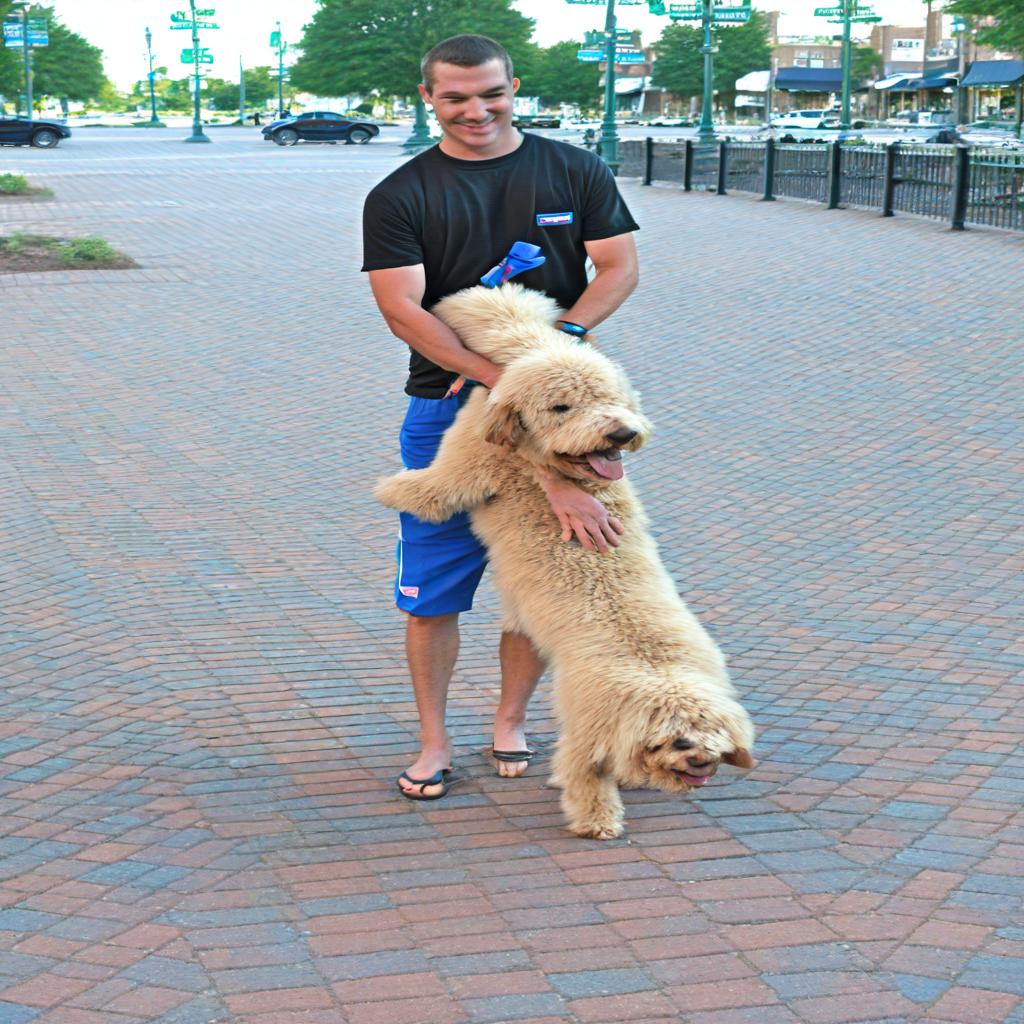}}
        & \raisebox{-0.5\height}{\includegraphics[width=.118\linewidth]{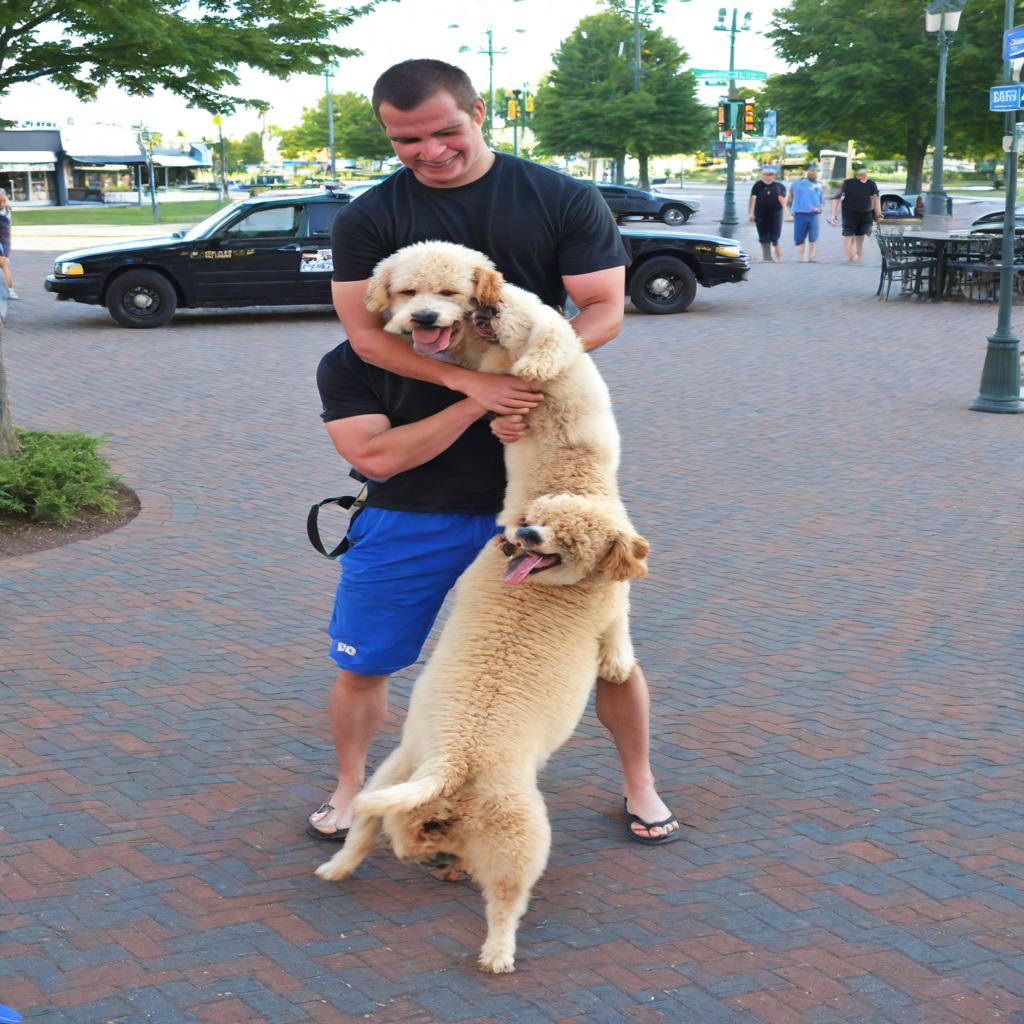}}
        & \raisebox{-0.5\height}{\includegraphics[width=.118\linewidth]{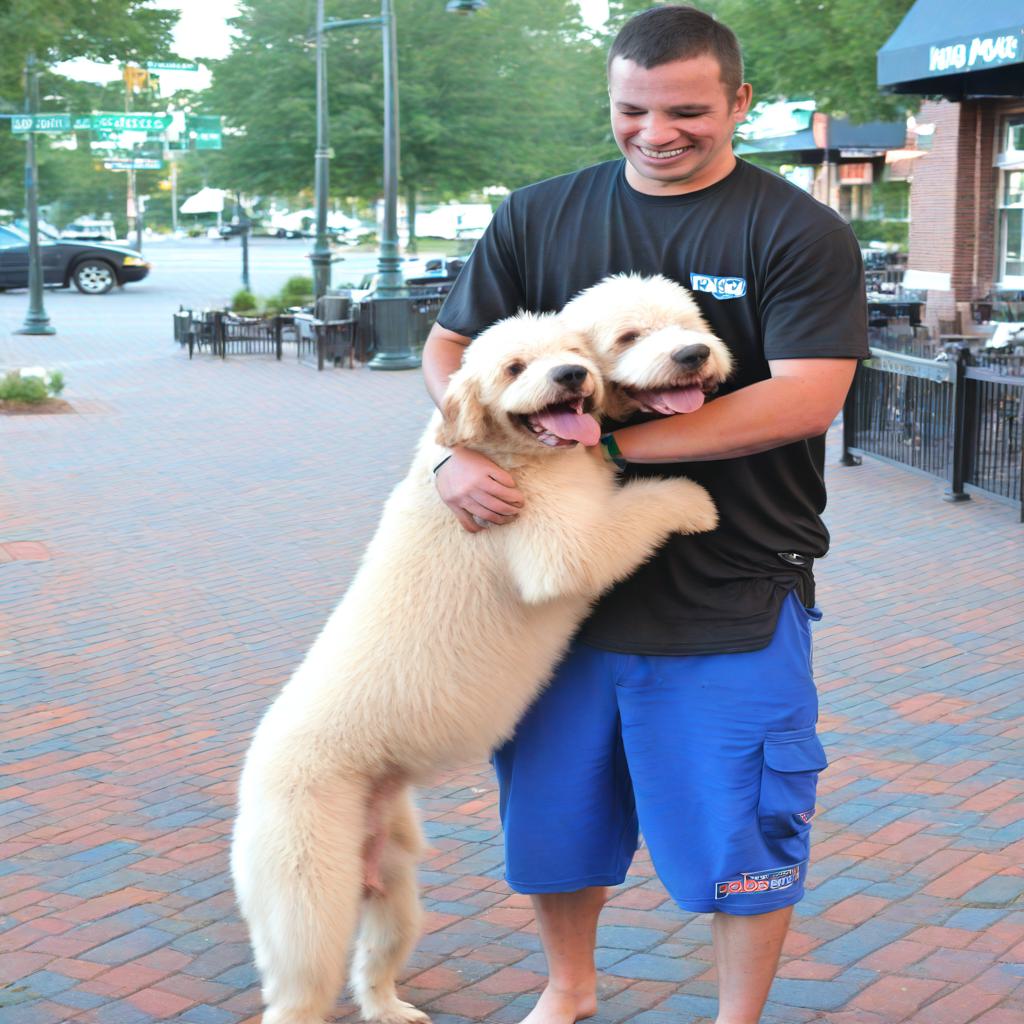}}
        & \raisebox{-0.5\height}{\includegraphics[width=.118\linewidth]{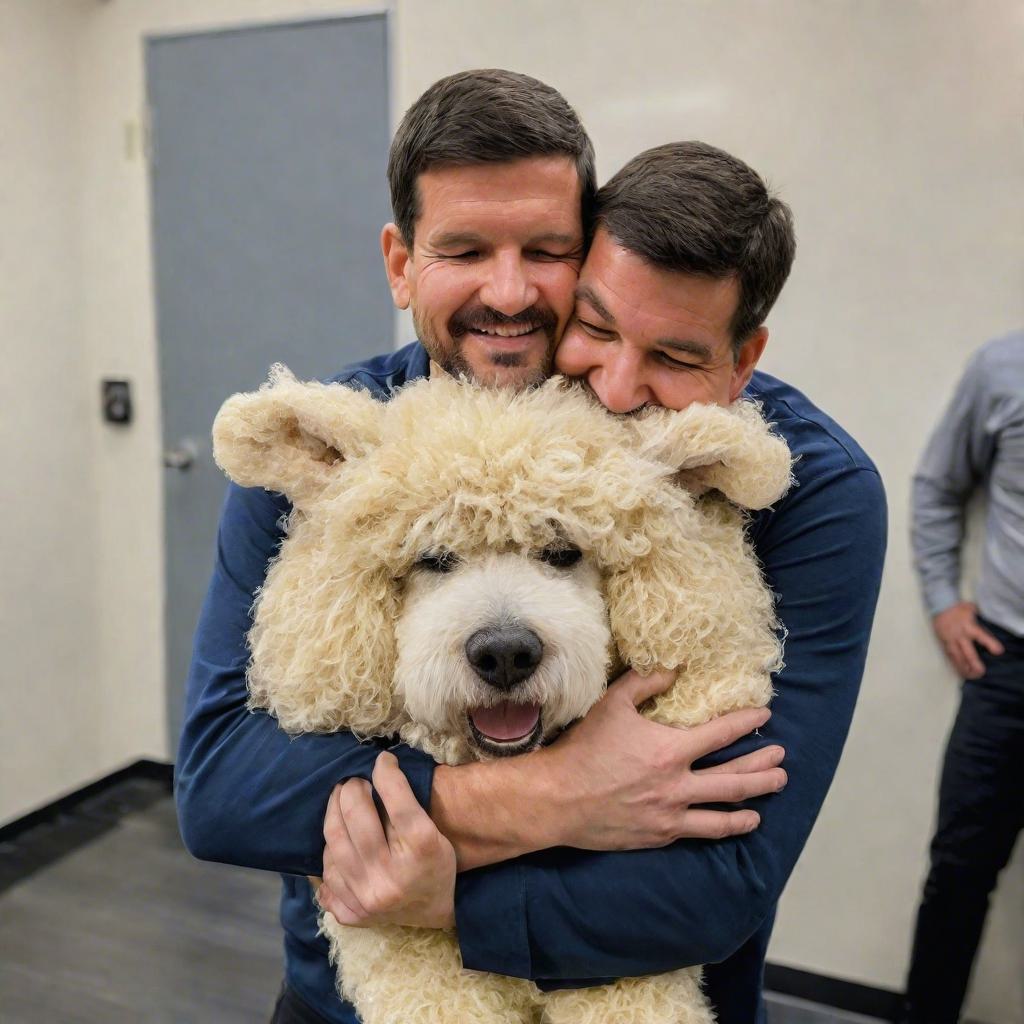}}
        & \raisebox{-0.5\height}{\includegraphics[width=.118\linewidth]{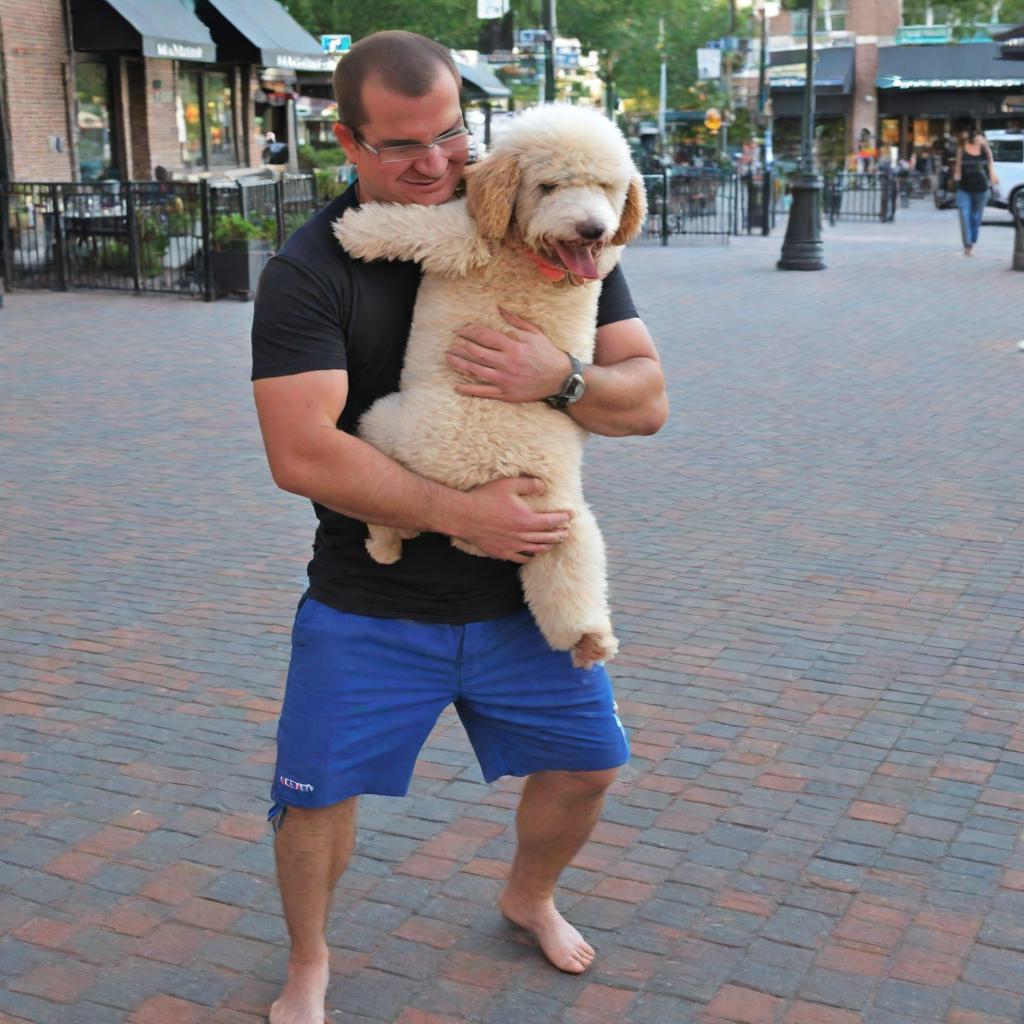}}\\
        \\[-1.2em]
        \multicolumn{8}{c}{walk dog \textrightarrow{} hug dog}\\
        
        \raisebox{-0.5\height}{\includegraphics[width=.118\linewidth]{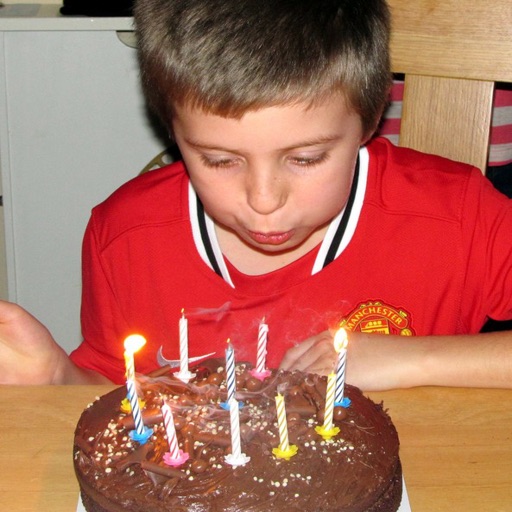}}
        & \raisebox{-0.5\height}{\includegraphics[width=.118\linewidth]{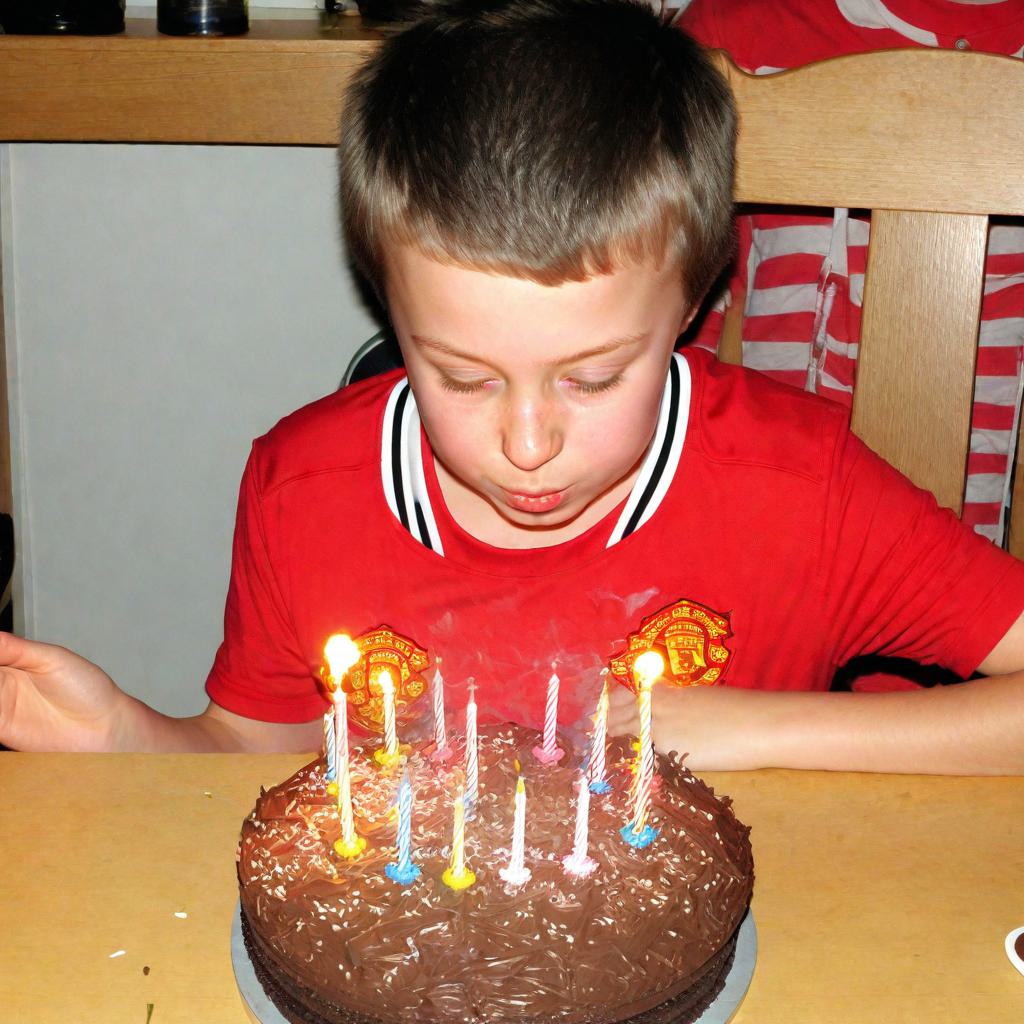}}
        & \raisebox{-0.5\height}{\includegraphics[width=.118\linewidth]{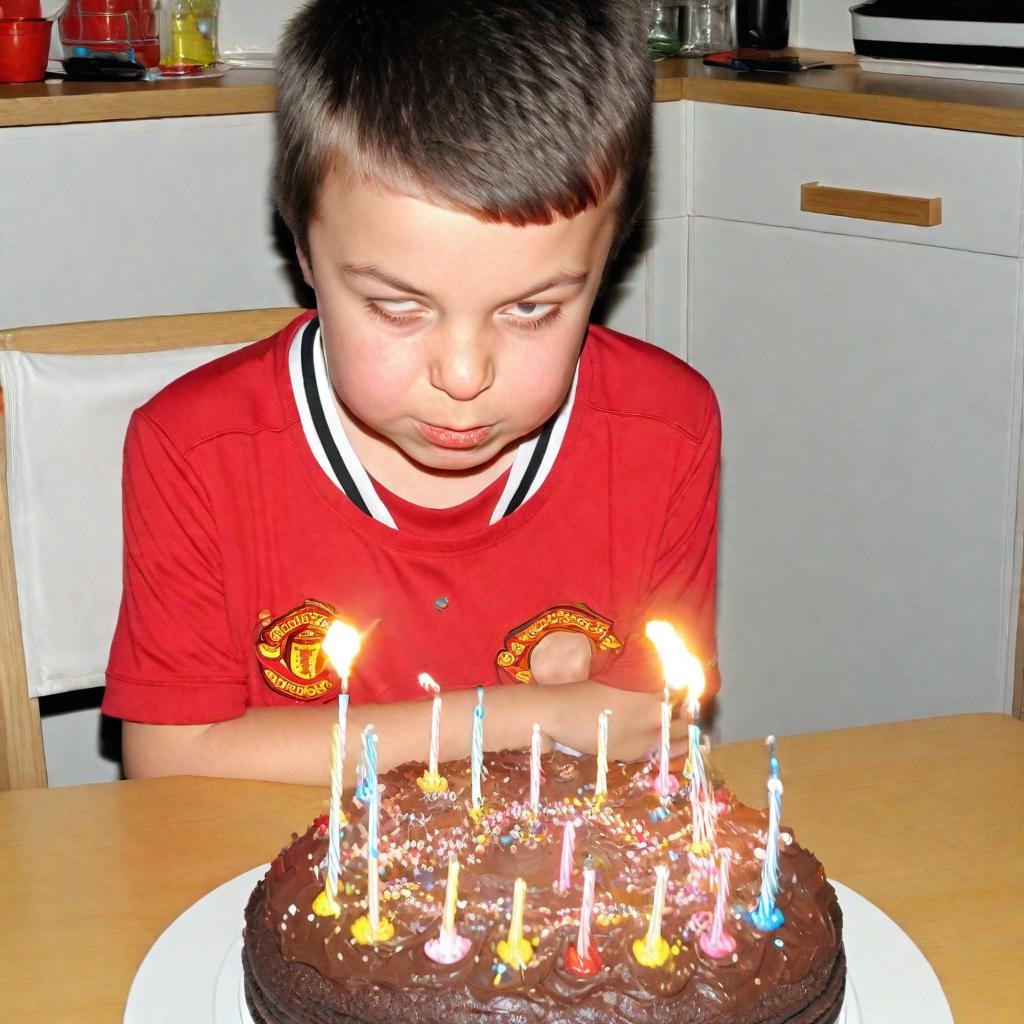}}
        & \raisebox{-0.5\height}{\includegraphics[width=.118\linewidth]{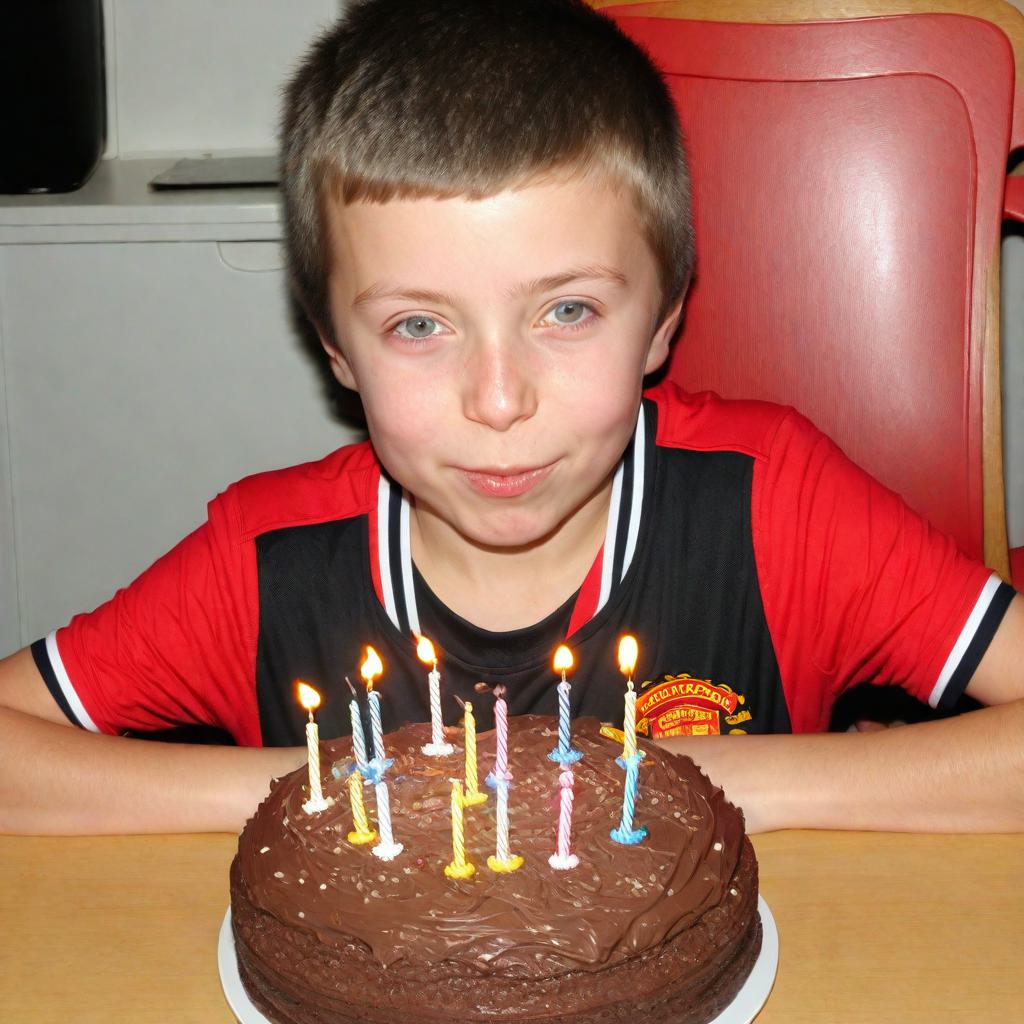}}
        & \raisebox{-0.5\height}{\includegraphics[width=.118\linewidth]{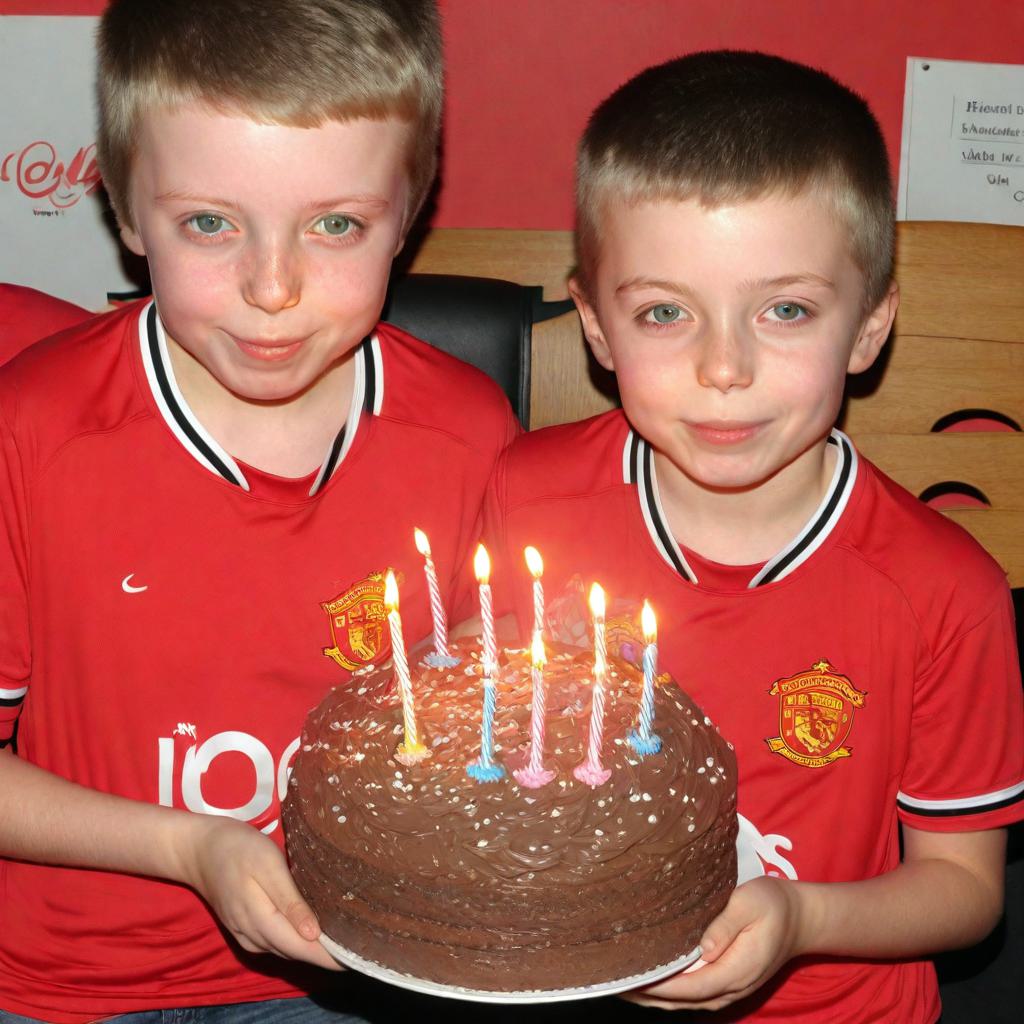}}
        & \raisebox{-0.5\height}{\includegraphics[width=.118\linewidth]{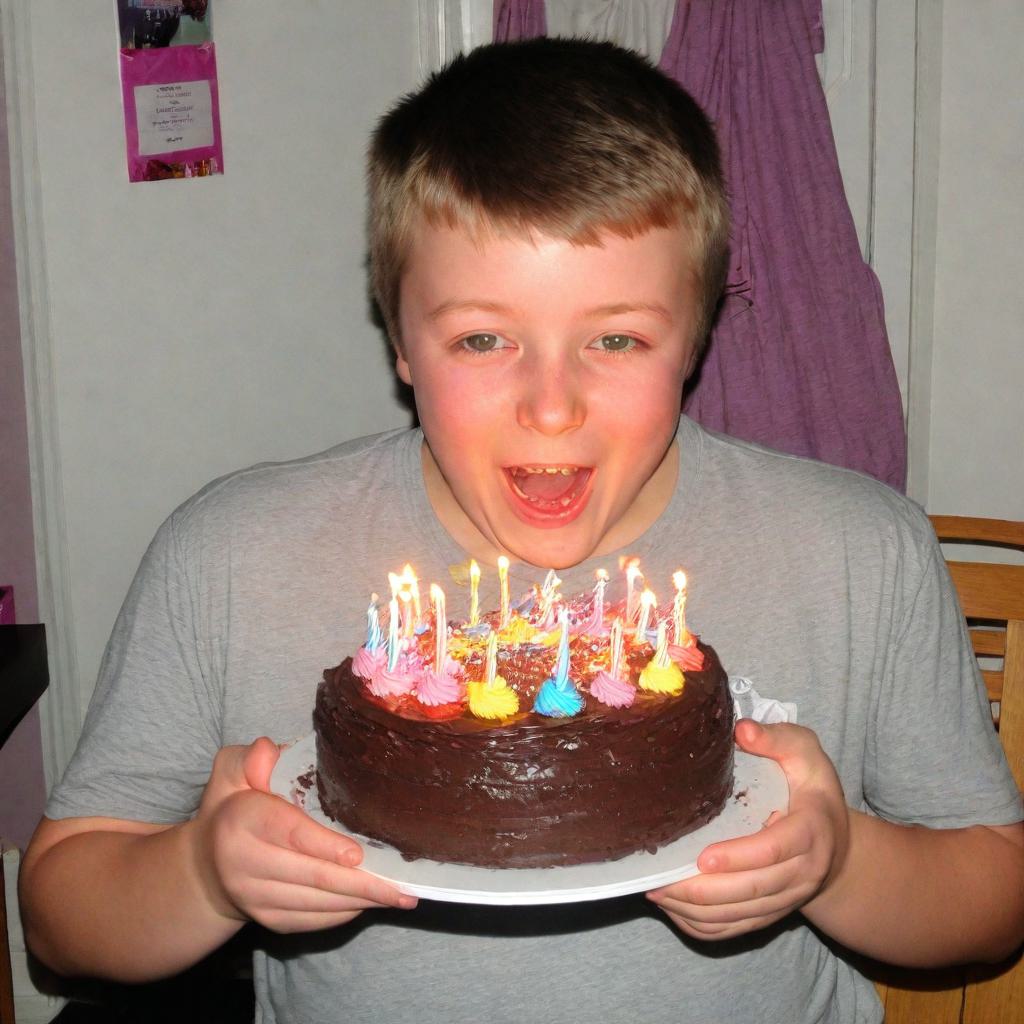}}
        & \raisebox{-0.5\height}{\includegraphics[width=.118\linewidth]{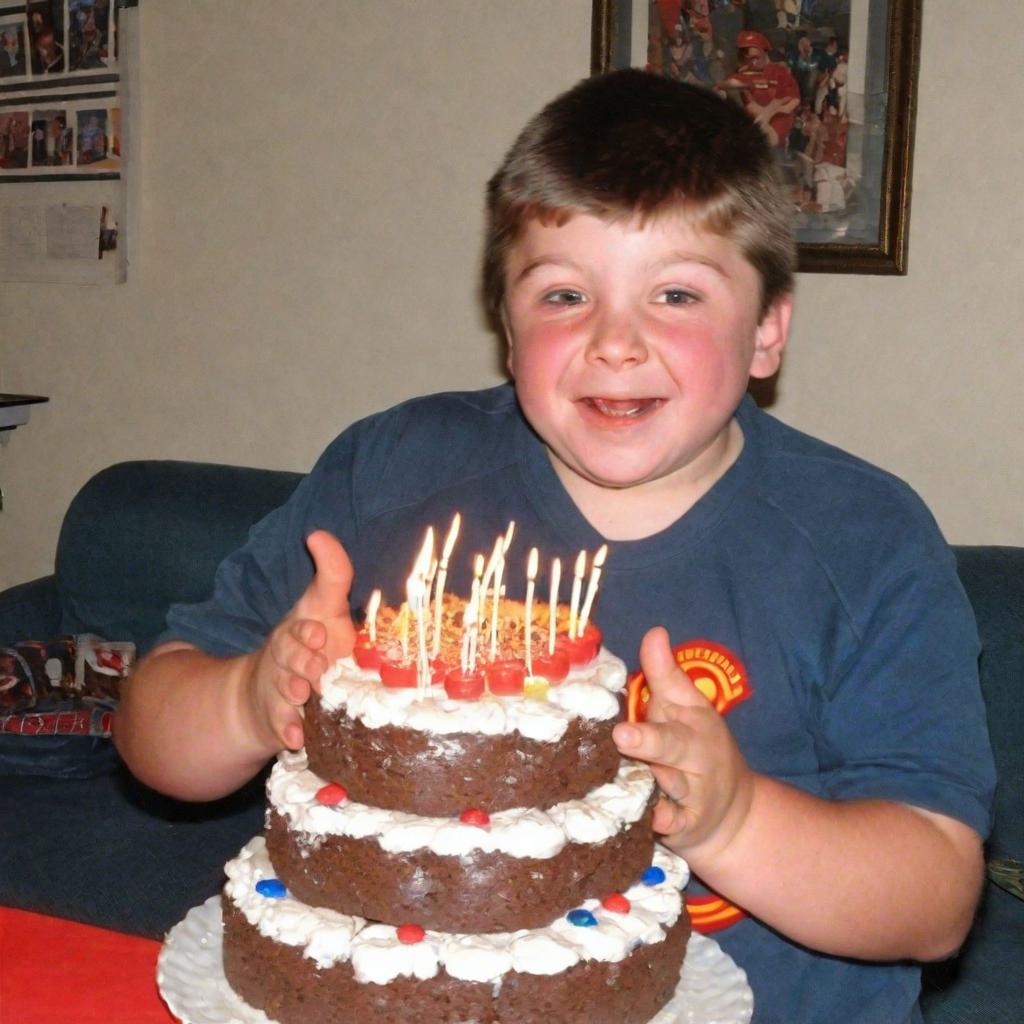}}
        & \raisebox{-0.5\height}{\includegraphics[width=.118\linewidth]{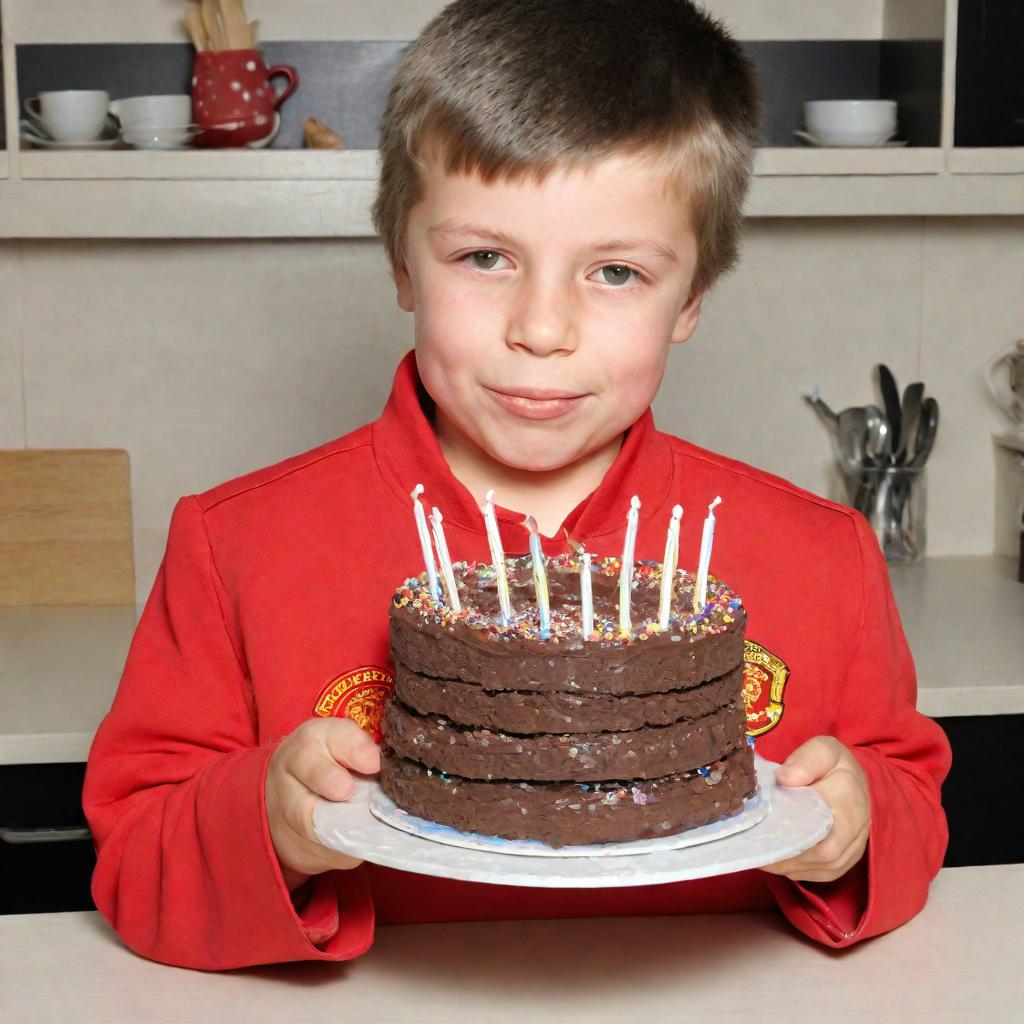}}\\
        \\[-1.2em]
        \multicolumn{8}{c}{blow cake \textrightarrow{} hold cake}\\
        
        \raisebox{-0.5\height}{\includegraphics[width=.118\linewidth]{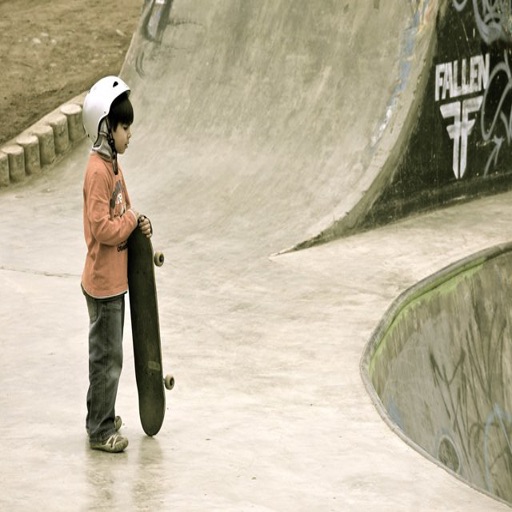}}
        & \raisebox{-0.5\height}{\includegraphics[width=.118\linewidth]{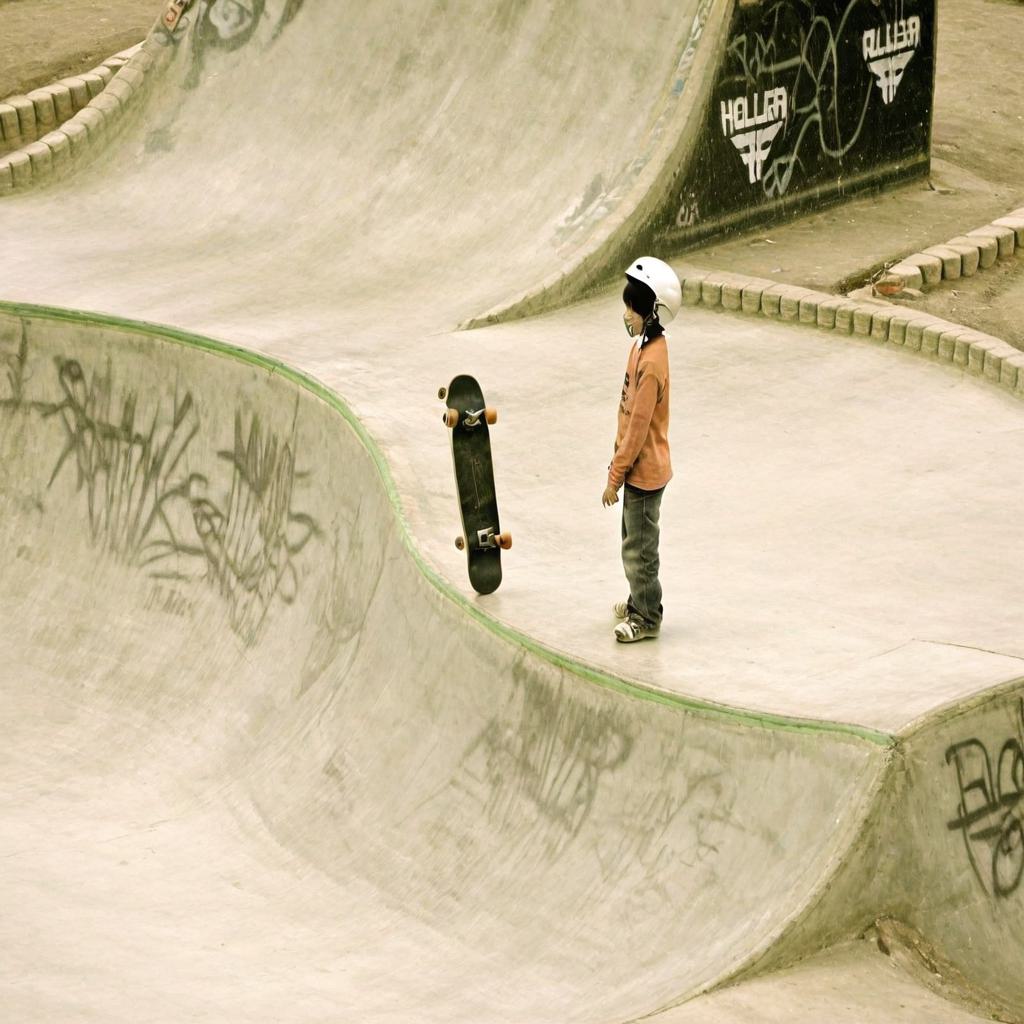}}
        & \raisebox{-0.5\height}{\includegraphics[width=.118\linewidth]{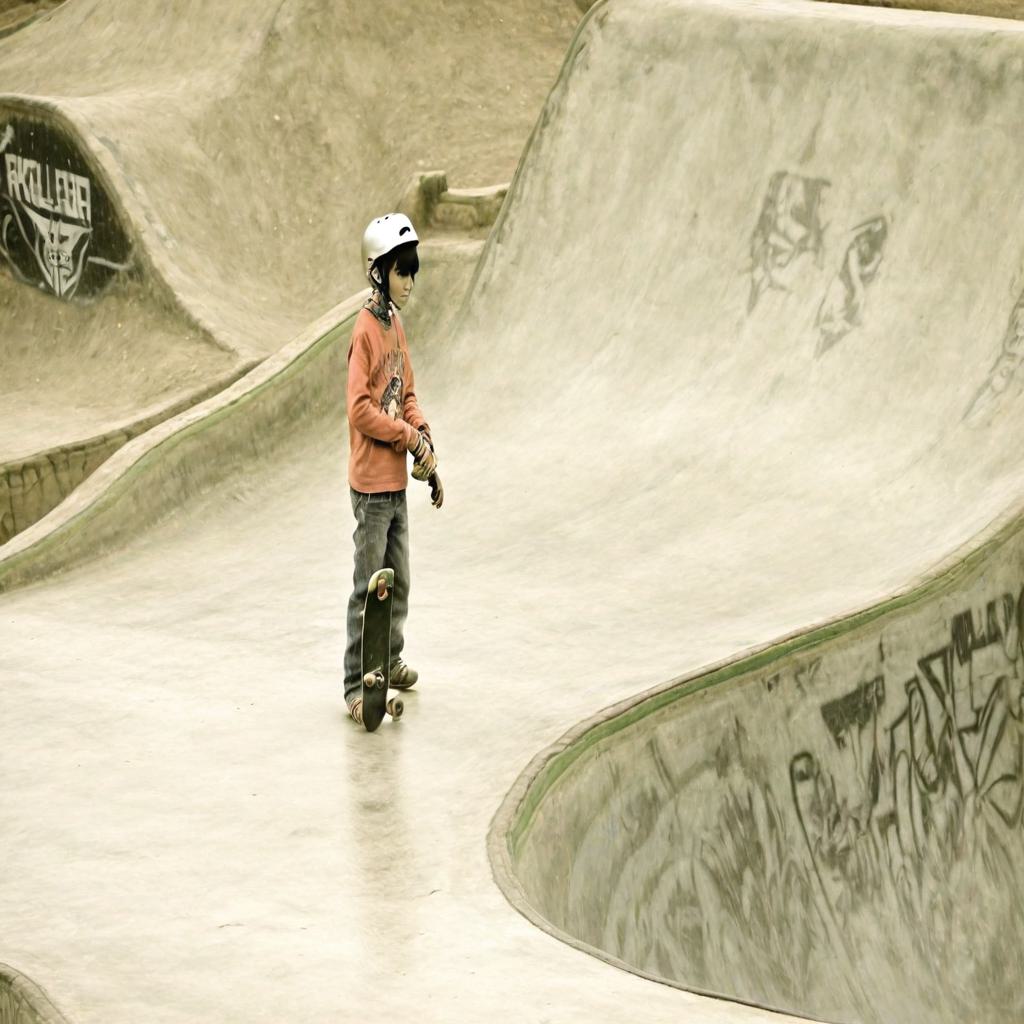}}
        & \raisebox{-0.5\height}{\includegraphics[width=.118\linewidth]{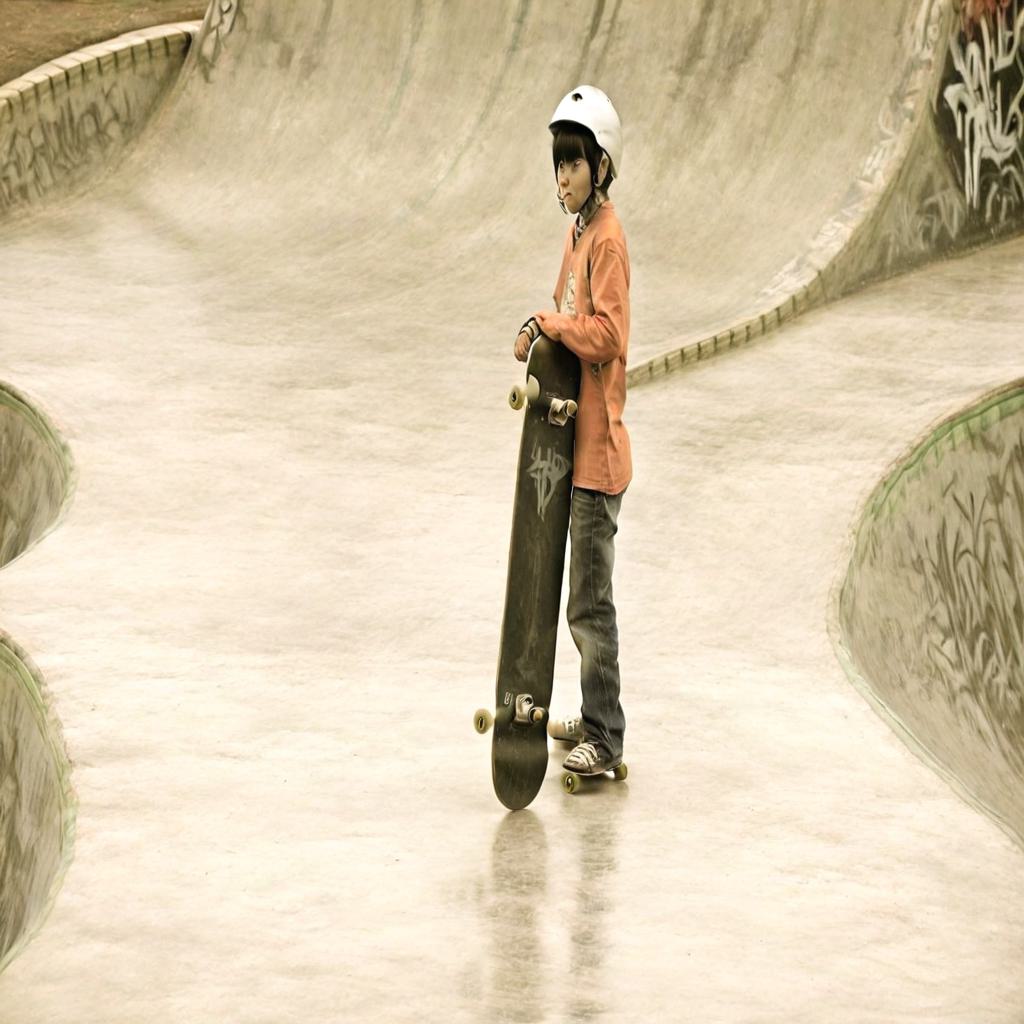}}
        & \raisebox{-0.5\height}{\includegraphics[width=.118\linewidth]{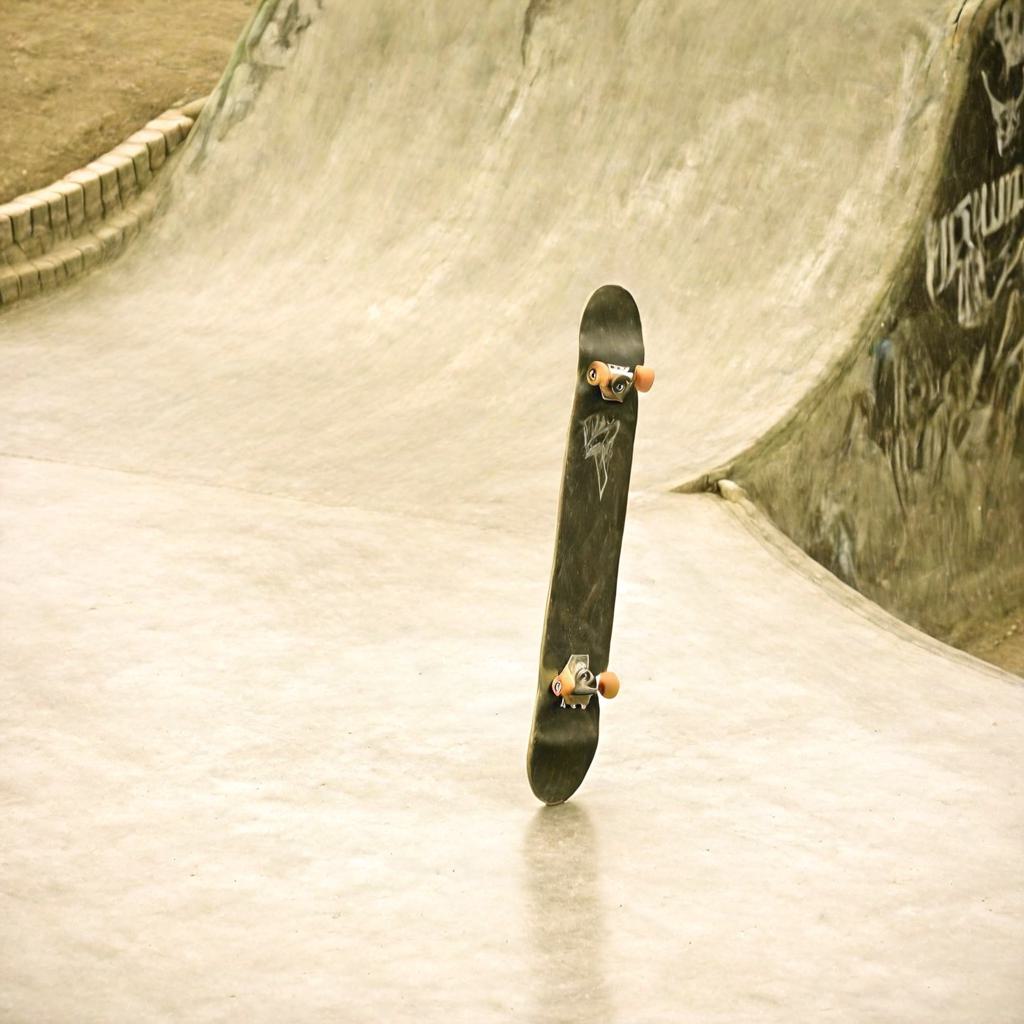}}
        & \raisebox{-0.5\height}{\includegraphics[width=.118\linewidth]{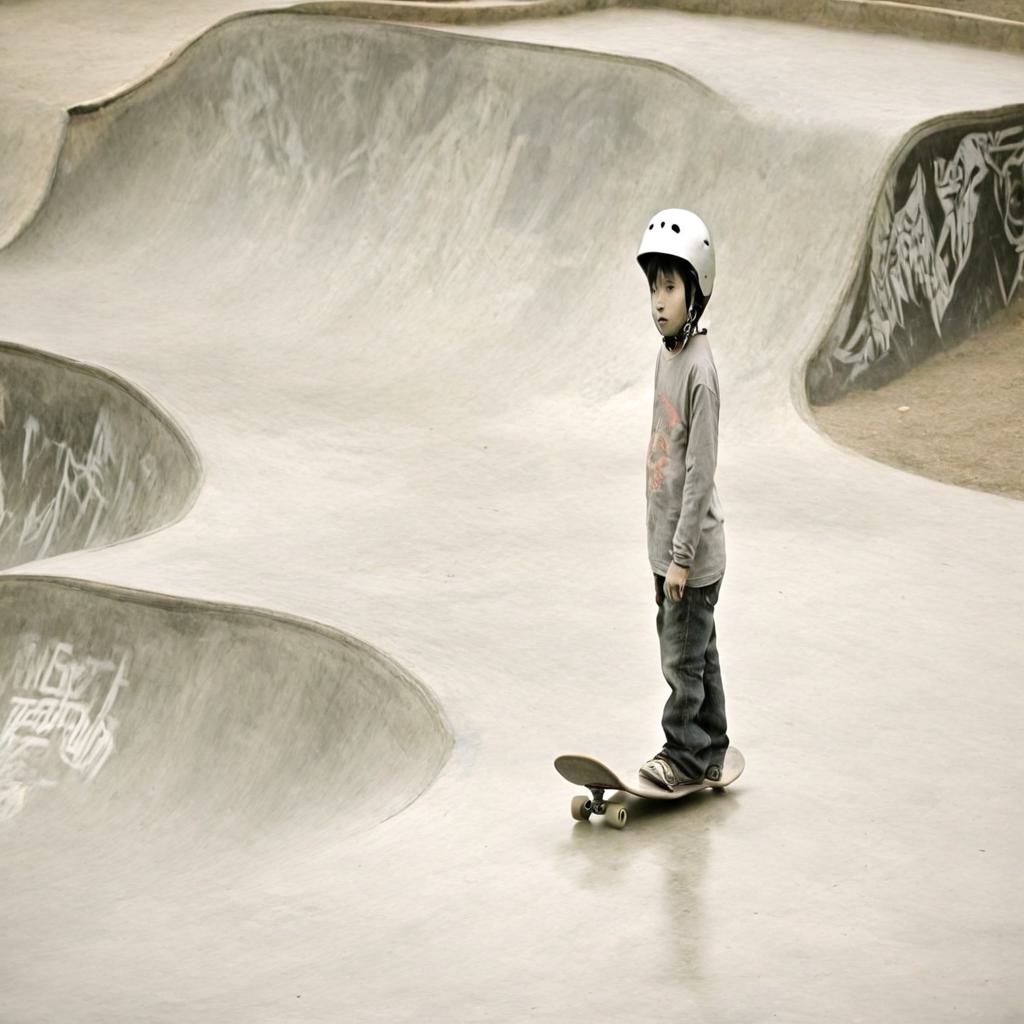}}
        & \raisebox{-0.5\height}{\includegraphics[width=.118\linewidth]{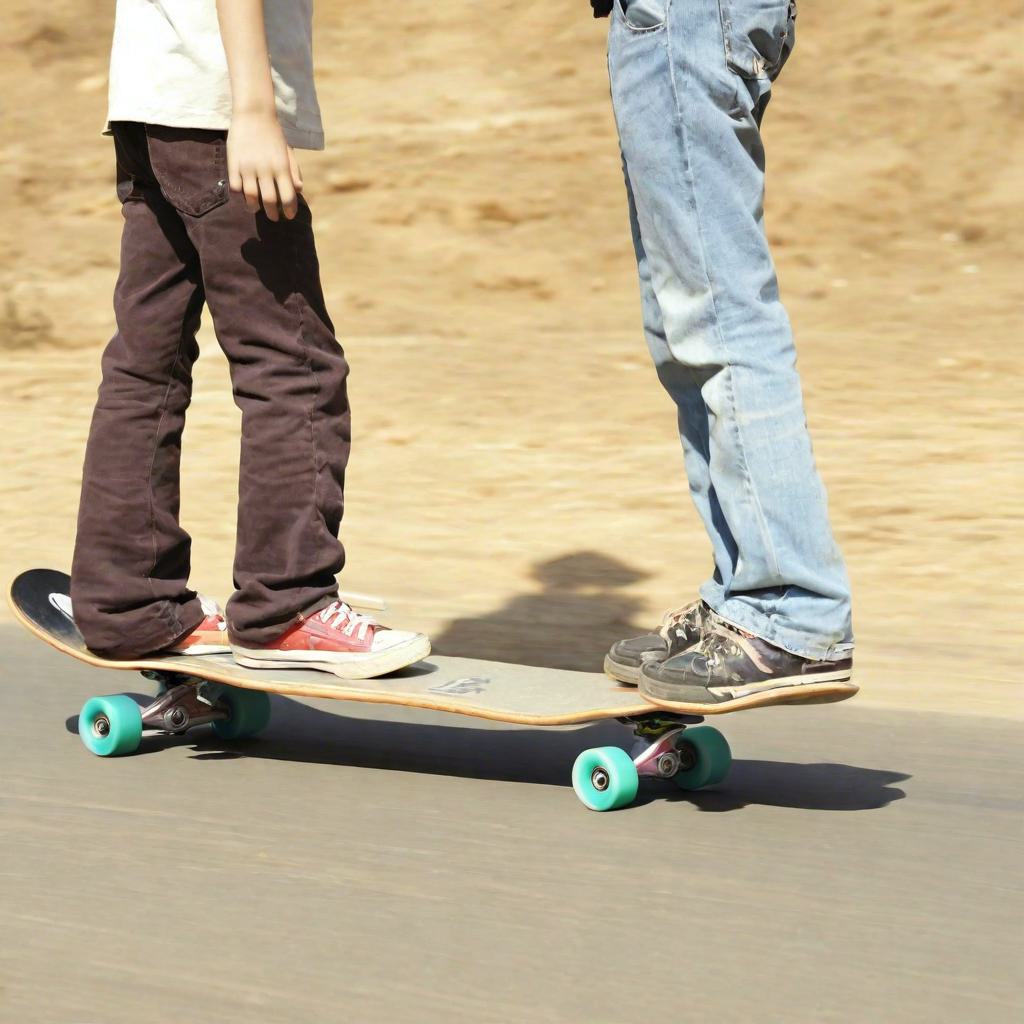}}
        & \raisebox{-0.5\height}{\includegraphics[width=.118\linewidth]{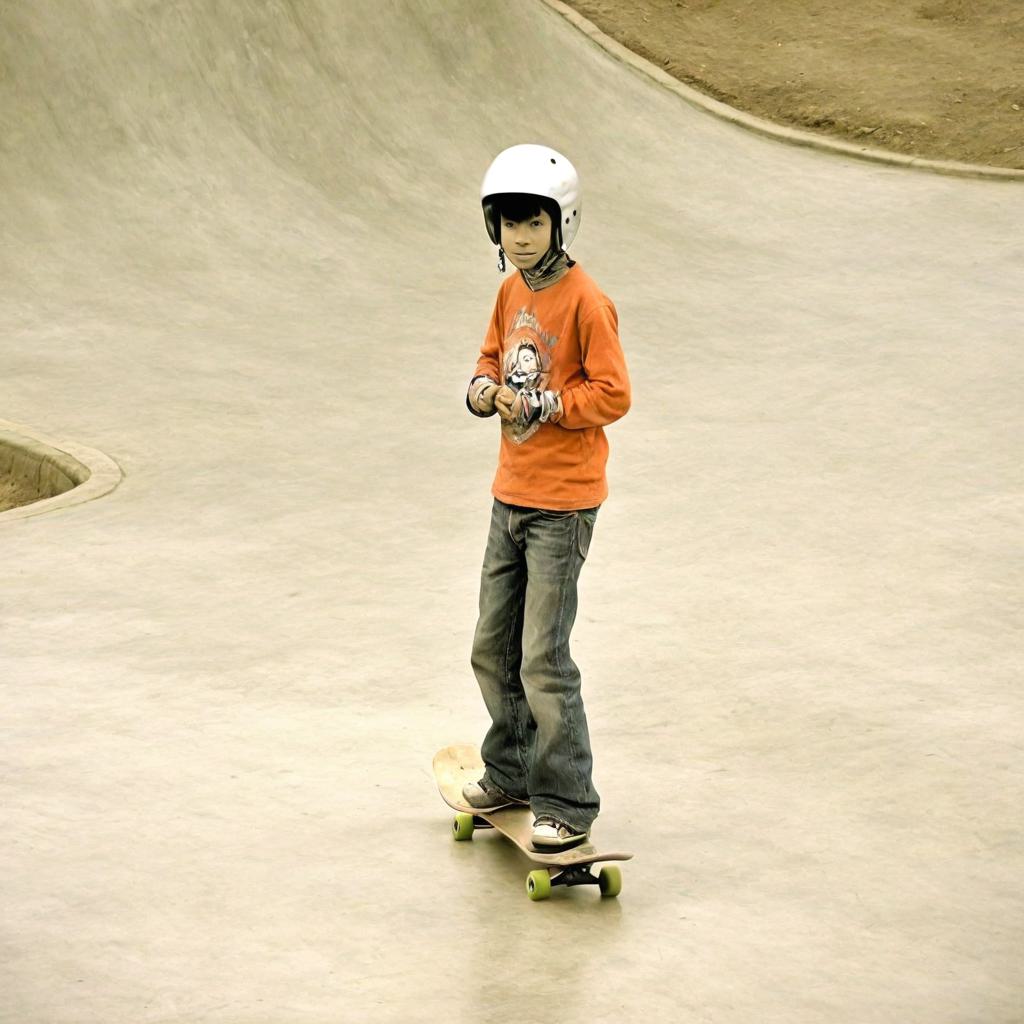}}\\
        \\[-1.2em]
        \multicolumn{8}{c}{hold skateboard \textrightarrow{} ride skateboard}\\
    \end{tabular}
    \caption{Ablation study comparing the effects of various model components on HOI editing performance. Each row shows a source image and the edited outputs for a given interaction edit (e.g., walk dog → hug dog).} 
    \label{fig:qualitative_abl}
\end{figure}

%% file: assets/figure/qkv_ablation.tex
\begin{table}[t]
\caption{Ablation on which cross-attention projections to freeze during Selective Inversion. \checkmark~indicates the projection is \emph{trainable}; $\times$ indicates it is \emph{frozen}. Freezing only $W_Q$ (our Selective Inversion) yields the best overall trade-off.}
\label{tab:qkv-ablation}
\centering
\begin{tabular}{l||ccc||c||ccc}
\hline \hline
\multirow{2}{*}{Configuration} & \multirow{2}{*}{$W_Q$} & \multirow{2}{*}{$W_K$} & \multirow{2}{*}{$W_V$} & \multirow{2}{*}{\text{{\rmfamily\scshape ei}}} & HOI         & Identity & Background \\
                               &                        &                        &                        &                                                          & Editability &          &            \\ \hline \hline
None frozen (Full Inversion)   & \checkmark             & \checkmark             & \checkmark             & 0.496                                                    & 0.449       & 0.555    & 0.634 \\
Freeze $W_K$                   & \checkmark             & $\times$               & \checkmark             & 0.558                                                    & 0.502       & 0.628    & 0.596 \\
Freeze $W_V$                   & \checkmark             & \checkmark             & $\times$               & 0.528                                                    & 0.489       & 0.574    & 0.509 \\
Freeze $W_K, W_V$              & \checkmark             & $\times$               & $\times$               & 0.553                                                    & \textbf{0.540} & 0.566 & 0.515 \\ \hline
Freeze $W_Q$ (Ours)            & $\times$               & \checkmark             & \checkmark             & \textbf{0.573}                                           & 0.514       & \textbf{0.649} & \textbf{0.641} \\ \hline \hline
\end{tabular}%
\end{table}

%% file: assets/figure/rank-reg-sweep.tex
\begin{figure}[!htbp]
    \centering
    \captionsetup[subfigure]{justification=centering}
    \begin{subfigure}[b]{0.25\textwidth}
        \centering
        \includegraphics[width=\textwidth]{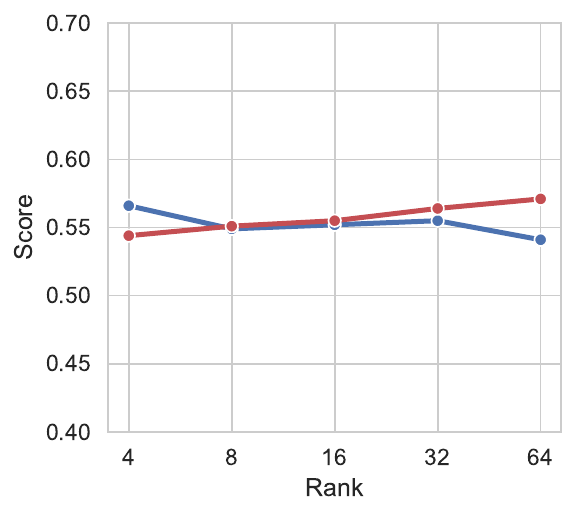}
        \caption{Editability-Identity (\text{\textsc{ei}})}
        \label{fig:rank_reg_sweep_f1}
    \end{subfigure}%
    \begin{subfigure}[b]{0.25\textwidth}
        \centering
        \includegraphics[width=0.983\textwidth]{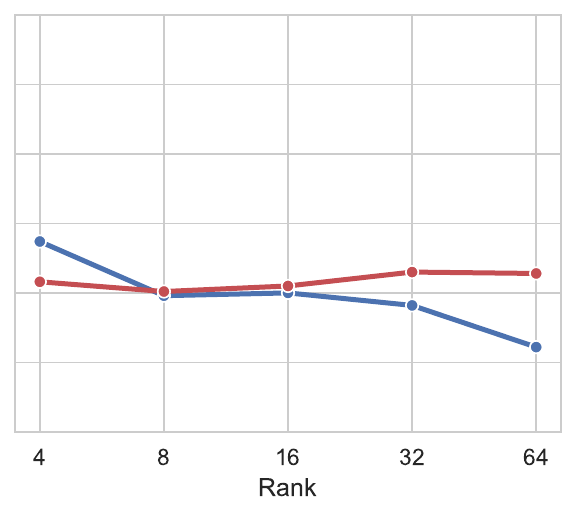}
        \caption{HOI Editability}
        \label{fig:rank_reg_sweep_hoi}
    \end{subfigure}%
    \begin{subfigure}[b]{0.25\textwidth}
        \centering
        \includegraphics[width=0.983\textwidth]{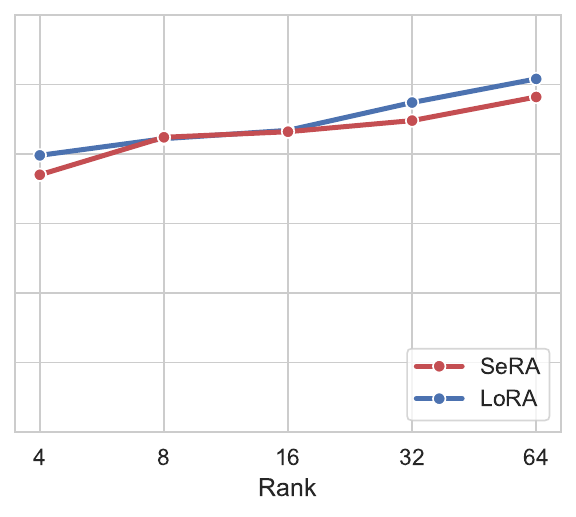}
        \caption{Identity Consistency}
        \label{fig:rank_reg_sweep_id}
    \end{subfigure}%
    \begin{subfigure}[b]{0.25\textwidth}
        \centering
        \includegraphics[width=0.960\textwidth]{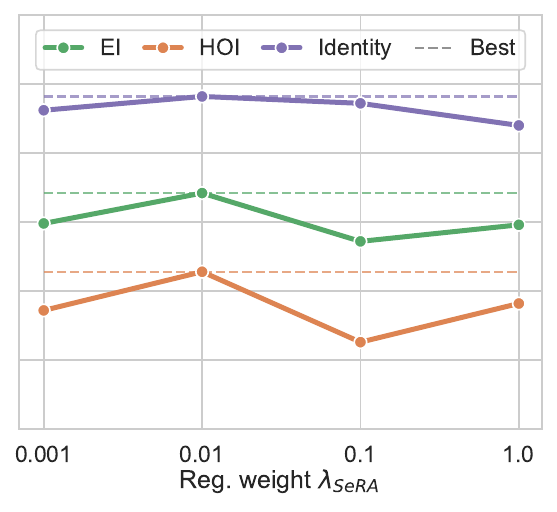}
        \caption{Regularization strength}
        \label{fig:rank_reg_sweep_reg}
    \end{subfigure}
    \caption{Sensitivity of {InteractEdit} to rank and SeRA regularization on \iebench. Panels (a)--(c) sweep the adapter rank $r \in \{4, 8, 16, 32, 64\}$ for both vanilla LoRA (blue) and SeRA (red), comparing (a) Editability-Identity (EI), (b) HOI Editability, and (c) Identity Consistency. 
    Panel (d) sweeps the SeRA uncertainty-loss weight $\lambda_{\textrm{SeRA}} \in \{0.001, 0.01, 0.1, 1.0\}$ on a log axis, with the per-metric best values marked as dashed lines: all three metrics peak at the default $\lambda_{\textrm{SeRA}} = 0.01$.}
    \label{fig:rank_reg_sweep}
\end{figure}

%% file: assets/figure/per-action.tex
\begin{table}[t]
  \centering
  \caption{Per-action disaggregation of HOI Editability on \iebenchext, sorted by HE (ascending). $N$ is the number of (image, edit) pairs. HE: HOI Editability.}
  \label{tab:he-per-action}
  \begin{tabular}{l r c c c}
    \toprule
    Target verb & $N$ & \text{{\rmfamily\scshape ei}} & HE & Identity \\
    \midrule
    groom & 6 & 0.000 & 0.000 & 0.207 \\
    open & 3 & 0.000 & 0.000 & 0.222 \\
    fill & 4 & 0.143 & 0.100 & 0.248 \\
    make & 14 & 0.140 & 0.114 & 0.180 \\
    stand on & 3 & 0.172 & 0.133 & 0.242 \\
    feed & 24 & 0.209 & 0.175 & 0.259 \\
    cut & 20 & 0.213 & 0.210 & 0.217 \\
    drink with & 7 & 0.230 & 0.229 & 0.231 \\
    sip & 4 & 0.203 & 0.250 & 0.172 \\
    wash & 26 & 0.279 & 0.323 & 0.245 \\
    walk & 15 & 0.297 & 0.333 & 0.268 \\
    catch & 12 & 0.321 & 0.417 & 0.261 \\
    smell & 6 & 0.276 & 0.433 & 0.203 \\
    eat & 28 & 0.282 & 0.471 & 0.201 \\
    kick & 5 & 0.329 & 0.520 & 0.240 \\
    kiss & 18 & 0.311 & 0.567 & 0.214 \\
    eat at & 1 & 0.447 & 0.600 & 0.356 \\
    lie on & 14 & 0.458 & 0.614 & 0.365 \\
    hit & 10 & 0.354 & 0.620 & 0.247 \\
    jump & 7 & 0.502 & 0.657 & 0.406 \\
    sit on & 11 & 0.497 & 0.673 & 0.394 \\
    pick up & 8 & 0.279 & 0.675 & 0.176 \\
    hold & 32 & 0.343 & 0.750 & 0.222 \\
    sit at & 1 & 0.436 & 0.800 & 0.300 \\
    carry & 4 & 0.374 & 0.850 & 0.240 \\
    dribble & 12 & 0.362 & 0.900 & 0.226 \\
    hug & 12 & 0.431 & 0.950 & 0.279 \\
    ride & 8 & 0.539 & 0.975 & 0.372 \\
    throw & 10 & 0.437 & 0.980 & 0.281 \\
    read & 1 & 0.406 & 1.000 & 0.255 \\
    \midrule
    \textbf{Average} & 326 & \textbf{0.332} & \textbf{0.498} & \textbf{0.250} \\
    \bottomrule
  \end{tabular}
\end{table}

%% file: assets/figure/per-object.tex
\begin{table}[t]
  \centering
  \caption{Per-object disaggregation of HOI Editability on \iebenchext, sorted by HE (ascending). Columns as in \Cref{tab:he-per-action}.}
  \label{tab:he-per-object}
  \begin{tabular}{l r c c c}
    \toprule
    Object & $N$ & \text{{\rmfamily\scshape ei}} & HE & Identity \\
    \midrule
    bottle & 6 & 0.189 & 0.167 & 0.219 \\
    cup & 16 & 0.207 & 0.200 & 0.214 \\
    banana & 10 & 0.269 & 0.240 & 0.307 \\
    cake & 48 & 0.221 & 0.296 & 0.177 \\
    chair & 9 & 0.292 & 0.311 & 0.274 \\
    dog & 26 & 0.290 & 0.369 & 0.239 \\
    broccoli & 8 & 0.289 & 0.400 & 0.226 \\
    horse & 54 & 0.319 & 0.411 & 0.261 \\
    cat & 20 & 0.296 & 0.420 & 0.228 \\
    pizza & 8 & 0.259 & 0.450 & 0.182 \\
    surfboard & 3 & 0.395 & 0.467 & 0.342 \\
    dining table & 2 & 0.447 & 0.700 & 0.328 \\
    skateboard & 15 & 0.510 & 0.733 & 0.391 \\
    sports ball & 60 & 0.373 & 0.743 & 0.249 \\
    snowboard & 4 & 0.493 & 0.750 & 0.368 \\
    bench & 12 & 0.551 & 0.783 & 0.425 \\
    donut & 12 & 0.292 & 0.800 & 0.178 \\
    sandwich & 3 & 0.327 & 0.800 & 0.206 \\
    couch & 2 & 0.408 & 0.800 & 0.274 \\
    book & 6 & 0.422 & 0.867 & 0.279 \\
    teddy bear & 2 & 0.390 & 1.000 & 0.243 \\
    \midrule
    \textbf{Average} & 326 & \textbf{0.332} & \textbf{0.498} & \textbf{0.250} \\
    \bottomrule
  \end{tabular}
\end{table}

%% file: sec/5_conclusion.tex
\section{Conclusion}
This paper introduces a novel reference-free interaction editing task and proposes \textsc{InteractEdit} to tackle its challenges. HOI editing remains a formidable problem, requiring interaction modifications while preserving identity. Existing methods struggle to balance HOI editability and identity consistency, as some overfit to the source image and overwrite pretrained interaction priors, while others fail to effectively leverage these priors, leading to suboptimal edits. To address this, \textsc{InteractEdit} employs Selective-Rank Adaptation (SeRA) to regularize inversion and mitigate overfitting, alongside a \emph{Selective Inversion} strategy to preserve target interaction knowledge while adapting appearance from the source image. Extensive quantitative and qualitative evaluations demonstrate that our method significantly outperforms state-of-the-art approaches, setting a new benchmark for HOI editing.

\subsection{Failure Cases and Limitations}\label{subsec:failure-limitations}
\input{assets/figure/failure}
While \textsc{InteractEdit} advances reference-free HOI editing, we identify four main failure modes, three of which are illustrated in Fig.~\ref{fig:failure}.
First, \emph{very small subjects or objects} (Fig.~\ref{fig:failure_smallobj}) occupy only a handful of tokens in the cross-attention maps, so the identity tokens receive weak supervision over that region and the inverted appearance is poorly reconstructed in the edited output.
Second, \emph{rare verb--object compositions} (Fig.~\ref{fig:failure_rarehoi}) that are underrepresented in the pretrained diffusion model cannot be reliably synthesised: because our framework is reference-free, it relies entirely on the base model's generative prior for the target interaction, and rare compositions such as \textit{make a cake} fall outside this prior.
Third, \emph{extreme non-rigid changes} (Fig.~\ref{fig:failure_extremeviewpointchange}), such as transitions between drastically different viewpoints or body poses (e.g., front-view \textit{eat} to side-view \textit{cut}), push the inverted tokens beyond the appearance distribution they were optimised on, leading to subtle drift in subject or object identity.
Fourth, \emph{simultaneous multi-HOI editing} is not supported by our current design: inversion is performed per subject--object--background triplet, so scenes containing multiple interacting pairs must be edited one pair at a time rather than jointly in a single pass.
We view these limitations as natural directions for future work, including higher-resolution attention conditioning for small instances, lightweight retrieval-based priors for rare compositions, and a multi-HOI inversion formulation for joint multi-HOI editing.


%% file: assets/figure/failure.tex
\begin{figure}[htbp]
	\centering
    \captionsetup[subfigure]{justification=centering}
	\begin{subfigure}[b]{0.32\textwidth}
		\centering
		\includegraphics[width=\textwidth]{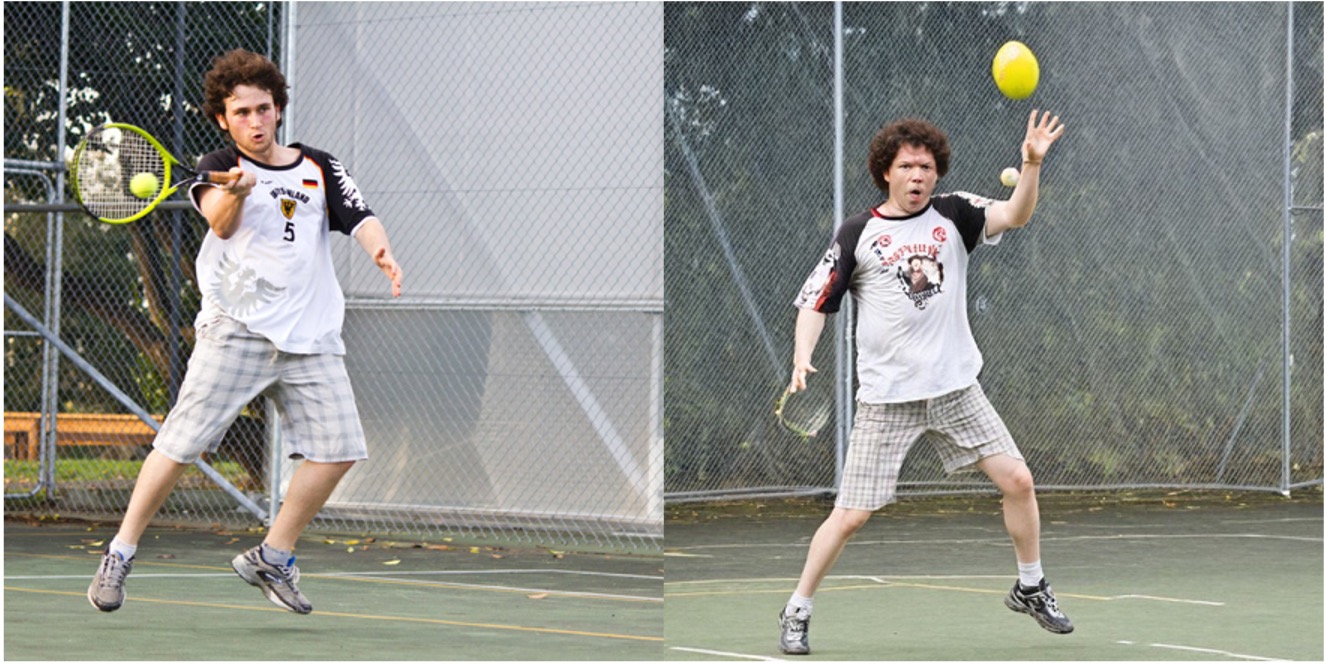}
		\caption{Very small object\\(hit->catch tennis ball)}
		\label{fig:failure_smallobj}
	\end{subfigure}%
	\hfill
	\begin{subfigure}[b]{0.32\textwidth}
		\centering
		\includegraphics[width=\textwidth]{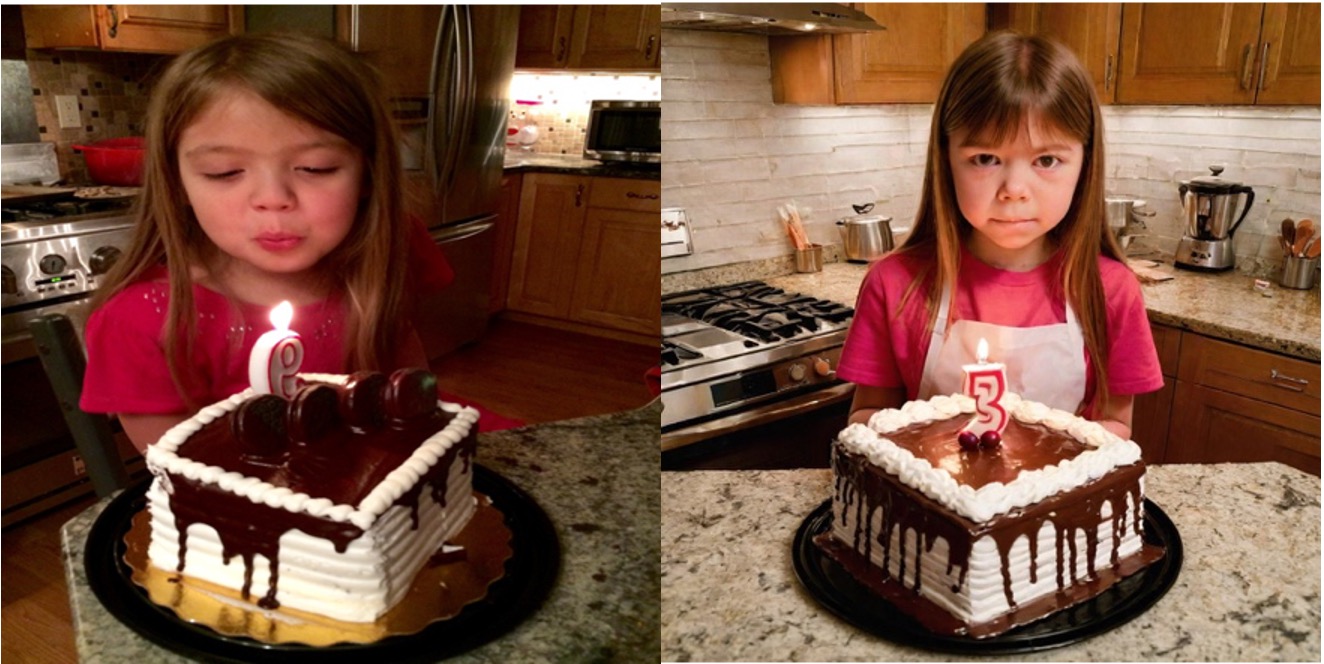}
		\caption{Rare verb--object composition\\(\textit{blow}$\rightarrow$\textit{make} cake)}
		\label{fig:failure_rarehoi}
	\end{subfigure}%
	\hfill
	\begin{subfigure}[b]{0.32\textwidth}
		\centering
		\includegraphics[width=\textwidth]{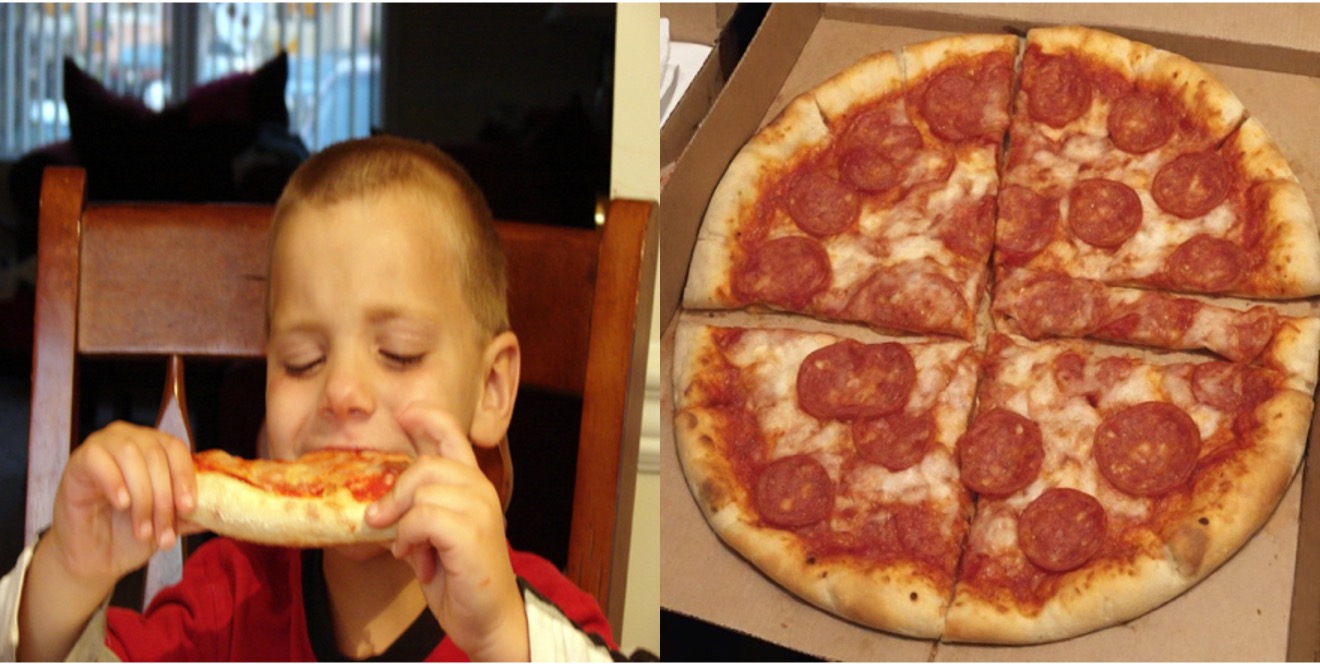}
		\caption{Extreme viewpoint change\\(\textit{eat}$\rightarrow$\textit{cut} pizza)}
		\label{fig:failure_extremeviewpointchange}
	\end{subfigure}
	\caption{Representative failure cases of {InteractEdit}. (a) When the target object occupies only a few pixels (e.g., a tennis ball), the limited spatial resolution of cross-attention maps provides weak supervision for the identity tokens, resulting in degraded object reconstruction. (b) For rare verb--object compositions that are underrepresented in the pretrained diffusion model (e.g., \textit{make a cake}), the reference-free setting cannot reliably synthesize the target interaction. (c) When the target interaction requires very large viewpoint or pose changes (e.g., from a front-facing \textit{eat} to a side-view \textit{cut} pose), the inverted tokens may fail to generalize, causing subtle drift in subject or object appearance.}
	\label{fig:failure}
\end{figure}

%% file: sec/X_suppl.tex

  
  


\input{sec/B_implementations}

\section{CLIP-T Sensitivity Analysis}\label{sec:clip-score}
To systematically evaluate the sensitivity of CLIP-T to actions (verbs) versus objects (nouns), we conducted an analysis on 1,911 images sampled from the HICO-DET test set. To avoid confounding factors, we filtered the dataset to include only images containing a single HOI instance. For each image, we generated mismatched prompts in two ways: (i) by replacing the ground-truth action with all other possible verbs applicable to the same object (verb mismatch), and (ii) by replacing the ground-truth object with all other possible nouns for the same action (noun mismatch). This yielded a total of 50,239 mismatched prompts: 36,872 object mismatches and 13,367 action mismatches. We then computed the cosine similarity between each mismatched prompt and the image embedding using CLIP, and compared it against the similarity score for the correct image–prompt pair. The score difference reflects CLIP-T’s ability to discriminate mismatches. Our analysis shows that verb mismatches produce extremely small margins clustered near zero, whereas noun mismatches yield much broader margins, confirming that CLIP is primarily sensitive to objects but weak at distinguishing actions.

\section{Theoretical Proofs}
\subsection{Capacity Control and Prior Preservation}

\begin{lemma}[Logit-Subspace Invariance under Selective Inversion]
\label{lem:logit_invariance}
Let the pretrained query matrix be \(W_Q^{(0)}\in\mathbb{R}^{d\times d}\) and define
the query-feature map \(\Phi(X)=X W_Q^{(0)}\).
If, during inversion, \(W_Q\) is kept fixed
\(\bigl(W_Q\equiv W_Q^{(0)}\bigr)\) while \(W_K\) is
updated arbitrarily, then for every input \(X\)
the attention-score matrix
\[
S(X) \;=\; \tfrac{1}{\sqrt d}\,\Phi(X)\, (X W_K)^{\!\top}
\]
has its \emph{row-space} contained in
\(\operatorname{row}\!\bigl(\Phi(X)\bigr)\),
which is exactly the row-space learned by the pretrained model.
Consequently, all edited images lie in the same
interaction-prior subspace as the original network.
\end{lemma}

\begin{proof}
Because \(\Phi(X)=X W_Q^{(0)}\) is fixed,
every row of \(S(X)\) is a linear combination of the rows of
\(\Phi(X)\) (scalar weights given by the corresponding columns of
\(X W_K\)), hence
\(\operatorname{row}(S(X))\subseteq\operatorname{row}(\Phi(X))\),
independent of the choice of \(W_K\).
\end{proof}

\section{More Qualitative Results}\label{sec:more-qualitative}
In \cref{fig:qualitative_3,fig:qualitative_5}, we provide a more visual comparison on same source images as in \cref{fig:qualitative} for methods that are not in the main paper. 
\cref{fig:qualitative_4} extends the comparison to additional source images. Compared to other methods, our method successfully enables interaction edits while maintaining the identity. 

\begin{figure}
    \centering
    \includegraphics[width=\linewidth]{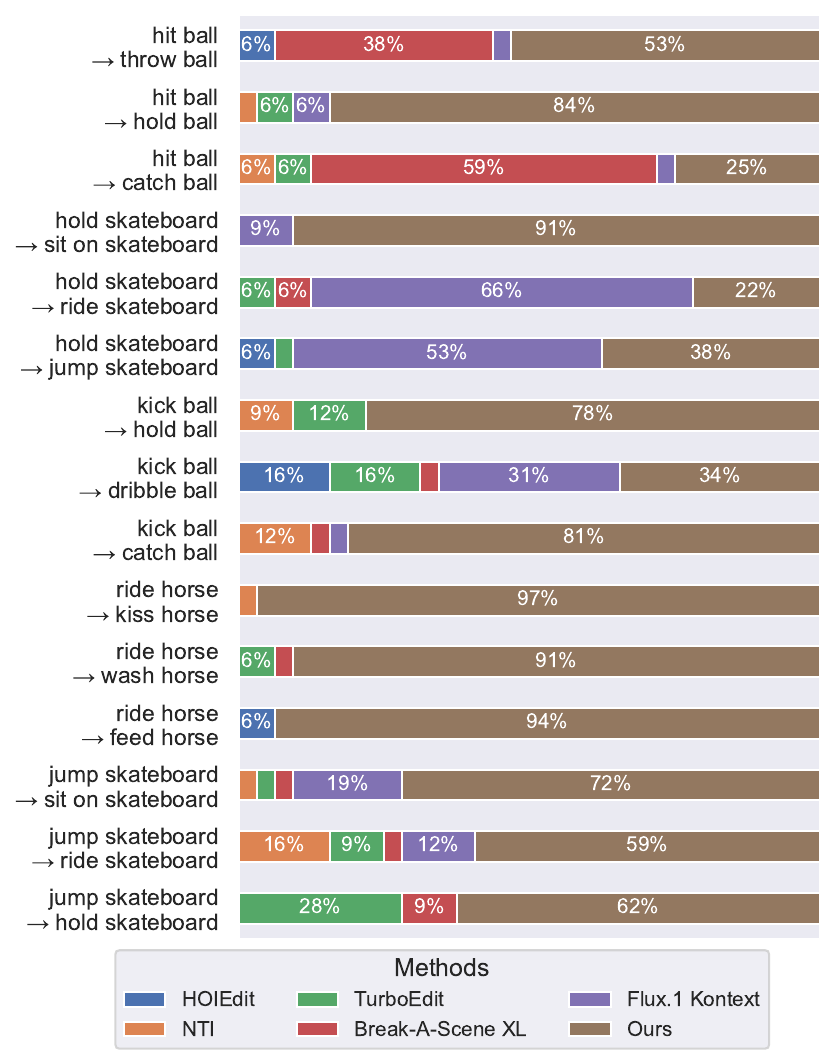}
    \caption{User preferences for each interaction edit.}
    \label{fig:userstudy-edits}
\end{figure}

\input{assets/figure/ab_qual_5}
\input{assets/figure/ab_qual_3}
\input{assets/figure/ab_qual_4}

\section{User Study Details}\label{sec:userstudy-details}

To assess user preferences, we conducted an online survey where participants were instructed to evaluate the edits based on three criteria: (1) adherence to the target interaction—how well the edit reflects the intended action, (2) identity consistency—whether the subject and object retain their original appearance, and (3) realism—how natural and visually plausible the edit appears.

In our user study, participants evaluated edits ranging from subtle action transitions to significant pose and structural modifications. Complex edits such as \textit{jump $\rightarrow$ hold/ride/sit on skateboard} require adjusting both the subject’s pose and the spatial relationship with the object, making them particularly challenging for existing methods. Similarly, \textit{ride $\rightarrow$ feed/wash/kiss horse} involve close-contact interactions that demand precise alterations to the subject’s positioning while preserving identity consistency.

\cref{fig:userstudy-edits} provides insights into user preferences for different interaction edits. Across 12 out of 15 interaction edits, our method is the most preferred. In some more challenging edits, such as \textit{hold skateboard $\rightarrow$ sit on skateboard} and \textit{ride horse $\rightarrow$ kiss/wash/feed horse}, our method significantly outperforms others, due to its ability to modify interaction while preserving identity.

Edits involving a sports ball, such as \textit{kick $\rightarrow$ catch/dribble/hold ball} assess a model’s ability to modify hand and foot placement while maintaining a natural transition between actions. Similarly, \textit{hold $\rightarrow$ jump/ride/sit on skateboard} require adapting the subject’s posture while ensuring the skateboard remains a consistent object in the scene. Finally, hand-object interaction edits like \textit{hit $\rightarrow$ catch/hold/hit ball} evaluate whether the model can effectively modify the action without introducing artifacts or unnatural grasp on the object. These diverse edits allow us to assess how well different methods handle a range of interaction modifications, from subtle changes to full-body pose adjustments.

%% file: sec/B_implementations.tex
\section{Implementation Details}\label{sec:implementation-details}
\noindent\textbf{Training Setup} Our model was trained on NVIDIA 6000 ADA 48GB, with baseline evaluations partially conducted on NVIDIA A100 SXM 80GB. 
In InteractEdit, both stages utilize Adam optimizer \cite{adam2014} with a batch size of 1 and weight decays of 1e-4. During the inference stage, we set the number of denoising steps to $T=50$.

\input{assets/figure/cost}

\subsection{Resources Analysis and Computational Efficiency}\label{subsec:resource-analysis}
To analyze the resource usage of each method, we evaluate the time and memory consumption, as detailed in \cref{tab:cost}. All experiments are conducted using a single NVIDIA A100 SXM4 80GB GPU, and the analysis is based on the resource usage required to complete \iebench. For most image editing methods, the total time taken is the sum of the time spent on source image inversion (28 inversions) and generating the edited images (1000 edits). 

However, some methods deviate from this workflow:
\begin{itemize}
    \item \textbf{HOIEdit}: Requires training an interaction token for each action. For IEBench, with 18 different target actions, the training process takes a total of 58,320 seconds.
    \item \textbf{Imagic}: Performs inversion using both the target prompt and source images, requiring 100 inversions for \iebench.
    \item \textbf{DAC}: Involves two types of inversion: Induction-1 (once per source image, 28 inversions) and Induction-2 (per source-target pair, 100 inversions, see \texttt{Others} column in \cref{tab:cost}).
    \item \textbf{InstructPix2Pix}: Requires extensive pretraining on a large instruction-image pair dataset. The pretraining dataset\footnote{\code{timbrooks/instructpix2pix-clip-filtered}}, contains 313010 edit instruction-image pairs. InstructPix2Pix was trained for 15,000 steps using 8x NVIDIA A100 GPUs with a batch size of 32 \footnote{\url{https://huggingface.co/diffusers/sdxl-instructpix2pix-768}}.
    \item \textbf{ReNoise}: Performs inversion for each generated image, resulting in 1000 inversions for IEBench.
    \item \textbf{TurboEdit}: Integrates inversion into the editing process.
\end{itemize}

Due to the inherent properties of latent inversion, some methods (Null-Text Inversion with Prompt-to-Prompt, PnP Diffusion, MasaCtrl, HOIEdit, LEDITS++, and Edit-Friendly DDPM) initialize the editing from a fixed inverted noise rather than random noise, resulting in deterministic editing outcomes with no variation across multiple generations. Nevertheless, to ensure a fair comparison with other methods, we report the time required to generate 10 edits, maintaining consistency with the evaluation protocol.

Our proposed method, InteractEdit, achieves comparable computational efficiency to existing methods while significantly enhancing HOI editability. Unlike HOIEdit, which relies on costly interaction token training, InteractEdit operates in a zero-shot manner without requiring interaction inversion. Furthermore, InteractEdit requires less than 15GB of GPU memory, making it runnable for mainstream GPUs.

\subsection{Baseline Implementations}\label{subsec:baseline-implementations}
To facilitate fair comparisons in the HOI editing task, we re-implemented existing methods as part of our benchmark. This re-implementation aims to standardize evaluation practices and support future research efforts in this area. The re-implemented methods will be made open-source alongside our benchmark. In this section, we provide detailed descriptions of how each method was implemented. 

\noindent\textbf{Null-Text Inversion \cite{mokady2023nti}} is implemented with Prompt-to-Prompt \cite{hertz2022p2p}, following the official code \footnote{\url{https://github.com/google/prompt-to-prompt}}. Null-Text Inversion enables real-image inversion, and the editing process is performed with Prompt-to-Prompt. For inversion, we use a null inversion prompt (``"), while for editing, the target prompt follows the format: ``a photo of [subject] [target interaction] [object]". We use Stable Diffusion v1.5 with a guidance scale of 7.5 and 50 sampling steps.

\noindent\textbf{PnP Diffusion \cite{tumanyan2023pnp}} is implemented following the official diffusers code \footnote{\url{https://github.com/MichalGeyer/pnp-diffusers}}. We adhere to the default hyperparameters: $\tau_f=0.8,\tau_A=0.5,\text{timesteps}=50$ and use Stable Diffusion 2.1. For inversion, we use a null inversion prompt (``"), while for editing, the target prompt follows the format: ``a photo of [subject] [target interaction] [object]".

\noindent\textbf{MasaCtrl \cite{cao2023masactrl}} is implemented following the official code \footnote{\url{https://github.com/TencentARC/MasaCtrl}}. We use Stable Diffusion v1.4 as we are unable to reproduce the method on Stable Diffusion XL. We follow the default hyperparameters: $S=4,L=10$. For inversion, we use a null inversion prompt (``"), while for editing, the target prompt follows the format: ``a photo of [subject] [target interaction] [object]".

\noindent\textbf{HOIEdit \cite{xu2025hoiedit}} is implemented following the official code \footnote{\url{https://github.com/Kenneth-Wong/hoiedit}}. We use Stable Diffusion v1.4 and follow the default hyperparameters: $S=4, L=10$. Following the paper, for inversion, we use a null inversion prompt (``"), while for editing, the target prompt follows the format: ``[subject] \textless{}R\textgreater{} [object]", which yields the best outcome in preserving identity while accurately modifying the interaction, where \textless{}R\textgreater{} is trained interaction token using ReVersion \cite{huang2023reversion}. To train the interaction token for all 25 actions in \iebench, we randomly select 10 images as the training images.

\noindent\textbf{Imagic \cite{kawar2023imagic}} is implemented following the community-provided code \footnote{\url{https://github.com/huggingface/diffusers/blob/main/examples/community/imagic_stable_diffusion.py}}. We use Stable Diffusion v1.5 and adopt the default hyperparameters from the community code: $\eta=1.2$, Although this value differs from the official implementation, we find it produces better results in our experiments. For editing, the target prompt is formatted as: ``a photo of [subject] [target interaction] [object]".

\noindent\textbf{CDS \cite{Nam2024cds}} is implemented following the official code \footnote{\url{https://github.com/HyelinNAM/ContrastiveDenoisingScore}}. We use Stable Diffusion v1.5 and follow the default hyperparameters: $\ell_\text{con}=3.0, \ell_\text{DDS}=1.0$. For inversion, we use a source prompt which follows the format: ``a photo of [subject] [source interaction] [object]", while for editing, the target prompt follows the format: ``a photo of [subject] [target interaction] [object]".

\noindent\textbf{DAC \cite{song2024dac}} is implemented following the official code \footnote{\url{https://github.com/xuesong39/DAC}}. We use Stable Diffusion v2.1 and follow the default hyperparameters: annealing of 0.8, and use a guidance scale of 4.0 and 30 inference steps. For abduction 1, which performs the inversion, we use a source prompt which follows the format: ``a photo of [subject] [source interaction] [object]", while for abduction 2, which related to the editing, we use the target prompt follows the format: ``a photo of [subject] [target interaction] [object]".

\noindent\textbf{LEDITS++ \cite{brack2024ledits++}} is implemented following the official code \footnote{\url{https://github.com/huggingface/diffusers/tree/main/src/diffusers/pipelines/ledits_pp}}. We use Stable Diffusion XL and follow the default parameters: edit guidance scale of [5.0, 10.0] and edit threshold of [0.9, 0.85] with edit direction of [reverse, forward] for source and target concepts. Since LEDITS++ requires the invert prompt to be the target concept, we let it be "[source interaction]", while the target prompt is "[target interaction]".

\noindent\textbf{InstructPix2Pix \cite{brooks2022instructpix2pix}} is implemented following the community-provided code \footnote{\url{https://github.com/huggingface/diffusers/tree/main/examples/instruct_pix2pix}}. We use Stable Diffusion XL based pretrained weight \code{diffusers/sdxl-instructpix2pix-768} and adopt the following hyperparameters: image guidance of 1.5, guidance scale of 3.0 and 100 inference steps. The target prompt is formatted as: ``Make [subject] [target interaction] [object]", as the method requires the prompt to be given as an instruction.

\noindent\textbf{ReNoise \cite{garibi2024renoise}} is implemented following the official code \footnote{\url{https://github.com/garibida/ReNoise-Inversion}}. We use Stable Diffusion XL Turbo and use the following hyperparameters: 4 inversion steps, 4 inference steps, 9 renoise steps, $\lambda_\text{pair}=20.0, \lambda_\text{patch-KL}=0.065$. For inversion, we use a source prompt which follows the format: ``a photo of [subject] [source interaction] [object]", while for editing, the target prompt follows the format: ``a photo of [subject] [target interaction] [object]".

\noindent\textbf{TurboEdit \cite{deutch2024turboedit}} is implemented following the official code \footnote{\url{https://github.com/GiilDe/turbo-edit}}. We use Stable Diffusion XL Turbo and use the following hyperparameters: pseudo-guidance scale $w=1.5$ and 4 denoising steps. For inversion, we use a source prompt which follows the format: ``a photo of [subject] [source interaction] [object]", while for editing, the target prompt follows the format: ``a photo of [subject] [target interaction] [object]". 

\noindent\textbf{SVDiff \cite{han2023svdiff}} is implemented following the community-provided code \footnote{\url{https://github.com/mkshing/svdiff-pytorch}}. We use Stable Diffusion v1.5. For inversion, we use a source prompt which follows the format: ``a photo of [subject] [source interaction] [object]", while for editing, the target prompt follows the format: ``a photo of [subject] [target interaction] [object]". 

\noindent\textbf{Edit Friendly DDPM \cite{huberman2024editfriendly}} is implemented following the official code \footnote{\url{https://github.com/inbarhub/DDPM_inversion}}. We use Stable Diffusion v1.5 and use the following default hyperparameters: strength $=15, T_\text{skip}=36, \eta=1$ and 100 inference steps. For inversion, we use a source prompt which follows the format: ``a photo of [subject] [source interaction] [object]", while for editing, the target prompt follows the format: ``a photo of [subject] [target interaction] [object]". 

\noindent\textbf{DDS \cite{hertz2023dds}} is implemented following the official code \footnote{\url{https://github.com/google/prompt-to-prompt/blob/main/DDS_zeroshot.ipynb}}. We use Stable Diffusion v2.1 and use the default hyperparameters. For inversion, we use a source prompt which follows the format: ``a photo of [subject] [source interaction] [object]", while for editing, the target prompt follows the format: ``a photo of [subject] [target interaction] [object]". 

\noindent\textbf{InfEdit \cite{xu2023infedit}} is implemented following the official code \footnote{\url{https://github.com/sled-group/InfEdit}}. Following the paper, we use Stable Diffusion v1.5 (\code{SimianLuo/LCM\_Dreamshaper\_v7}) and use the following default hyperparameters: source guidance scale of 1, target guidance scale of 2, cross replace steps of 0.3, self replace steps of 0.3, target blend threshold of 0.3, source blend threshold of 0.3, and inference steps of 15. For inversion, we use a source prompt which follows the format: ``a photo of [subject] [source interaction] [object]", while for editing, the target prompt follows the format: ``a photo of [subject] [target interaction] [object]". 

\noindent\textbf{Break-A-Scene \cite{avrahami2023breakascene}} is implemented following the official code \footnote{\url{https://github.com/google/break-a-scene}}. We use Stable Diffusion v2.1 and use the following default hyperparameters: $\lambda_\text{attn}=0.01$. We fine-tune 400 steps in stage 1 and 800 steps in stage 2 with the default learning rate of 5e-4 and 2e-6, which we found to work well empirically. For fair comparison, we re-implement \textbf{Break-A-Scene XL} which is based on Stable Diffusion XL (\code{SG161222/RealVisXL\_V4.0}). We use $\lambda_\text{attn}=0.01$ and fine-tune 1000 steps in stage 1 and 200 steps in stage 2, with a learning rate of 3e-4 and 2e-6, respectively, which we found to work well empirically.

\noindent \textbf{Mix-of-Show \cite{gu2024mixofshow}} is implemented using the official code \footnote{\url{https://github.com/TencentARC/Mix-of-Show}} and default settings. For inversion, the subject and object are each represented by two tokens, which are processed independently and then fused via gradient fusion. For editing, the prompt is ``a photo of [subject tokens] [target interaction] [object tokens]''.

\noindent \textbf{DisenDiff \cite{zhang2024disendiff}} is implemented following the official code \footnote{\url{https://github.com/Monalissaa/DisenDiff}} and default settings. During inversion, we use the default prompt format: ``\textless{}new1\textgreater{} [subject] and \textless{}new2\textgreater{} [object]'', while during editing, we use ``a photo of \textless{}new1\textgreater{} [subject] [target interaction] \textless{}new2\textgreater{} [object]''.

\noindent \textbf{OmniGen \cite{xiao2025omnigen}} is implemented using diffusers library. There is two prompting strategies: single image and part-splitting. Single image prompt uses only one source image and follows the format: ``Edit \textless{}img\textgreater{}\textless{}\textbar{}image\_1\textbar\textgreater{}\textless{}/img\textgreater{} to change the interaction of the [subject] to [target interaction] the [object].'' The part-splitting approach splits the image into subject, object and background, and use the following prompt: ``A [subject] \textless{}img\textgreater{}\textless{}\textbar{}image\_1\textbar\textgreater{}\textless{}/img\textgreater{} [target interaction] the [object] \textless{}img\textgreater{}\textless{}\textbar{}image\_2\textbar\textgreater{}\textless{}/img\textgreater{} at \textless{}img\textgreater{}\textless{}\textbar{}image\_3\textbar\textgreater{}\textless{}/img\textgreater{}.'' We observe that part-splitting offers better interaction editability, and report scores based on this setting.

\noindent \textbf{HiDream-E1 \cite{cai2025hidream}} is implemented following the official code \footnote{\url{https://github.com/HiDream-ai/HiDream-E1}}. Following the default setting, input image is resized to 768$\times$768 and we adopt the prompt format:``the [subject] is now [target interaction] [object]''. This prompt is then further refined using OpenAI GPT-4o, conditioned on the source image, using the prompt engineering protocol provided in the official code.

\noindent \textbf{Flux.1 Kontext \cite{labs2025flux1kontext}} is implemented using the diffusers library and model weight \footnote{\code{black-forest-labs/FLUX.1-Kontext-dev}}. We follow the default hyperparameters: guidance scale of 2.5 and prompt format of ``the [subject] is now [target interaction] [object]''.

\noindent\textbf{FireFlow \cite{deng2024fireflow}} is implemented following the official code \footnote{\specialurl{https://github.com/HolmesShuan/FireFlow-Fast-Inversion-of-Rectified-Flow-for-Image-Semantic-Editing.git}}. Following the default setting, we adopt the source prompt format:``a photo of [subject] [source interaction] [object]'' and the target prompt format of ``a photo of [subject] [target interaction] [object]''. We use guidance scale of 3, \code{flux-dev} base model, \code{add\_q} editing strategy, \code{fireflow} sampling strategy, 8 inference steps, \code{end\_layer\_index} at 17, and use default setting on editing strategy.

%% file: assets/figure/cost.tex
\begin{table*}
\centering
\caption{Comparison between InteractEdit and existing baselines in terms of resource cost. Total time taken is computed as the total required time to complete IEBench, while average time taken is averaged over the number of image generated. 
$^\dagger$ indicates methods requiring pretraining, pretraining time is not included in this table. $^*$ “Others” refers to the total time taken for training interaction tokens. $^\ddagger$ “Others” refers to Induction-2, which requires 100 times. For further details, refer to \cref{subsec:resource-analysis}.}
\begin{tabular}{l|l|ccc|c|c|cc}
\hline
                           &            & \multicolumn{3}{c|}{Time Taken (s)}                                    & \multicolumn{1}{l|}{Total} & \multicolumn{1}{l|}{Average} & \multicolumn{2}{c}{GPU Memory (GB)}      \\ \cline{3-5} \cline{8-9} 
Method                     & Base       & \multicolumn{1}{c|}{Inversion} & \multicolumn{1}{c|}{Editing} & Others & (min)                      & (s)                          & \multicolumn{1}{c|}{Inversion} & Editing \\ \hline
NTI                        & SDv1.5     & \multicolumn{1}{c|}{18.81}     & \multicolumn{1}{c|}{4.56}    &        & 85                         & 5.09                         & \multicolumn{1}{c|}{11.14}     & 11.15   \\
PnP                        & SDv2.1     & \multicolumn{1}{c|}{50.16}     & \multicolumn{1}{c|}{}        &        & 130                        & 7.81                         & \multicolumn{1}{c|}{3.84}      & 7.97    \\
MasaCtrl                   & SDv1.4     & \multicolumn{1}{c|}{3.68}      & \multicolumn{1}{c|}{10.36}   &        & 174                        & 10.46                        & \multicolumn{1}{c|}{6.07}      & 12.48   \\
HOIEdit                    & SDv1.4     & \multicolumn{1}{c|}{46.93}     & \multicolumn{1}{c|}{14.07}   & 81000  & 1606                       & 96.38                        & \multicolumn{1}{c|}{11.07}     & 20.57   \\
Imagic                     & SDv1.5     & \multicolumn{1}{c|}{244.98}    & \multicolumn{1}{c|}{3.81}    &        & 472                        & 28.31                        & \multicolumn{1}{c|}{18.83}     & 19.93   \\
CDS                        & SDv1.5     & \multicolumn{1}{c|}{-}         & \multicolumn{1}{c|}{34.19}   &        & 570                        & 34.19                        & \multicolumn{1}{c|}{-}         & 10.23   \\
DAC                        & SDv2.1     & \multicolumn{1}{c|}{330.52}    & \multicolumn{1}{c|}{5.07}    & 335.88 & 799                        & 47.91                        & \multicolumn{1}{c|}{16.43}     & 9.71    \\
LEDITS++                   & SDXL       & \multicolumn{1}{c|}{4.13}      & \multicolumn{1}{c|}{4.46}    &        & 76                         & 4.58                         & \multicolumn{1}{c|}{9.45}      & 9.45    \\
InstructPix2Pix $^\dagger$ & SDXL       & \multicolumn{1}{c|}{-}         & \multicolumn{1}{c|}{74.44}   &        & 1241                       & 74.44                        & \multicolumn{1}{c|}{-}         & 12.36   \\
ReNoise                    & SDXL Turbo & \multicolumn{1}{c|}{4.99}      & \multicolumn{1}{c|}{0.36}    &        & 89                         & 5.35                         & \multicolumn{1}{c|}{8.38}      & 9.36    \\
TurboEdit                  & SDXL Turbo & \multicolumn{1}{c|}{-}         & \multicolumn{1}{c|}{1.31}    &        & 22                         & 1.31                         & \multicolumn{1}{c|}{-}         & 23.88   \\
SVDiff                     & SDv1.5     & \multicolumn{1}{c|}{330.01}    & \multicolumn{1}{c|}{22.4}    &        & 527                        & 31.64                        & \multicolumn{1}{c|}{15.89}     & 12.65   \\
Edit Friendly DDPM         & SDv1.5     & \multicolumn{1}{c|}{9.07}      & \multicolumn{1}{c|}{5.79}    &        & 101                        & 6.04                         & \multicolumn{1}{c|}{6.32}      & 6.95    \\
DDS                        & SDv2.1     & \multicolumn{1}{c|}{-}         & \multicolumn{1}{c|}{9.23}    &        & 154                        & 9.23                         & \multicolumn{1}{c|}{-}         & 7.32    \\
InfEdit                    & SDv1.5     & \multicolumn{1}{c|}{-}         & \multicolumn{1}{c|}{3.92}    &        & 65                         & 3.92                         & \multicolumn{1}{c|}{-}         & 15.58   \\
Break-A-Scene              & SDv2.1     & \multicolumn{1}{c|}{274.3}     & \multicolumn{1}{c|}{1.42}    &        & 152                        & 9.10                         & \multicolumn{1}{c|}{27.79}     & 4.09    \\
Break-A-Scene XL           & SDXL       & \multicolumn{1}{c|}{502.15}    & \multicolumn{1}{c|}{5.34}    &        & 323                        & 19.40                        & \multicolumn{1}{c|}{65.84}     & 14.55   \\
Mix-of-Show                & SDv1.4     & \multicolumn{1}{c|}{332}       & \multicolumn{1}{c|}{1.82}    &        & 185                        & 11.12                        & \multicolumn{1}{c|}{10.10}     & 3.44    \\
DisenDiff                  & SDv1.5     & \multicolumn{1}{c|}{373}       & \multicolumn{1}{c|}{4.4}     &        & 247                        & 14.83                        & \multicolumn{1}{c|}{9.687}     & 10.66   \\
OmniGen                    & -          & \multicolumn{1}{c|}{-}         & \multicolumn{1}{c|}{142.9}   &        & 2382                       & 142.9                        & \multicolumn{1}{c|}{-}         & 10.98   \\
HiDream-E1                 & HiDream-L1 & \multicolumn{1}{c|}{-}         & \multicolumn{1}{c|}{101}     &        & 1683                       & 101                          & \multicolumn{1}{c|}{-}         & 63.23   \\
Flux.1 Kontext             & Flux.1     & \multicolumn{1}{c|}{-}         & \multicolumn{1}{c|}{59}      &        & 983                        & 59                           & \multicolumn{1}{c|}{-}         & 36.65   \\ \hline
InteractEdit               & SDXL       & \multicolumn{1}{c|}{466.27}    & \multicolumn{1}{c|}{5.92}    &        & 316                        & 18.98                        & \multicolumn{1}{c|}{14.11}     & 14.74   \\ \hline
\end{tabular}
\label{tab:cost}
\end{table*}

%% file: assets/figure/ab_qual_5.tex
\begin{figure}
    \centering
    \setlength{\tabcolsep}{1pt} 
    \renewcommand{\arraystretch}{1} 
    \resizebox{0.9\linewidth}{!}{
    \begin{tabular}{ccccccccc}
        \footnotesize Source & \footnotesize SVDiff & \footnotesize CDS & \footnotesize DisenDiff & \footnotesize Mix-of-Show & \footnotesize HiDream-E1 & \footnotesize FastEdit & \footnotesize MSDiffusion & \footnotesize Ours\\
        
          \includegraphics[width=.104\linewidth]{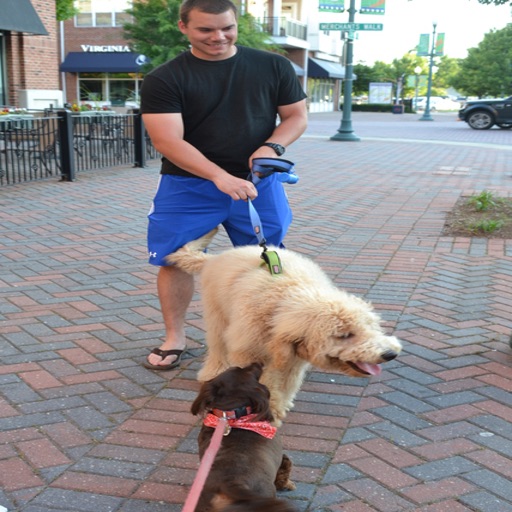}
        & \includegraphics[width=.104\linewidth]{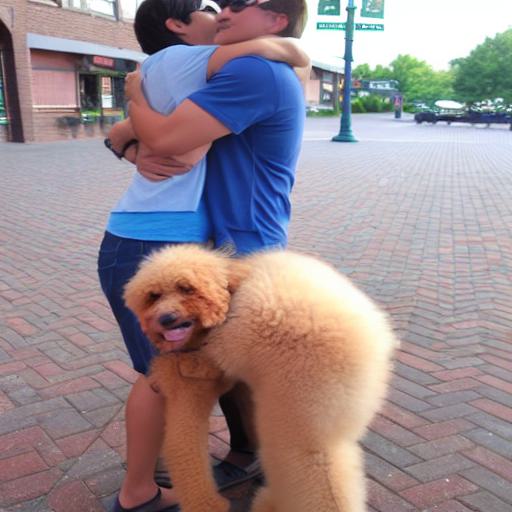}
        & \includegraphics[width=.104\linewidth]{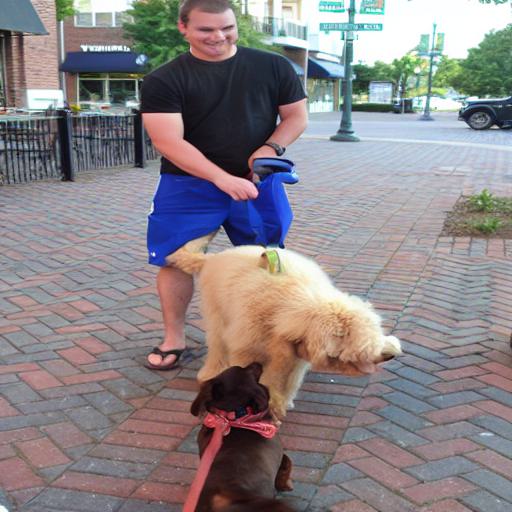}
        & \includegraphics[width=.104\linewidth]{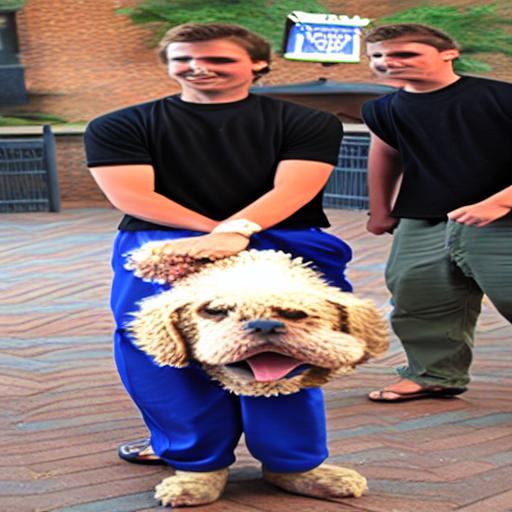}
        & \includegraphics[width=.104\linewidth]{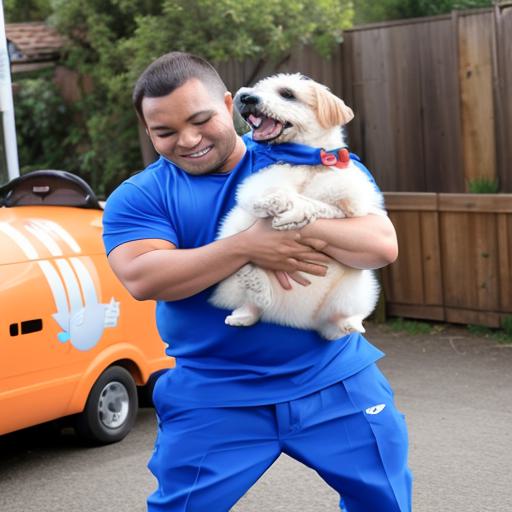}
        & \includegraphics[width=.104\linewidth]{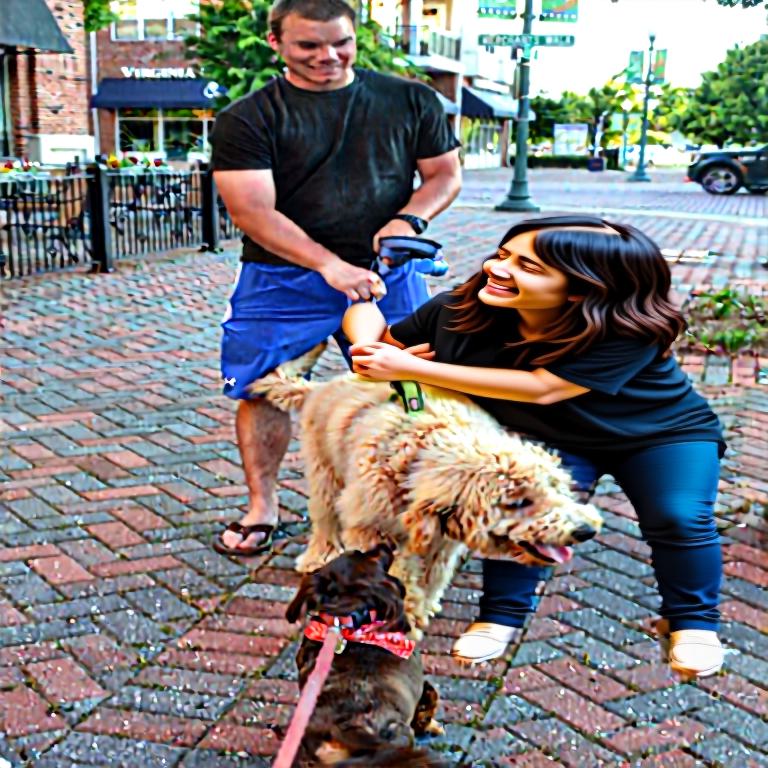}
        & \includegraphics[width=.104\linewidth]{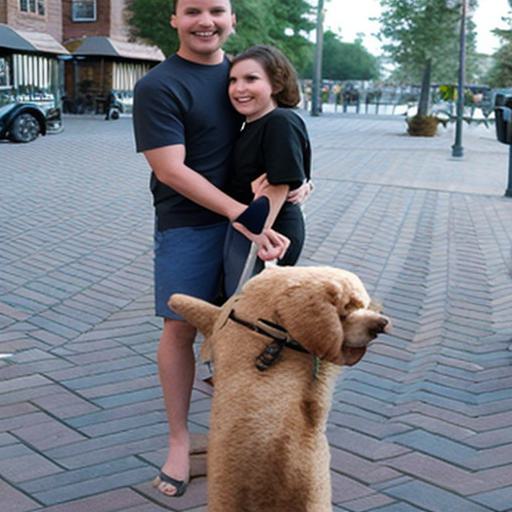}
        & \includegraphics[width=.104\linewidth]{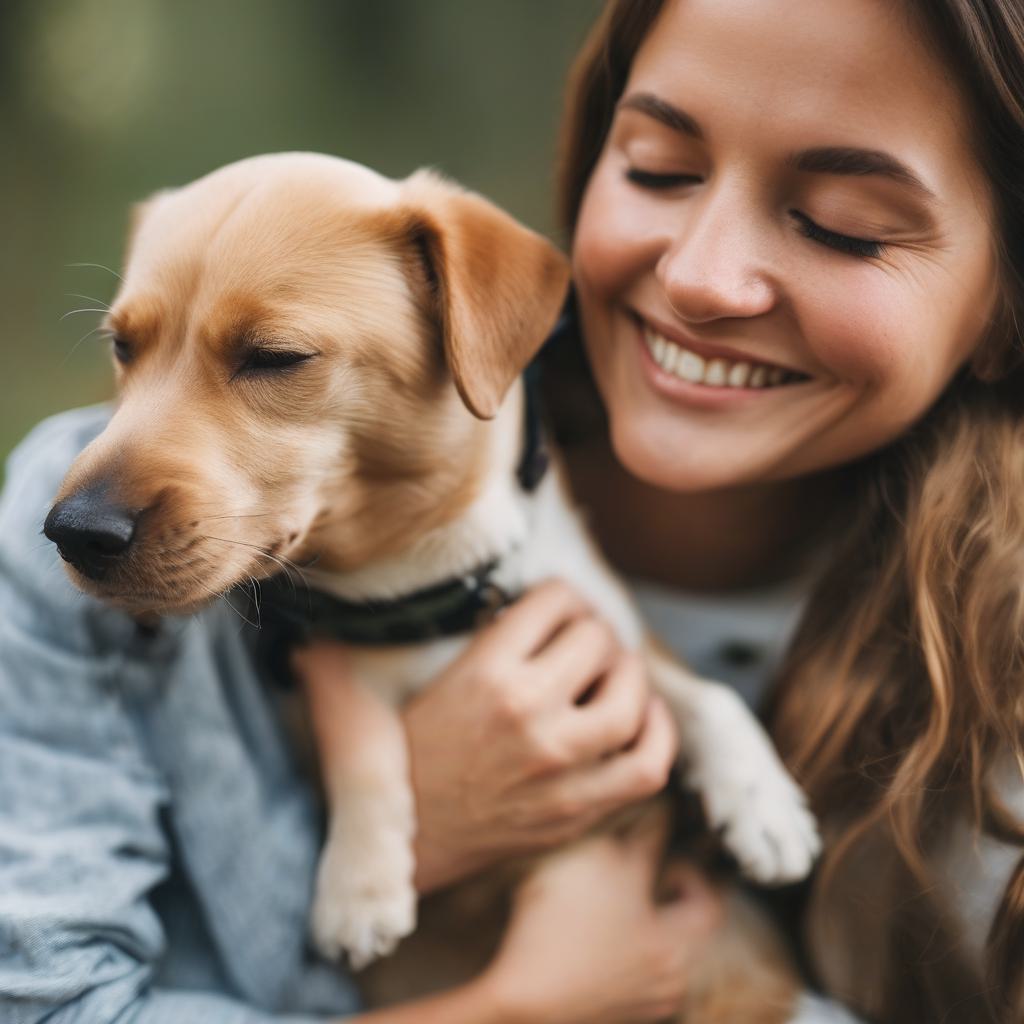}
        & \includegraphics[width=.104\linewidth]{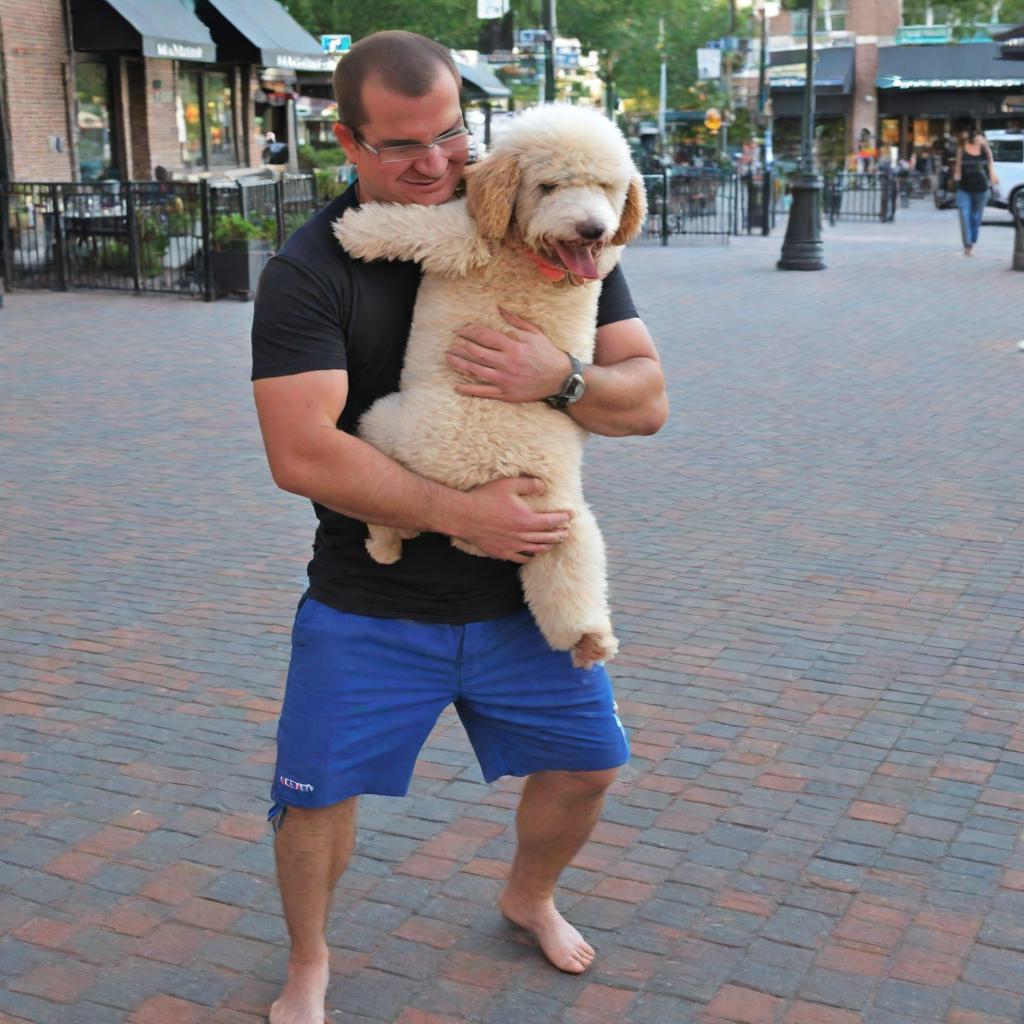}\\
        \\[-1.2em]
        walk dog & \multicolumn{8}{c}{hug dog}\\
        \includegraphics[width=.104\linewidth]{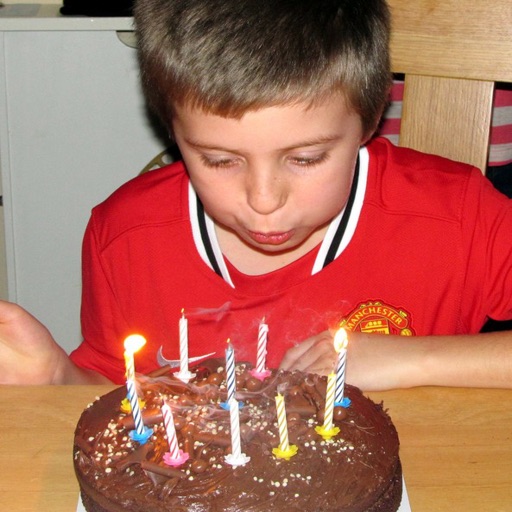}
        & \includegraphics[width=.104\linewidth]{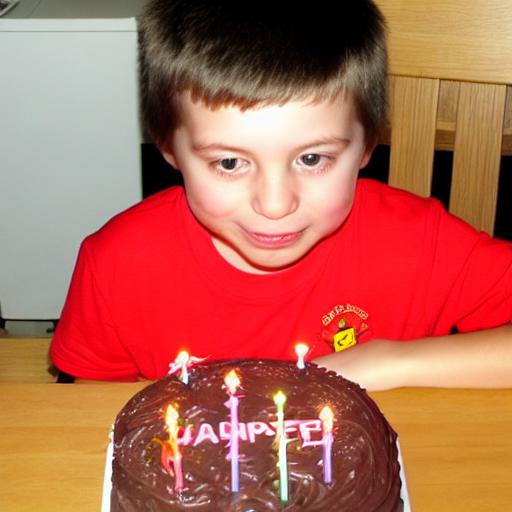}
        & \includegraphics[width=.104\linewidth]{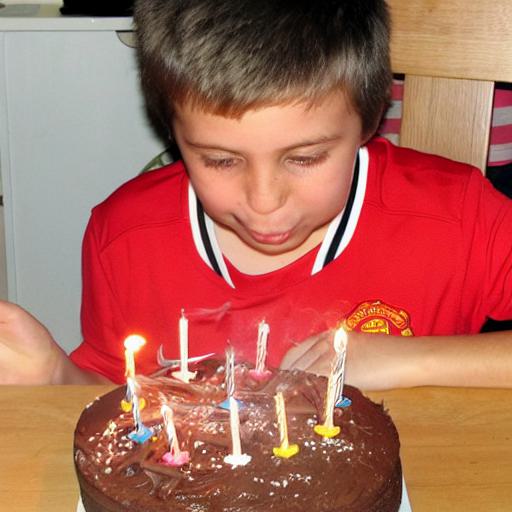}
        & \includegraphics[width=.104\linewidth]{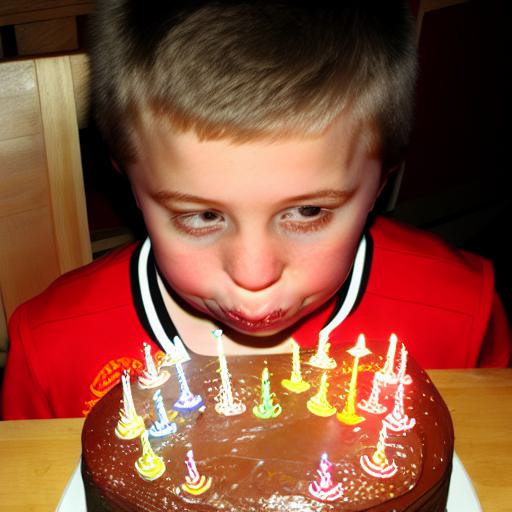}
        & \includegraphics[width=.104\linewidth]{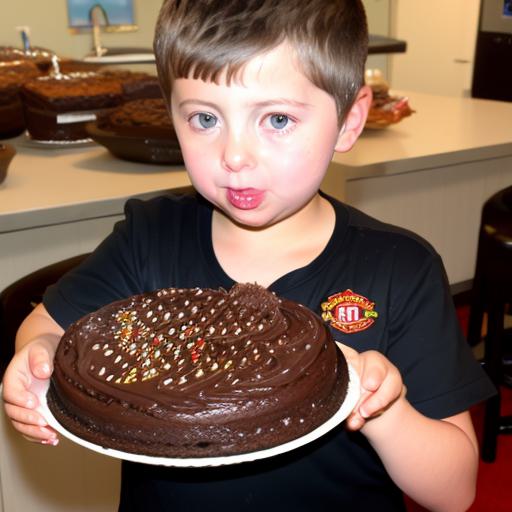}
        & \includegraphics[width=.104\linewidth]{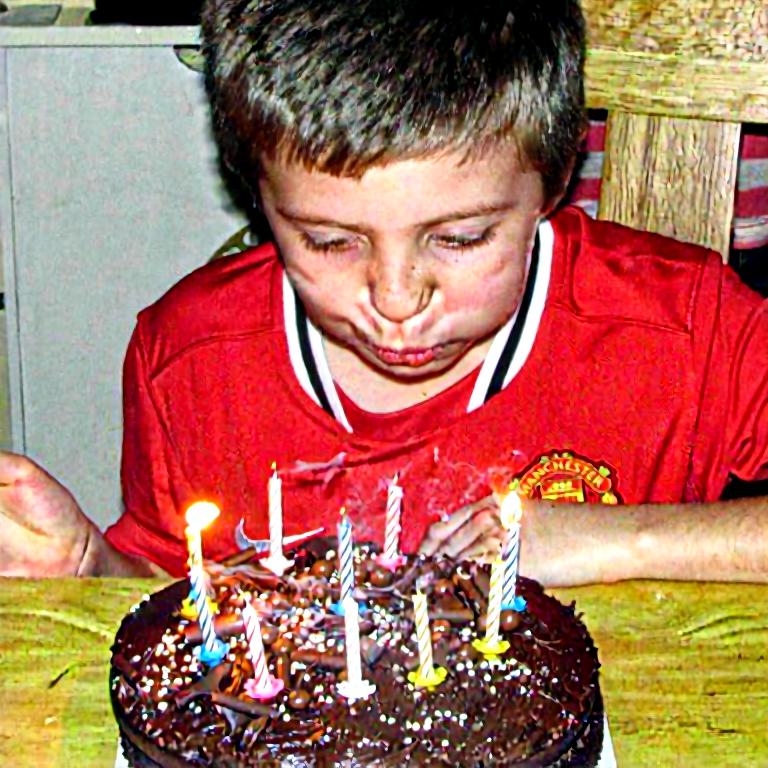}
        & \includegraphics[width=.104\linewidth]{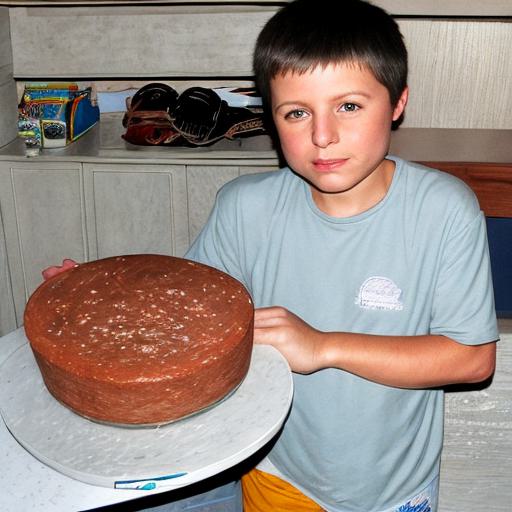}
        & \includegraphics[width=.104\linewidth]{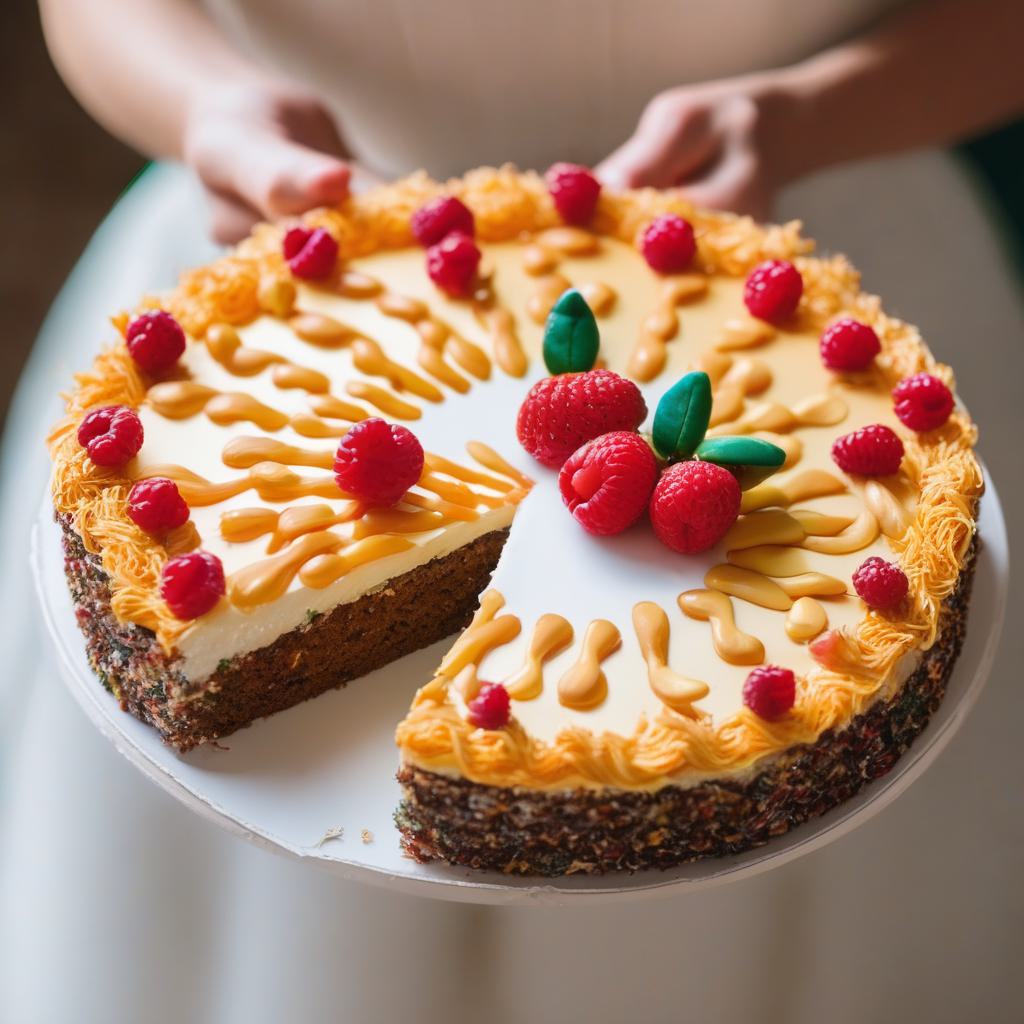}
        & \includegraphics[width=.104\linewidth]{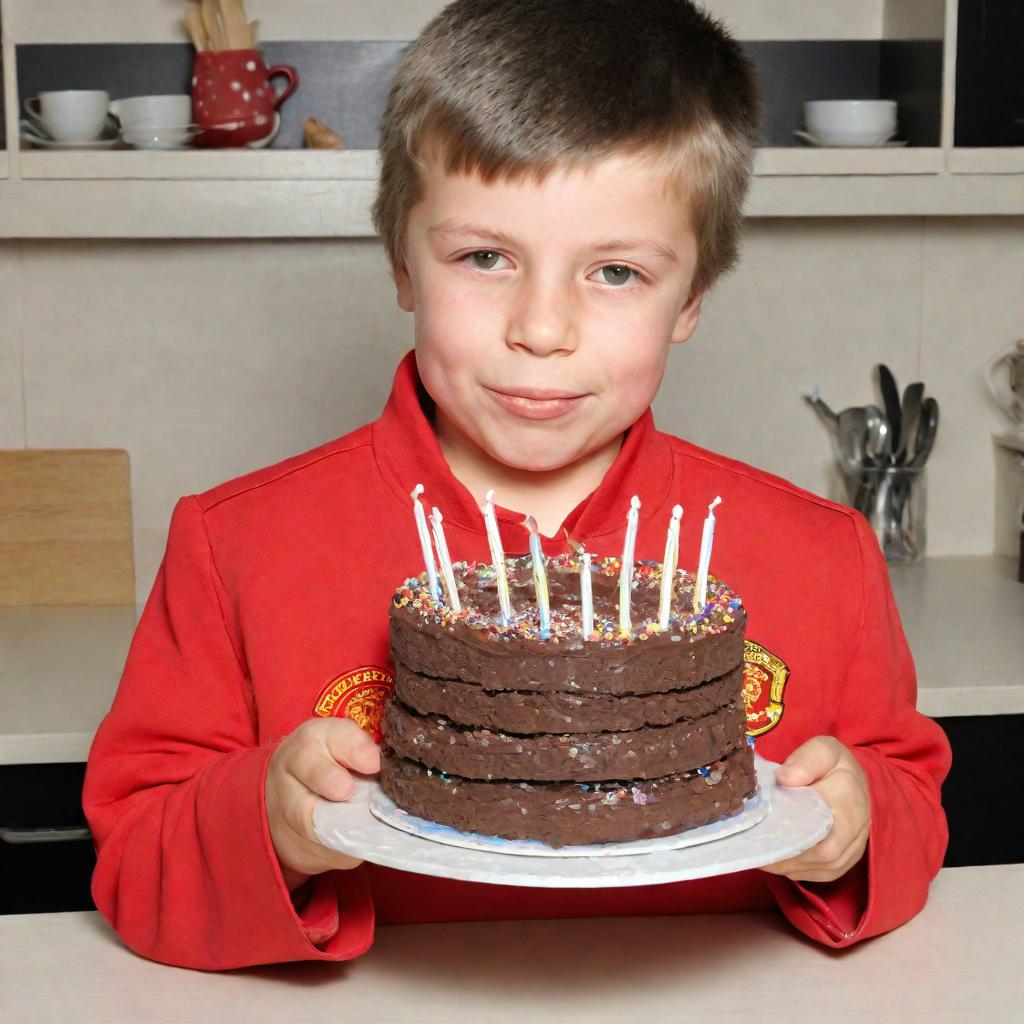}\\
        \\[-1.2em]
        & \multicolumn{8}{c}{hold cake}\\
        & \includegraphics[width=.104\linewidth]{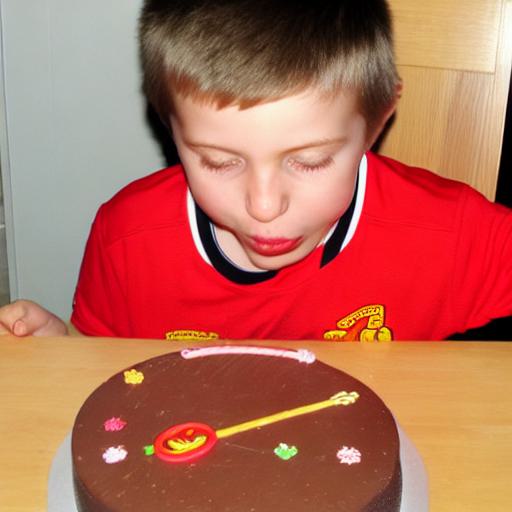}
        & \includegraphics[width=.104\linewidth]{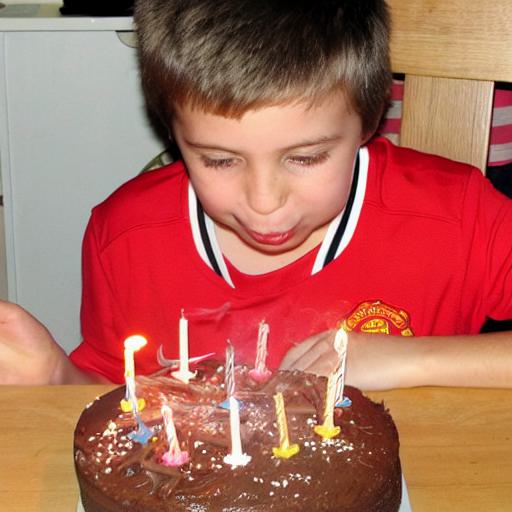}
        & \includegraphics[width=.104\linewidth]{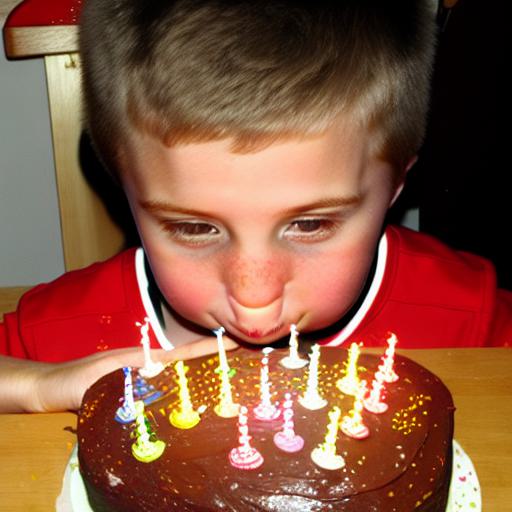}
        & \includegraphics[width=.104\linewidth]{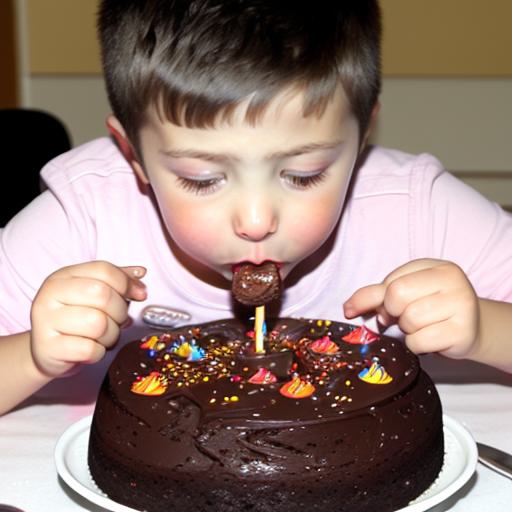}
        & \includegraphics[width=.104\linewidth]{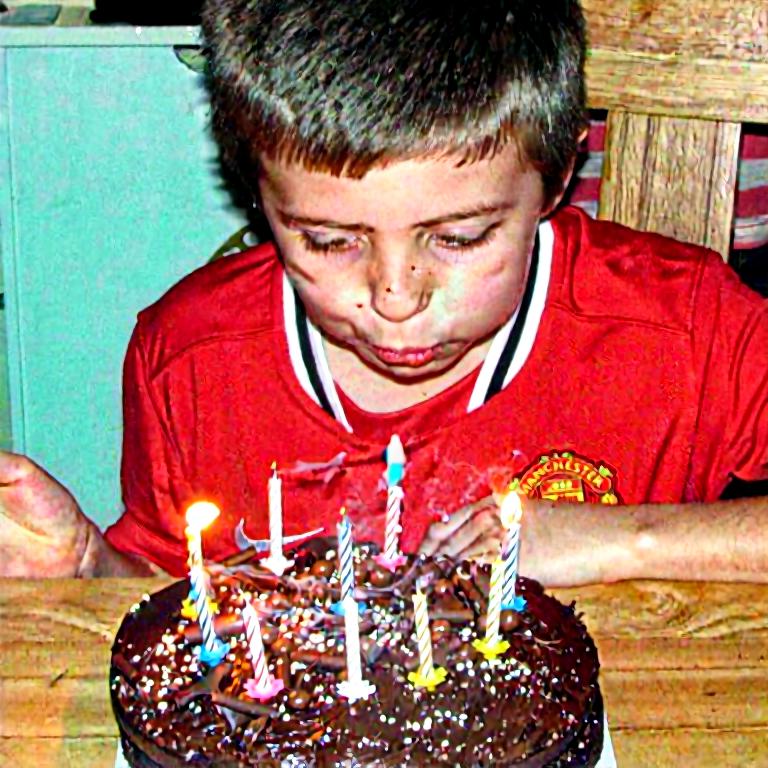}
        & \includegraphics[width=.104\linewidth]{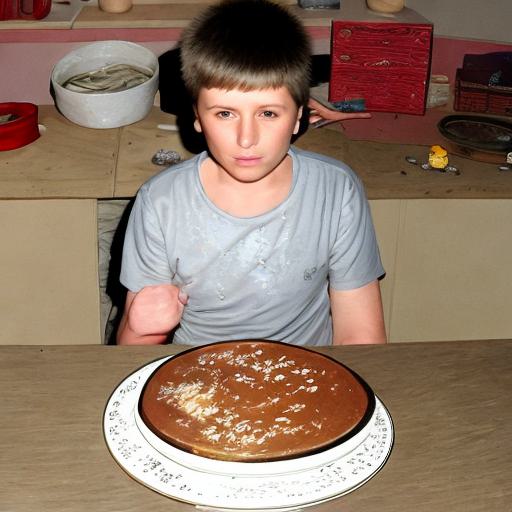}
        & \includegraphics[width=.104\linewidth]{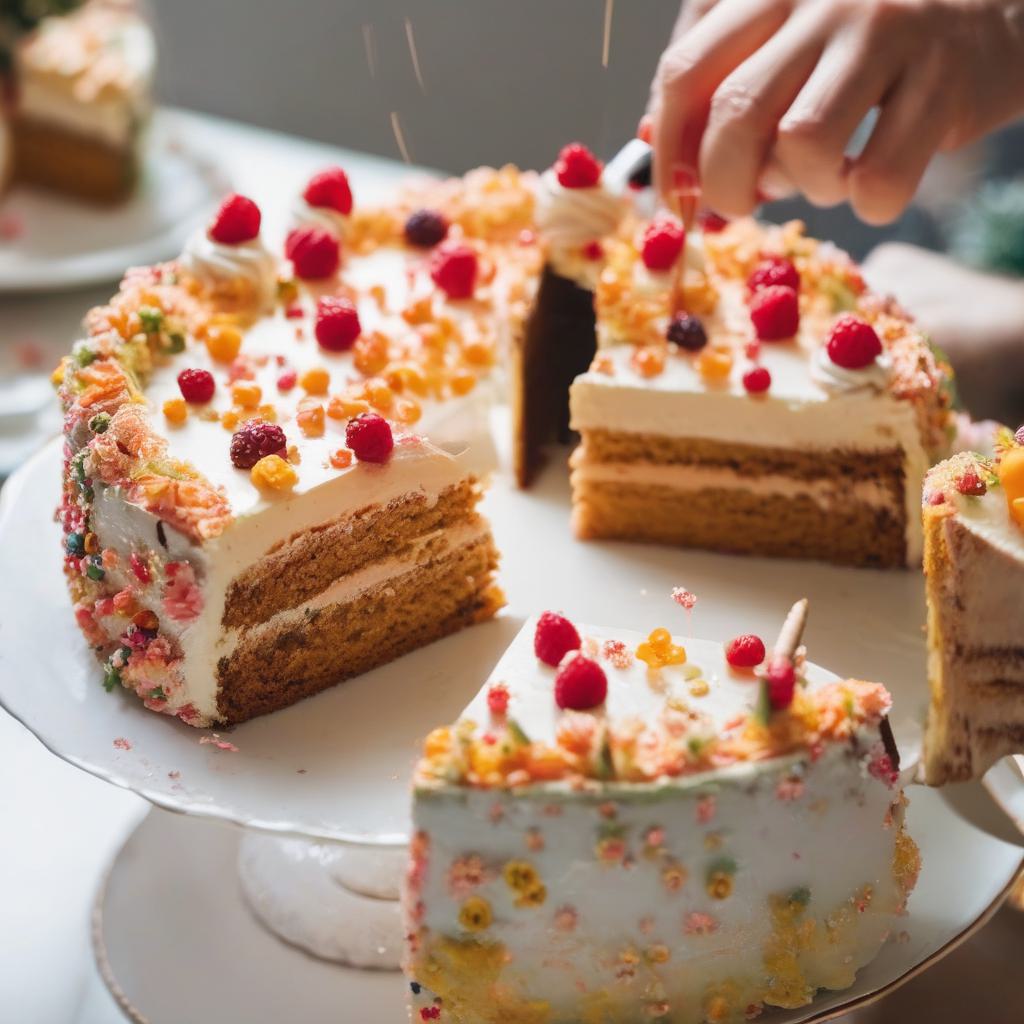}
        & \includegraphics[width=.104\linewidth]{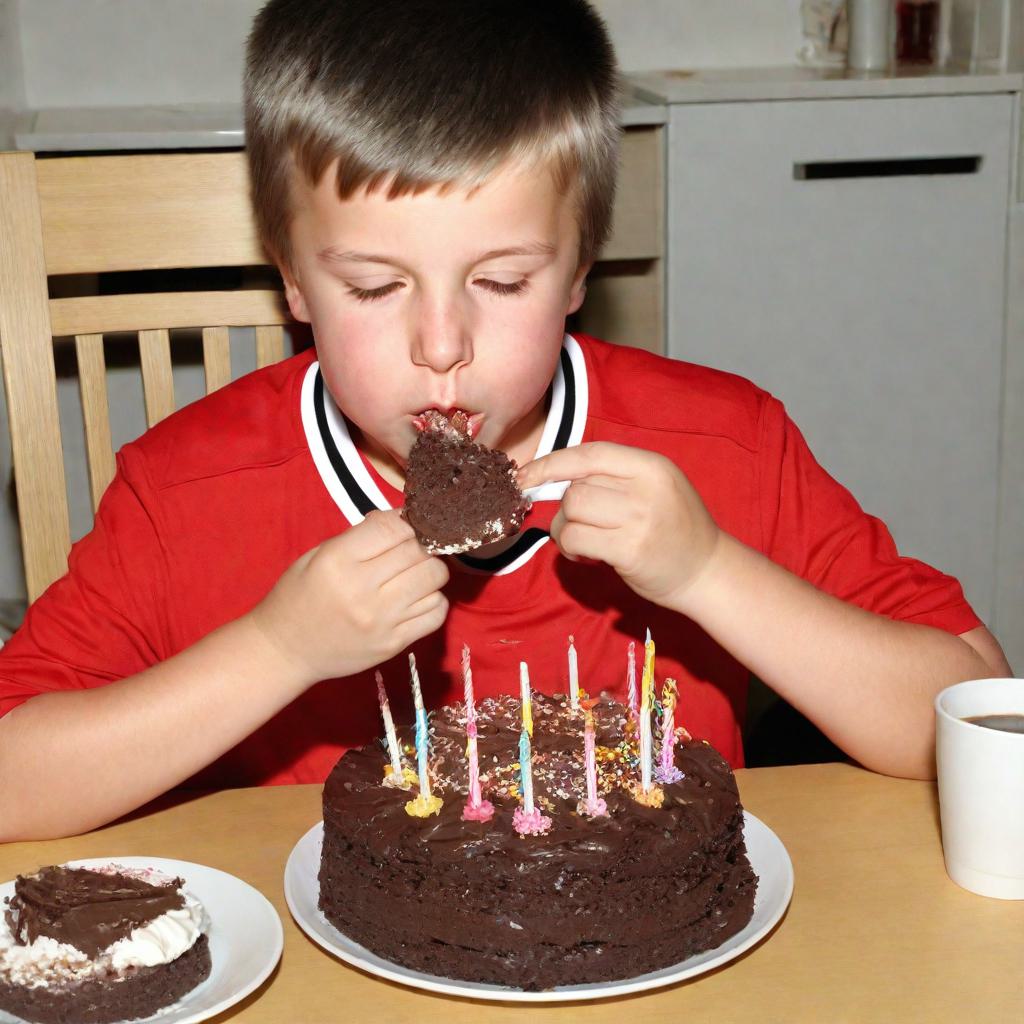}\\
        \\[-1.2em]
        blow cake & \multicolumn{8}{c}{eat cake}\\
        & \includegraphics[width=.104\linewidth]{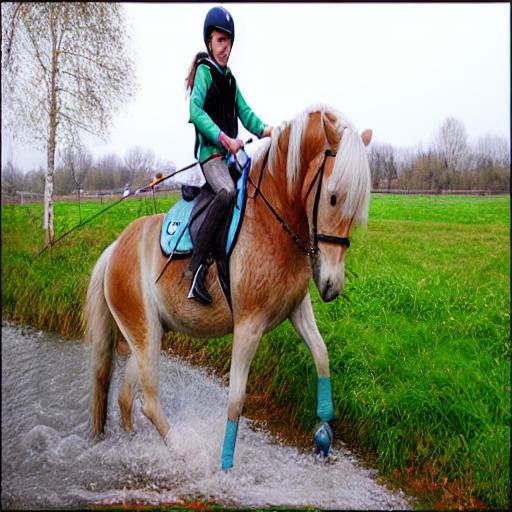}
        & \includegraphics[width=.104\linewidth]{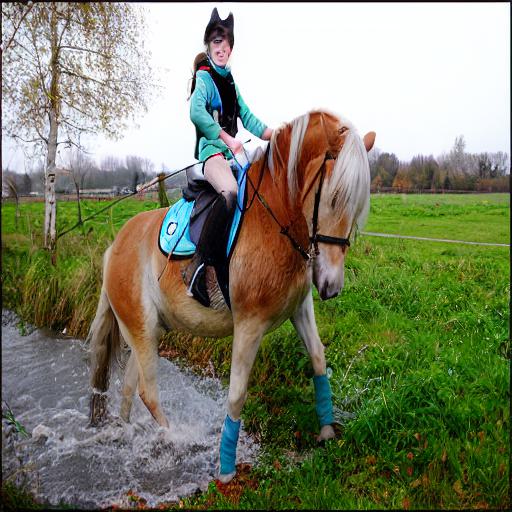}
        & \includegraphics[width=.104\linewidth]{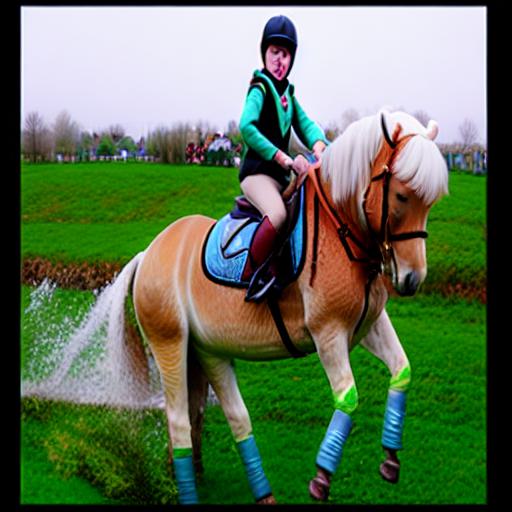}
        & \includegraphics[width=.104\linewidth]{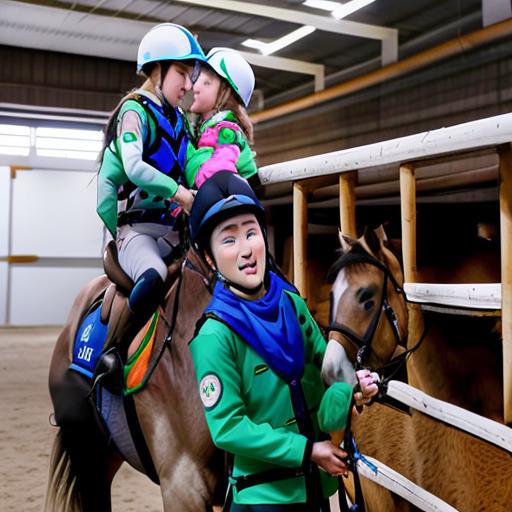}
        & \includegraphics[width=.104\linewidth]{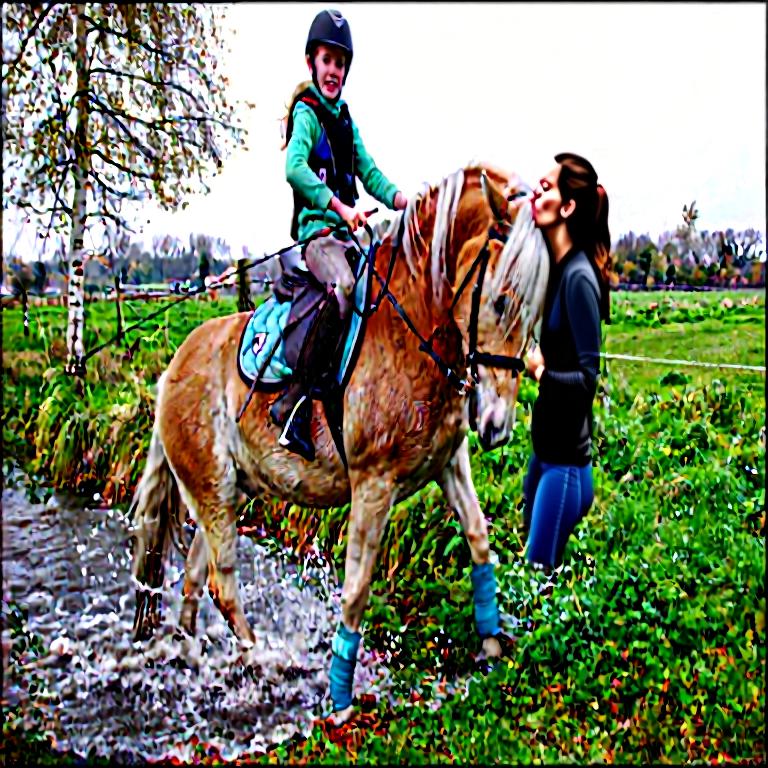}
        & \includegraphics[width=.104\linewidth]{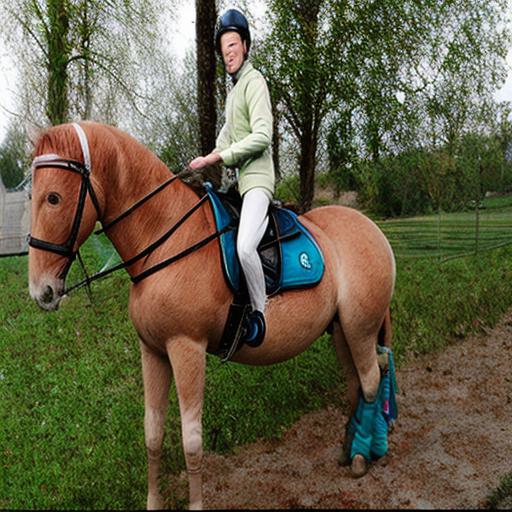}
        & \includegraphics[width=.104\linewidth]{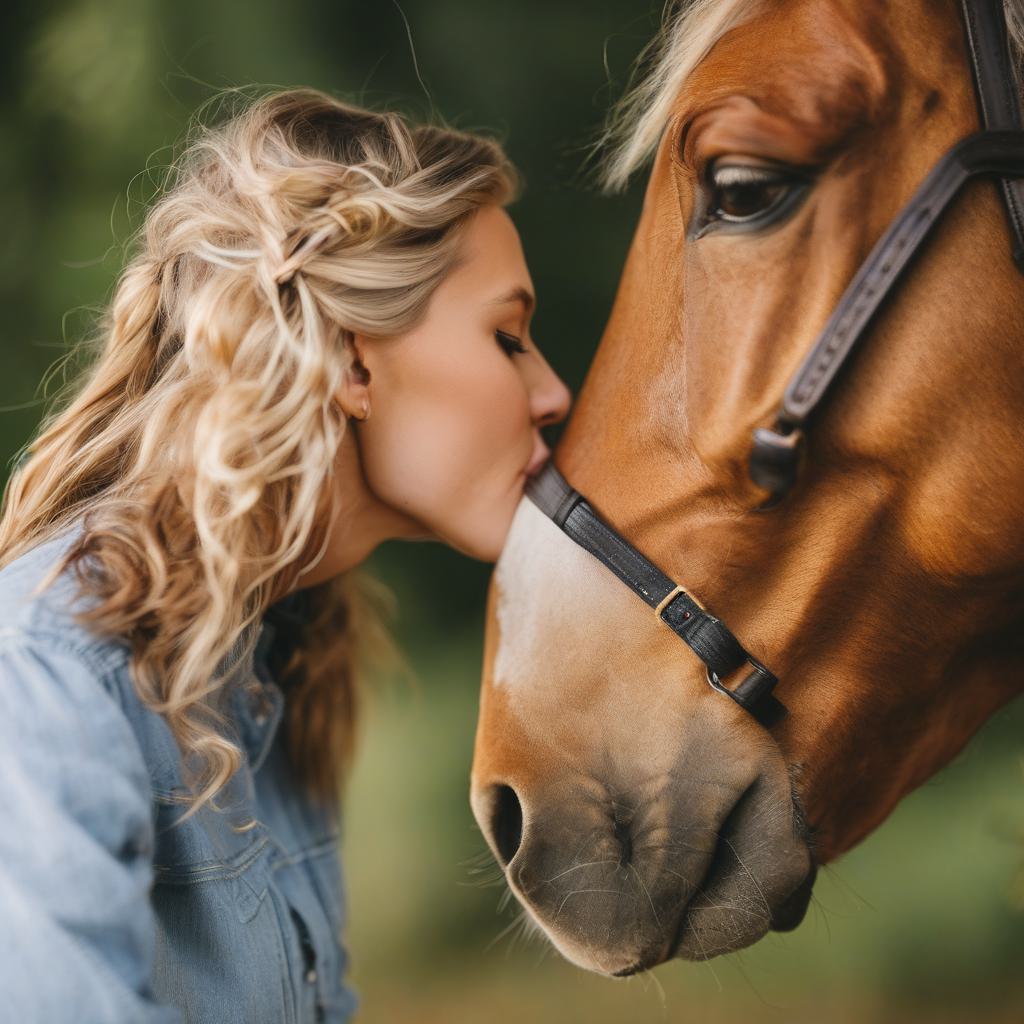}
        & \includegraphics[width=.104\linewidth]{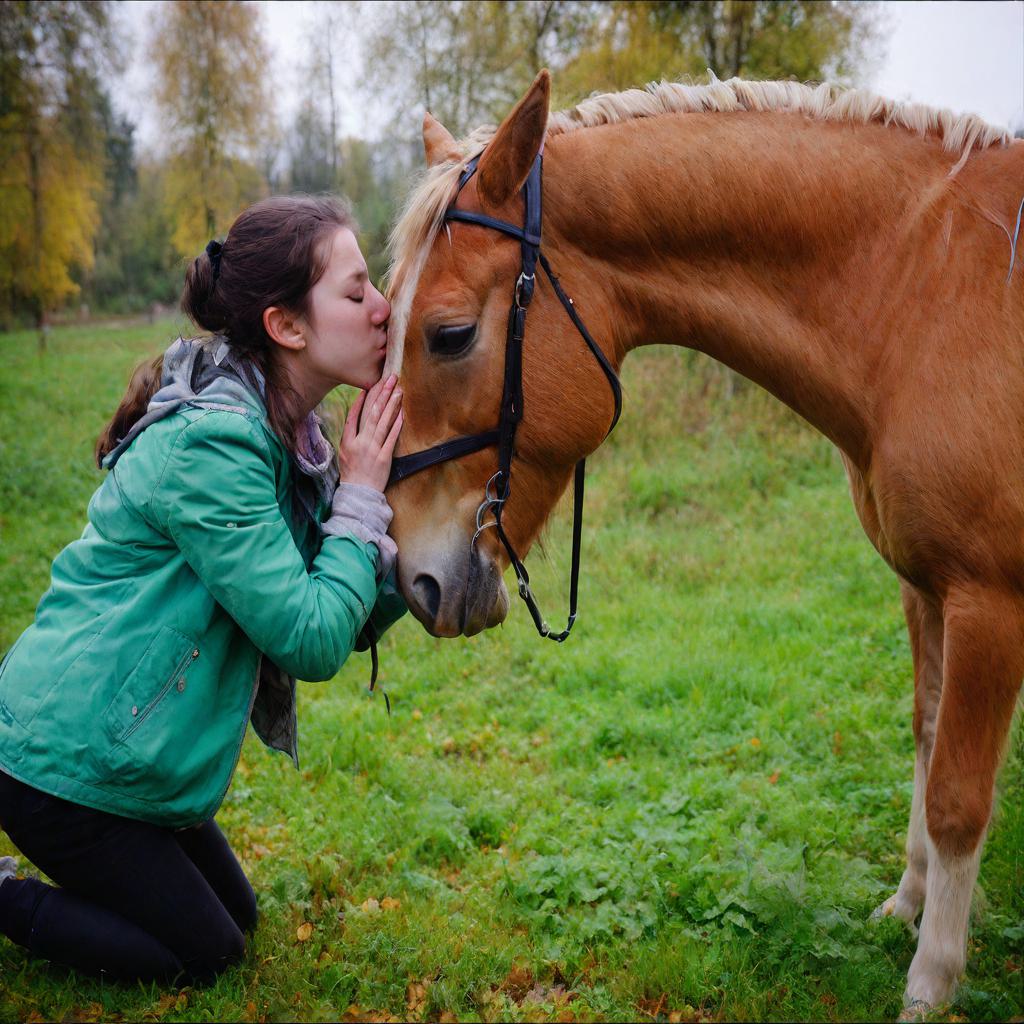}\\
        \\[-1.2em]
        & \multicolumn{8}{c}{kiss horse}\\
          \includegraphics[width=.104\linewidth]{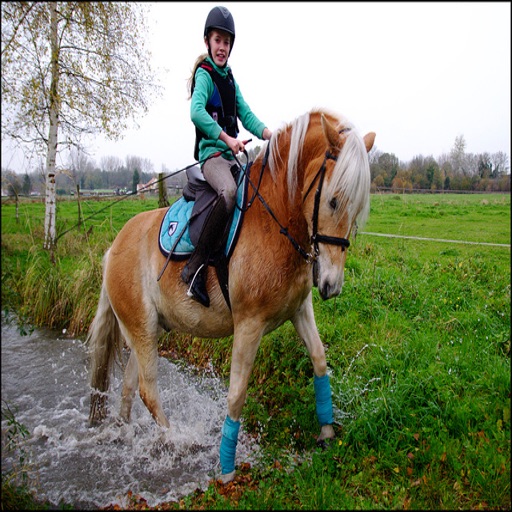}
        & \includegraphics[width=.104\linewidth]{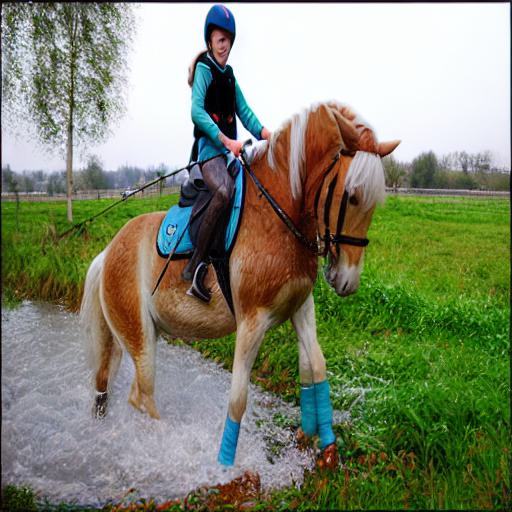}
        & \includegraphics[width=.104\linewidth]{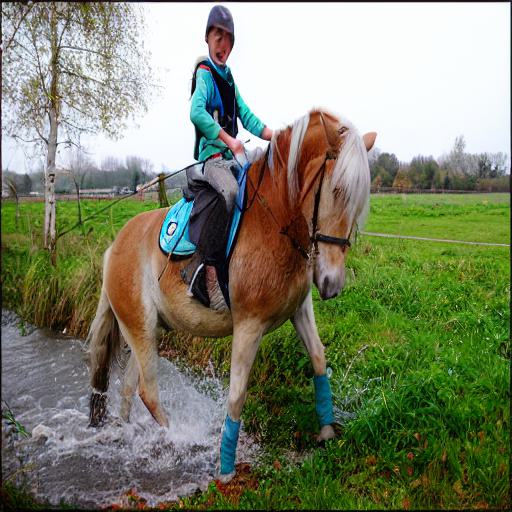}
        & \includegraphics[width=.104\linewidth]{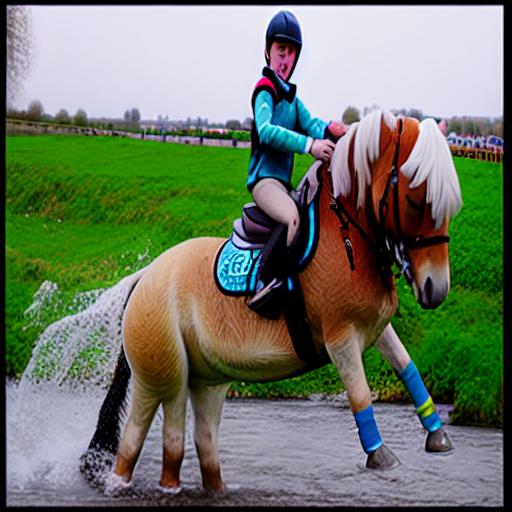}
        & \includegraphics[width=.104\linewidth]{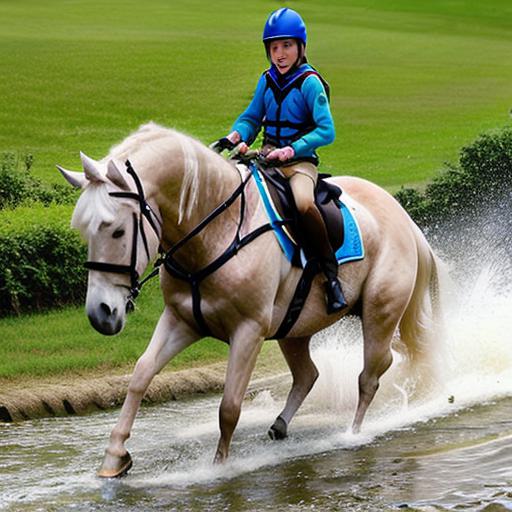}
        & \includegraphics[width=.104\linewidth]{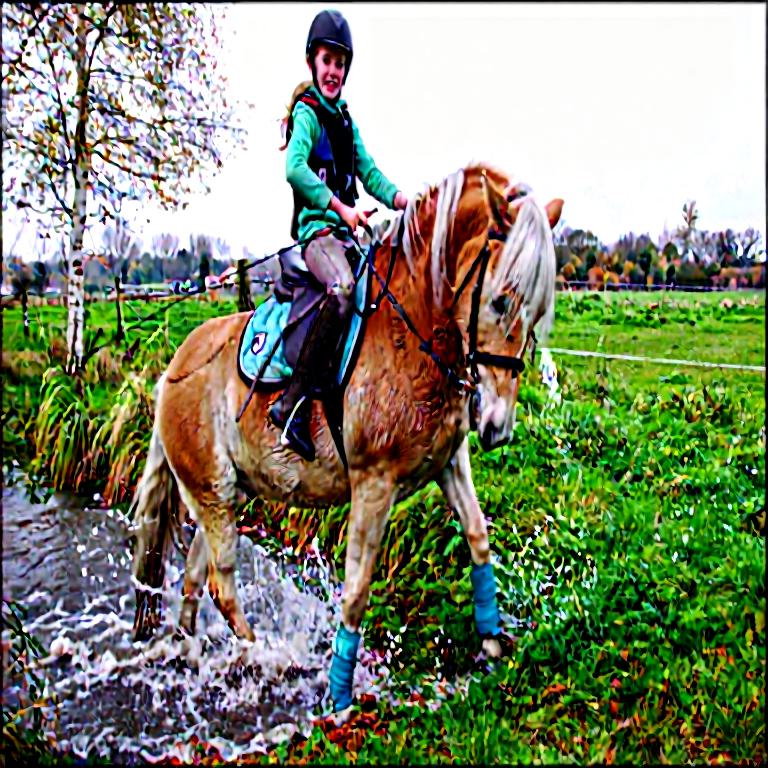}
        & \includegraphics[width=.104\linewidth]{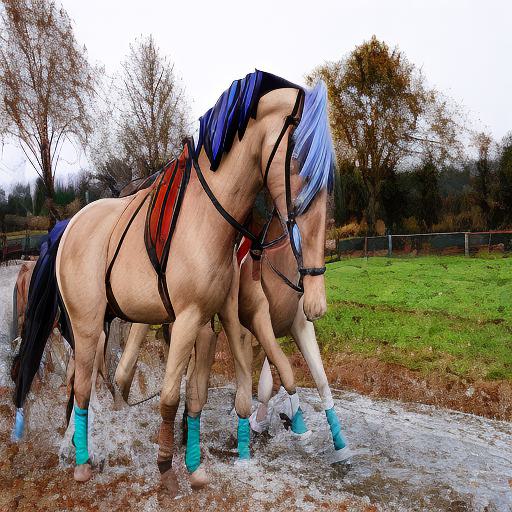}
        & \includegraphics[width=.104\linewidth]{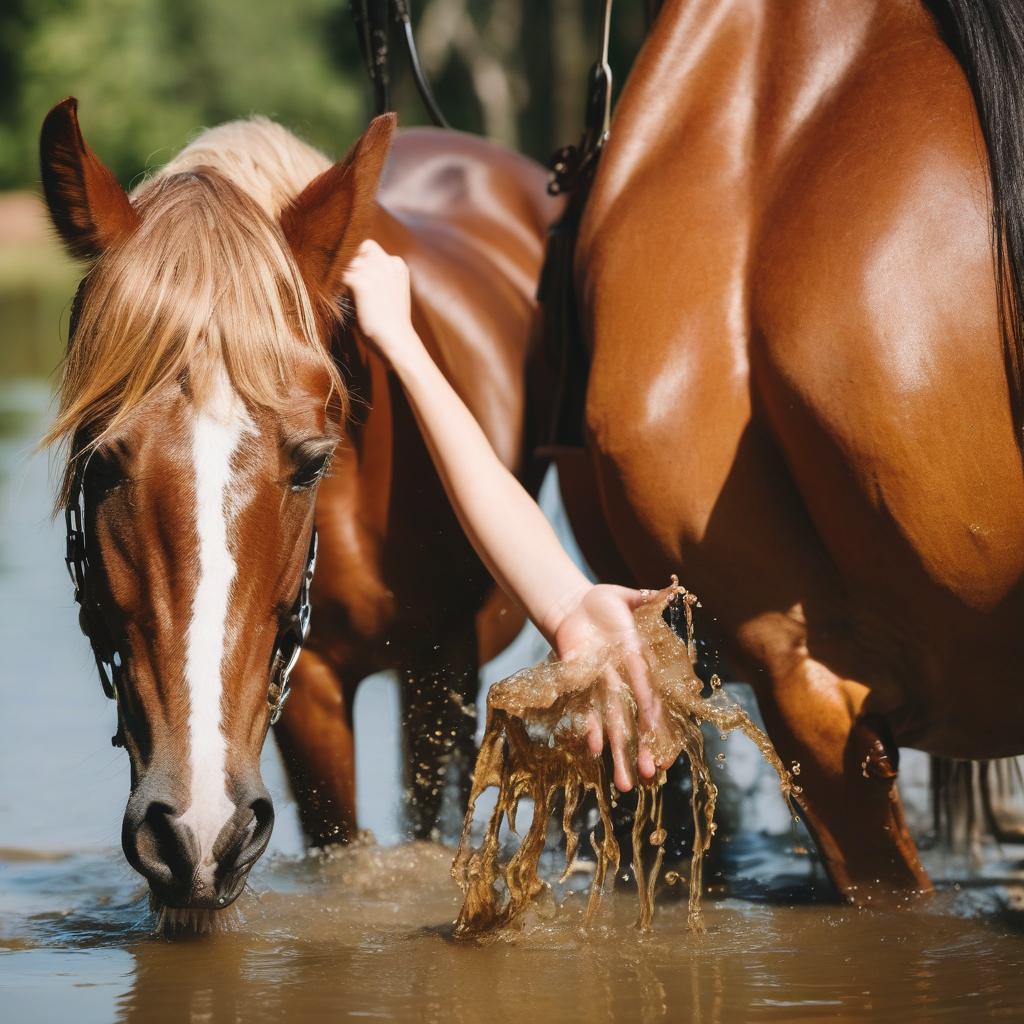}
        & \includegraphics[width=.104\linewidth]{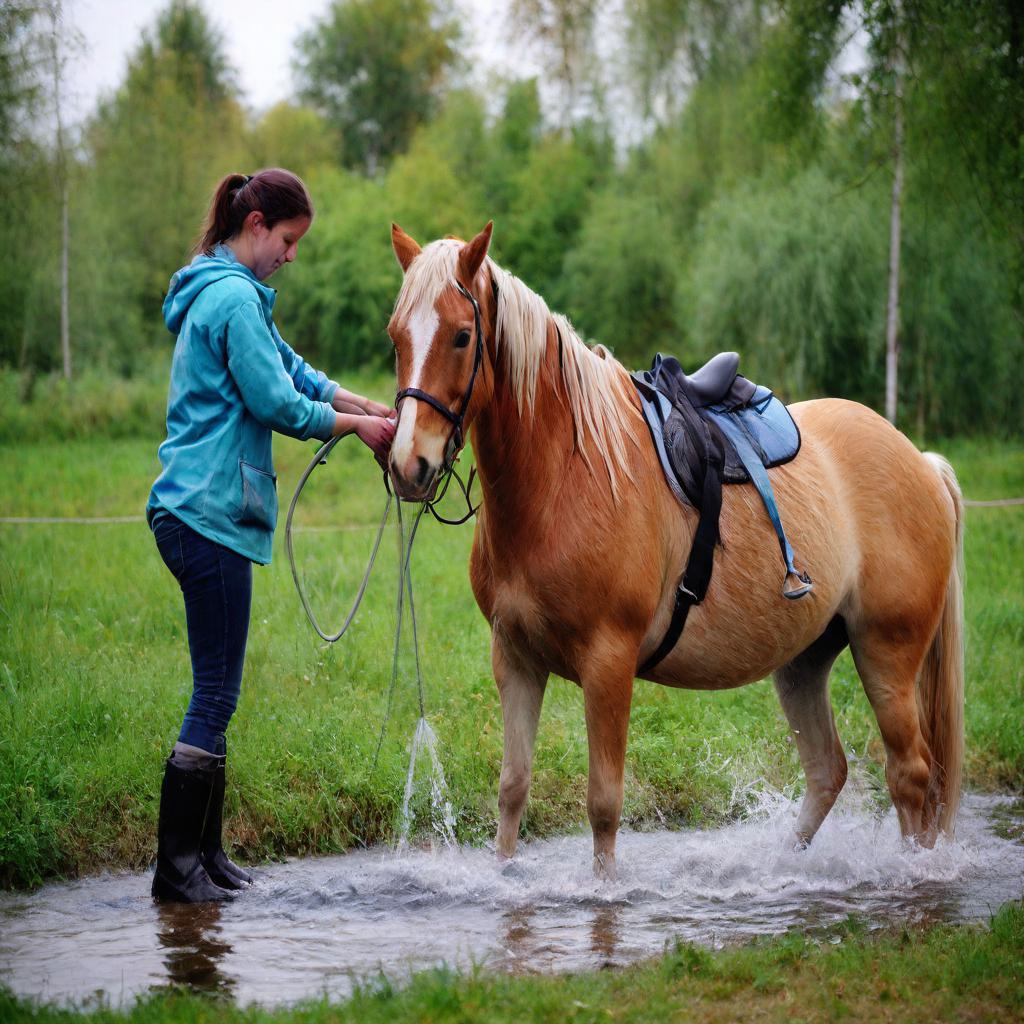}\\
        \\[-1.2em]
        & \multicolumn{8}{c}{wash horse}\\
        & \includegraphics[width=.104\linewidth]{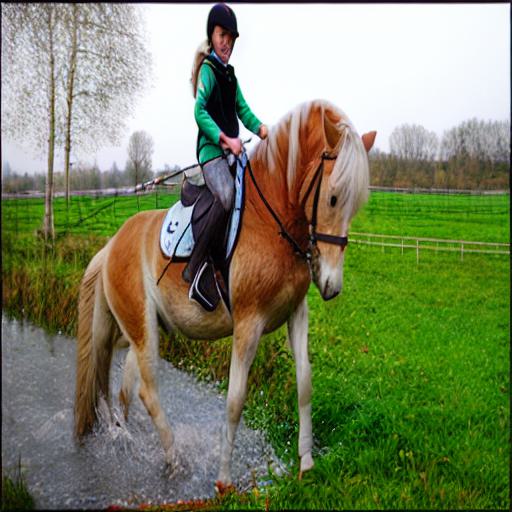}
        & \includegraphics[width=.104\linewidth]{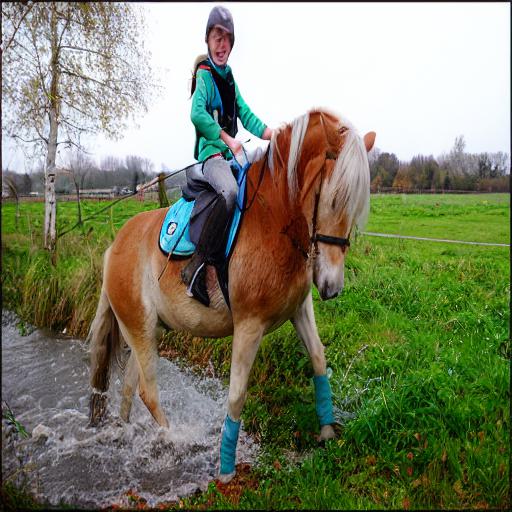}
        & \includegraphics[width=.104\linewidth]{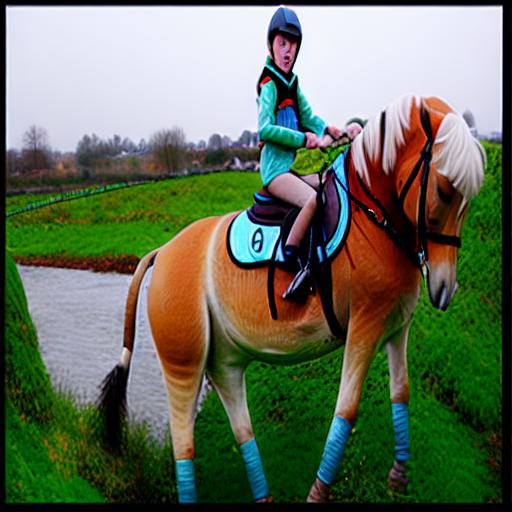}
        & \includegraphics[width=.104\linewidth]{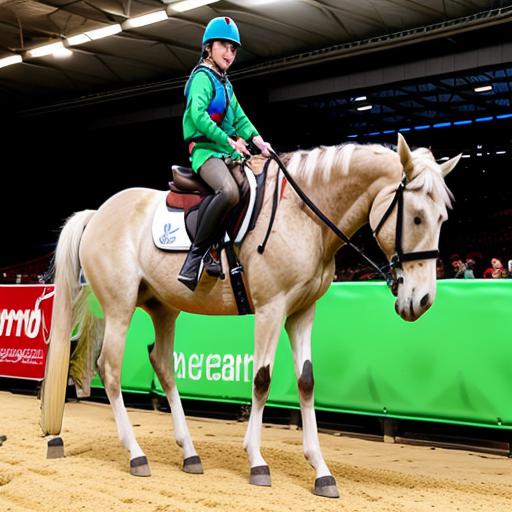}
        & \includegraphics[width=.104\linewidth]{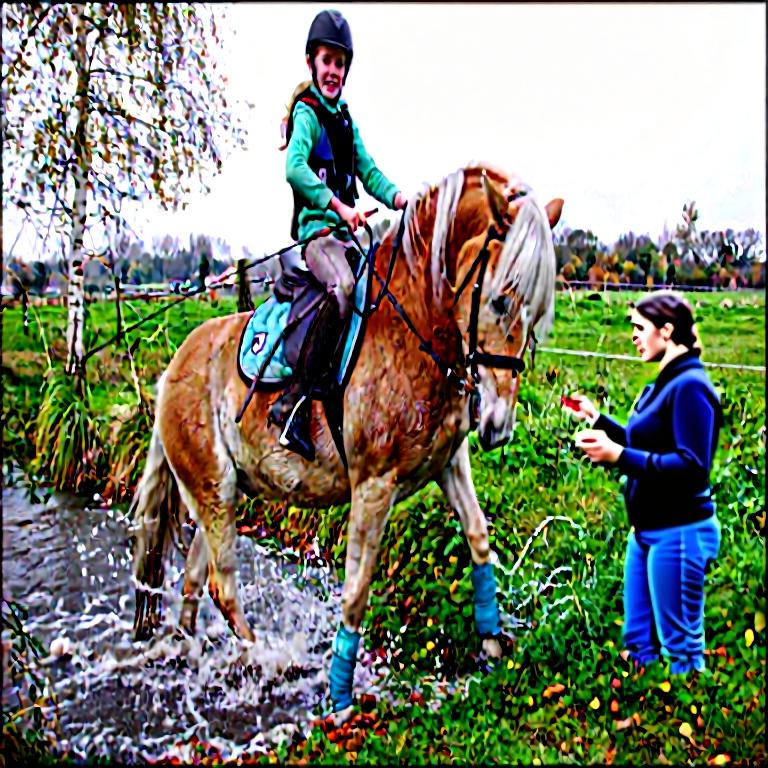}
        & \includegraphics[width=.104\linewidth]{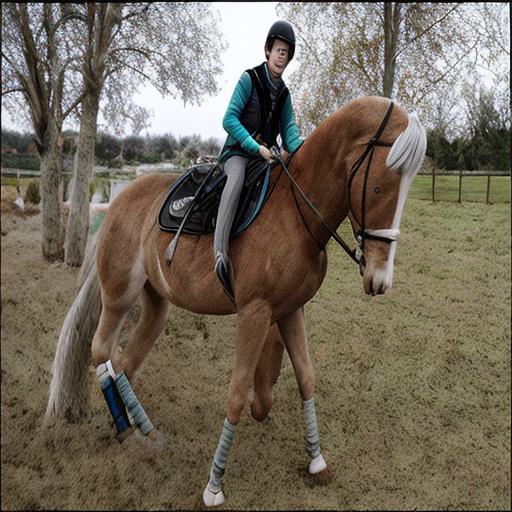}
        & \includegraphics[width=.104\linewidth]{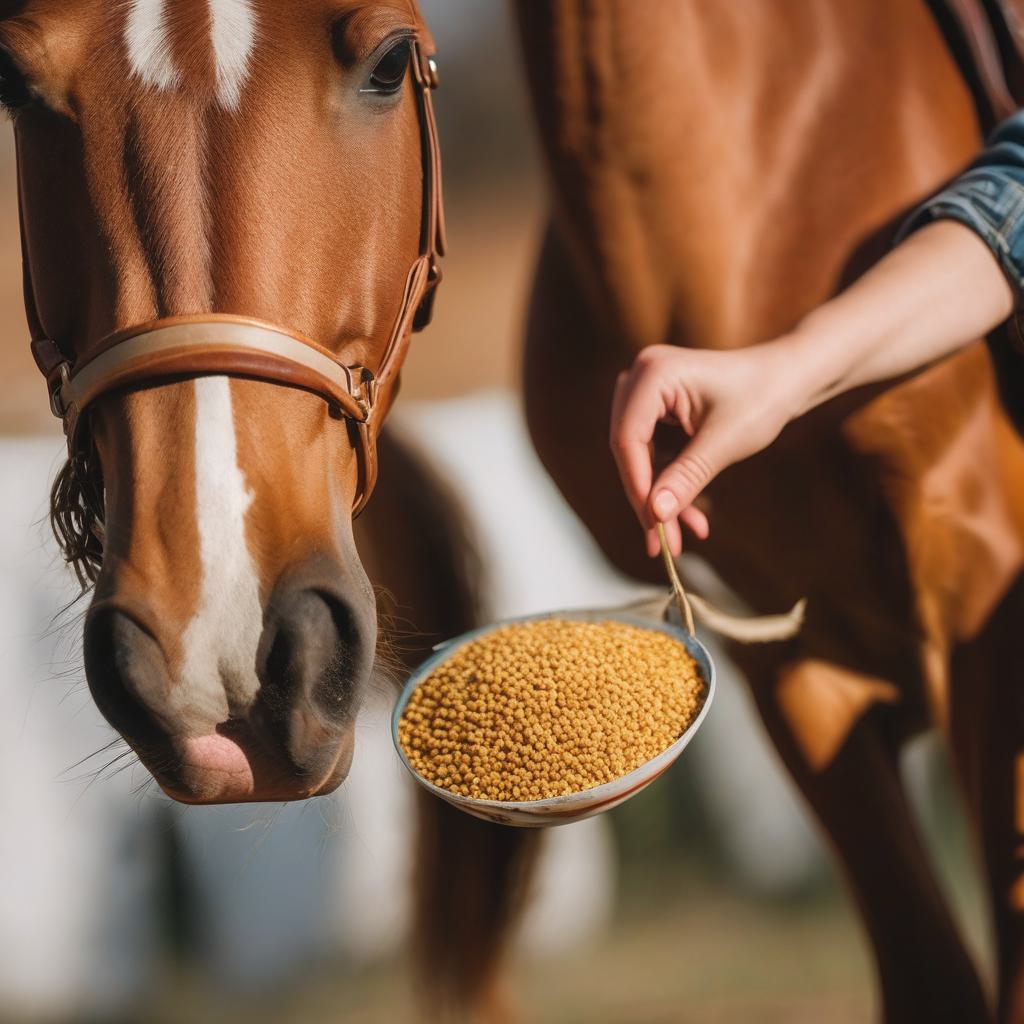}
        & \includegraphics[width=.104\linewidth]{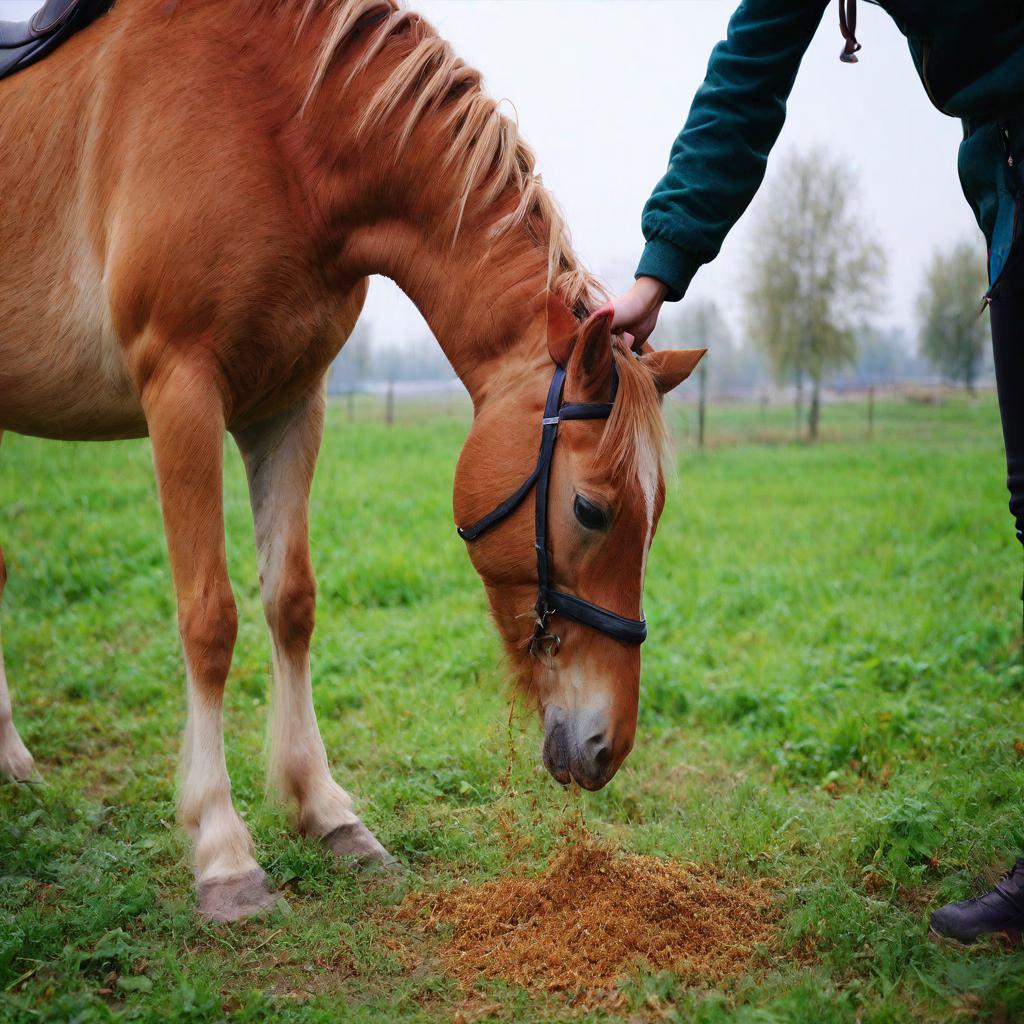}\\
        ride horse & \multicolumn{8}{c}{feed horse}\\
    \end{tabular}
    }
    \caption{Qualitative comparison with more existing baselines. The Source column shows the source image and its original interaction.}
    \label{fig:qualitative_5}
\end{figure}

%% file: assets/figure/ab_qual_3.tex
\begin{figure}
    \centering
    \setlength{\tabcolsep}{1pt} 
    \renewcommand{\arraystretch}{1} 
    \resizebox{0.9\linewidth}{!}{
    \begin{tabular}{ccccccccc}
        \footnotesize Source & \footnotesize HOIEdit & \footnotesize LEDITS++ & \footnotesize DAC & \footnotesize FireFlow & \footnotesize Ours\\
        
          \includegraphics[width=.150\linewidth]{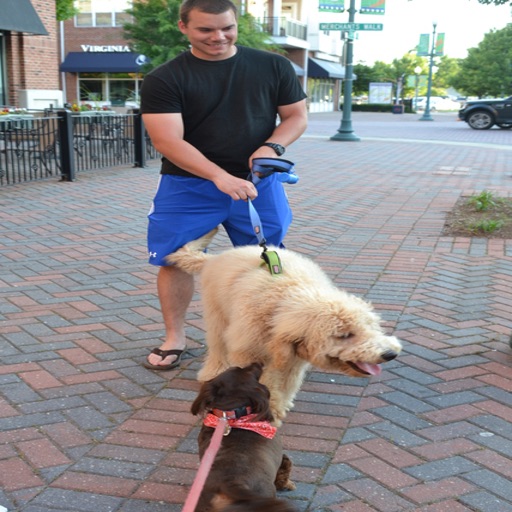}
        & \includegraphics[width=.150\linewidth]{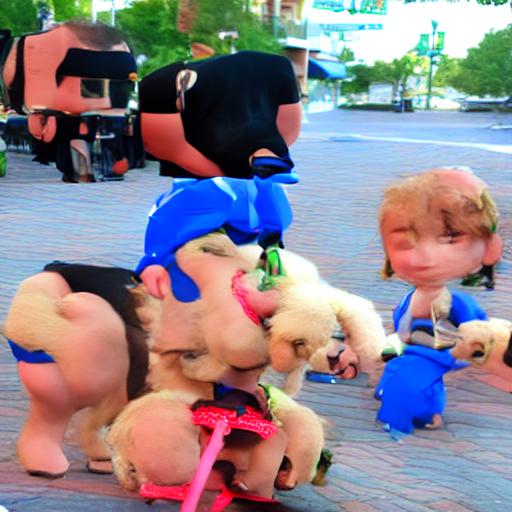}
        & \includegraphics[width=.150\linewidth]{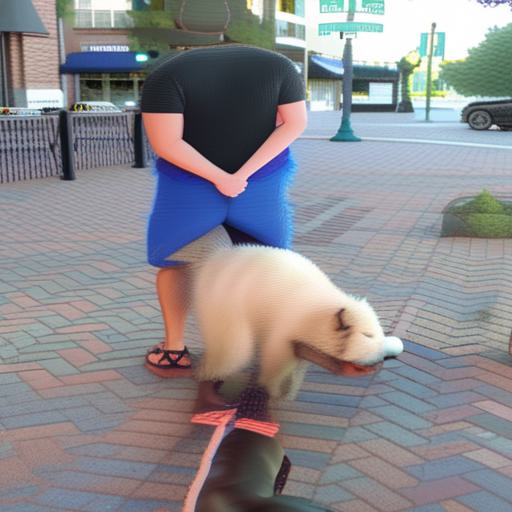}
        & \includegraphics[width=.150\linewidth]{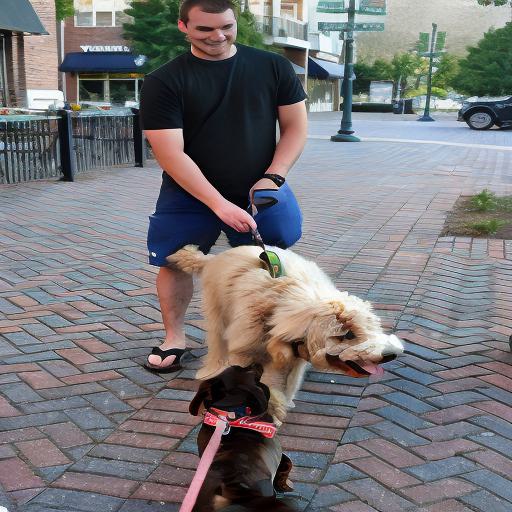}
        & \includegraphics[width=.150\linewidth]{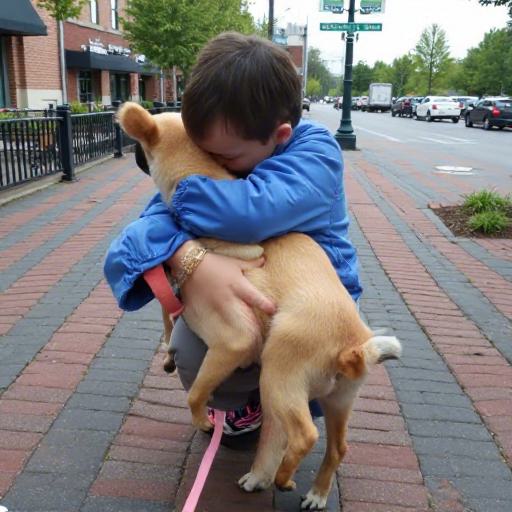}
        & \includegraphics[width=.150\linewidth]{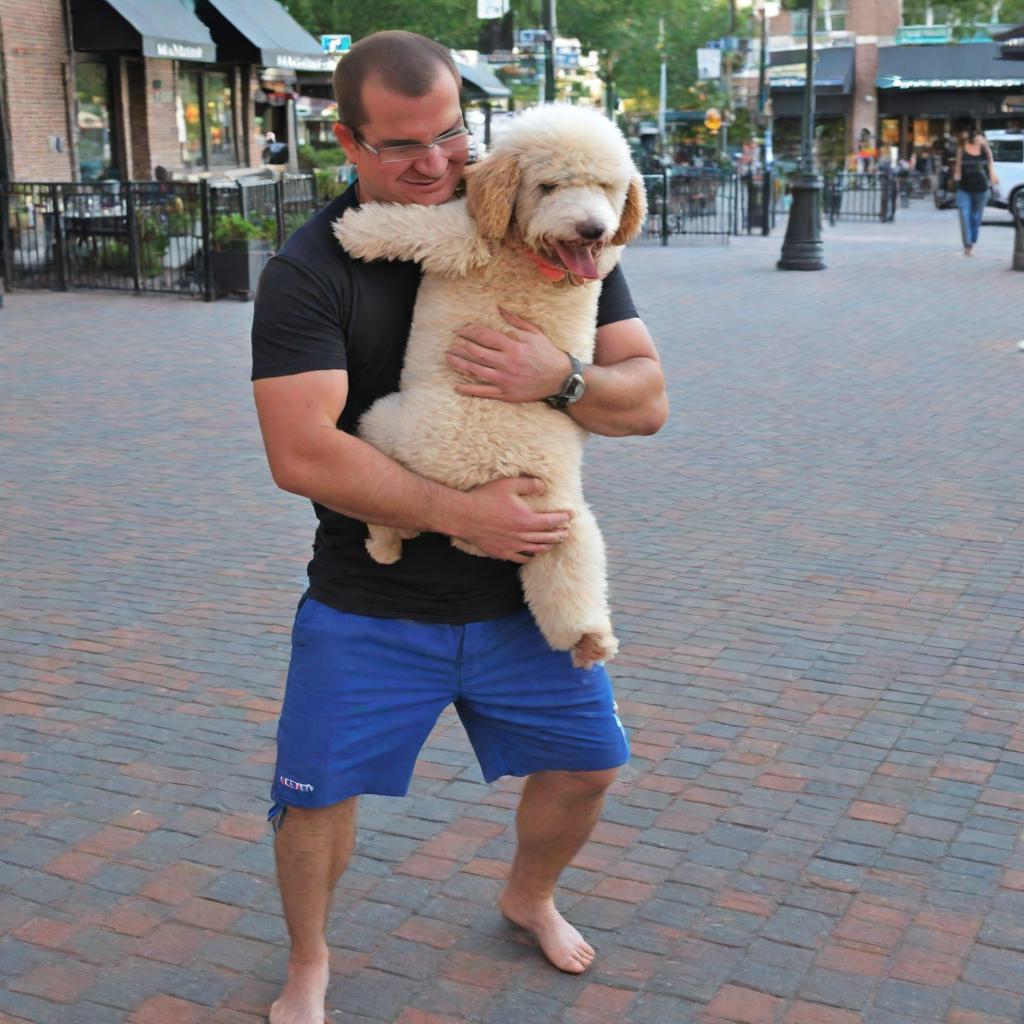}\\
        \\[-1.2em]
        walk dog & \multicolumn{5}{c}{hug dog}\\
        \includegraphics[width=.150\linewidth]{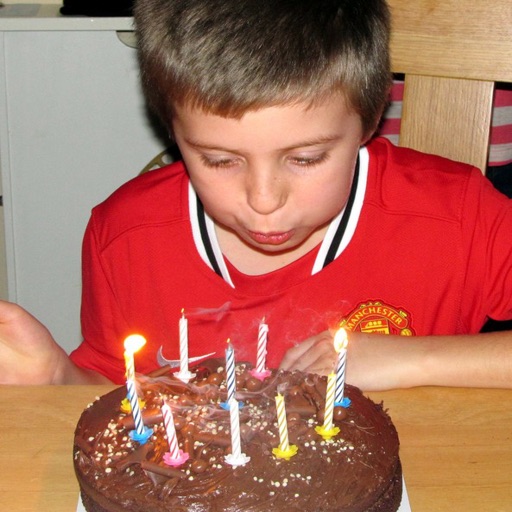}
        & \includegraphics[width=.150\linewidth]{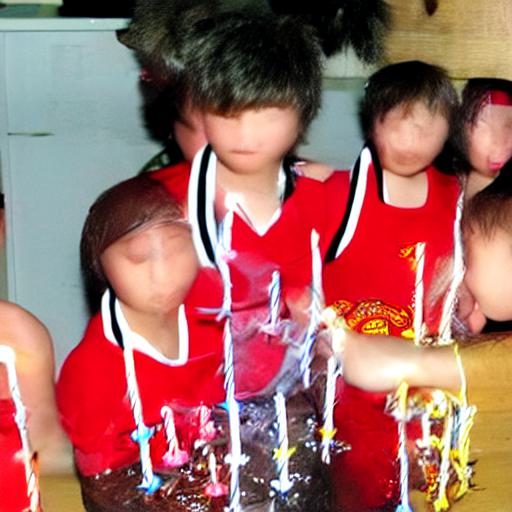}
        & \includegraphics[width=.150\linewidth]{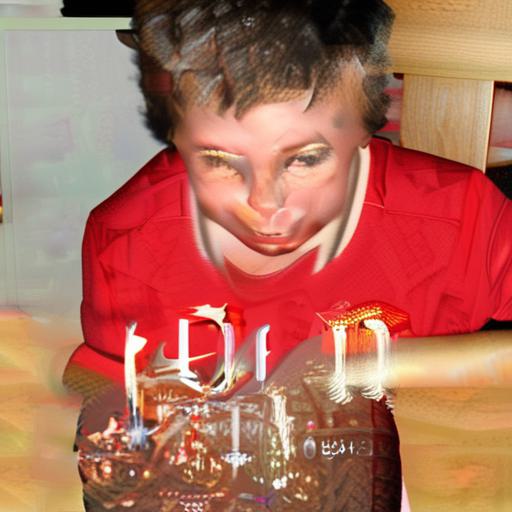}
        & \includegraphics[width=.150\linewidth]{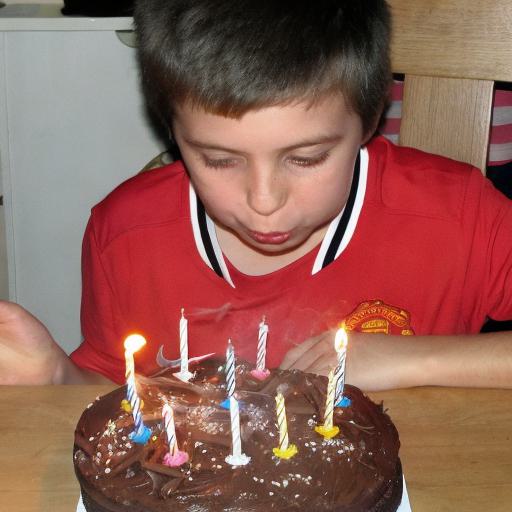}
        & \includegraphics[width=.150\linewidth]{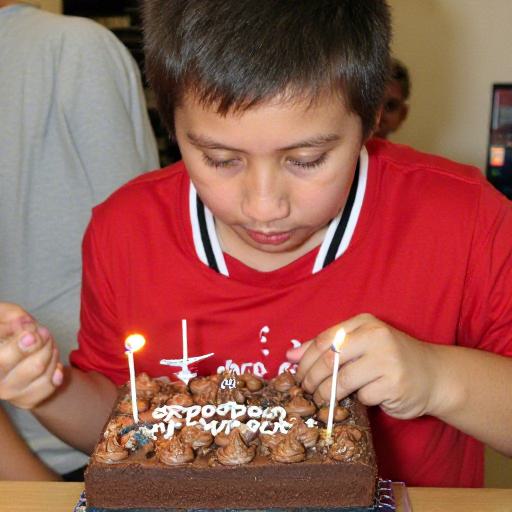}
        & \includegraphics[width=.150\linewidth]{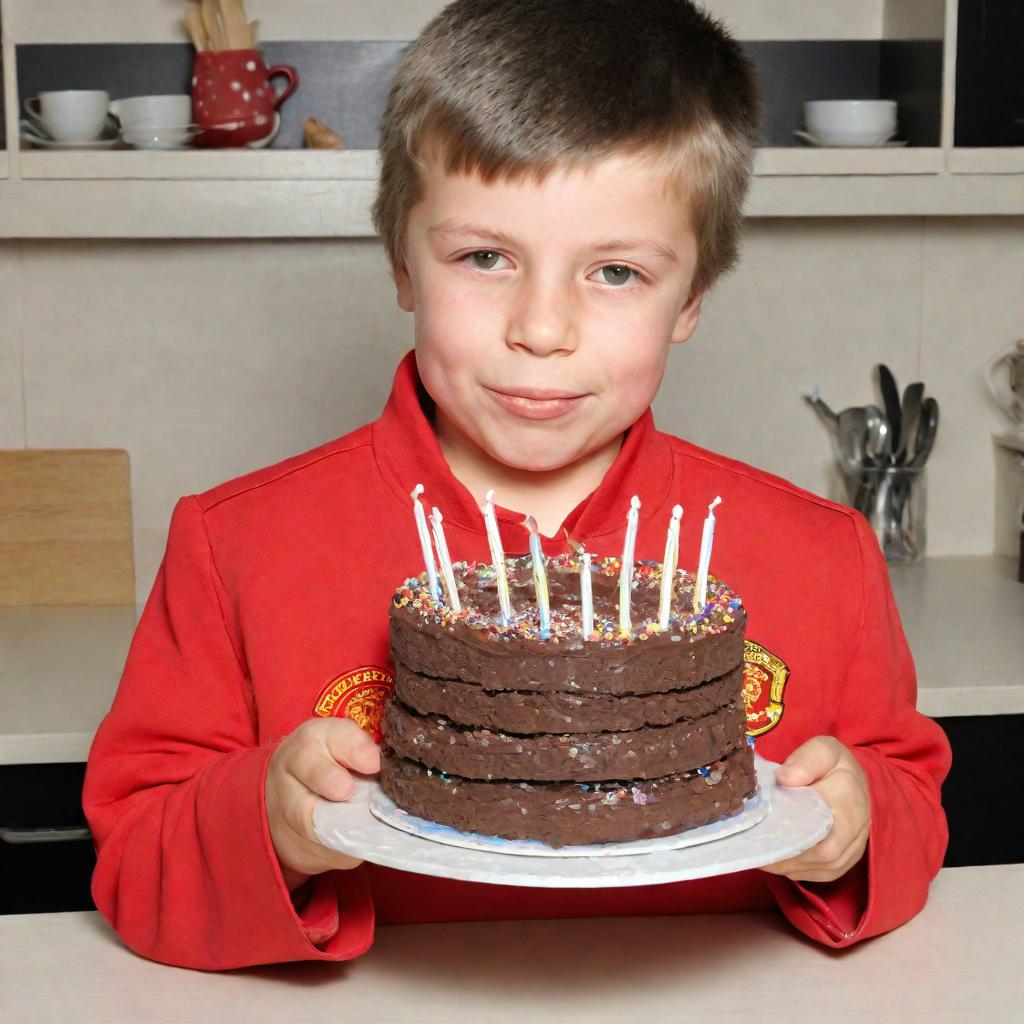}\\
        \\[-1.2em]
        & \multicolumn{5}{c}{hold cake}\\
        
        & \includegraphics[width=.150\linewidth]{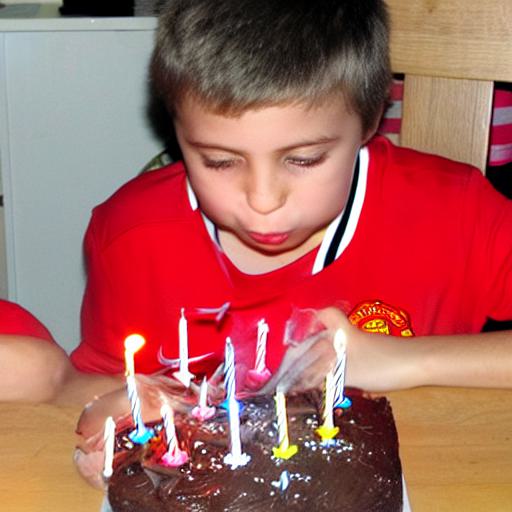}
        & \includegraphics[width=.150\linewidth]{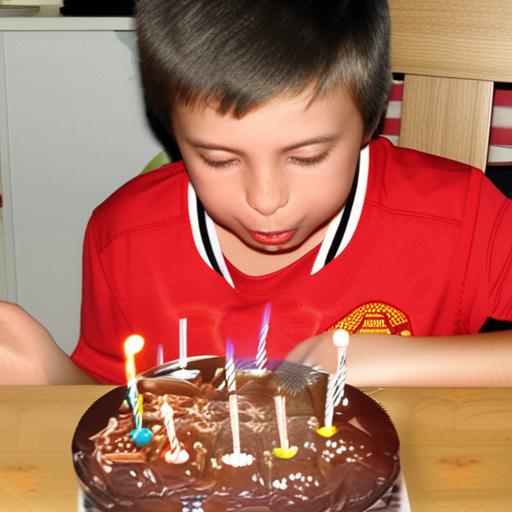}
        & \includegraphics[width=.150\linewidth]{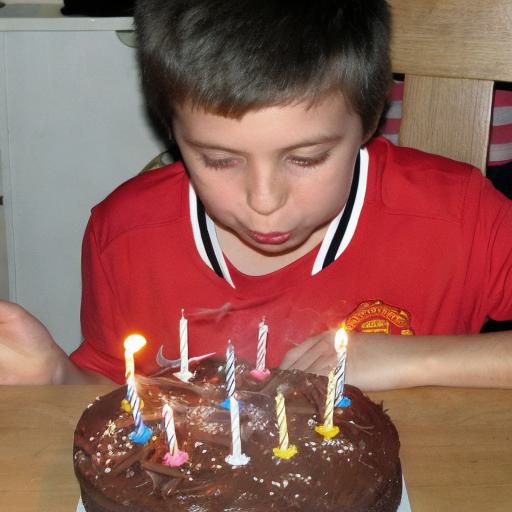}
        & \includegraphics[width=.150\linewidth]{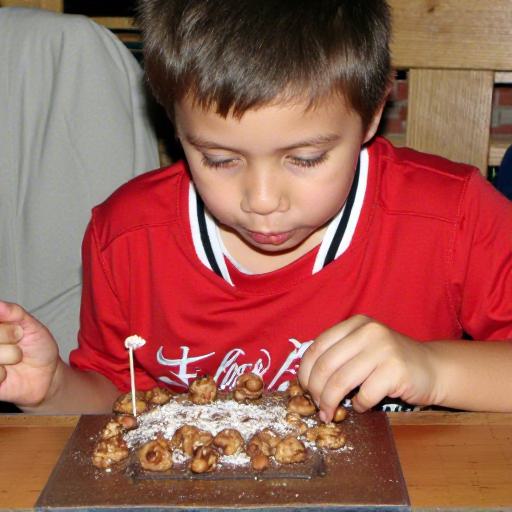}
        & \includegraphics[width=.150\linewidth]{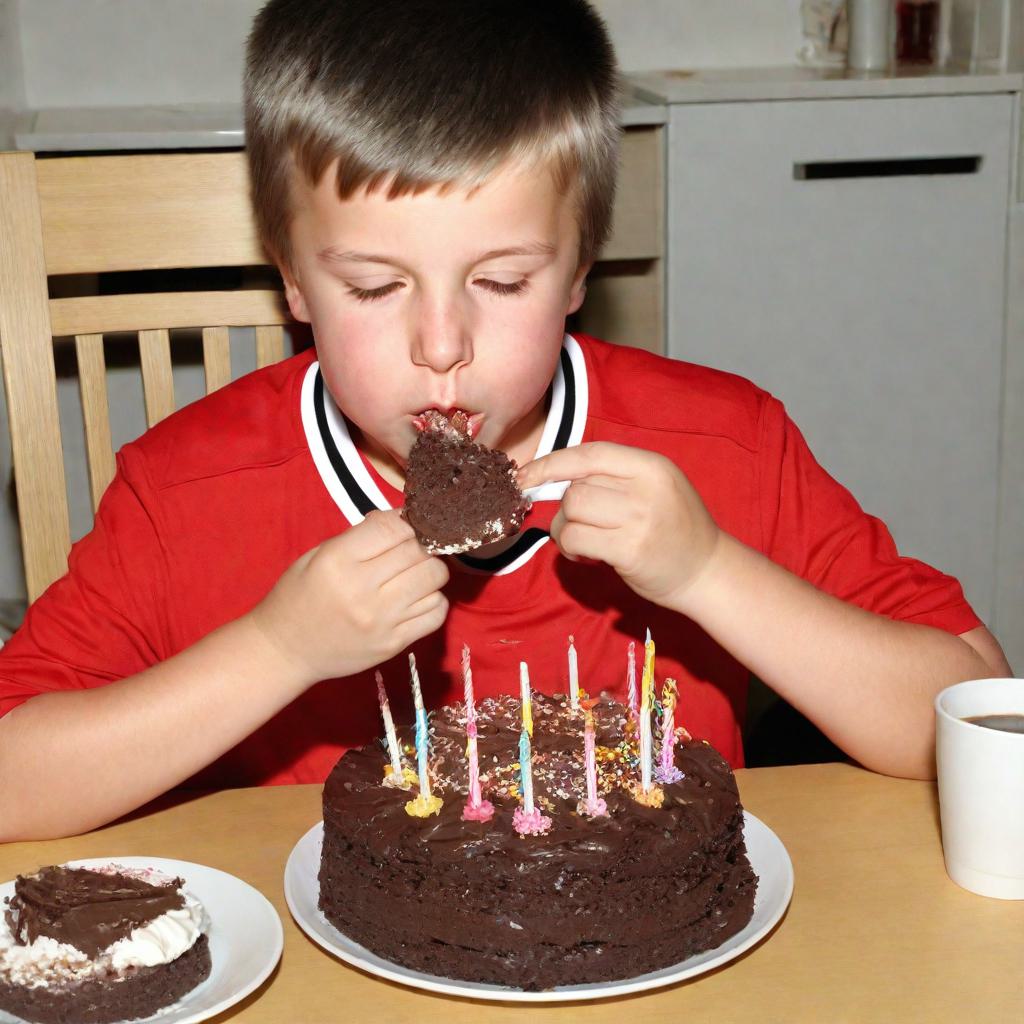}\\
        \\[-1.2em]
        blow cake & \multicolumn{5}{c}{eat cake}\\
        
        & \includegraphics[width=.150\linewidth]{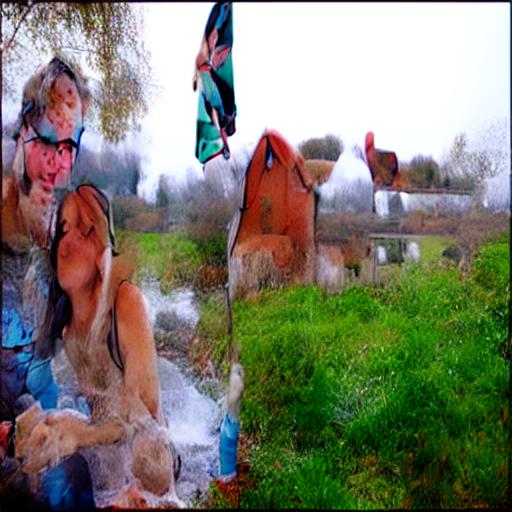}
        & \includegraphics[width=.150\linewidth]{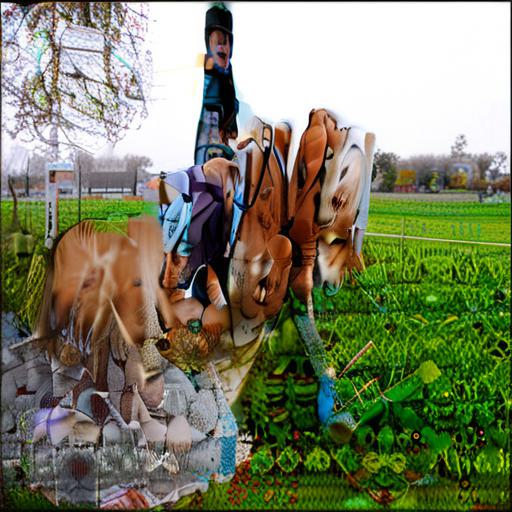}
        & \includegraphics[width=.150\linewidth]{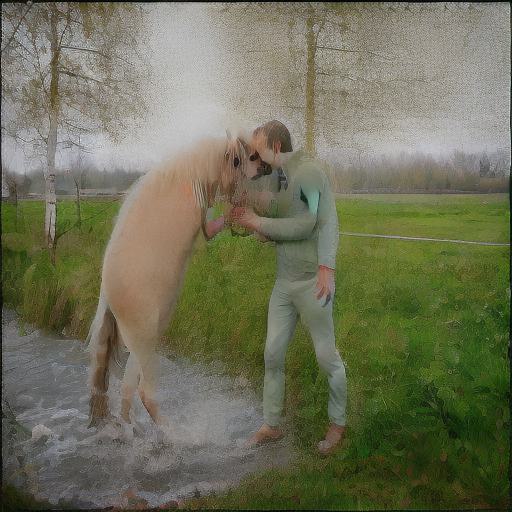}
        & \includegraphics[width=.150\linewidth]{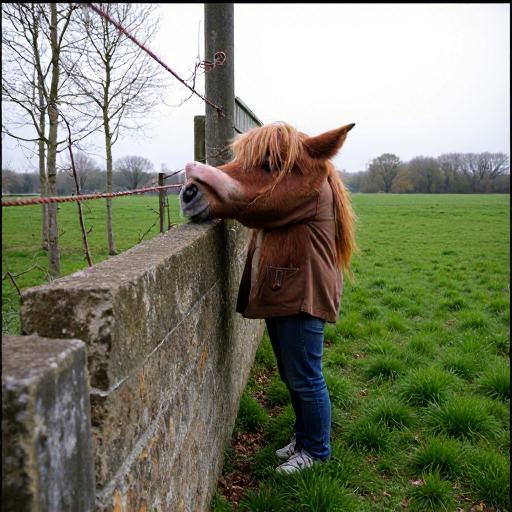}
        & \includegraphics[width=.150\linewidth]{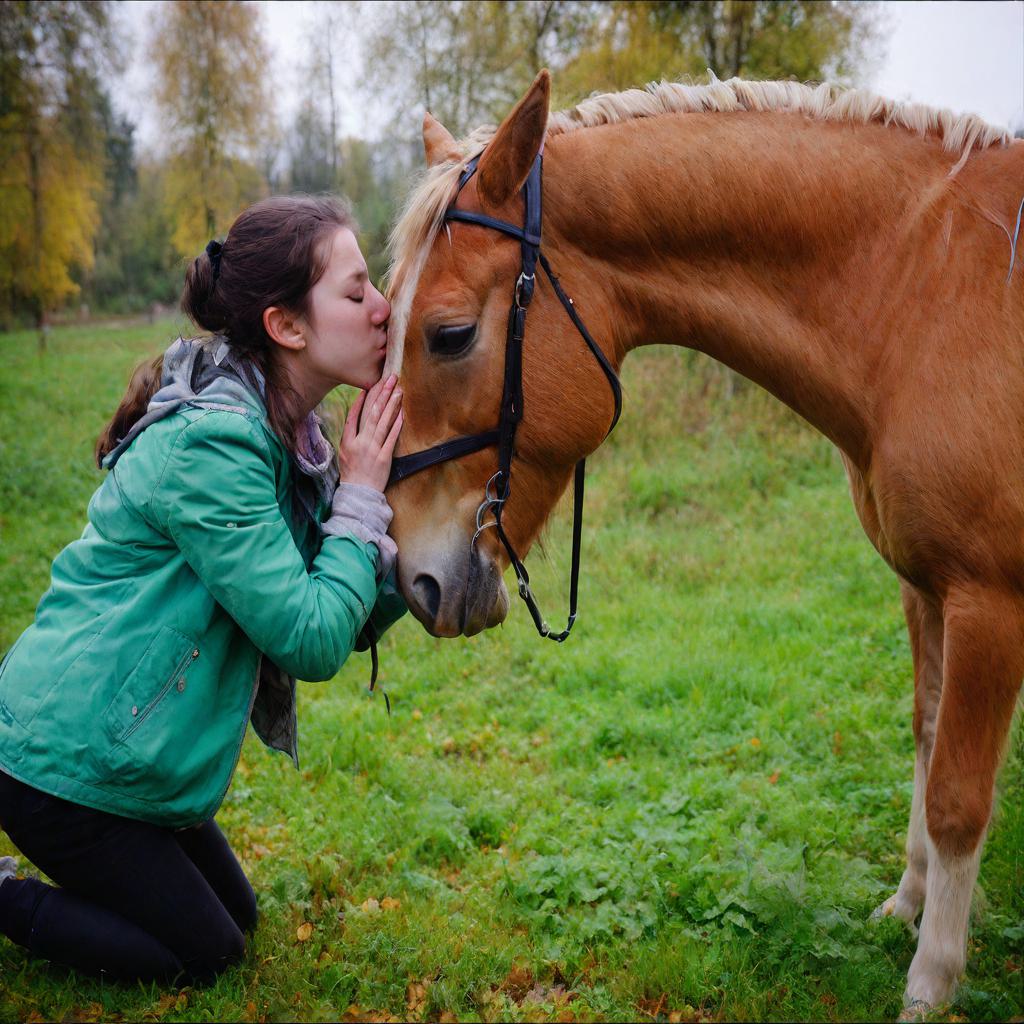}\\
        \\[-1.2em]
        & \multicolumn{5}{c}{kiss horse}\\

          \includegraphics[width=.150\linewidth]{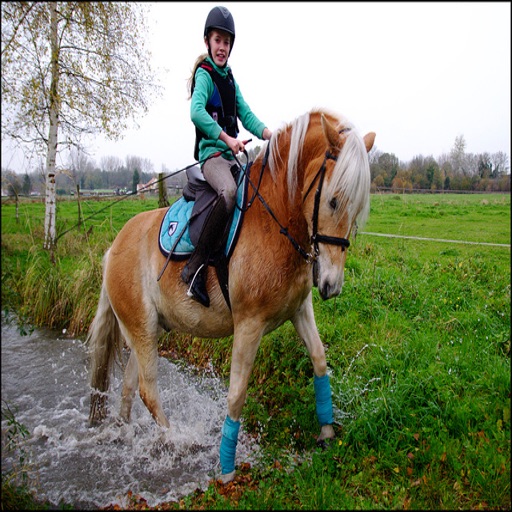}
        & \includegraphics[width=.150\linewidth]{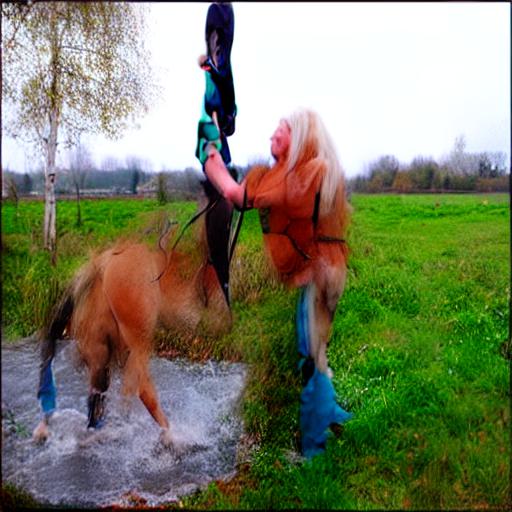}
        & \includegraphics[width=.150\linewidth]{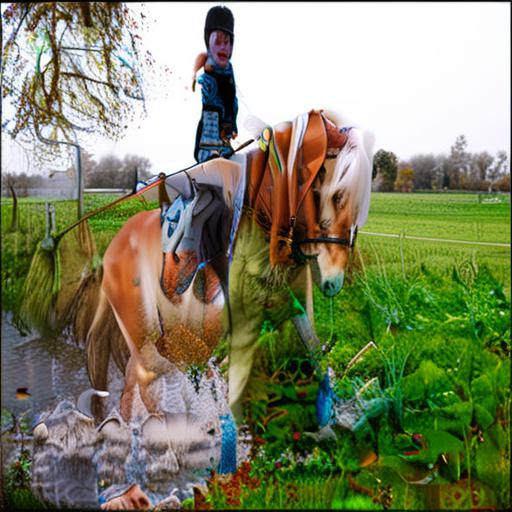}
        & \includegraphics[width=.150\linewidth]{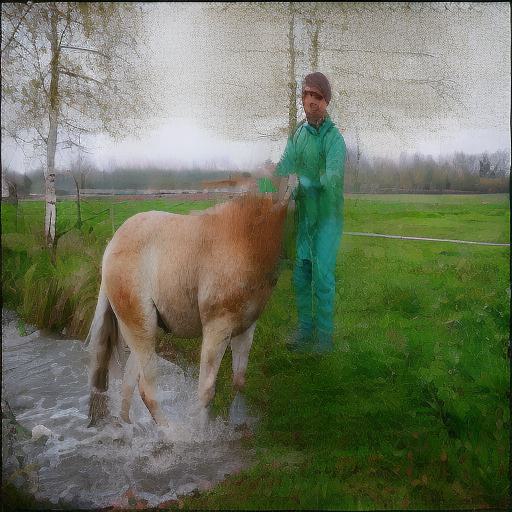}
        & \includegraphics[width=.150\linewidth]{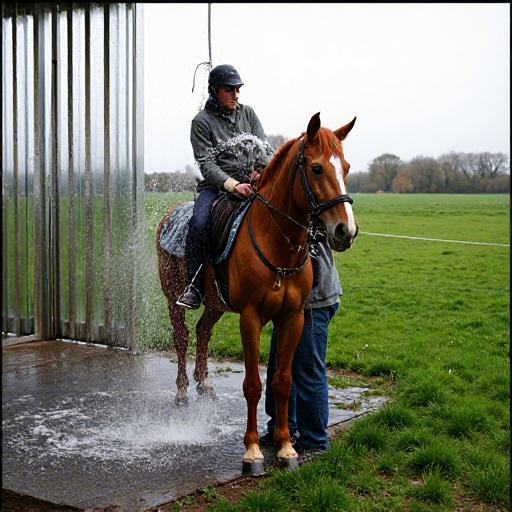}
        & \includegraphics[width=.150\linewidth]{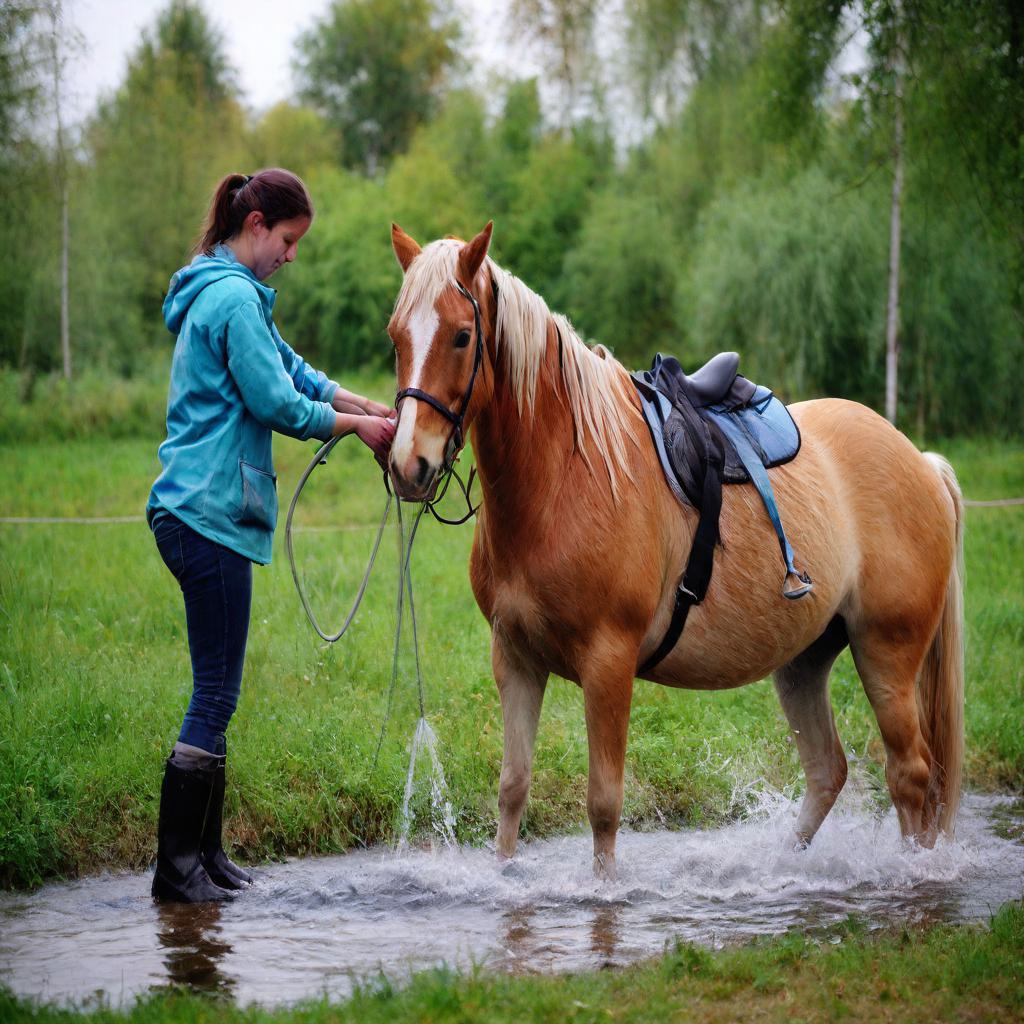}\\
        \\[-1.2em]
        & \multicolumn{5}{c}{wash horse}\\
        
        & \includegraphics[width=.150\linewidth]{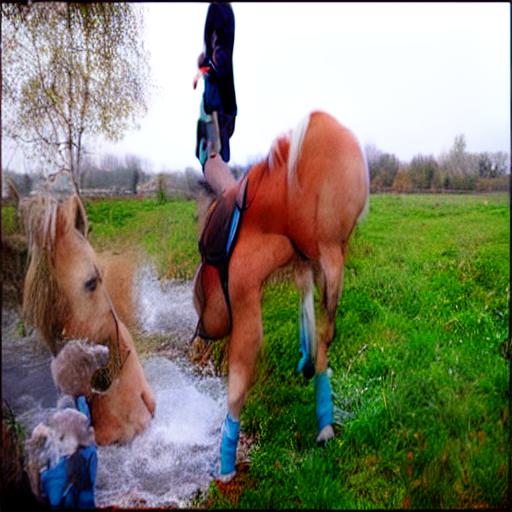}
        & \includegraphics[width=.150\linewidth]{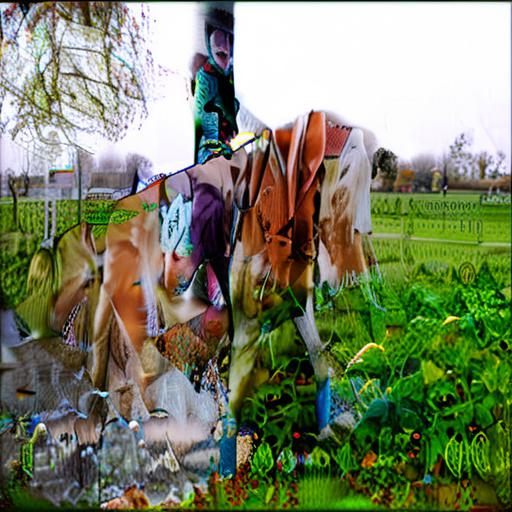}
        & \includegraphics[width=.150\linewidth]{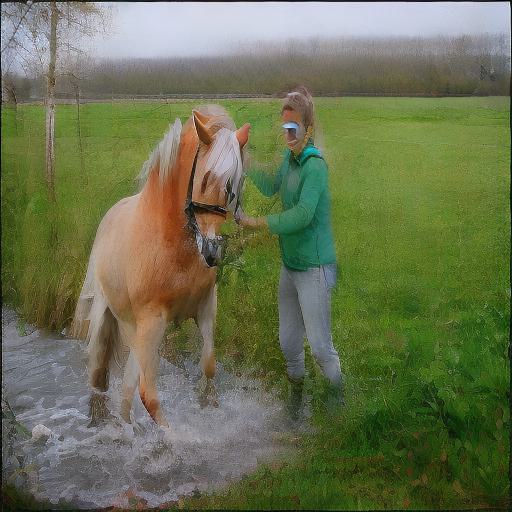}
        & \includegraphics[width=.150\linewidth]{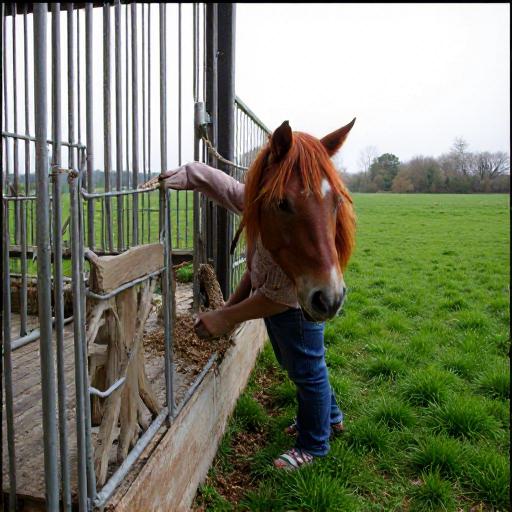}
        & \includegraphics[width=.150\linewidth]{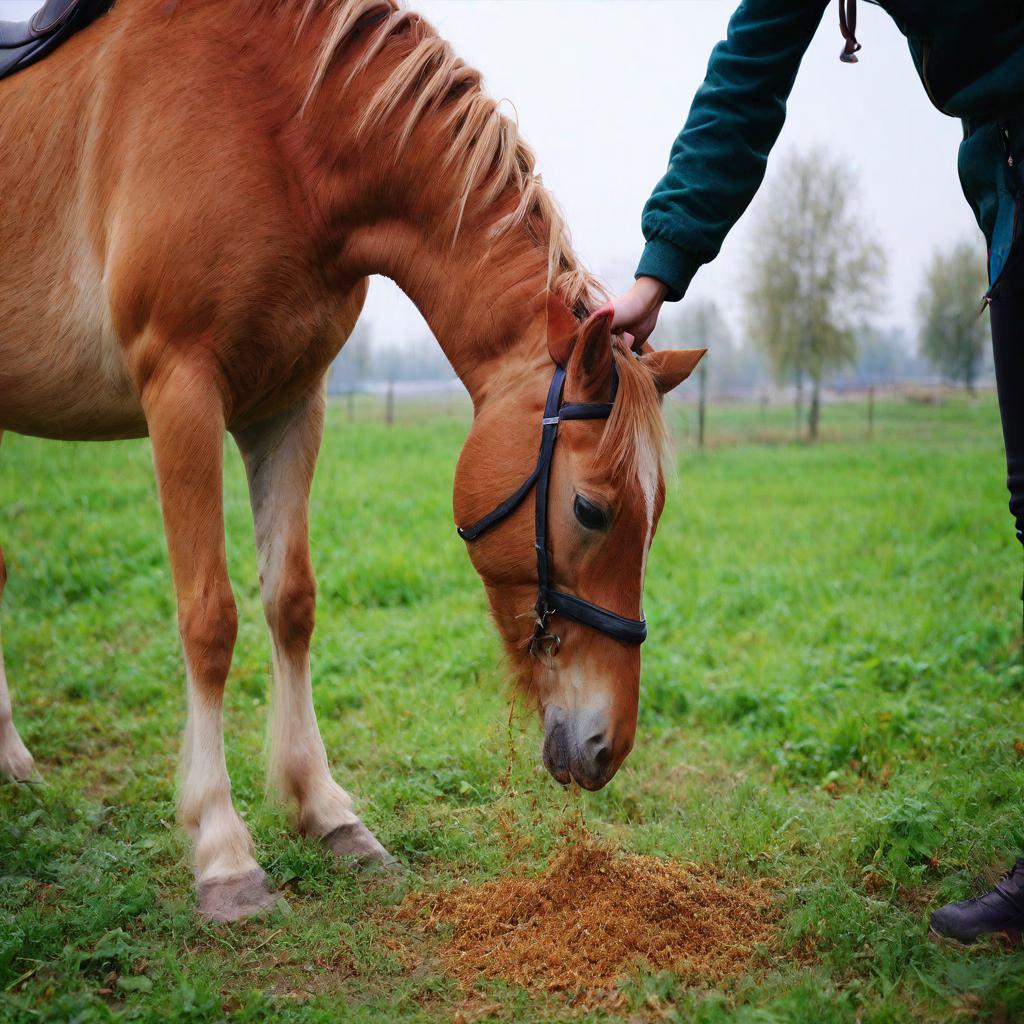}\\
        ride horse & \multicolumn{5}{c}{feed horse}\\
    \end{tabular}
    }
    \caption{Qualitative comparison with more existing baselines.}
    \label{fig:qualitative_3}
\end{figure}

%% file: assets/figure/ab_qual_4.tex
\begin{figure*}
    \centering
    \setlength{\tabcolsep}{1pt} 
    \renewcommand{\arraystretch}{1} 
    \begin{tabular}{ccccccccc}
        \footnotesize Source & \footnotesize P2P & \footnotesize PnP & \footnotesize SVDiff & \footnotesize InfEdit & \footnotesize CDS & \footnotesize TurboEdit & \footnotesize \resizebox{.104\linewidth}{!}{Break-A-Scene} & \footnotesize Ours\\
        & \includegraphics[width=.104\linewidth]{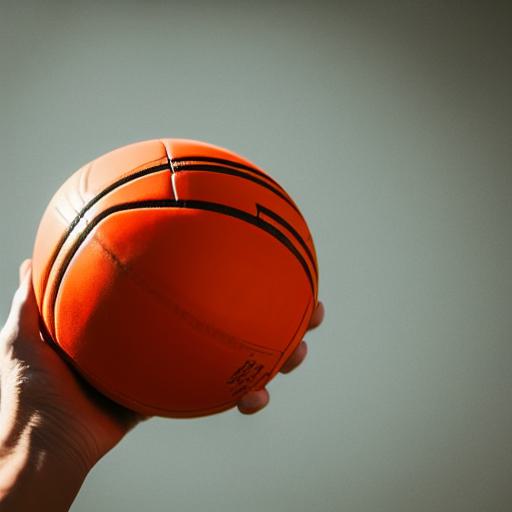}
        & \includegraphics[width=.104\linewidth]{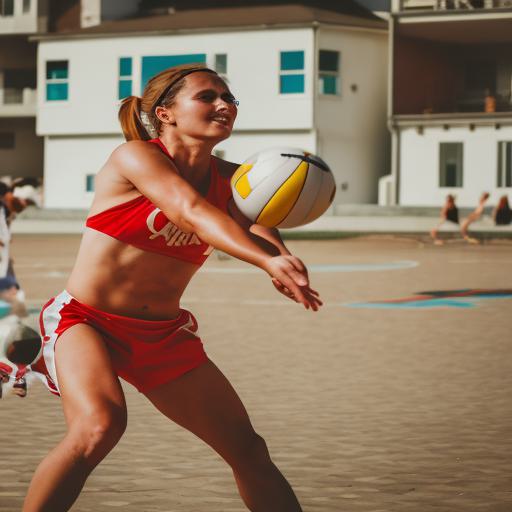}
        & \includegraphics[width=.104\linewidth]{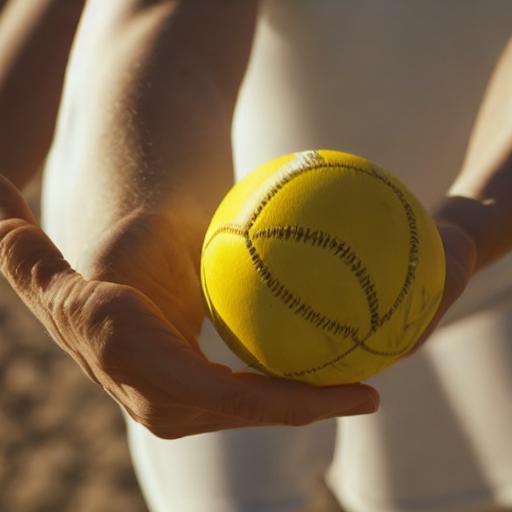}
        & \includegraphics[width=.104\linewidth]{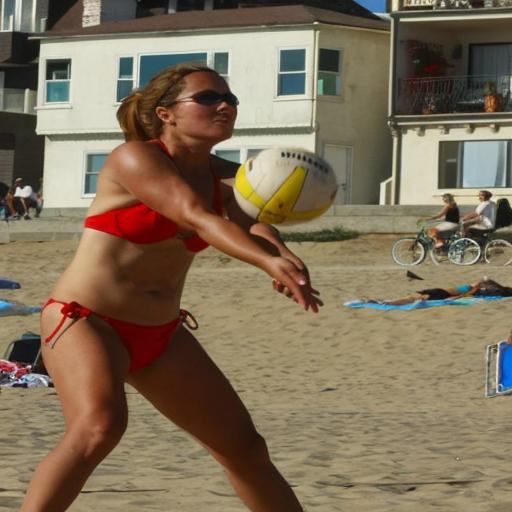}
        & \includegraphics[width=.104\linewidth]{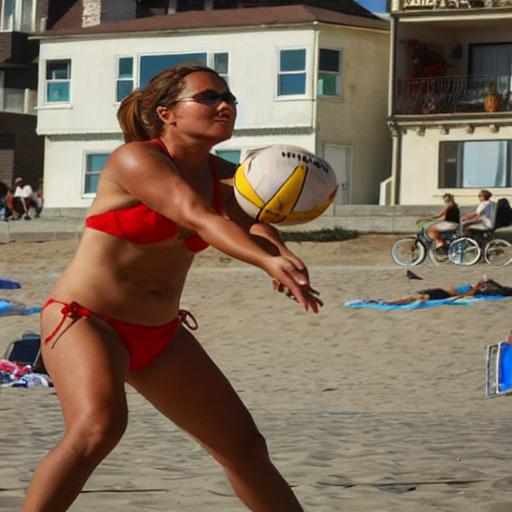}
        & \includegraphics[width=.104\linewidth]{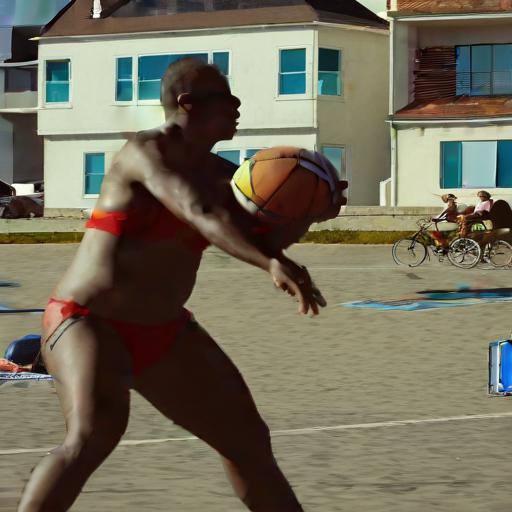}
        & \includegraphics[width=.104\linewidth]{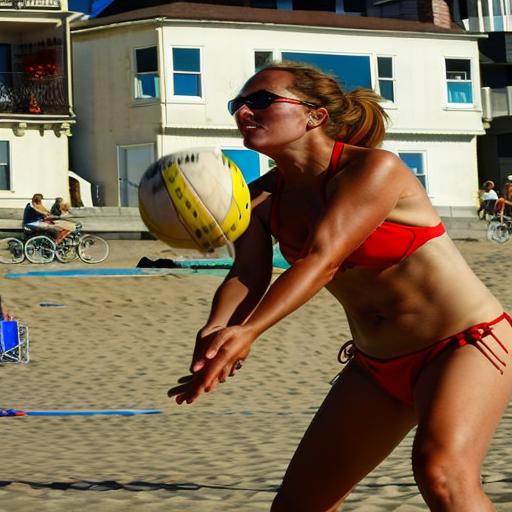}
        & \includegraphics[width=.104\linewidth]{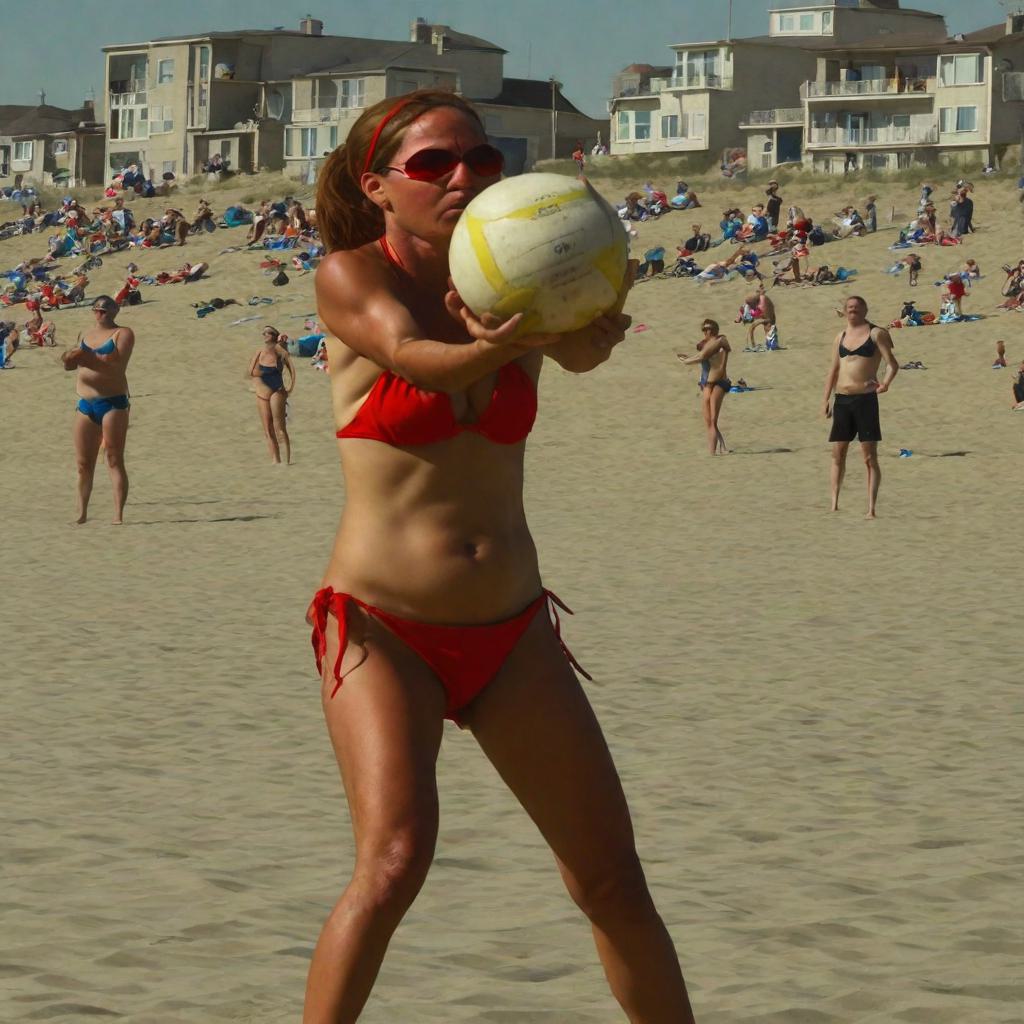}\\
        \\[-1.2em]
        & \multicolumn{8}{c}{hold ball}\\
          \includegraphics[width=.104\linewidth]{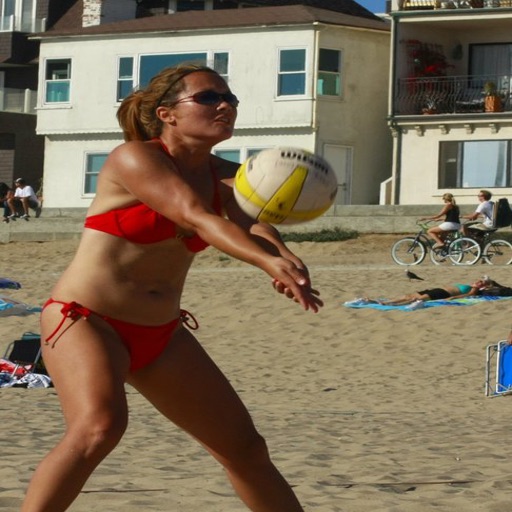}
        & \includegraphics[width=.104\linewidth]{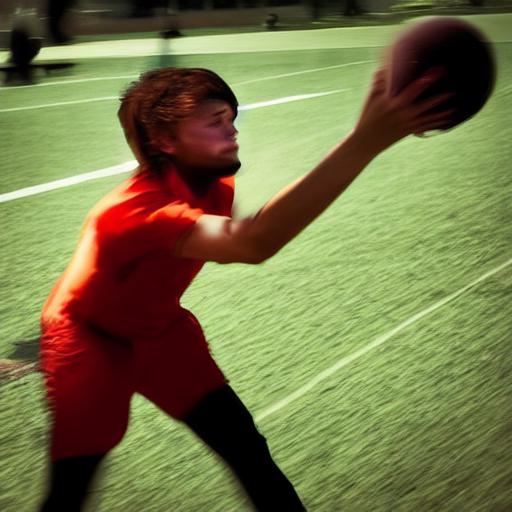}
        & \includegraphics[width=.104\linewidth]{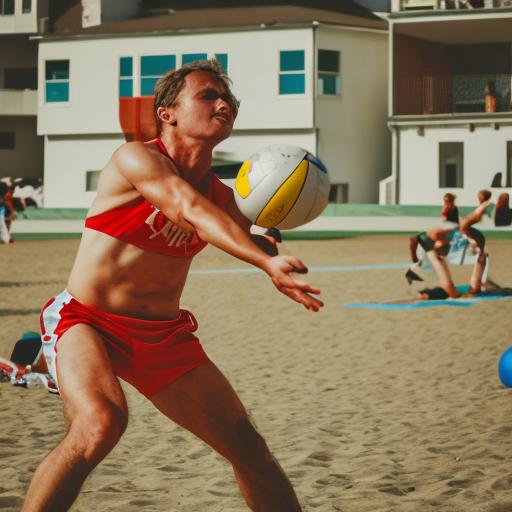}
        & \includegraphics[width=.104\linewidth]{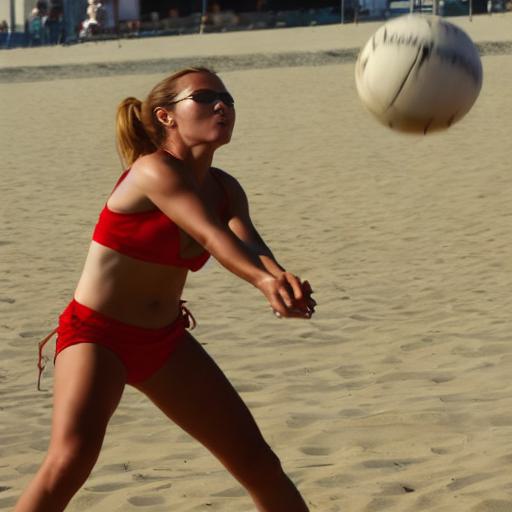}
        & \includegraphics[width=.104\linewidth]{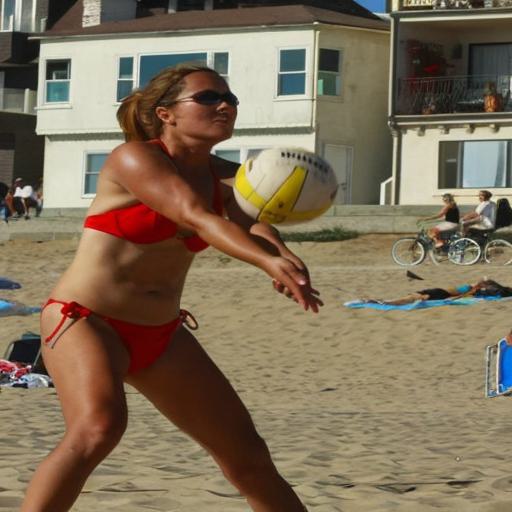}
        & \includegraphics[width=.104\linewidth]{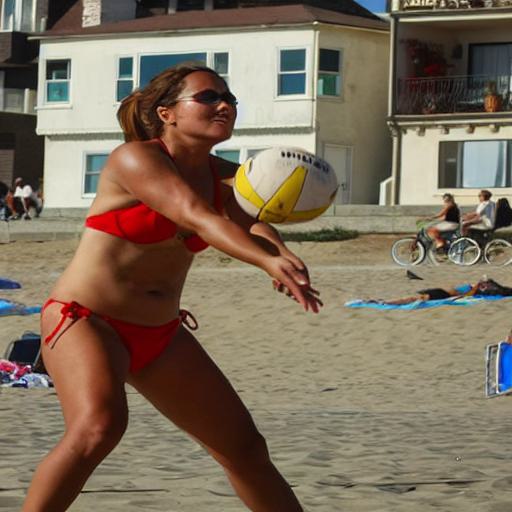}
        & \includegraphics[width=.104\linewidth]{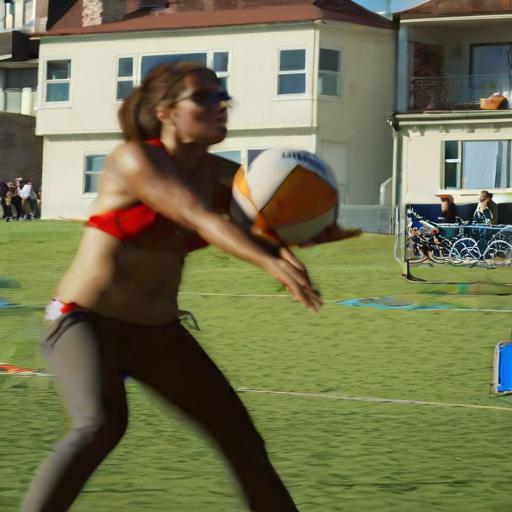}
        & \includegraphics[width=.104\linewidth]{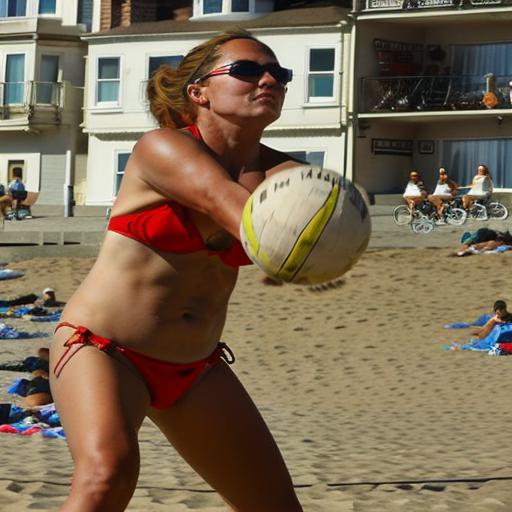}
        & \includegraphics[width=.104\linewidth]{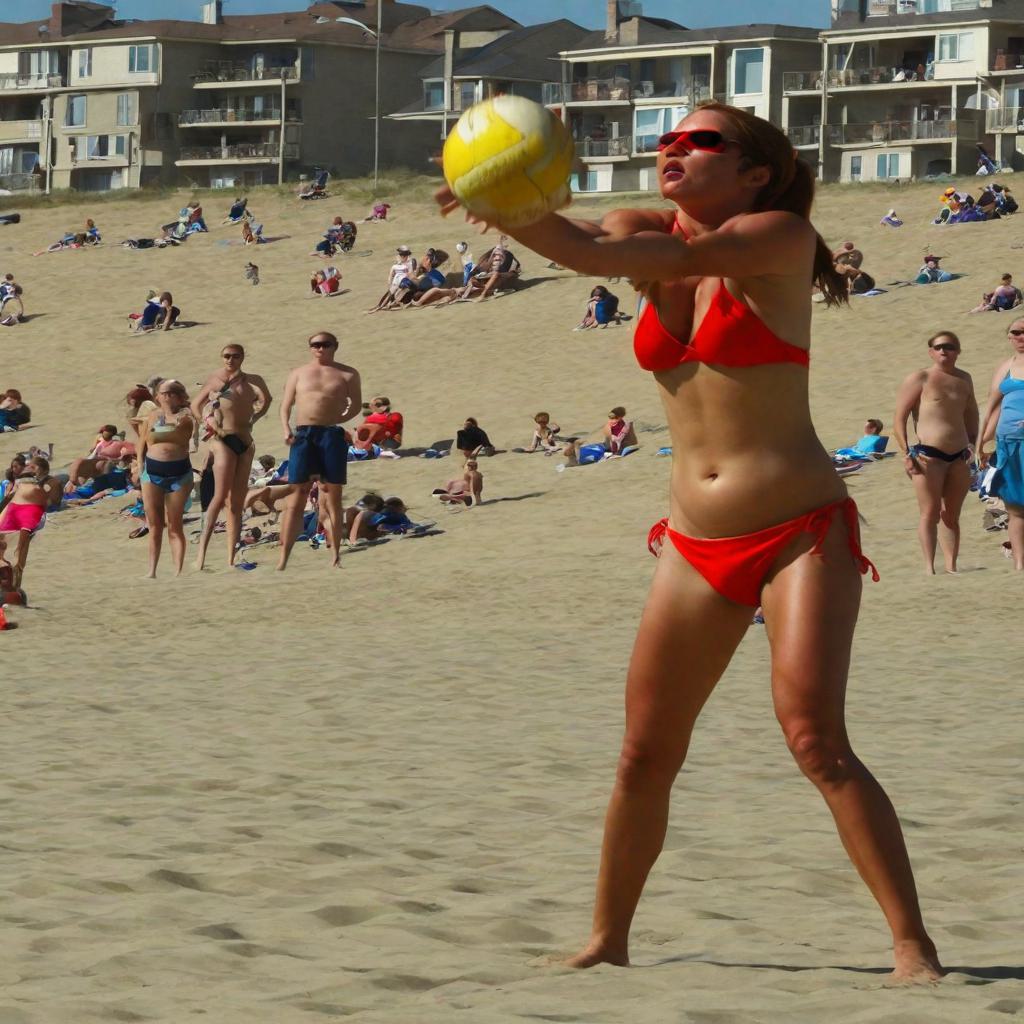}\\
        \\[-1.2em]
        & \multicolumn{8}{c}{catch ball}\\
        & \includegraphics[width=.104\linewidth]{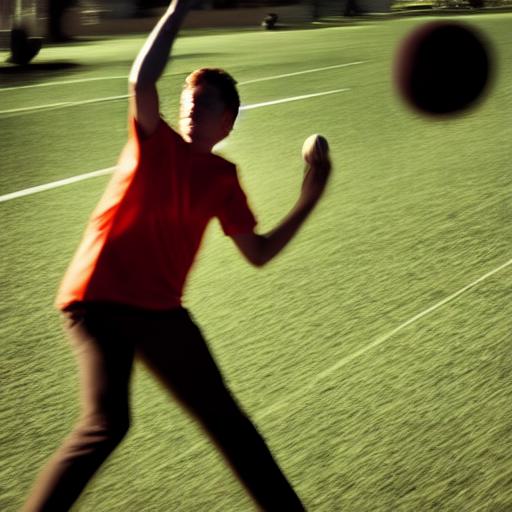}
        & \includegraphics[width=.104\linewidth]{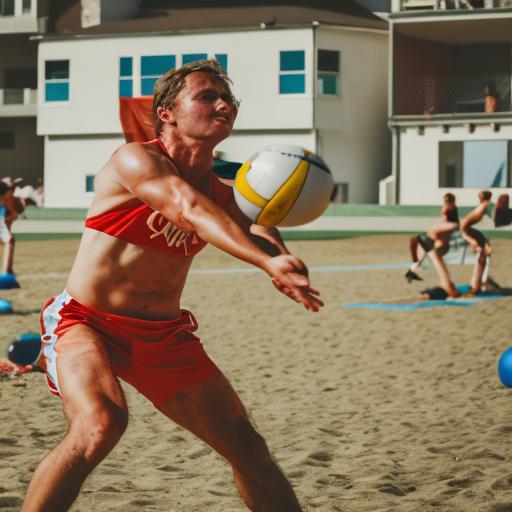}
        & \includegraphics[width=.104\linewidth]{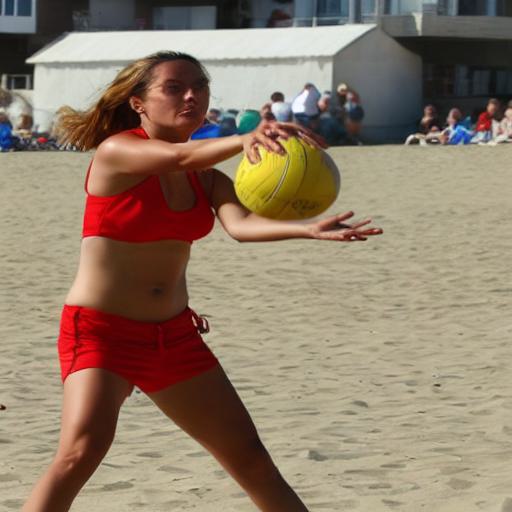}
        & \includegraphics[width=.104\linewidth]{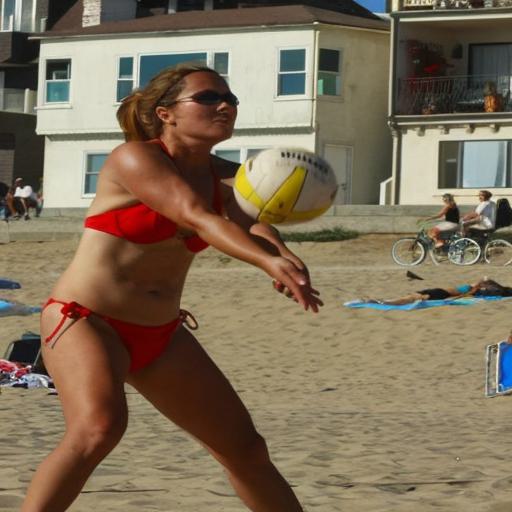}
        & \includegraphics[width=.104\linewidth]{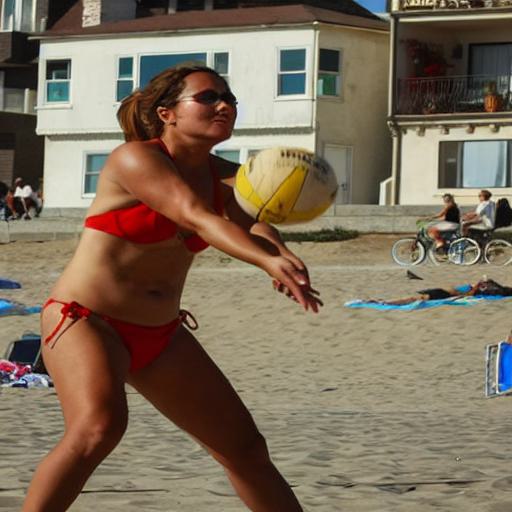}
        & \includegraphics[width=.104\linewidth]{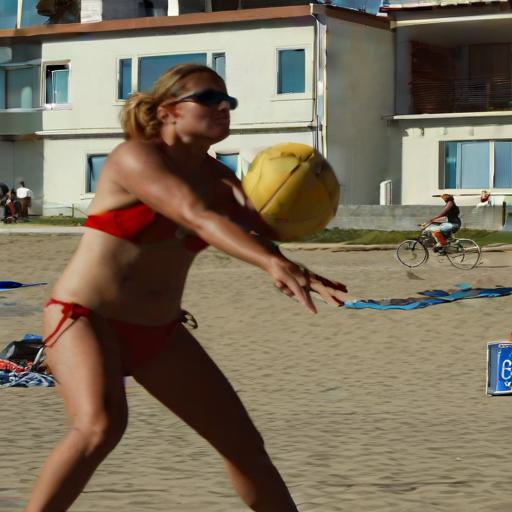}
        & \includegraphics[width=.104\linewidth]{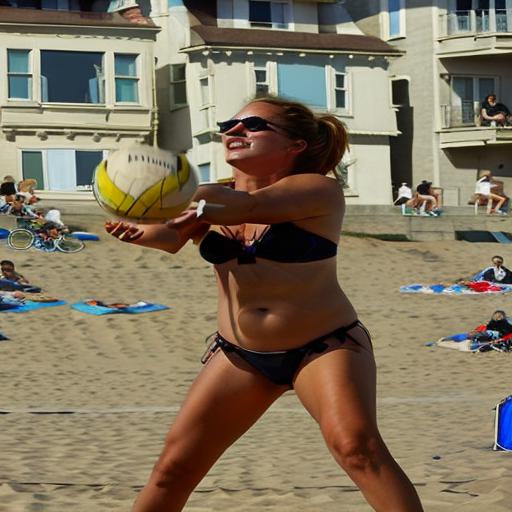}
        & \includegraphics[width=.104\linewidth]{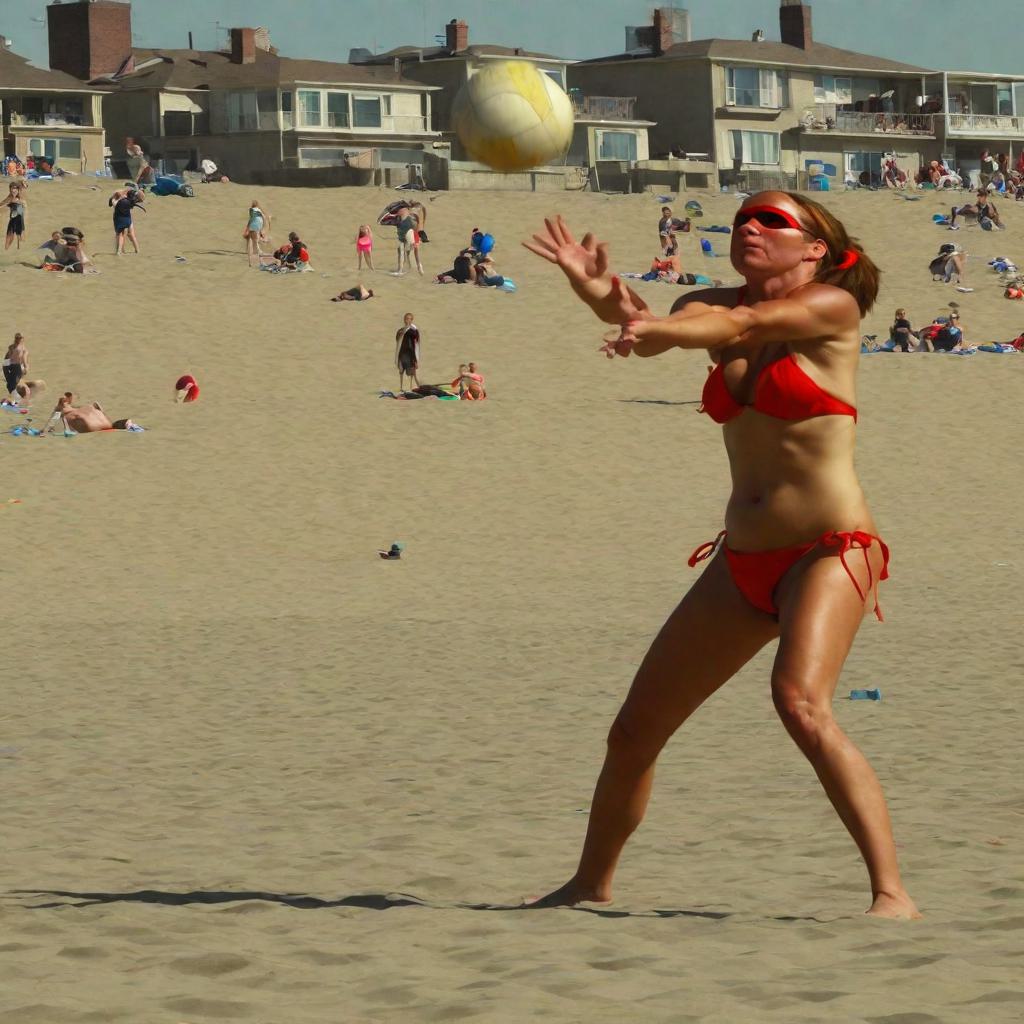}\\
        \\[-1.2em]
        hit ball & \multicolumn{8}{c}{throw ball}\\
        
          \includegraphics[width=.104\linewidth]{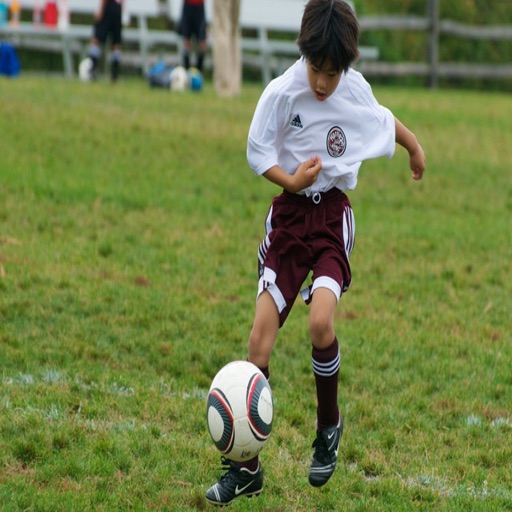}
        & \includegraphics[width=.104\linewidth]{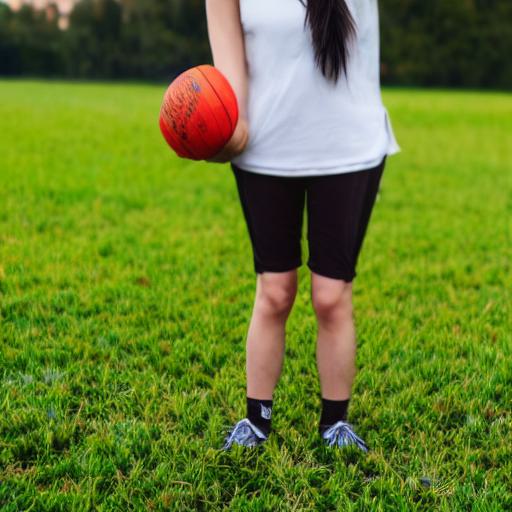}
        & \includegraphics[width=.104\linewidth]{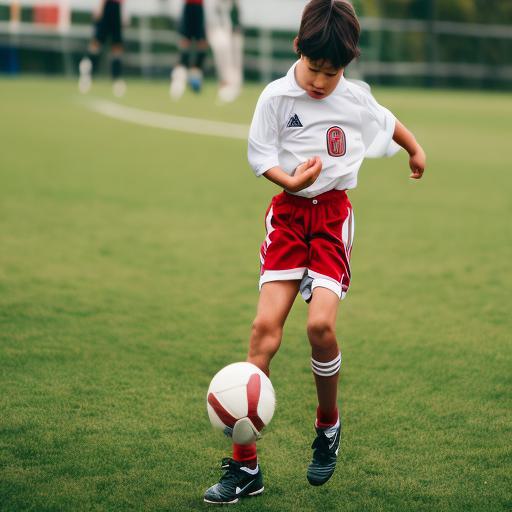}
        & \includegraphics[width=.104\linewidth]{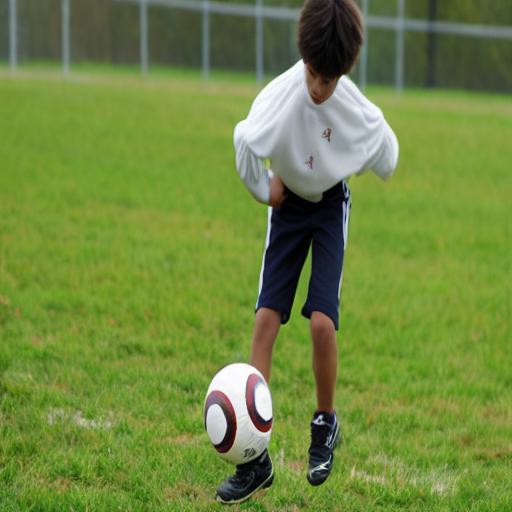}
        & \includegraphics[width=.104\linewidth]{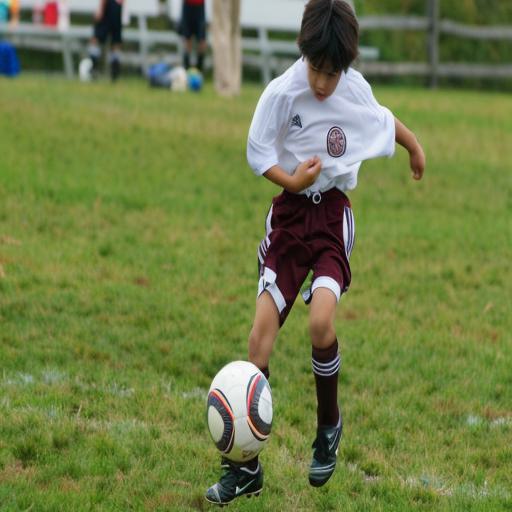}
        & \includegraphics[width=.104\linewidth]{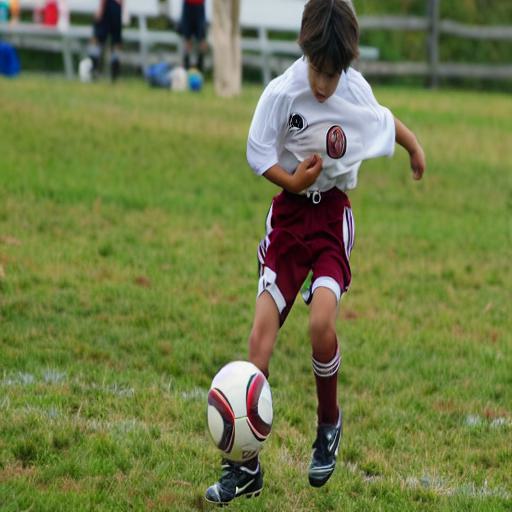}
        & \includegraphics[width=.104\linewidth]{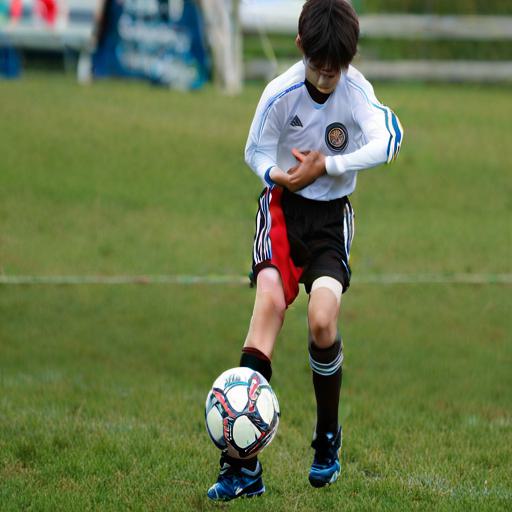}
        & \includegraphics[width=.104\linewidth]{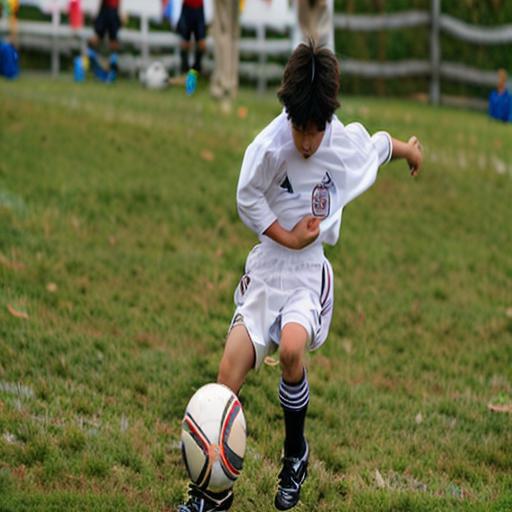}
        & \includegraphics[width=.104\linewidth]{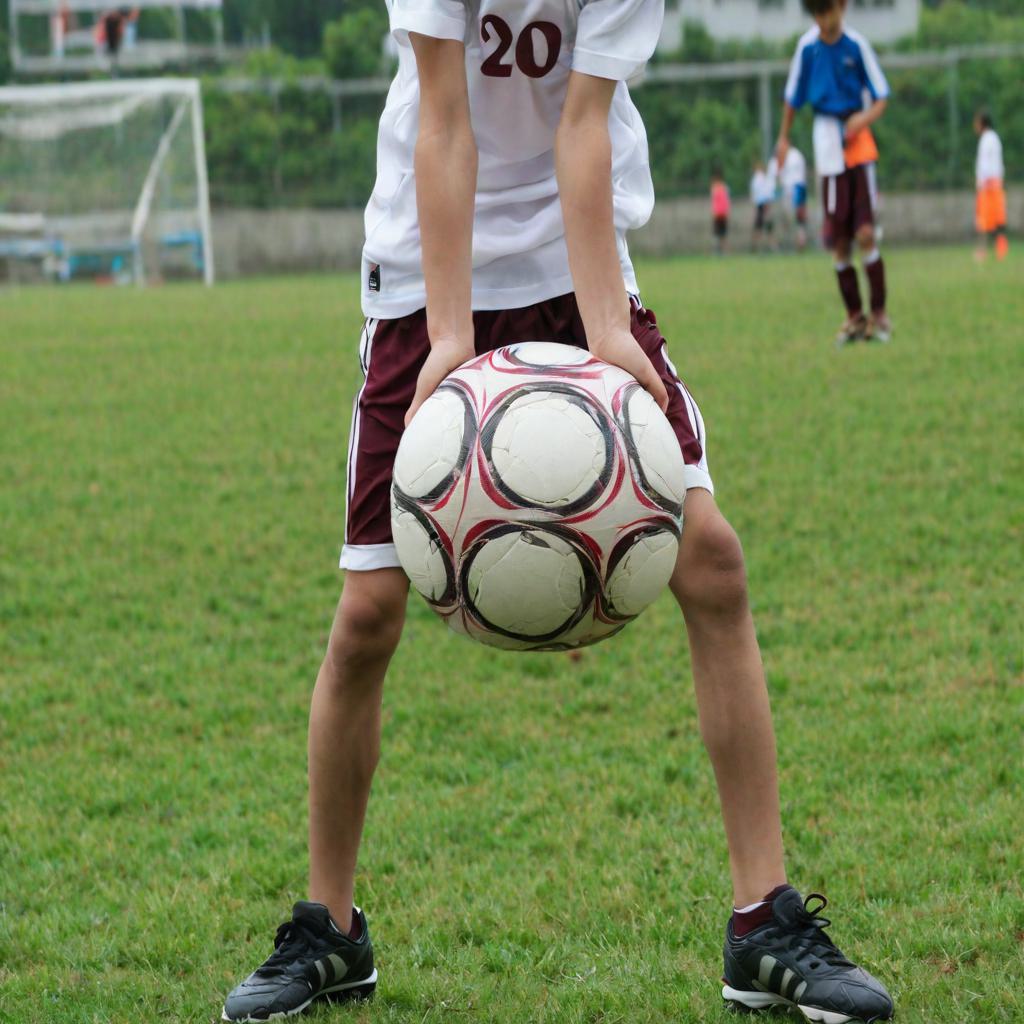}\\
        & \multicolumn{8}{c}{hold ball}\\
        \\[-1.2em]
        & \includegraphics[width=.104\linewidth]{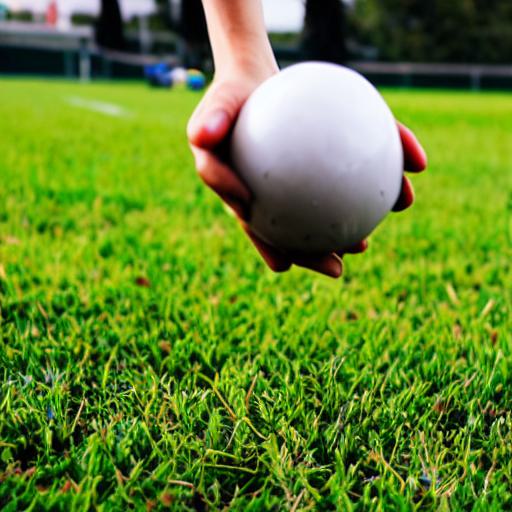}
        & \includegraphics[width=.104\linewidth]{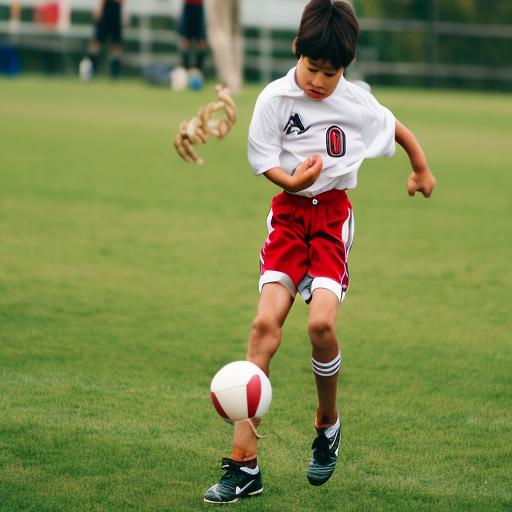}
        & \includegraphics[width=.104\linewidth]{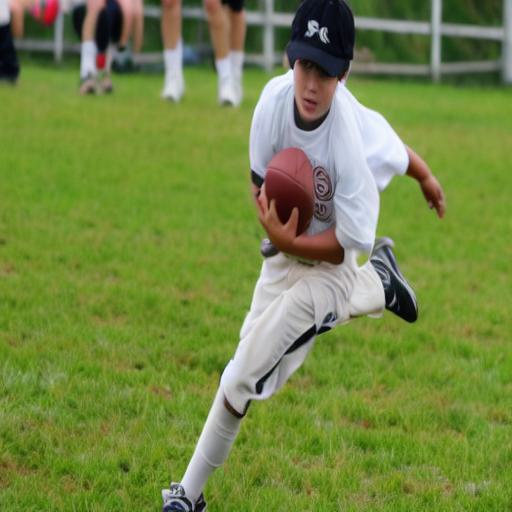}
        & \includegraphics[width=.104\linewidth]{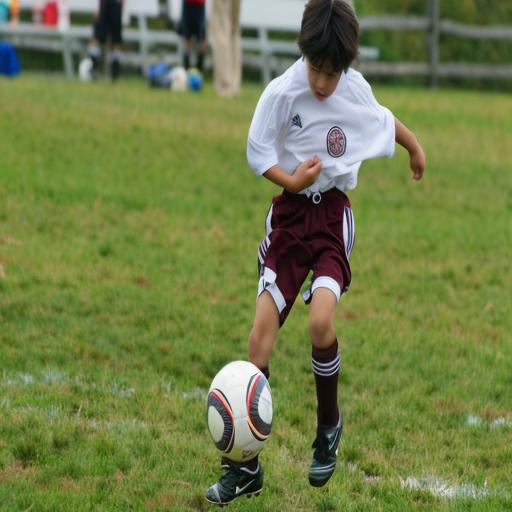}
        & \includegraphics[width=.104\linewidth]{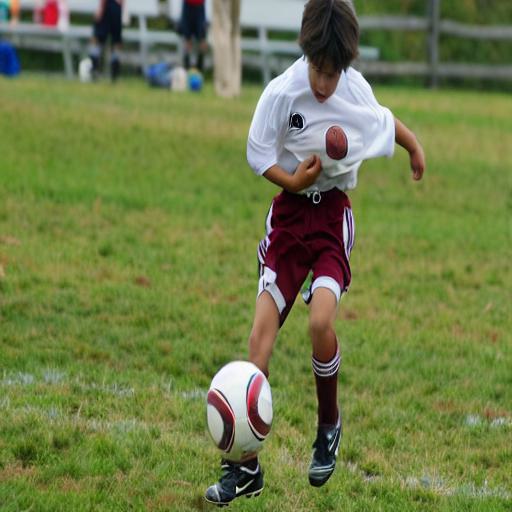}
        & \includegraphics[width=.104\linewidth]{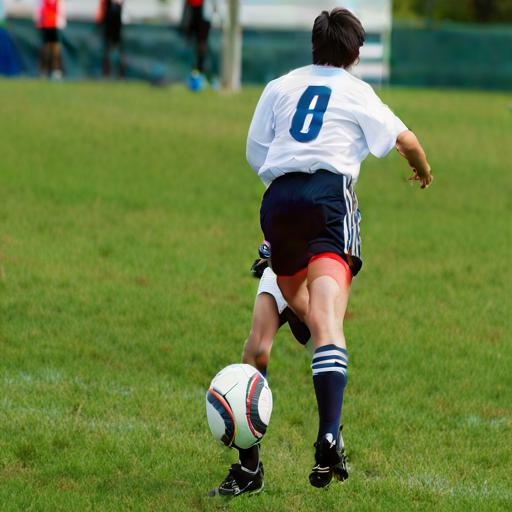}
        & \includegraphics[width=.104\linewidth]{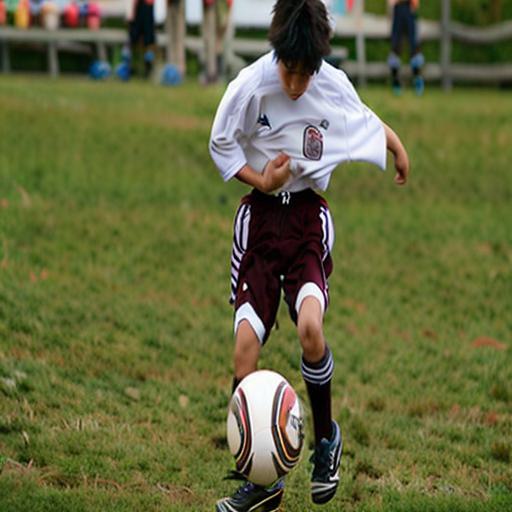}
        & \includegraphics[width=.104\linewidth]{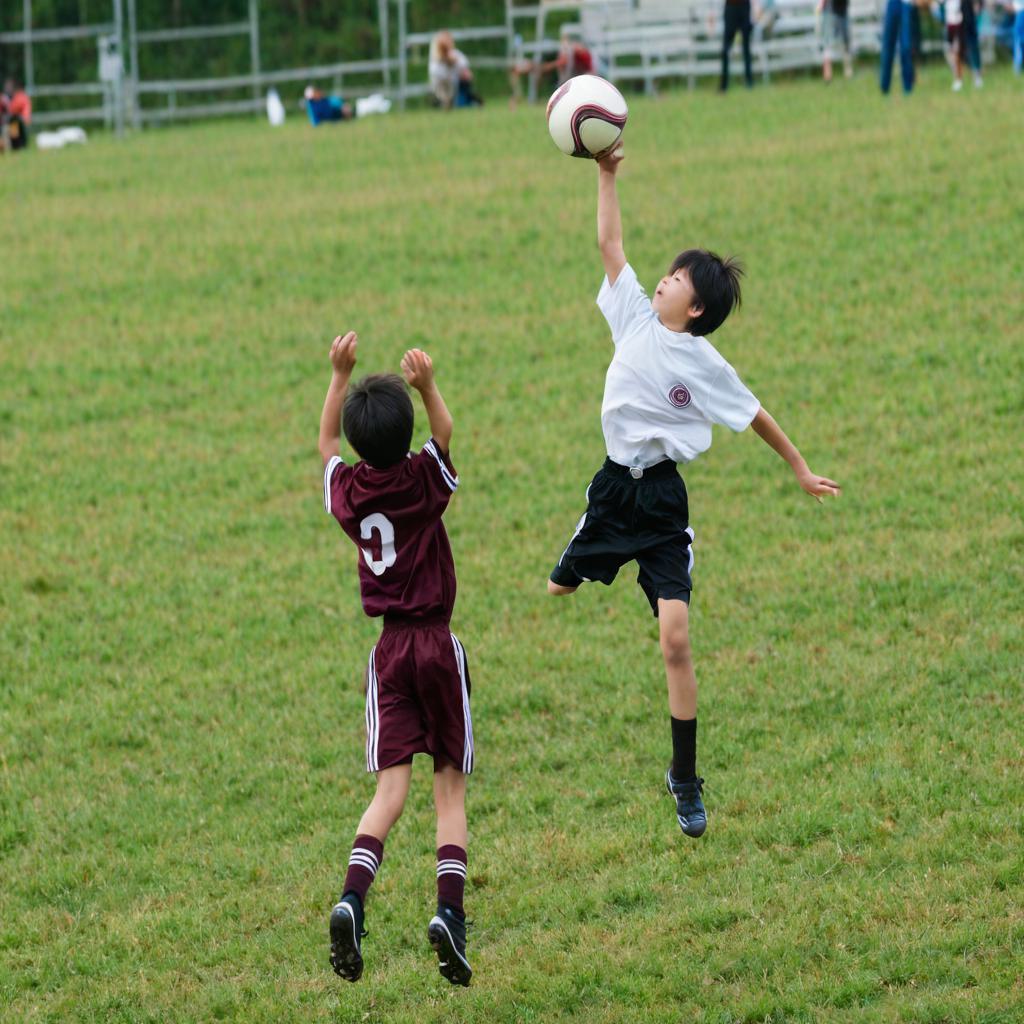}\\
        kick ball & \multicolumn{8}{c}{catch ball}\\
        \\[-1.2em]
          \includegraphics[width=.104\linewidth]{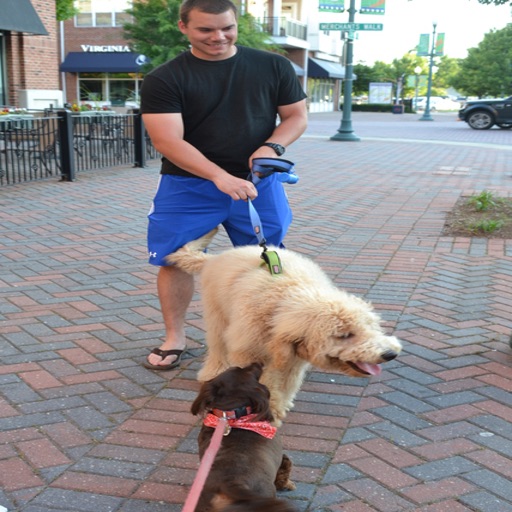}
        & \includegraphics[width=.104\linewidth]{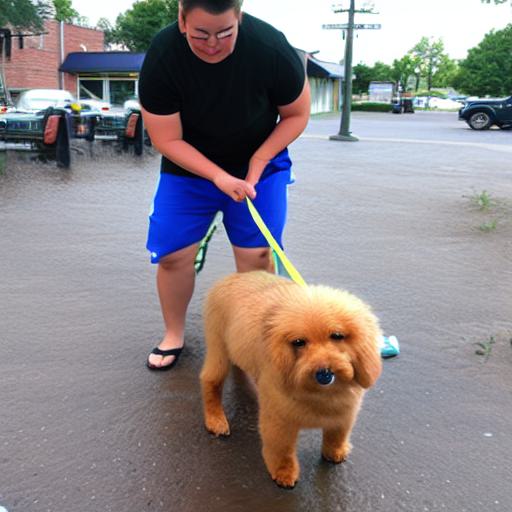}
        & \includegraphics[width=.104\linewidth]{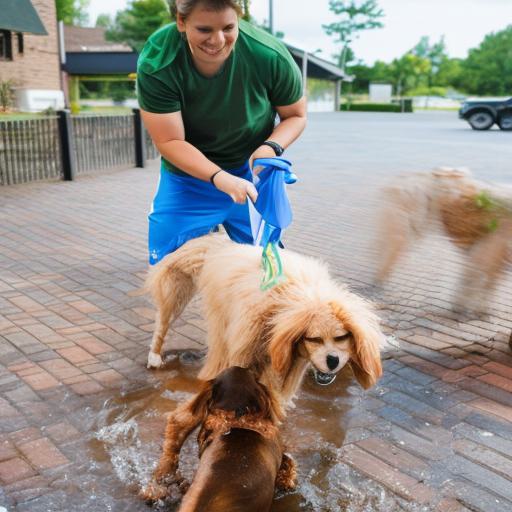}
        & \includegraphics[width=.104\linewidth]{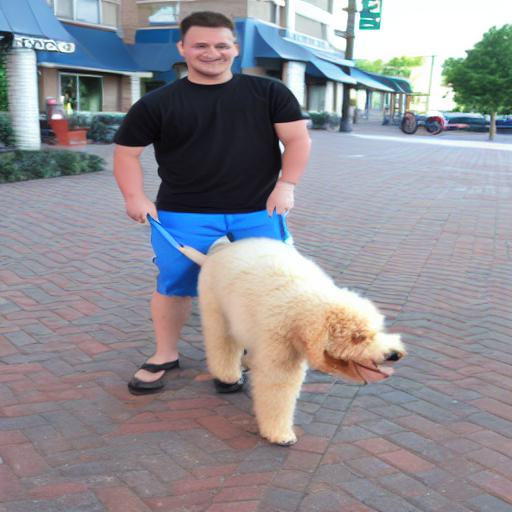}
        & \includegraphics[width=.104\linewidth]{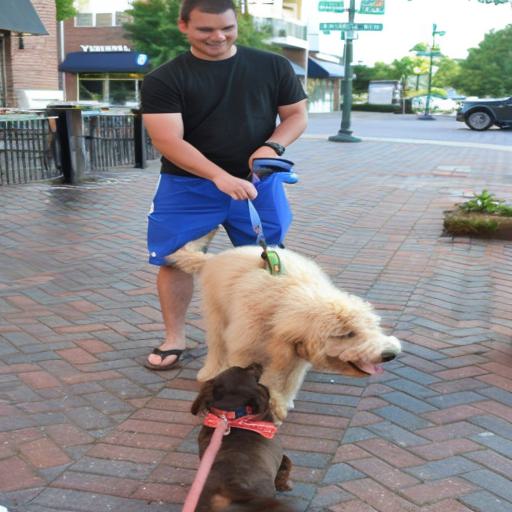}
        & \includegraphics[width=.104\linewidth]{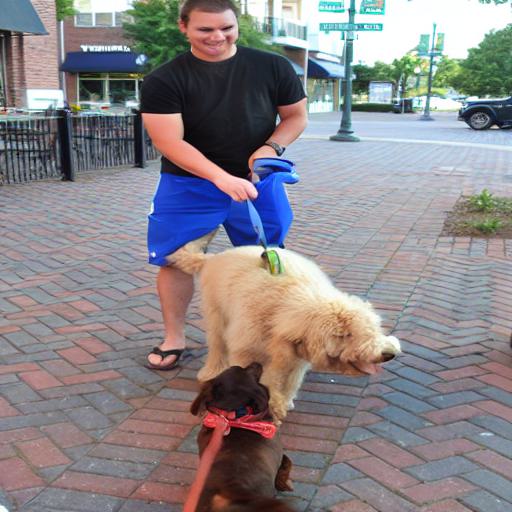}
        & \includegraphics[width=.104\linewidth]{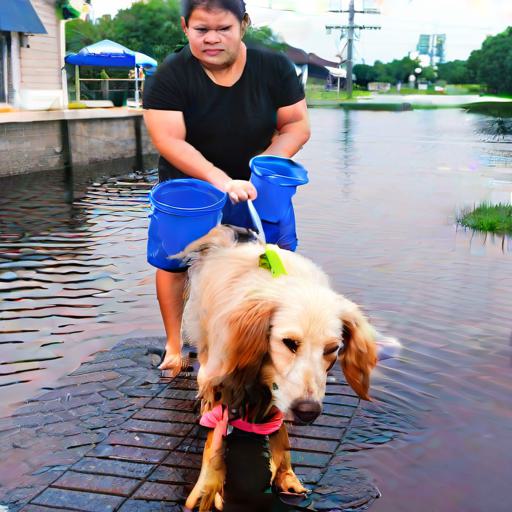}
        & \includegraphics[width=.104\linewidth]{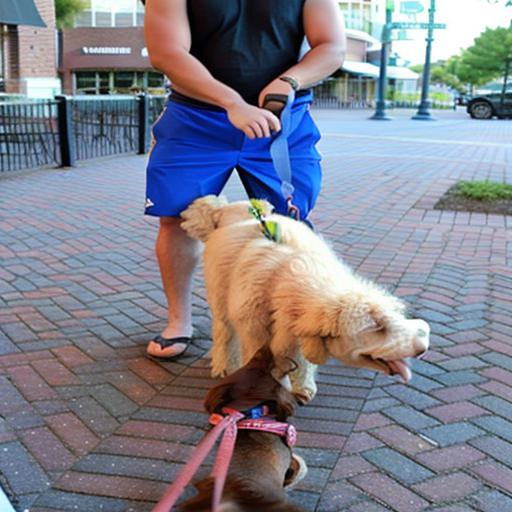}
        & \includegraphics[width=.104\linewidth]{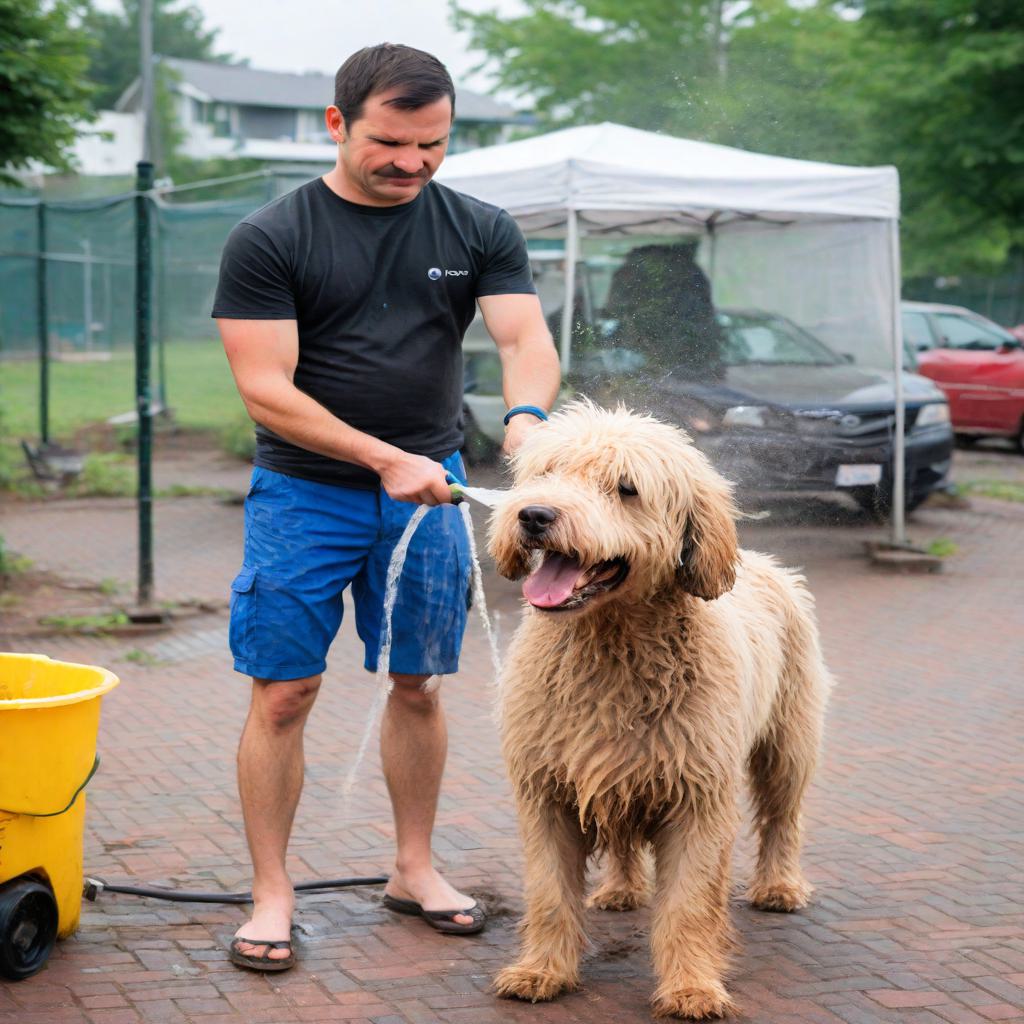}\\
        walk dog & \multicolumn{8}{c}{wash dog}\\
        \includegraphics[width=.104\linewidth]{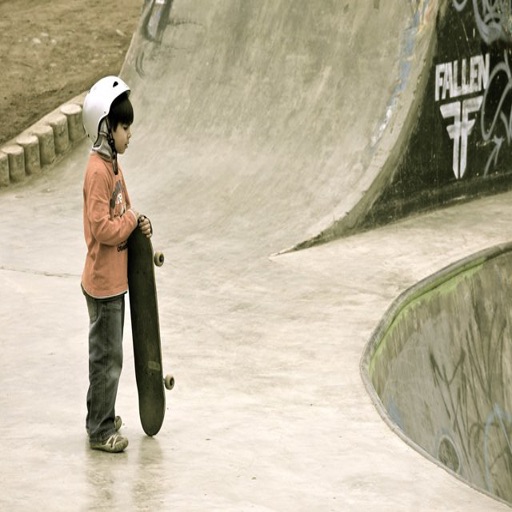}
        & \includegraphics[width=.104\linewidth]{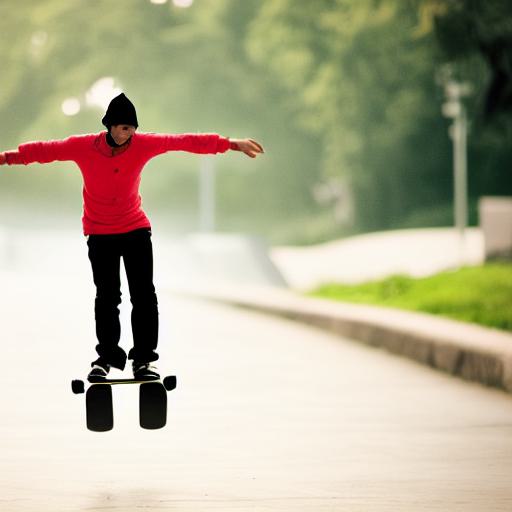}
        & \includegraphics[width=.104\linewidth]{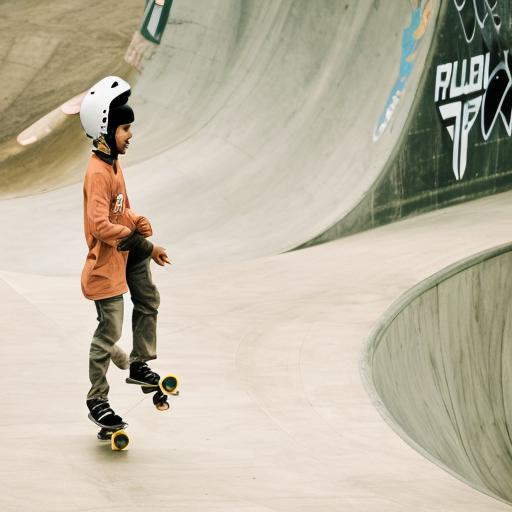}
        & \includegraphics[width=.104\linewidth]{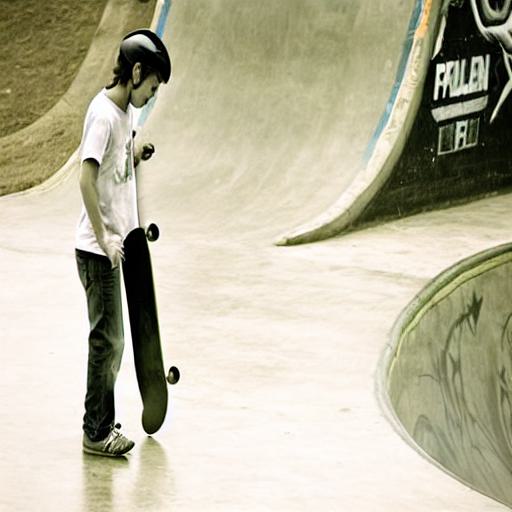}
        & \includegraphics[width=.104\linewidth]{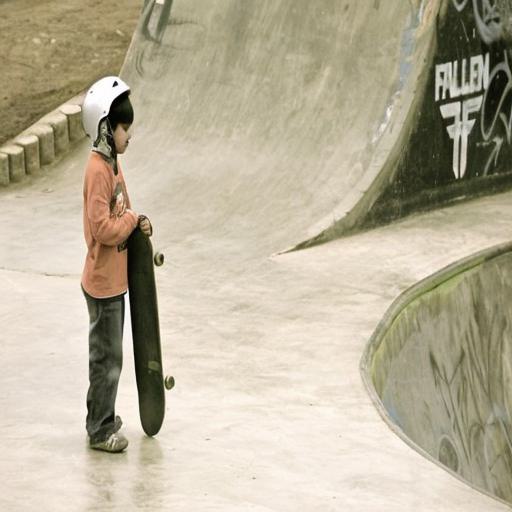}
        & \includegraphics[width=.104\linewidth]{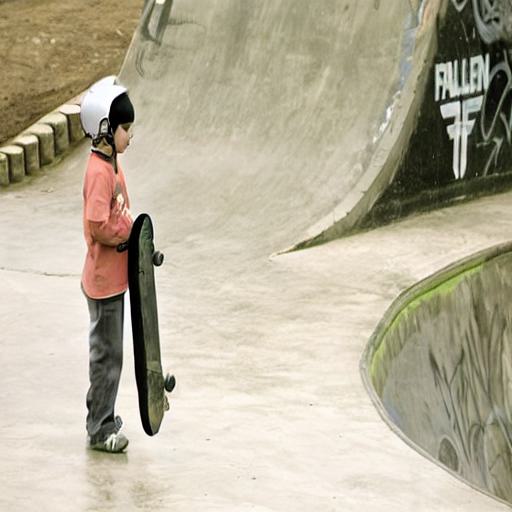}
        & \includegraphics[width=.104\linewidth]{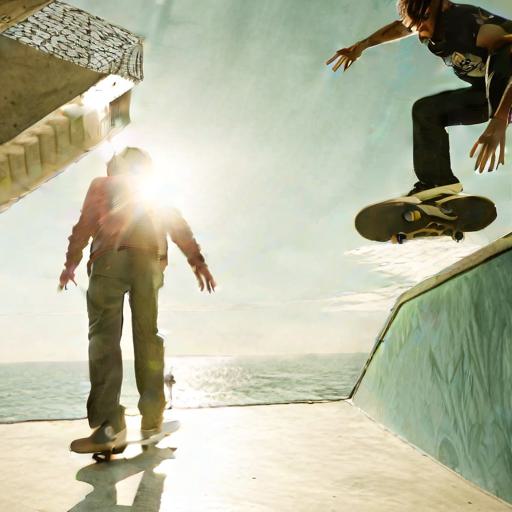}
        & \includegraphics[width=.104\linewidth]{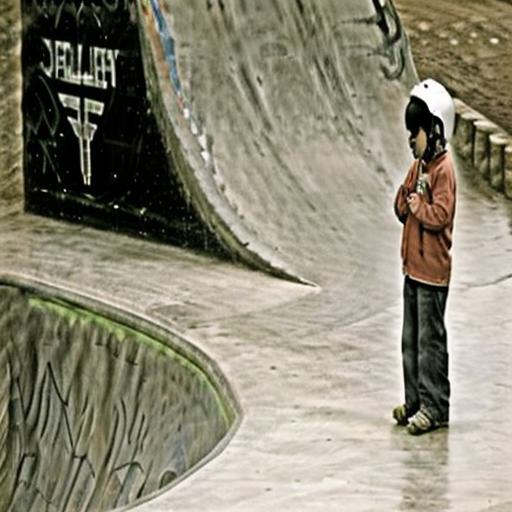}
        & \includegraphics[width=.104\linewidth]{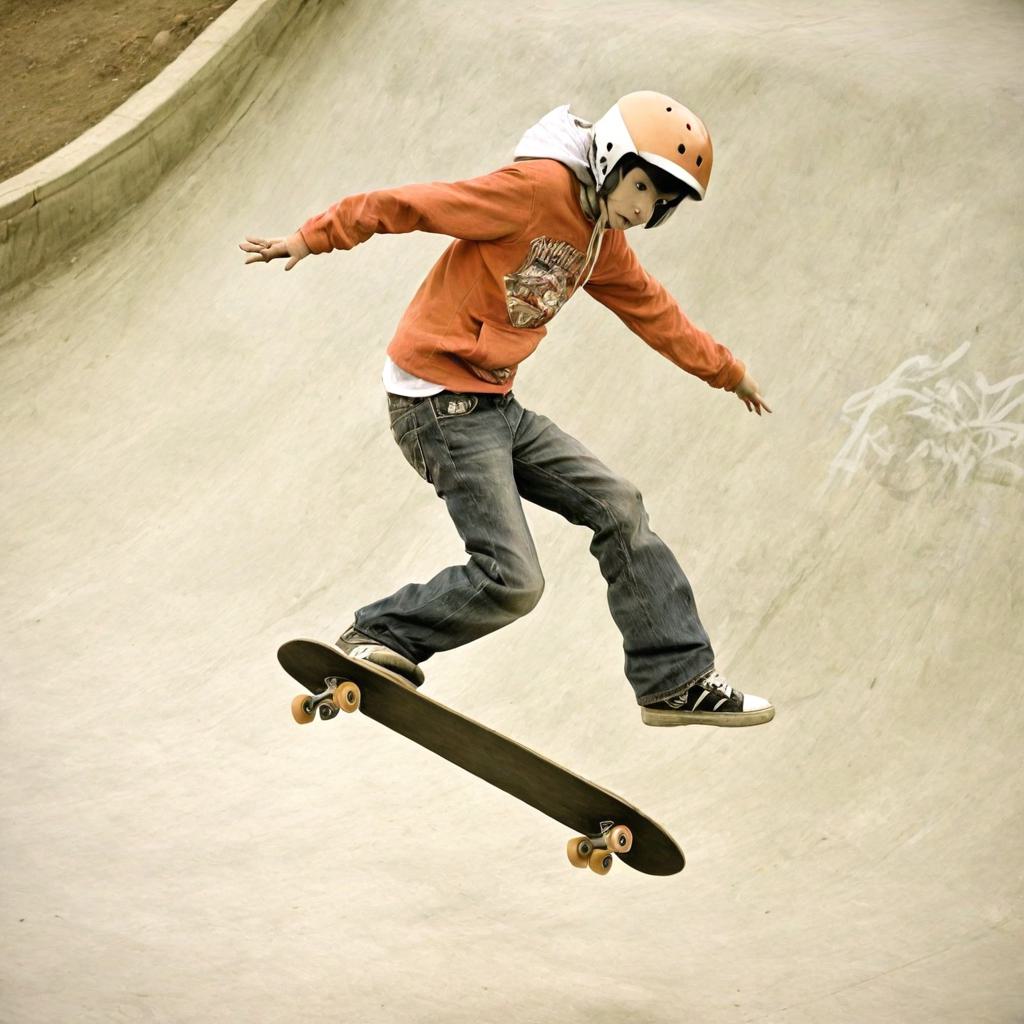}\\
        \\[-1.2em]
        & \multicolumn{8}{c}{jump skateboard}\\
        & \includegraphics[width=.104\linewidth]{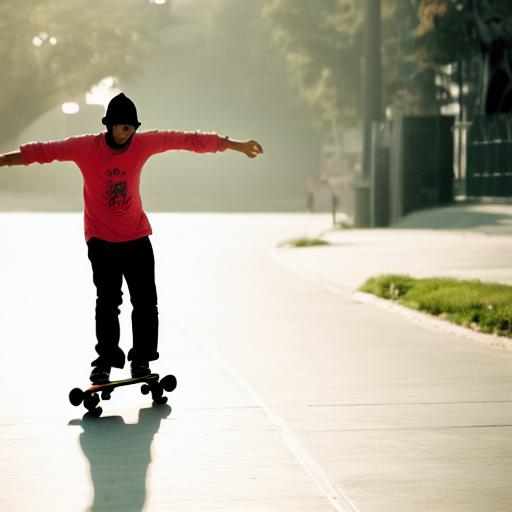}
        & \includegraphics[width=.104\linewidth]{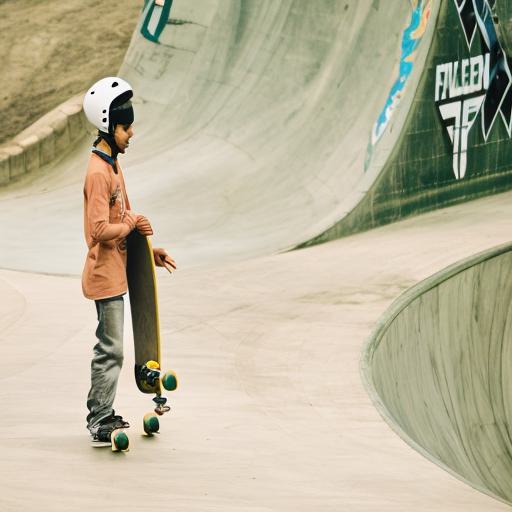}
        & \includegraphics[width=.104\linewidth]{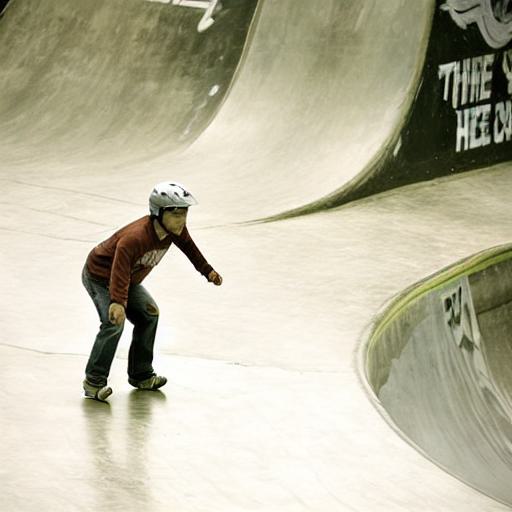}
        & \includegraphics[width=.104\linewidth]{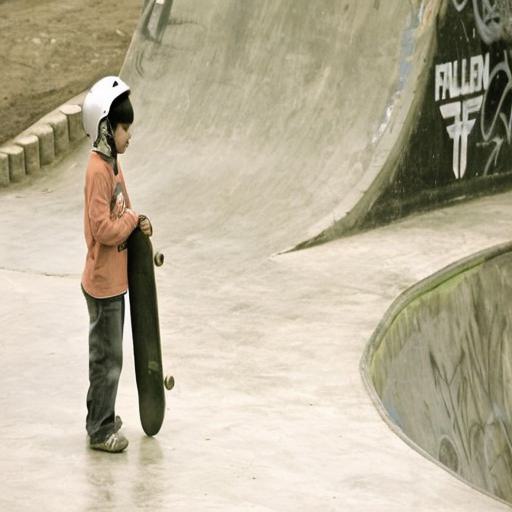}
        & \includegraphics[width=.104\linewidth]{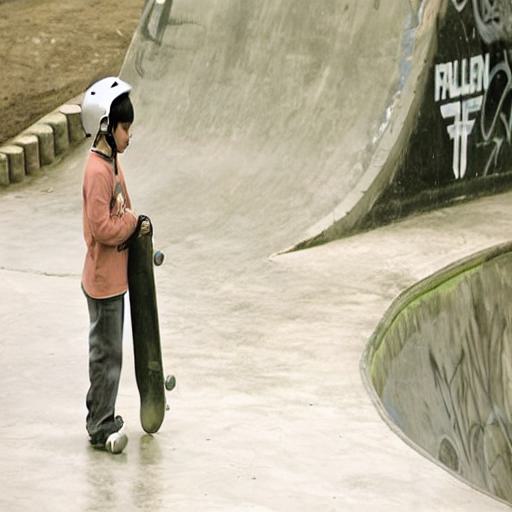}
        & \includegraphics[width=.104\linewidth]{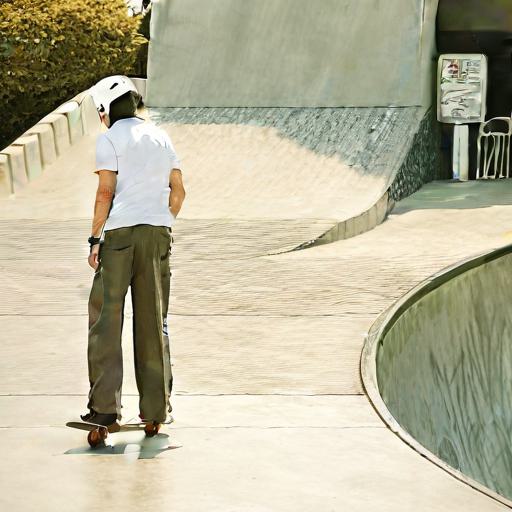}
        & \includegraphics[width=.104\linewidth]{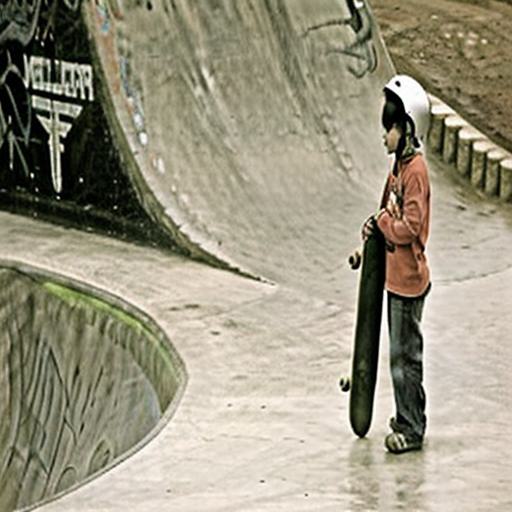}
        & \includegraphics[width=.104\linewidth]{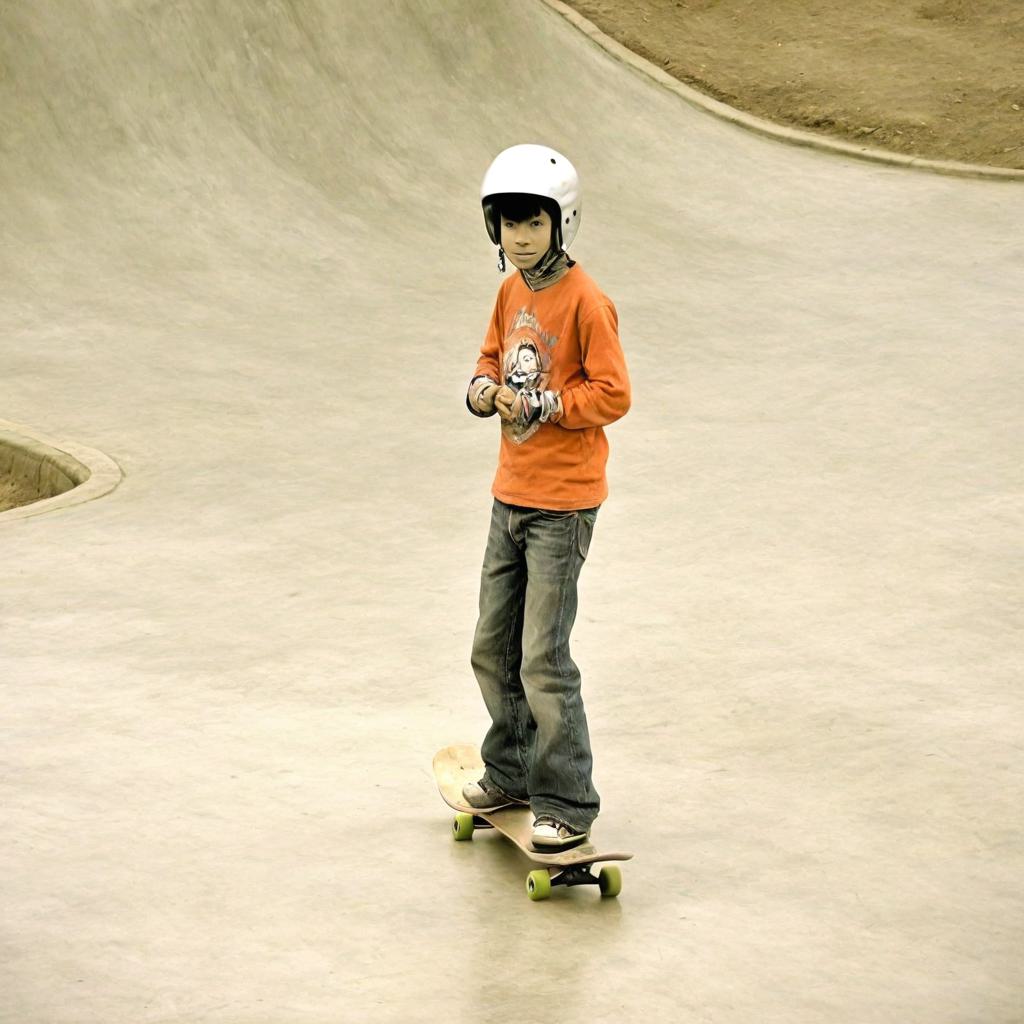}\\
        \\[-1.2em]
        hold skateboard & \multicolumn{8}{c}{ride skateboard}\\
    \end{tabular}
    \caption{More qualitative comparison. Our method demonstrates the best HOI editability.}
    \label{fig:qualitative_4}
\end{figure*}